\documentclass{aastex631}
\let\splitbox\undefined 

\pdfoutput=1

\usepackage{ulem}

\usepackage{float}
\usepackage{amsmath}
\usepackage{rotating}
\usepackage{tikz}
\usepackage{booktabs}
\usepackage{amssymb}
\usepackage{graphicx}
\usepackage{adjustbox}
\usepackage{tabularx}
\usepackage{natbib}

\begin{document}

\title{Gamma-Ray Burst Light Curve Reconstruction: A Comparative Machine and Deep Learning Analysis}

\author{A. Manchanda$^*$}
\affiliation{Centre for Astrophysics and Supercomputing, Swinburne University of Technology, Victoria 3122, Australia.}
\affiliation{Department of Physics, Indian Institute of Technology, Hyderabad, Telangana, 502284, India.}

\author{A. Kaushal$^*$}
\affiliation{Department of Computer Science and Engineering, UIET-H, Panjab University, Punjab, 146001, India}

\author{M. G. Dainotti$^*,^+$}
\affiliation{National Astronomical Observatory of Japan,2 Chome-21-1 Osawa, Mitaka, Tokyo 181-8588, Japan.}
\affiliation{Department of Astronomical Sciences, The Graduate University for Advanced Studies, SOKENDAI, Shonankokusaimura, Hayama, Miura District, Kanagawa 240-0115, Japan.}
\affiliation{Nevada Center for Astrophysics, University of Nevada, 4505 Maryland Parkway, Las Vegas, NV 89154, USA.}
\affiliation{Space Science Institute, 4765 Walnut Street, Suite B, Boulder, CO 80301, USA.}

\author{K. Gupta}
\affiliation{Department of Computer Science, PSIT College, Kanpur, Uttar Pradesh, 209305, India.}

\author{A. Deepu}
\affiliation{Department of Ocean Engineering and Naval Architecture, Indian Institute of Technology, Kharagpur, West Bengal, 721302, India.}

\author{S. Naqi}
\affiliation{B. Tech, Information Technology, KNIT Sultanpur, Uttar Pradesh, 228118, India.}

\author{J. Felix}
\affiliation{B.Tech, Industrial and Systems Engineering, Indian Institute of Technology Kharagpur, 721302, India.}

\author{N. Indoriya}
\affiliation{Department of Electrical Engineering and Computer Science, IISER Bhopal, Madhya Pradesh, 462066, India.}

\author{S. P. Magesh}
\affiliation{B. Tech, Electrical engineering, Indian Institute of Technology, Madras, 600036, India.}

\author{H. Gupta}
\affiliation{MIT WPU School of Computer Science \& Engineering, Pune, 411038, India.}

\author{A. Madhan}
\affiliation{BE CSE, Visveswaraya Technological University, Belagavi, Karnataka, 590018, India.}

\author{D. H. Hartmann}
\affiliation{Department of Physics and Astronomy, Clemson University, Clemson, SC 29634, USA.}      

\author{A. Pollo}
\affiliation{Astronomical Observatory of Jagiellonian University in Kraków, Orla 171, 30-244 Kraków.}
\affiliation{National Centre for Nuclear Research, 02-093 Warsaw, Poland.}

\author{M. Bogdan}
\affiliation{Department of Mathematics, University of Wroclaw, Wroclaw, 50-384, Poland.}

\author{J. X. Prochaska}
\affiliation{University of California, Santa Cruz, 1156 High Street, Santa Cruz, CA 95064, USA.}

\author{N. Fraija}
\affiliation{National Autonoma University of Mexico, Circuito Interior, 04510 Mexico City, Mexico City.}

\author{D. Debnath}
\affiliation{B.Tech, Electronics \& Communication Engineering, NIT Agartala, Tripura, 799046, India.}

\def\thefootnote{*}\footnotetext{These authors contributed equally to this work}
\def\thefootnote{+}\footnotetext{this author is the corresponding author}
 



\begin{abstract}

Gamma-Ray Bursts (GRBs), observed at high-z, are probes of the evolution of the Universe and can be used as cosmological tools. Thus, we need correlations with small dispersion among key parameters. To reduce such a dispersion, we mitigate gaps in light curves (LCs), including the plateau region, key to building the two-dimensional Dainotti relation between the end time of plateau emission ($T_a$) and its luminosity ($L_a$). We reconstruct LCs using nine models: Multi-Layer Perceptron (MLP), Bi-Mamba, Fourier Transform, Gaussian Process–Random Forest Hybrid (GP-RF), Bidirectional Long Short-Term Memory (Bi-LSTM), Conditional GAN (CGAN), SARIMAX-based Kalman filter, Kolmogorov–Arnold Networks (KANs), and Attention U-Net. These methods are compared to the Willingale model (W07) over a sample of 521 GRBs.
MLP and Attention U-Net outperform other methods, with MLP reducing the plateau parameter uncertainties by 37.2\% for $\log T_a$, 38.0\% for $\log F_a$, and 41.2\% for $\alpha$ (the post-plateau slope in the W07 model), achieving the lowest 5-fold cross-validation (CV) mean squared error (MSE) of 0.0275. Attention U-Net achieved the lowest uncertainty of parameters, a 37.9\% reduction in $\log T_a$, a 38.5\% reduction in $\log F_a$ and a 41.4\% reduction in $\alpha$, but with a higher MSE of 0.134. 
Although Attention U-Net achieves the largest uncertainty reduction, the MLP attains the lowest test MSE while maintaining comparable uncertainty performance, making it the more reliable model. The other methods yield MSE values ranging from 0.0339 to 0.174.
These improvements in parameter precision are needed to use GRBs as standard candles, investigate theoretical models, and predict GRB redshifts through machine learning. 

\end{abstract}

\keywords{$\gamma$-ray bursts--- statistical methods---machine - learning ---cosmology: cosmological parameters--- light curve reconstruction}


\section{Introduction} 
\label{sec:intro}

GRBs are brief and highly luminous astrophysical phenomena detectable at remarkable distances \citep[for review, see][]{2015PhR...561....1K}, with observations extending up to redshift $z=9.4$ \citep{Cucchiara2011}. This property makes GRBs exceptional cosmological tools for probing the universe's early evolution. A thorough analysis of GRBs also yields essential information about Population III stars, the first generation formed during the epoch of re-ionization.
GRB emission is usually observed in two episodes: the prompt and the afterglow. The prompt phase is interpreted by internal shell collisions or magnetic reconnection and is characterized by the short duration and high energy. Hence, this is detected predominantly in $\gamma$ rays and X-rays and occasionally in optical wavelengths \citep{Vestrand2005Natur,Blake2005Natur,Beskin2010ApJ,2012MNRAS.421.1874G,2014Sci...343...38V}. During the afterglow phase, which follows the primary episode, the relativistic jet impacts the circumstellar environment, transferring part of its energy. The afterglow is characterized by a long duration and energies that span a broad spectrum, including X-ray, optical, and sometimes radio bands \citep{costa1997,vanParadijs1997,Piro1998,Gehrels2009ARA&A}. 


The Neil Gehrels Swift Observatory (Swift, \cite{Gehrels2004ApJ...611.1005G}) is vital for detecting the temporal properties of GRBs. The Swift Burst Alert Telescope (BAT), which operates in the range of 15-150 keV \citep{barthelmy2005burst}, plays a crucial role in quickly detecting prompt emission. It enables a rapid follow-up of the afterglow through the X-ray Telescope (XRT) instrument, which covers the 0.3-10 keV range \citep{burrows2005swift}, and the Ultra-Violet Optical Telescope (UVOT 170 - 600 nm, \citep{Roming2005}). Moreover, Swift's rapid multi-wavelength afterglow follow-up has revealed new characteristics in GRB LCs \citep{Tagliaferri2005,Nousek2006,Troja2007}. 

Most X-ray LCs exhibit rapid decay in flux after the prompt episode, occasionally followed by flares and/or a plateau \citep{Zhang2006,OBrien2006,Nousek2006,Sakamoto2007,Liang2007,willingale2007testing,Dainotti2008,dainotti2010a,Dainotti2016,dainotti2017a, dereli2024unraveling}. The plateau observed in GRB LCs can be modeled using a Broken Power-Law (BPL) \citep{Zhang2006, Zhang2007ApJ...655L..25Z,Racusin2009}, a smooth BPL, or the phenomenological model proposed by (W07, \cite{willingale2007testing}). The W07 model determines critical parameters such as the time at the end of the plateau, \(T_a\), the corresponding flux, \(F_a\), and the temporal index after the plateau, \(\alpha_a\). On the other hand, the BPL model provides $T_a$, $F_a$, and the slope of the LC during the plateau, \(\alpha_1\) and after the plateau, \(\alpha_2\). Section ~\ref{section:sampling} details the W07 and the parameters.

The plateau phase is frequently interpreted using the magnetar model \citep{Zhang2001,Rowlinson2014, Rea2015, Stratta2018}, which attributes the emission to dipole radiations generated by the rotational energy of a newly formed Neutron Star (NS). According to this model, the plateau ends when the NS reaches its critical spin-down timescale. Uncertainties in determining $T_a$ are often related to the magnetar spin period and uncertainties of the magnetic field. Therefore, precise measurements of $T_a$ are essential to verify the validity of this model. The plateau phase, exhibiting more consistent features across various GRBs, such as length and flatness, has attracted attention because of its potential to establish relevant correlations with the plateau parameters and their application as cosmological tools. \citep{Dainotti2008, dainotti2010a, dainotti2011a,dainotti2013determination, dainotti2015, dainotti2017a, Tang2019ApJS..245....1T, Wang:2021hcx, Zhao2019ApJ...883...97Z, Liang2007, Li2018b} have explored the luminosity at the end of the plateau, $L_{X,a}$ versus its rest-frame time $T^{*}_{X,a}$ (known as the Dainotti relation or the 2D L-T relation) \footnote{the rest-frame time is denoted with an asterisk}. The 2D relation has also been identified in the optical plateau emissions \citep{dainotti2020b,2022ApJS..261...25D}. Within the theoretical magnetar framework, \cite{Rowlinson2014} showed that the X-ray Dainotti relation is reproduced with a slope for $L_{a,X}$-$T^{*}_{a, X}$ of $-1$. This correlation has been applied in cosmological research, such as the development of the GRB Hubble diagram, which extends to redshifts greater than $z>8$ \citep{cardone2009updated,cardone2010constraining,postnikov2014nonparametric,dainotti2013}. 

The 2D L-T relation has been further expanded by incorporating the peak prompt luminosity, $L_{X, peak}$, resulting in the Dainotti 3D relation \citep{Dainotti2016,dainotti2017a,dainotti2020a, 2022ApJS..261...25D}. This 3D relation has also been successfully applied to constrain cosmological parameters \citep{dainotti2023b, dainotti2022g, dainotti2022b, 2022MNRAS.512..439C, 2022MNRAS.510.2928C}.
Importantly, \cite{dainotti2022g} demonstrated that reducing the uncertainties associated with the plateau emission parameters by 47.5\% could achieve the same precision for the cosmological value of $\Omega_M$ quoted in \cite{Betoule} in just 8 years as we have calculated in \citep{dainotti2022c,narendra2024grb}, if we consider the optical sample and the addition of GRBs for which the redshift was inferred \citep{giovanna2024inferring, dainotti2024optical}. This improvement could be realized immediately, compared to the precision requiring 22 more years of observations under current rates and parameter uncertainties \citep{dainotti2022g}, which highlights the significant potential of a more reliable Light Curve Reconstruction (LCR) approach, as it could substantially accelerate progress to reach the same precision achieved by SNe Ia \citep [for details, see] []{dainotti2020x}.

In addition, GRB LCs with temporal gaps present significant challenges in testing theoretical models, such as the standard fireball model \citep{panaitescu2000analytic, piran1999gamma}. This model is typically evaluated using closure relations \citep{willingale2007testing,evans2009methods,racusin2009jet,kumar2010external, srinivasaragavan2020investigation, dainotti2021closure, ryan2020gamma, tak2019closure}, which involve the temporal ($\alpha$) and spectral ($\beta$) index of the afterglow. The value of $\alpha$ can correspond to $\alpha_1$ or $\alpha_2$, depending on the segment of the LC analyzed, while $\beta$ is measured for the same time window as $\alpha$. 
An accurate assessment of these factors is essential to evaluate the fireball model and categorize GRBs according to their morphology. 

However, gaps due to satellite orbital periods, lack of fast follow-up studies, and factors such as meteorological turbulence and instrumental errors complicate these measurements. To address these challenges, further improvement in understanding GRBs requires extensive data coverage and the development of a reliable taxonomy of GRB classes. Such efforts could considerably improve studies on GRB populations, their cosmological evolution, emission mechanisms, and/or progenitors.
Lastly, another valuable application involves the utilization of reconstructed LCs to train ML models to estimate the redshifts of GRBs \citep{bargiacchi2025high}.

Thus, here we tackle the LCR, which provides a novel approach to address the challenge of temporal gaps in LCs and it is a continuation of a previous effort by some of us \citep{dainotti2023stochastic}. 
The paper introduced a stochastic reconstruction technique using existing models and GPs, reducing uncertainties for key GRB parameters.

Recent advances in machine learning have further pushed the boundaries of LCR. For example, \cite{demianenko2023understanding} examined the application of neural network-based methods, including Bayesian neural networks, multilayer perceptrons, and normalization flows, to obtain approximate findings of a single LC. These methods were tested on the LCs of Supernovae type 1a taken from the Zwicky Transient Facility Bright Transient Survey and simulated PLAsTiCC. These methods demonstrated that even with limited observations, neural networks could significantly enhance the quality of the reconstruction in comparison to state-of-the-art models. Moreover, they are computationally efficient, outperforming GPs in terms of speed, and were found to be effective for subsequent tasks such as peak identification and transient classification. 

Building on these advancements, this work further utilizes deep learning techniques to improve the GRB LCR. In particular, we utilized Bi-Mamba, Multi-Layer Perceptron (MLP), Fourier transform, ensemble Gaussian Processes (GP) with Random Forest (RF) models, Bi-directional Long Short-Term Memory (Bi-LSTM), Conditional Generative Adversarial Networks (CGAN), SARIMAX-based Kalman filter, Kolmogorov-Arnold Networks (KANs), and the Hybrid model of U-Net with Attention Mechanism (Attention U-Net). This study highlights the strengths and limitations of these models and performs a comparative analysis to assess their performance.

This paper is divided into the following sections: Sec. \S\ref{section:methods} details on the dataset and the various models employed to reconstruct GRB LCs.
The uncertainty, performance, and outliers in Sec. \S\ref{section:results}. Sec. \S\ref{section:conclusion} provides the synopsis and conclusions on the observed efficacy of each model.

\section{Methodology}\label{section:methods}

\subsection{Dataset and the Willingale model}\label{section:sampling}

This research analyzed a dataset comprising 521 GRBs.
The initial sample comprises 455 GRBs detected by Swift between January 2005 and August 2019, as compiled by \cite{2020ApJ...903...18S,dainotti2020a}. This dataset is then enlarged by additional 14 GRBs observed by Fermi-LAT, as presented in \cite{dainotti2024analysis} and a further 78 GRBs detected by Swift between 2019 and 2023, as reported by \cite{narendra2024grb}. After accounting for two overlapping GRBs between the original and extended samples, the final dataset consists of 521 unique GRBs. The dataset includes 230 GRBs with known redshifts and 315 without redshift information, as described in \cite{giovanna2024inferring, dainotti2024optical, narendra2024grb}. All GRBs in the sample, including flares (about 31.5\% of our sample), exhibit temporal gaps in their LCs arising from various factors.
To address irregular gaps in GRB LCs, we implemented a gap-aware reconstruction in log-time space. A minimum gap threshold of 0.05 was imposed. The number of reconstructed points was adaptively set as a fraction of the original LC size: 5\% for LCs with more than 500 points, 10\% for those with 250–500 points, 30\% for 100–250 points, and 40\% for fewer than 100 points (minimum of 20 points). Extremely sparse LCs (about 5\% of the sample with $<$26 points) with very large gaps, reconstruction was not feasible. Thus, the dataset was reduced to 521 GRBs.


The W07 function is to model the LCs of GRBs, defined in Eq. \ref{eqn1} and introduced by \cite{willingale2007testing}:

\begin{equation}
f(t) = \left \{
\begin{array}{ll}
\displaystyle{F_i \exp{\left ( \alpha_i \left( 1 - \frac{t}{T_i} \right) \right )} \exp{\left (
- \frac{t_i}{t} \right )}} {\hspace{1cm}\rm for} \ \ t < T_i, \\
\\
\displaystyle{F_i \left ( \frac{t}{T_i} \right )^{-\alpha_i}
\exp{\left ( - \frac{t_i}{t} \right )}} {\hspace{1cm}\rm for} \ \ t \ge T_i. \\
\end{array}
\right .
\label{eqn1}
\end{equation}

Here, the symbols $T_i$ and $F_i$ refer to time and flux, respectively, in our case, at the end of the plateau emission,
 $\alpha_i$ is the temporal index parameter after $T_i$. The parameter $t_i$ corresponds to the onset of the rise phase, and the afterglow emission is represented by $t_a$. We have shown in previous papers that the onset is usually compatible with 0 \citep{willingale2007testing}.

The LCs analyzed in this study have had the prompt emission segment removed because of its significant variability and the challenges in modeling it effectively. Consequently, the data set is limited to the plateau and afterglow phases. 

\begin{figure*}
\begin{center}
\includegraphics[width=.45\textwidth]{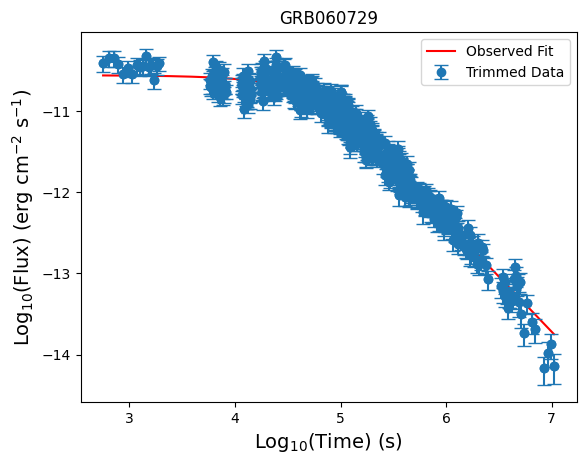}
\includegraphics[width=.45\textwidth]{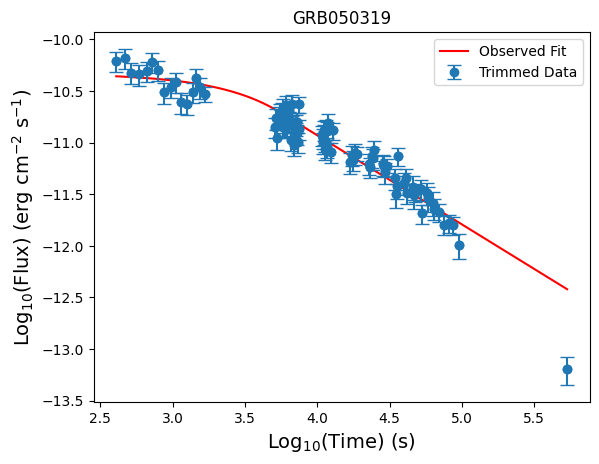}
\includegraphics[width=.45\textwidth]{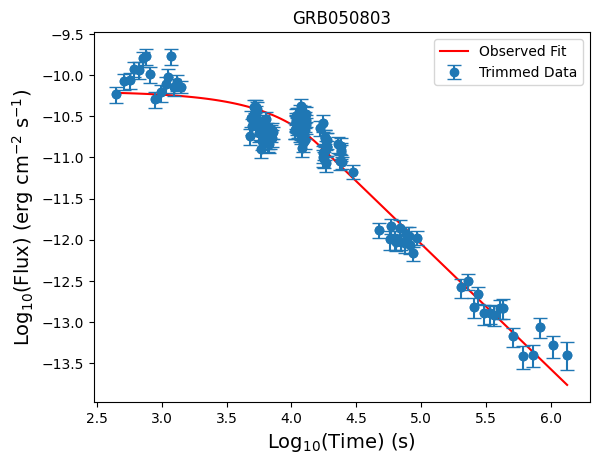}
\includegraphics[width=.45\textwidth]{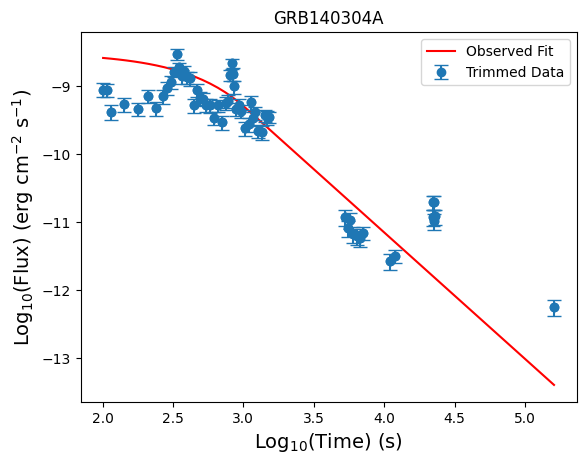}  

\includegraphics[width=.45\textwidth]{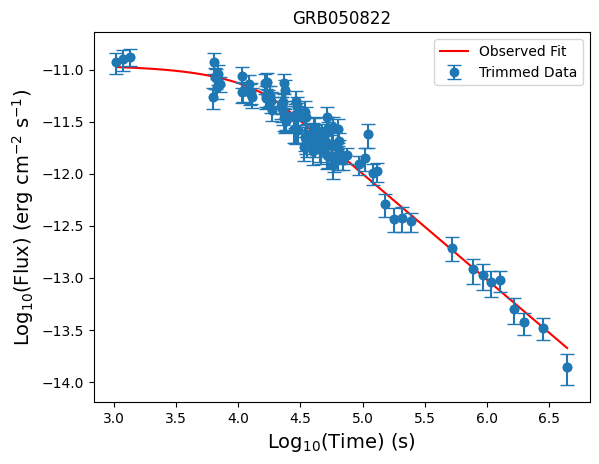}
\includegraphics[width=.42\textwidth]{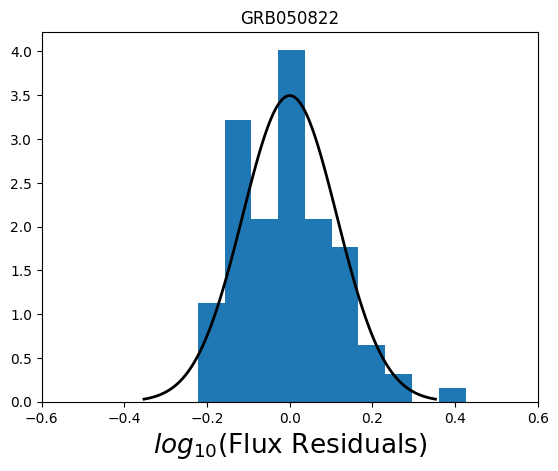}  

\end{center}
    \caption{Row 1 and Row 2 describe the GRB LCs divided into four classes depending on the afterglow feature: i) Good GRBs (Row 1, left); ii) GRB LCs with a break towards the end (Row 1, right); iii) Flares or Bumps in the afterglow (Row 2, left); iv) Flares or Bumps with a Double Break towards the end of the LC (Row 3, right). Row 3's left plot shows the LC of GRB050822, starting from the plateau emission, and the best-fit W07 model is shown in red. The Row 3 right plot shows the log(flux) residual histogram, and the best-fit Gaussian distribution is displayed in black.}
    \label{fig:categories}
\end{figure*}

The GRBs were subsequently categorized into four distinct classes based on the features of their afterglow emission, using the classification methodology established in \citealt{dainotti2023stochastic}:
\begin{itemize}
 \item \textbf{Good GRBs:} The afterglow is a good approximation with the W07 model, represents 55\% of the data set.
 \item \textbf{Flares/Bumps:} GRBs in this category show flares and bumps throughout the afterglow phase, accounting for 24\% of the total.
 \item \textbf{Break:} GRBs with a single break observed towards the end of the LC, which makes up 13\% of the sample.
 \item \textbf{Flares/Bumps + Double Break:} GRBs that display a combination of flares/bumps and a double break constitute 7.5\% of the dataset.
\end{itemize}
The first four panels of Fig.\ref{fig:categories} illustrate the LCs for the four identified classes and their corresponding W07 model fits. 
Our analysis relies on the analysis of GRB afterglow LCs by comparing each LC with the W07 fitted model. The parameters for each GRB LC have been derived from the work of \cite{2020ApJ...903...18S}. The blue points represent the trimmed data, where the prompt emission phase has been removed.

\subsection{Reconstruction with the Willingale model}\label{section:willingale_theory}

In this section, we adhere to the methodology described in \cite{dainotti2023stochastic} to reconstruct the LCs using the W07 model. However, our approach further extends the analysis by including all classes of GRBs, allowing for a more extensive study of the model across different categories of GRBs.

The flux residual for each LC is calculated as the difference between the logarithmic flux of the original LC and the logarithmic flux value predicted by the W07 fit at a given time t (in log10 scale):

\begin{equation}
\log_{10} F_{\text{res}} = \log_{10} F_{\text{obs}}(t) - \log_{10} f(t).
\end{equation}

where $\log_{10} F_{\text{res}}$ is the flux computed as the residual, while $F_{obs}$ is the observed flux and f(t) is the functional form of W07 at the given time t.

We then perform the same steps described, generating histograms of the residuals for each GRB and fitting a Gaussian distribution, as a first approximation, to those residuals as presented in right bottom panel of Fig. \ref{fig:categories}. In our analysis, we followed the same methodology as outlined in \cite{dainotti2023stochastic}, where the fitting was performed using the Gaussian, since in most of the cases the Gaussian distribution is quoted to be the best-fit in the package FindDistribution. However, there are several cases in which other distributions may have fitted better, but to provide an uniform treatment for simplicity and to better compare our analysis with the one of \cite{dainotti2023stochastic}, we have adopted the same methodology. 
In future work, we will use in each case the best fit distribution for each GRB.
The reconstructed flux at each time point is calculated by adding noise, sampled from the residual distribution, to the flux predicted by the W07 model:

\begin{equation}
\log_{10} F_{\text{recon}}(t) = \log_{10} f(t) + (1 + n) \times \text{RVN}. 
\end{equation}

where $\log_{10} F_{\text{recon}}(t)$ represents the reconstructed flux at time t, $f(t)$ denotes the W07 flux at time t, n is the noise level, and RVN refers to the random variate sampled from the Normal (Gaussian) distribution.

We evaluated the reconstruction at 10\% and 20\% noise levels. This method ensures that the reconstructed data points are statistically consistent with the best-fit W07 model as the initial observed points.
We reconstruct time series using a logarithmic distribution via the \textit{geomspace} function and perform 100 times MCMC calculations for each GRB to guarantee the stability of the model.
The combined data (original and reconstructed points) is then refitted using a least-squares regression with the \textit{minimize()} function from the \textit{lmfit} library, providing the new fitting parameters and uncertainties of the reconstructed LCs.

\subsection{The Machine-Learning Approach}\label{section:ML}

Machine learning models are algorithms that learn from the data, recognize patterns, and give predictions on unseen data.
One of the most widely used ML techniques is the Artificial Neural Network (ANN) \citep{DeepLearning, ANN}. These are inspired by how the human brain (neurons) processes information. Just as the brain consists of interconnected neurons that help us think and learn, a neural network consists of layers of artificial neurons that process and learn from data. A neural network consists of the input, hidden, and output layers. The input layer is the first layer that receives the raw input data. In this layer, no computations are performed, as it is responsible for passing the data to the next layer. The output layer is the final layer that gives the output predictions of the hidden layer. The hidden layers consist of artificial neurons to analyze the data. Each connection between neurons has a weight, determining the input value's importance. A bias is an extra value added to fine-tune and adjust the predictions. At first, the weights and biases are initialized randomly, but they are adjusted to improve the prediction as the model trains. The model undergoes training over a series of epochs, progressively adjusting its weights and biases within each epoch to minimize the errors it commits.

Each neuron gives output as a weighted sum of the inputs (containing biases). However, if we only use this method, the network can learn only linear relationships, which limits its ability to capture complex patterns. To overcome this, we apply activation functions, which introduce non-linearity and allow the network to learn more advanced relationships in data. After applying the activation function, the neuron passes the result to the next layer, allowing the network to process data step by step until it reaches the output layer. The performance of neural networks greatly depends on hyperparameters, which are the configuration variables that must be fixed before training. These may include the number of hidden layers, neurons, epochs, etc. The technical terms are summarized in Table \ref{tab:ML_def}.

This study uses various machine-learning models to reconstruct the GRB LCs. The Multilayer Perceptron (MLP) is the most basic type of neural network. It is a specific type of ANN with fully connected layers, meaning that each neuron in one layer is connected to every neuron in the next layer. We have compared this with other advanced neural networks extending the basic neural network differently. For instance, Bi-LSTM is designed to work with sequences. A standard neural network treats every input independently, but Bi-LSTM “remembers” the past information to process the new ones. Sometimes, not all parts of the data are equally important. We introduce Attention U-Net, where the “attention” mechanisms help the network focus on the most relevant parts while ignoring less important details. By incorporating probability into the model parameters (weights and biases), we use Gaussian Processes (GP) to predict output and uncertainty. In most neural networks, how neurons process information is fixed, meaning that the transformation applied to the data remains the same throughout training. KAN improves this by adapting the activation functions during learning. Other methods, such as SARIMAX and Fourier, use statistical techniques to analyze time-series patterns.

\begin{table}[htbp]
\centering
\hspace*{-2.2cm}
\renewcommand{\arraystretch}{1.5}
\resizebox{1.2\textwidth}{!}{
\begin{tabular}{|l|c|}
    \hline
    \textbf{ML Terms}    &
    \textbf{Definition} \\ \hline

    Neural Network & ML model inspired by our brain, consists of layers of interconnected "neurons" to process information and learn patterns from the data.\\ \hline

    Hyperparameters & Variables that need to be optimized to control the training process. In our study, we have used \textit{GridSearchCV} and \textit{optuna} to optimize. \\ \hline
    
    Activation Function & It introduces non-linearity to the outputs of the neurons to learn complex data. \\ \hline

    ReLU (Rectified Linear Unit) & An activation function that outputs the input directly if it's positive, and zero otherwise, mathematically represented as max(0,x). \\ \hline

    LeakyReLU & Variant of ReLU introduces small, non-zero slope values for negative inputs. \\ \hline

    Sigmoid Activation Function & A mathematical function that maps input values to a range between 0 and 1. \\ \hline
    
    Loss Function & It quantifies the difference between the model's prediction and the actual values. \\ \hline

    MSE (Mean Squared Error) & The loss function that measures the average squared difference between predicted and true values. \\ \hline

    Optimizer & An algorithm that adjusts the model parameters (like weights and biases) to minimize the loss function. We use the Adam's optimizer. \\ \hline

    Backpropagation & Technique to adjust weights and biases to minimize the loss function. \\ \hline

    Learning Rate & A hyperparameter that controls the rate at which the model updates its parameters. \\ \hline

    Epoch & One full pass through the entire dataset during training, allowing the model to learn patterns and adjust the parameters \\ \hline

    EarlyStopping & A technique to prevent overfitting by stopping training when the validation loss stops improving after a set number of iterations. \\ \hline

    Patience & A parameter used in early stopping that defines how many epochs to wait after the last improvement before stopping training.\\ \hline
    
    He-Uniform Initializer & A weight initialization method that draws weights from a uniform distribution scaled based on the number of input neurons. \\ \hline
    
    Cross-Validation (CV) & Evaluates the model’s performance by dividing the dataset into multiple subsets for training and testing.\\ \hline

    Outliers & The GRBs displaying uncertainty greater than 100\% for a particular model when fitted with W07 fitting function. \\ \hline

    \end{tabular}
}
    \caption{Definition of Machine Learning Terms.}
    \label{tab:ML_def}
\end{table}

\vspace{5pt}

    
    
    

To ensure robustness and consistency, the proposed models share several common steps in both pre-processing and result calculation, as outlined below:
\begin{itemize}

    \item \textbf{Individual GRB handling:} Due to the distinct nature of each GRB's afterglow light curve, each GRB was trained and reconstructed individually rather than using a generalized model. This approach preserves each burst's unique temporal and flux evolution patterns. The Appendix(\ref{section:appendix}) explains this in more detail. For every model, the following process was applied: time values (inputs) and corresponding flux values (labels) for each GRB were log$_{10}$-transformed and used for training.

    \item \textbf{Input and Output:} The model is trained using time values—whose lengths vary depending on each GRB—paired with their corresponding target flux values. During reconstruction, the model is given missing time window (gap length greater than 0.03 in log time scale units) to predict the intermediate flux values.
    
    \item \textbf{Parameter Tuning:} We selected a subset of 16 GRBs for hyperparameter tuning, with 4 GRBs randomly chosen from each of the 4 GRB classes. For each of these 16 GRBs, we held out 20\% of the data as a validation set and used the remaining 80\% for training. Hyperparameter optimization was conducted individually for each GRB using Bayesian Optimization via the \textit{Optuna} module \citep{2019optuna}, with the goal of minimizing the validation loss on the 20\% hold-out set.

   Importantly, we did not use early stopping in this process or any other model training stage. Instead, the number of training epochs was itself treated as a tunable hyperparameter and optimized during the \textit{Optuna}-based tuning phase. For each GRB, the optimal number of epochs (along with other hyperparameters) was selected based on validation loss. Once optimal hyperparameters were obtained for each of the 16 GRBs, we computed their average across the 16 runs. For parameters requiring integer values, we rounded the averaged result to the nearest integer.
    
   This resulted in a fixed set of hyperparameters (including the number of epochs) for each model architecture. These were then used to train each model on the complete dataset of 521 GRBs. This tuning process was performed separately for each model type evaluated in the study.

  The following hyperparameters were tuned depending on the model architecture:

    \begin{itemize}
        \item \textbf{MLP:} Number of hidden layers, neurons per layer, number of epochs
        \item \textbf{Attention U-Net:} Number of filters per layer, kernel size, number of convolutional blocks, learning rate, batch size, number of epochs
        \item \textbf{Bi-LSTM:} Number of Bi-LSTM layers, number of hidden units, activation function, number of epochs, learning rate.
        \item \textbf{Bi-MAMBA:} d\_state, dropout rate, d\_expand, number of epochs, learning rate.
        \item \textbf{GP-RF:} Number of estimators, maximum tree depth, minimum samples for node splitting, minimum samples per leaf.
        \item \textbf{CGAN:} Generator and discriminator hidden dimensions, number of epochs, learning rate.
        \item \textbf{KAN:} Width, grid, number of epochs, learning rate.
    \end{itemize}

    \item \textbf{Min-Max Scaling:} For each GRB, the data is scaled to the [0, 1] range using standard min-max normalization (\ref{eq:minmax}). However, scaling is not applied to Attention U-Net and MLP models, as by performing some tuning in several GRBs showed a decrease in performance. Additionally, Attention U-Net inherently includes multiple Batch Normalization layers, while the MLP, being a relatively simpler network, is less susceptible to the weights of the model either increasing exponentially or going toward zero, namely exploding or vanishing gradients.
    
    \begin{equation}
    \label{eq:minmax}
        X_{i} = \frac{X_{i} - X_{min}}{X_{max} - X_{min}};  i \in D,
    \end{equation}
    where $X_{min}$ and $ X_{max}$ are the minimum and the maximum value in $D$.

    \item \textbf{Uncertainty Estimation and Confidence Interval:} The uncertainty in the reconstructed flux values was estimated by fitting a Gaussian distribution to the flux residuals, defined as the difference between the original data and the reconstructed data. This Gaussian distribution provides a probabilistic model of the noise. Random noise samples were drawn from this distribution and added to the model’s mean prediction to generate realistic flux variations. To account for variability, we performed Monte Carlo simulations with 1000 iterations. For each iteration, random noise was sampled for each reconstructed data point and added to the corresponding flux value. This process resulted in multiple realizations of the reconstructed LC, capturing the statistical uncertainty of the predictions. The mean of these realizations was computed to represent the final reconstructed flux. The 95\% confidence intervals were determined by calculating the 2.5th and 97.5th percentiles, denoted with P, of the realizations for each time point, providing a reliable measure of uncertainty for the reconstructed data:

    \begin{equation}
    \text{CI}_{95} = 
    \left[
    \text{Percentile}_{2.5}, 
    \text{Percentile}_{97.5}
    \right].
    \end{equation}
        
    \item \textbf{5-Fold Cross-Validation:} After fixing the hyperparameters for each model, 5-fold CV is used as an alternate metric to evaluate the performance of each model. Rather than splitting the entire dataset of 521 GRBs into separate training and testing sets, we apply CV to each GRB separately. For each GRB, its time-flux data points are first randomly shuffled and then partitioned into five approximately equal-sized, non-overlapping subsets (folds). In each of the five iterations, four folds (80\% of the data) are used to train the model, and the remaining fold (20\%) is used to validate it \citep{kohaviCV}. This ensures that all data points are used for both training and validation across the five folds. After training, the model predicts flux values at the time points in the validation fold.
    These predicted fluxes are inversely transformed back to the original scale before calculating the MSE, as it depends on the scale. The MSE loss is averaged across the folds and then across the 521 GRBs. CV was used solely to evaluate model performance, not for tuning model parameters or minimizing training error.

    \item \textbf{Final Model:} After evaluating model performance using 5-fold CV on each GRB independently, we proceed to train a final model for each GRB. Since the hyperparameters are fixed for every model, we train the final model using the entire available LC information (time-flux values) of each GRB, without any train-test split. This model is then used to reconstruct missing regions (gaps greater than 0.03 in log-time scale) in the LC of that GRB. Then, we use this result to calculate the reduction in parameter uncertainty.

\end{itemize}

\subsubsection{Gaussian Process}\label{sec:GP}

GP regression \citep{GP2003gaussian} is a probabilistic, non-parametric approach designed for regression tasks. The probabilistic nature of the GP model allows for the likelihood of the prediction instead of producing a singular prediction. The method relies on Bayesian inference, where the prior based on trends in the observed data is updated to compute a posterior that aligns with the data. The model prediction is bound by the confidence interval, which defines the region of likelihood of the prediction.



Let \(X\) be a set in \(\mathbb{R}^d\). A GP is a collection of random functions, $f(x)$, defined over the input space X set in \(\mathbb{R}^d\). For any set of points in \(x_1,....,x_n \in X\), the values \((f(x_1),...,f(x_n))\) follows a joint multivariate Gaussian distribution.

Therefore, we can characterize the GP by its mean function:

\begin{equation}
m(x) = \mathbb{E}[f(x)],
\end{equation}

And its covariance function:

\begin{equation} 
    \mathbf{K}(x,x') = \mathbb{E} \left[(f(x) - m(x))(f(x') - m(x'))\right].
\end{equation}

A key component of GP is the kernel, or covariance function $\mathbf{K}(x,x')$, which measures the similarity between data points. 
In machine learning, the most popular kernel function is
Gaussian Radial Basis Function (RBF) kernel, which is given by: 
\begin{equation}
K_{\gamma}(x, x') = \exp\left(-\frac{\|x - x'\|_2^2}{\gamma^2}\right),
\end{equation}
where \(\|x - x'\|_2^2\) denotes the squared Euclidean distance between \(x\) and \(x'\),
for \(x \in \mathbb{R}^d\) and \(x' \in \mathbb{R}^d\),
and the parameter $\gamma$ determines
the length scale of the associated hypothesis space of the functions. As $\gamma$ increases, the kernel and induced functions change less rapidly and thus become `smoother' (while they are always infinitely differentiable).

To capture observational uncertainty, RBF kernel is combined with the White Noise kernel, defined as: 
\begin{equation}
    \mathbf{K}(x,x') = \sigma^2 \delta_{x x'},
\end{equation}

where $\delta_{x x'}$ is the Kronecker delta function (1 if $x =x'$, 0 otherwise), and $\sigma^2$ represents the variance of the noise.

For our analysis, we utilize a Radial Basis Function kernel combined with a White Noise kernel because we assume that the flux noises are independent and identically normally distributed. The RBF captures the smooth, continuous relationships inherent in the data, and the White Kernel accounts for observational noise, ensuring reliability in the model predictions.

The formulation of the kernel is as follows:

\begin{equation}
    \mathrm{kernel} = \mathrm{RBF} \left(s_{1}=10, b_{1}=[10^{-2}, 10^{3}]\right) + \\ 
     \mathrm{WhiteKernel} \left(s_{2}=1, b_{2}=[10^{-5}, 10] \right).
\end{equation}

Here, $s_{1}$ represents the length\_scale, $b_{1}$ the length\_scale\_bounds, $s_{2}$ the noise\_level, and $b_{2}$ the noise\_level\_bounds.

To implement GPs, we utilized the \textit{GaussianProcessRegressor} from scikit-learn, enabling prediction with prior knowledge of the GP without its prior fitting. The standardized prior indicates that the mean is centered at zero. The GP regression model was fitted with the built-in \textit{fit} function to our data and the reconstruction of data points was performed using the in-built \textit{predict} function.

We have selected a 95\% confidence interval to reconstruct the LC using GP. Using the form of the function derived by the Gaussian Regressor (built-in function in Python), we avoid the distribution of the data points in the GPs by conducting 100 MCMC simulations of the reconstructed LC for each distinct LC. The value and its corresponding uncertainty were then randomly chosen.

\begin{figure}
    \centering
    \gridline{
        \fig{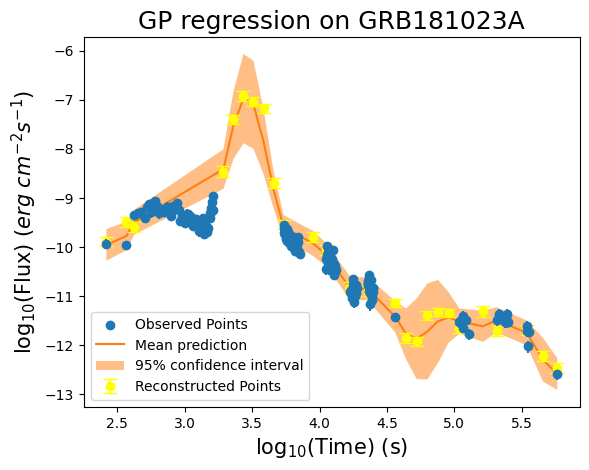}{0.45\textwidth}{(a) Without Tuning}
        \fig{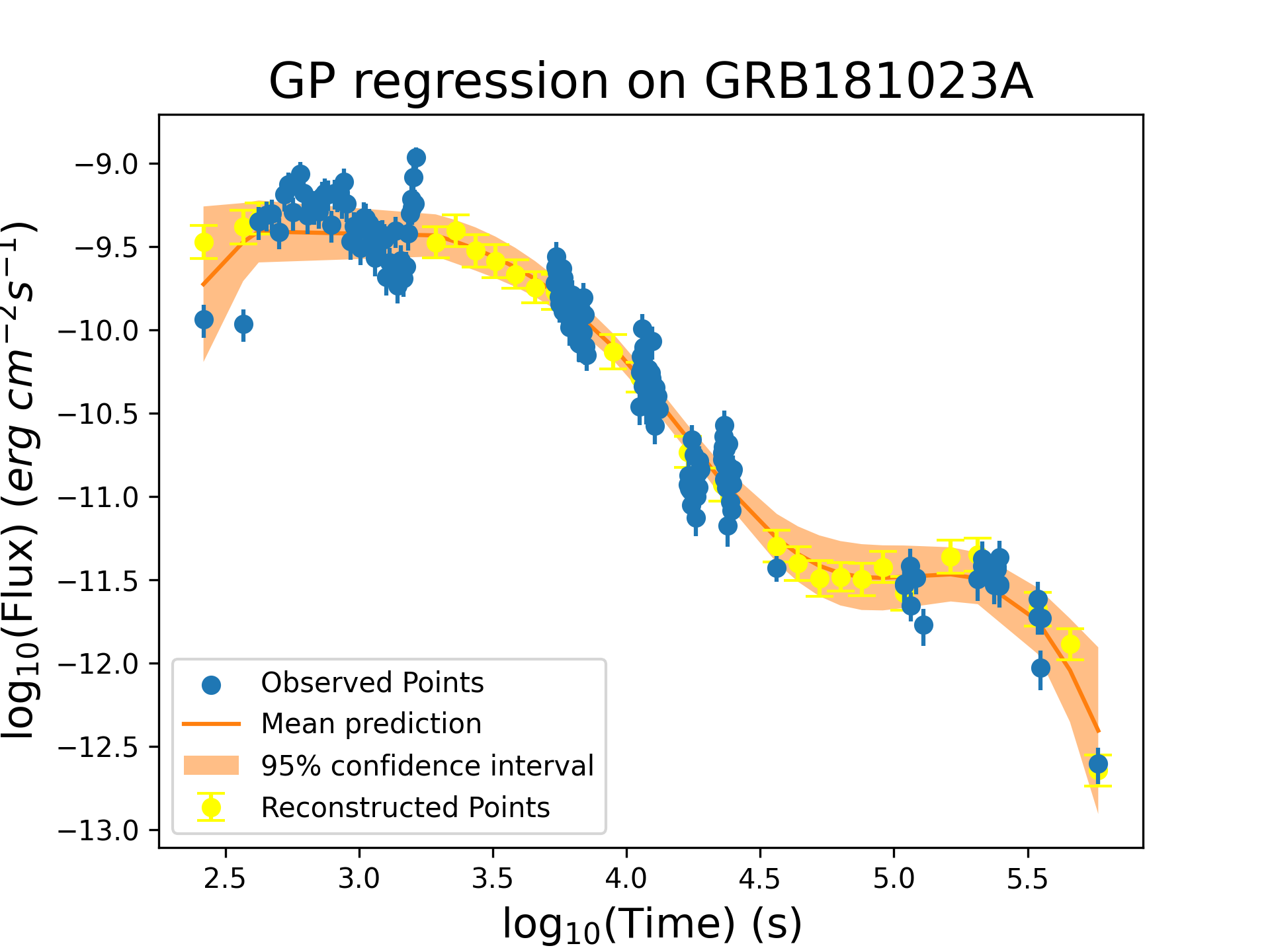}{0.45\textwidth}{(b) With Tuning }
    }
    \caption{Comparison of LCs before and after hyperparameter tuning.}
    \label{fig:comparison}
\end{figure}

In our data, out of the total sample of 521 GRBs, a tiny subset containing 9 GRBs with redshifts and 15 GRBs without redshifts, exhibited significant temporal gaps, defined as time intervals greater than 0.5 (in log$_{10}$ scale). As Fig. \ref{fig:comparison}a shows, the GP Regressor overfitted these data points in both the gaps, resulting in erratic predictions and extensive 95\% confidence intervals. This inconsistent behavior suggested the need to fine-tune two primary hyperparameters: \textit{alpha} and \textit{n\_restarts\_optimizer}, by employing \textit{GridSearchCV}; 
\textit{alpha} is a value added to the kernel matrix's diagonal during fitting. This hyperparameter helps avoid potential numerical issues, shown in Fig. \ref{fig:comparison}a, by ensuring the matrix remains positive and definite. Furthermore, \textit{alpha} can be interpreted as the additional Gaussian measurement noise variance on the training observations. The default value is 1e-10. A more considerable \textit{alpha} value is frequently required to avoid overfitting in cases where the data is noisy. 
\textit{n\_restarts\_optimizer} denotes the number of times the optimizer restarts to find the kernel parameters that maximize the log-marginal likelihood. This procedure guarantees a more reliable optimization of the kernel's parameters and aids in avoiding local minima.

\textit{GridSearchCV} uses 5-fold CV to assess model performance as it iteratively searches through a predefined set of hyperparameters. It exhaustively tests all possible combinations of the specified hyperparameters and selects the combination that results in the best performance according to a chosen metric. The \textit{n\_restarts\_optimizer} is set to 20 and the \textit{alpha} hyperparameter is set as 0.1.

After tuning the hyperparameters, we fit the GP Regressor to the data. The optimized model improved the reconstruction of the LCs, as shown in Fig. \ref{fig:comparison}b. It might appear that Fig. \ref{fig:comparison}a offers a better reconstruction than Fig. \ref{fig:comparison}b, as it reconstructs the flare in the first temporal gap. However, this is misleading as very large flares like this have been unusual in the sample of the 521 GRBs. Additionally, the confidence intervals in Fig. \ref{fig:comparison}a are significantly broader due to the larger temporal gap, reflecting high uncertainty. After hyperparameter optimization, the model shown in Fig. \ref{fig:comparison}b not only yields smoother and more stable predictions but also exhibits reduced uncertainty, as reflected in narrower confidence intervals. 

\subsubsection{Multi-layer Perceptron Model}

The MLP model offers a simple yet effective approach to reconstructing the LCs. By combining the best-fitting curve provided by the model and the noise provided by the Gaussian method, we get an excellent estimate of the flux values and reduction in uncertainty (as shown in \S\ref{section:results}). This method is helpful since it allows for as much accuracy as the other models while being more straightforward than the other methods.



The neural network was designed to capture the non-linear relationship between time and flux in the logarithmic space allowing for smooth reconstruction of LCs. The model has eight layers in total, including the input and output layers. The first layer is the input layer ($0^{th}$ layer) which accepts a singular $log_{10}(time)$ value. In every epoch the entire LC is passed through the model exactly once. The following six layers are the hidden layers ($1^{st}$ to $6^{th}$ layers), and each hidden layer has a varying number of neurons to ensure that the model learns the non-linear relationship between the flux and time. The hidden layers from the first to the last have 32, 64, 128, 64, 32 and 16 neurons each. Each layer uses the \textit{ReLU} activation function because it helps enhance the rate at which the network learns the non-linear relationship between the target and feature variables \citep{agarap2018deep}. The last layer is the output layer with a single neuron which makes the final prediction of the $log_{10}(flux)$ value corresponding to the $log_{10}(time)$ value which was passed into the input layer. The model is represented in Fig. \ref{fig:methods}a.

Each hidden layer performs the following operation:
\begin{equation}
\label{eq:mlp}
    \begin{split}
    \mathbf{Z^{(l)}} &= \mathbf{W}^{(l)} \cdot \mathbf{h}^{(l-1)} + \mathbf{b}^{(l)},\\
    \mathbf{h^{(l)}} &= \text{ReLU}(\mathbf{Z^{(l)}}),
    \end{split}
\end{equation}

where,
$\mathbf{W^{(l)}}$ and $\mathbf{b^{(l)}}$ are the weight and bias
matrices of the $l^{th}$ layer. 
$\mathbf{h^{(l-1)}}$ is the activation output of $(l-1)^{th}$ layer and $\mathbf{Z^{(l)}}$ is the pre-activation output of $l^{th}$ layer.

The kernel initialization of each hidden layer uses the \textit{He-Uniform} \citep{7410480} initializer method.  The \textit{He-Uniform} initializer initializes weights by sampling from a uniform distribution within the range:
\begin{equation}
    range = \left[ -\sqrt{\frac{6}{n_{in}}}, \sqrt{\frac{6}{n_{in}}} \right].
\end{equation}
here, $n_{in}$ is the number of input units (neurons) to the layer. This ensures that the weights are small enough to prevent exploding gradients while still being large enough to avoid vanishing gradients. It is particularly suited for \textit{ReLU} activation function \citep{7410480} since it does not saturate like sigmoid or tanh activation functions. The \textit{ReLU} activation function, defined as $max(0,x)$, outputs zero for negative values and keeps positive values unchanged, making it efficient for training deep networks.

The output layer gives the prediction as follows:
\begin{equation}
    log_{10}(flux)_{predicted} = \mathbf{W^{(out)} \cdot h^{(l-1)}} + \mathbf{b^{(out)}}.
\end{equation}

The model is trained using the Adam optimizer \citep{kingma2014adam} with 0.001 as the learning rate and MSE as the loss function.  The number of epochs is set to 5000. The model was implemented using the TensorFlow framework with Keras.

Next, we perform the 5-fold CV for all 521 GRBs individually and then report the average Mean Squared Error (MSE) across all GRBs as our performance metric in Table  \ref{tab:reconstruction_results}.

The predicted GRB LCs given by the MLP model would be smooth and unrealistic. To bring about the inherently noisy nature of the logarithmic values of the natural flux values, Gaussian noise was added to each $log_{10}(flux)$ prediction in the same manner as \ref{section:ML}.

\subsubsection{Attention U-Net}


U-Net is a convolutional neural network architecture designed primarily for biomedical image segmentation \citep{ronneberger2015u} but has been successfully adapted for one-dimensional data such as time series \citep{perslev2019u}. The architecture consists of a symmetric encoder-decoder structure, where the encoder compresses the input into a latent representation, and the decoder reconstructs the output from this latent space. Latent space is a lower-dimensional representation of the data where only the key features are retained and the rest discarded. \textit{Skip connections} between the encoder and decoder ensure the preservation of spatial details, which are critical for high-resolution reconstruction tasks.

For the LCR, we enhance the standard U-Net with an attention mechanism\citep{oktay2018attention}, which is a section of a neural network that enhances feature extraction and boosts performance on tasks involving complicated or sequential data by dynamically focusing on the most pertinent portions of the input data and giving distinct aspects varied degrees of priority as stated in \cite{2017arXiv170603762V}. The attention mechanism enables the model to focus on relevant features at each resolution level by weighting them based on their importance. This allows the network to capture both global and local structures effectively.

In our implementation of the Attention U-Net for GRB LCR, we employ 5-fold CV to ensure robust performance evaluation as explained in the starting of \S\ref{section:ML}. To handle the inherent variability in lightcurve lengths, all lightcurves are interpolated to a fixed length of 100 using uniform linear interpolation prior to training. This preprocessing ensures compatibility with the U-Net architecture, which requires fixed-size input tensors, without altering the temporal characteristics of the original lightcurves.

The attention mechanism is introduced through an \textit{Attention Block}, which is integrated into the decoder path of the U-Net. Each attention block computes the importance weights as follows: 
\begin{equation}
\psi_f = \sigma(\text{Conv1D}(f(\theta_x + \phi_g))), 
\end{equation} 
where $\theta_x$ represents the feature map from the encoder, $\phi_g$ is the corresponding upsampled feature map from the decoder, $f$ applies the \textit{ReLU} activation function, and $\sigma$ is the \textit{sigmoid} activation function (explained in Table \ref{tab:ML_def}). The calculated weights $\psi_f$ are then used to modulate the encoder features: \begin{equation} 
\text{Attention Output} = x \cdot \psi_f.
\end{equation}
where $x$ and $g$ represent the encoder and decoder features, respectively.

The encoder-decoder structure of the model forms the backbone of the architecture. The encoder's hierarchical features are extracted from the input LC data using convolutional layers with \textit{ReLU} activation and \textit{He-uniform} initialization. Mathematically, the output of each convolutional layer is given by: 

\begin{equation} 
h^{(l)} (t) = ReLU(W^{(l)} . h^{(l-1)}(t) + b^{(l)}),  
\end{equation} 
where $h^{(l)}(t)$ represents the output of the $l$-th layer, $W^{(l)}$ and $b^{(l)}$ are weight and bias matrices, and $t $ denotes the temporal index. Max-pooling operations reduce the temporal resolution at each stage, enabling the model to capture long-range temporal dependencies.

We implemented a 1D U-Net with attention mechanism. The network consisted of three downsampling blocks with ConvID (kernel size = 3, padding = same, activation = ReLU), which reduces the sequence length (temporal dimension) by a factor of 2, followed by a bottleneck. The bottleneck is the sequence with the shortest length. This stage captures the most compressed and high-level features of the input sequence, allowing the model to learn global patterns across the time series.  Then, we use three upsampling blocks with the UpSampling1D function to double the sequence length back to the original state. Each convolutional block used He-uniform initialization.  Attention blocks were inserted in the decoder path for feature refinement. These blocks help the network to selectively focus on the most relevant temporal features, enhancing the reconstruction quality. The model was trained using the Adam optimizer (learning rate = 0.001), with a batch size of 64 and for 1000 epochs. The input was log-transformed time values, and the output was the corresponding log-transformed flux values. 5-fold CV was used for evaluation, and final predictions were made using the full training set. The implementation of the Attention U-Net architecture was carried out using TensorFlow and Keras.

\begin{figure}
    \centering

    \gridline{
        \fig{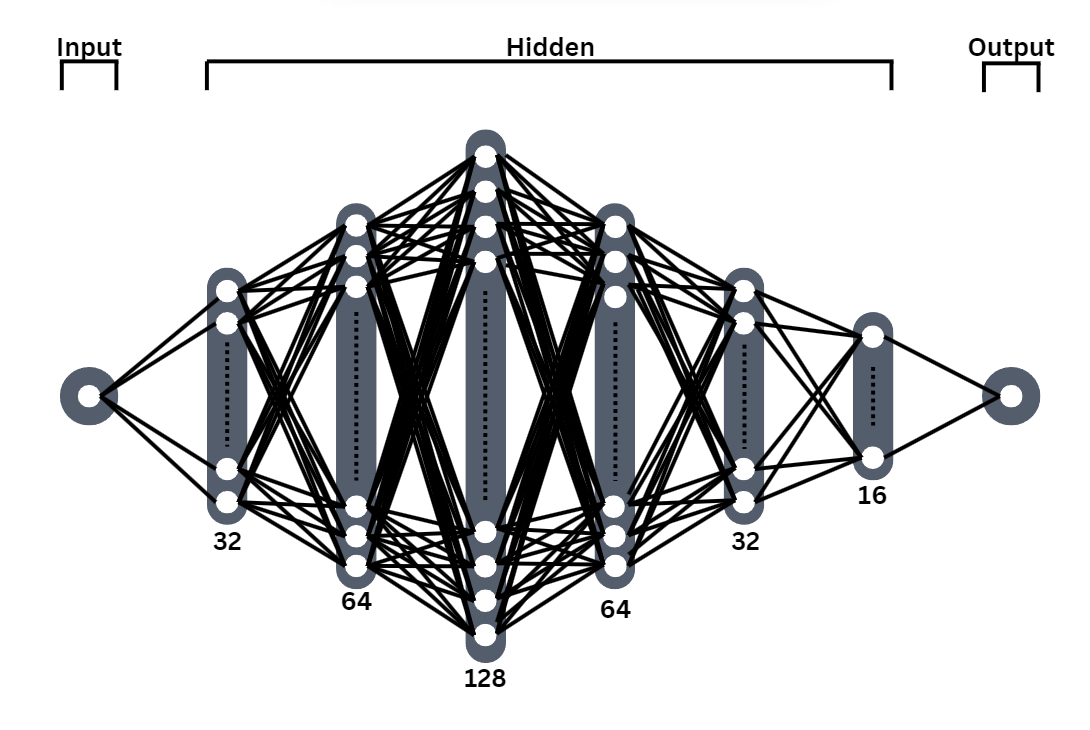}{0.32\textwidth}{(a) MLP}
        \fig{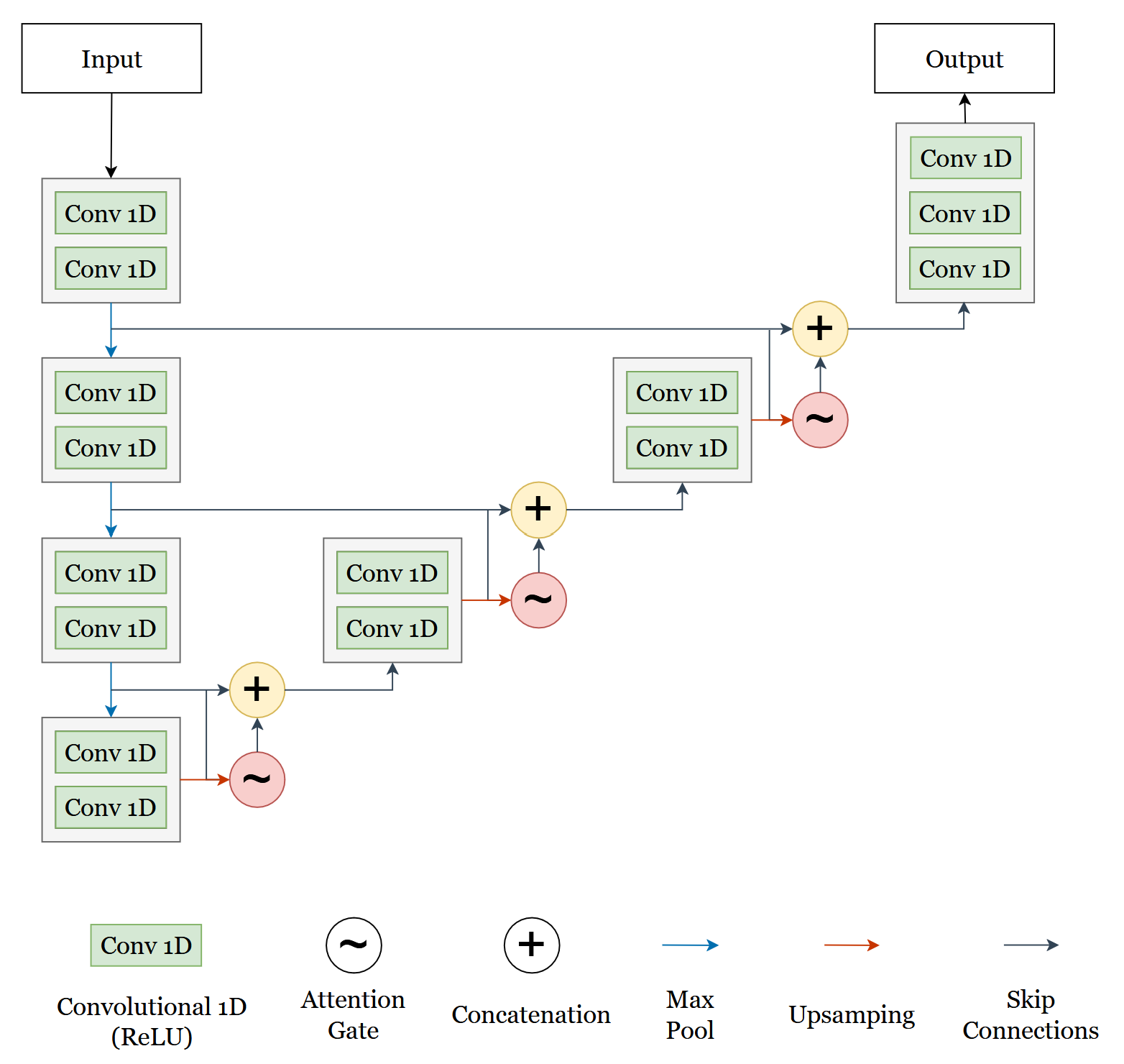}{0.32\textwidth}{(b) Attention U-Net}
        \fig{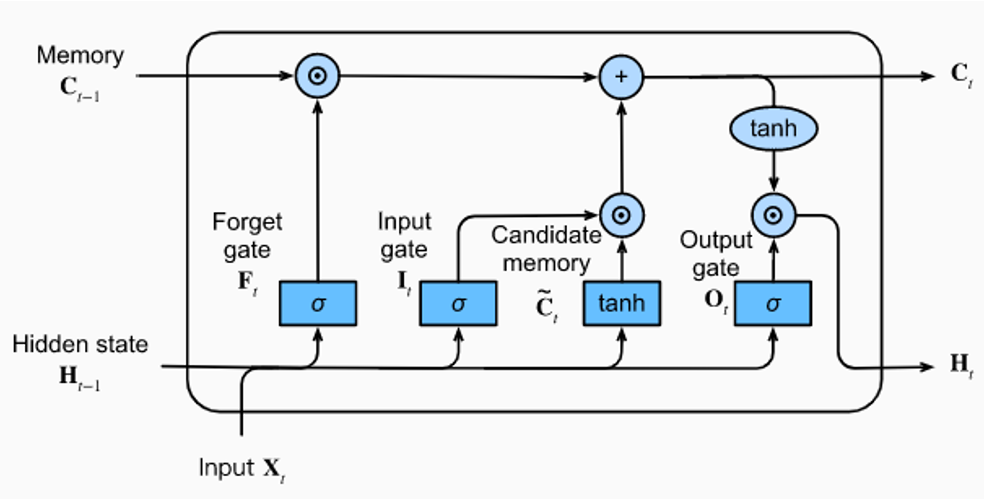}{0.32\textwidth}{(c) Bi-LSTM}
    }

    \gridline{
        \fig{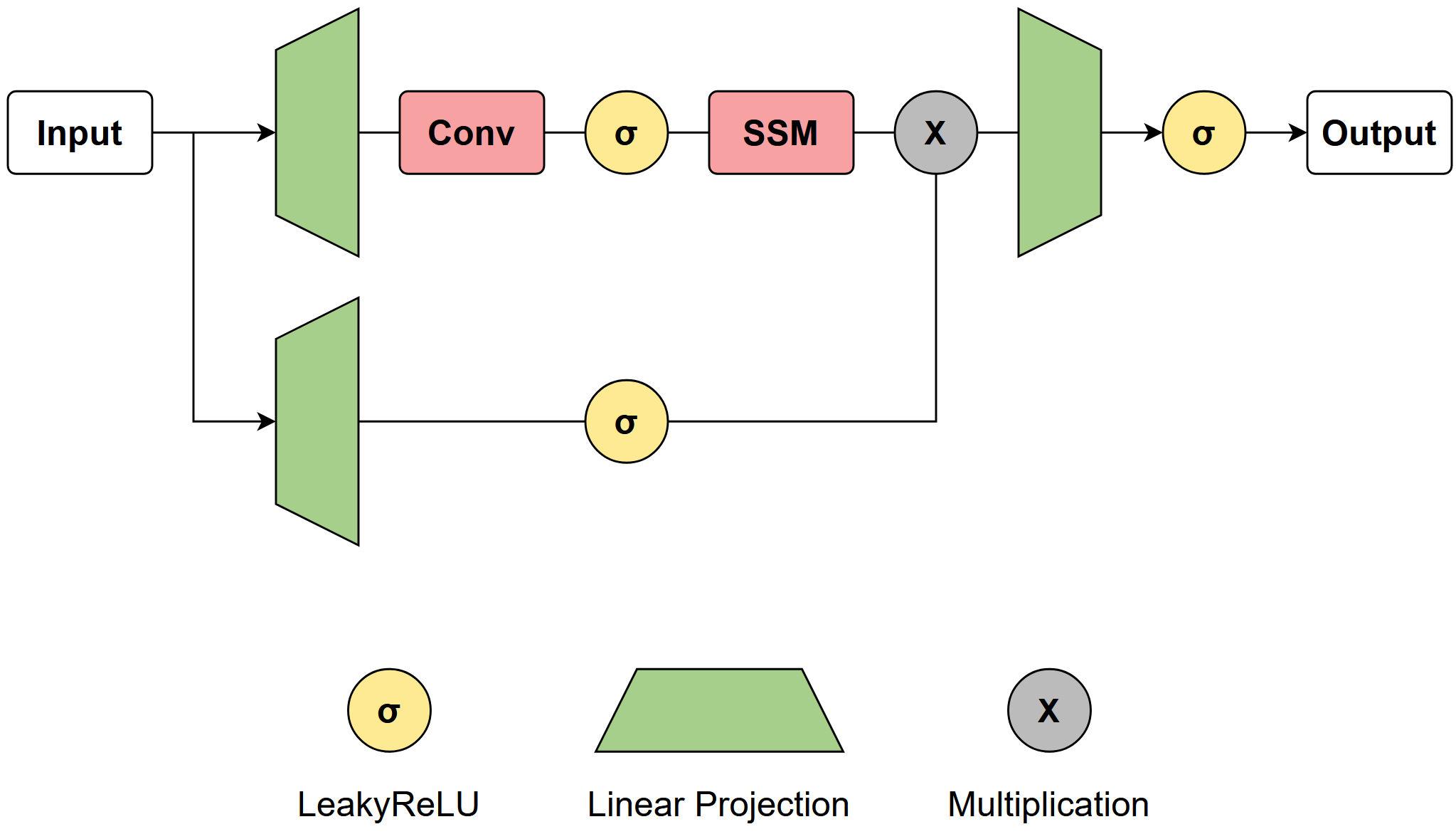}{0.32\textwidth}{(d) Bi-Mamba}
        \fig{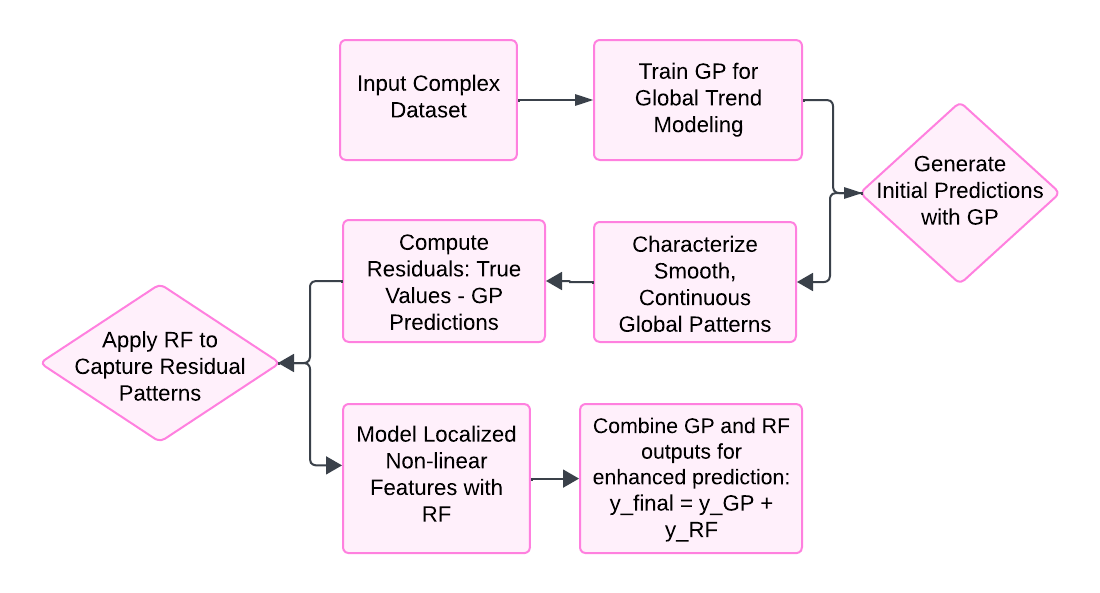}{0.32\textwidth}{(e) GP-RF}
        \fig{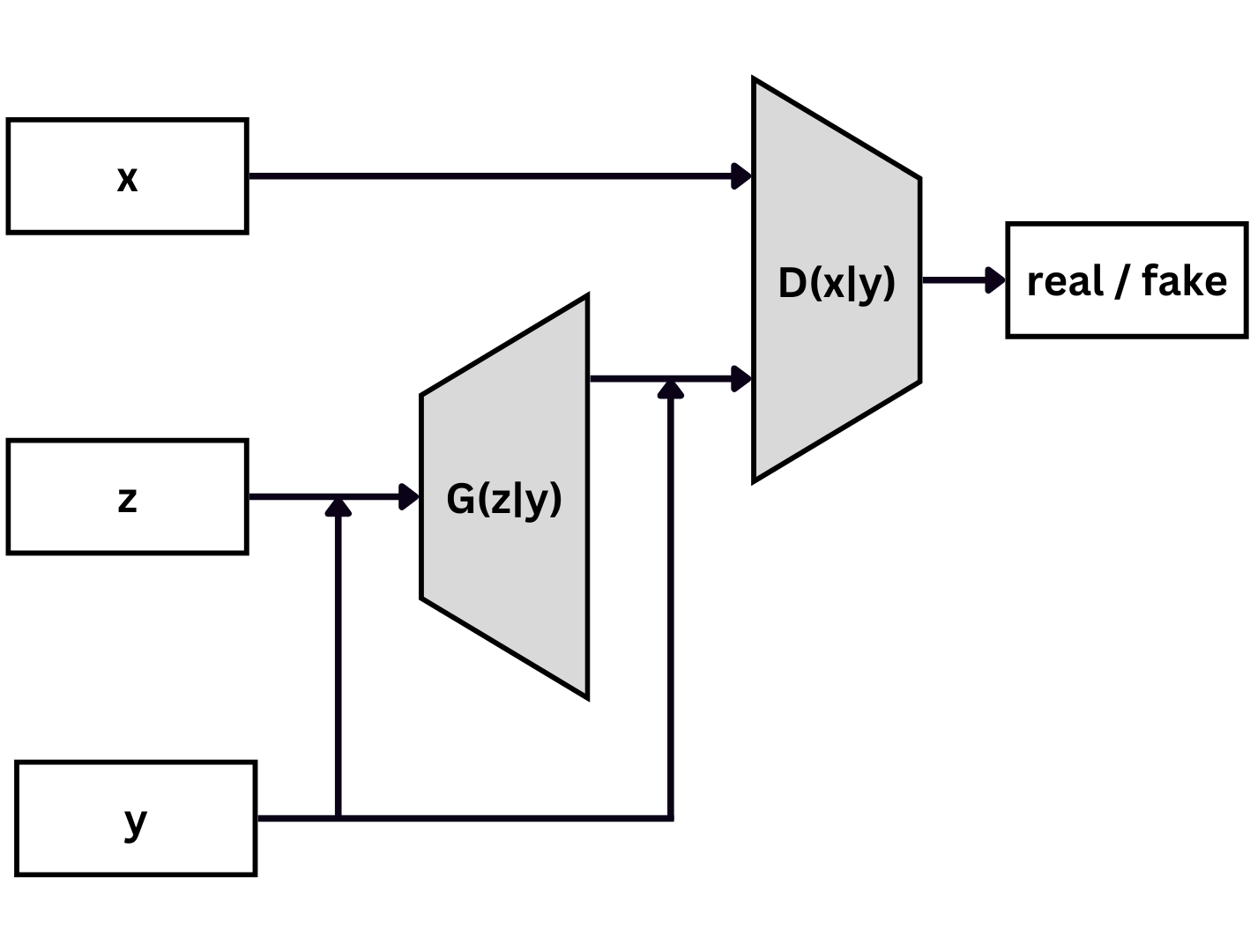}{0.32\textwidth}{(f) CGAN}
    }

    \gridline{
        \fig{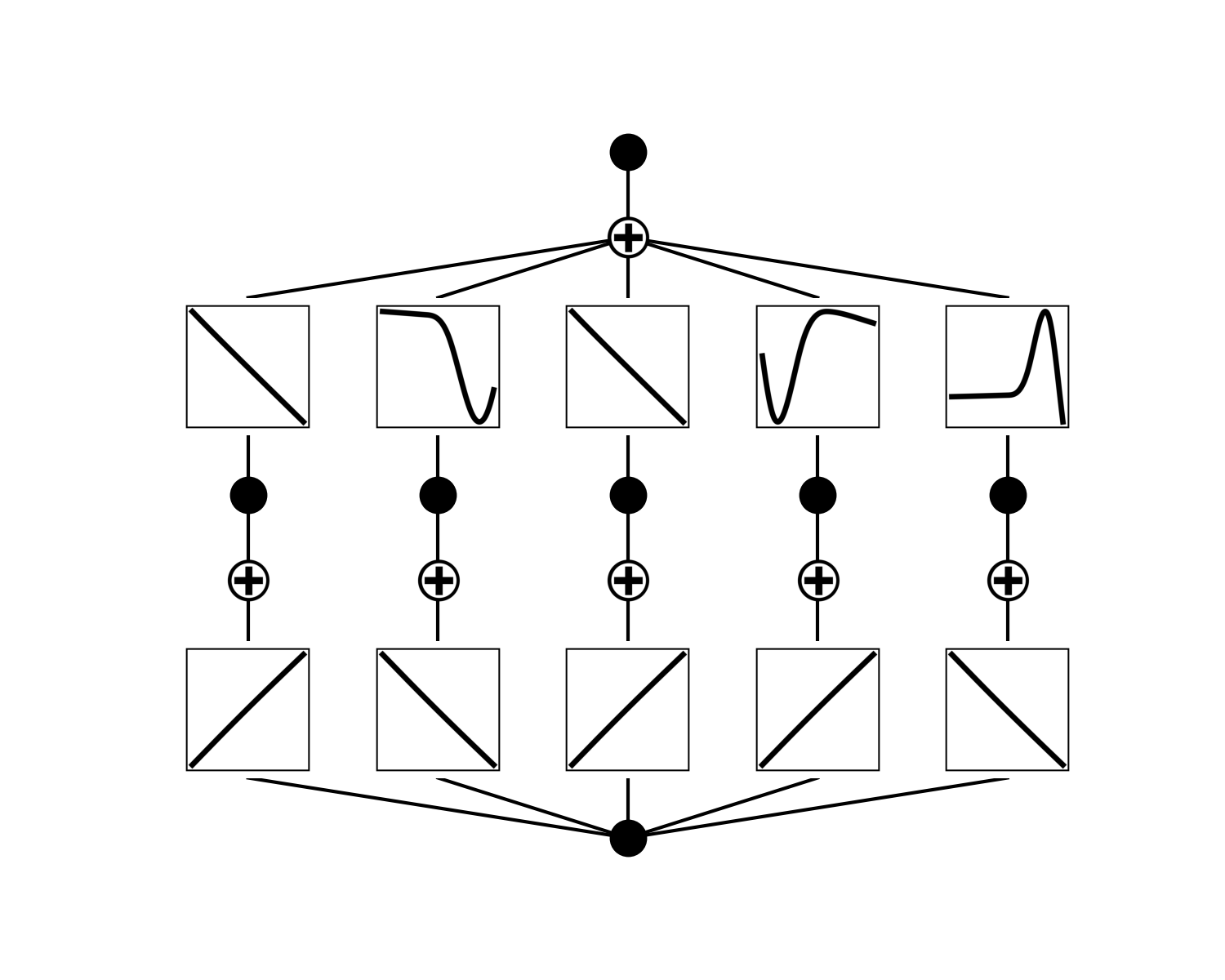}{0.32\textwidth}{(g) KAN}
        \fig{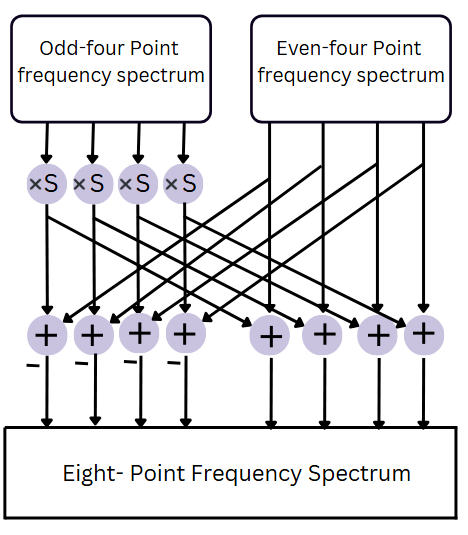}{0.32\textwidth}{(h) Fourier Transform}
        \fig{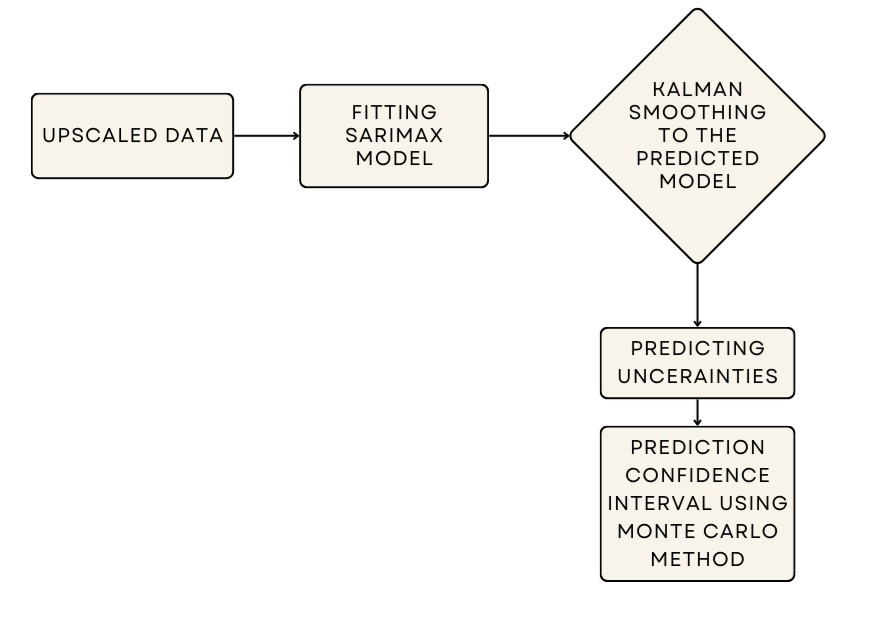}{0.32\textwidth}{(i) SARIMAX-based Kalman}}

    \caption{Architecture of nine models shown from top left to bottom right in each row: (a) MLP, (b) Attention U-Net, (c) Bi-LSTM, (d) Bi-MAMBA, (e) GP-RF, (f) CGAN, (g) KAN, (h) Fourier Transform, (i) SARIMAX-based Kalman . Architectures of Bi-Mamba, MLP, KAN, and Attention U-Net are actual representations of the models utilized.}
    \label{fig:methods}
\end{figure}

The data is transformed into a high-dimensional latent representation using two Conv1D layers with 256 filters, where 1D indicates that the convolution operates along a single spatial dimension (time, in our case), and filters refer to learnable kernels that extract features from the input light curve data.  The model is represented in Fig. \ref{fig:methods}b. This representation captures the most abstract features of the input sequence, which are essential for accurate reconstruction.

The decoder reconstructs the temporal sequence by progressively restoring the temporal resolution through upsampling layers, which increases the resolution of the input data.
{Skip connections} between the encoder and decoder layers ensure that fine-grained temporal details are preserved. To enhance the utility of skip connections, attention blocks are applied \citep{oktay2018attention}.

The model's last output layer is the Conv1D layer, which previously maps the reconstructed features to the previously anticipated flux values. It has a single filter. The complete model is trained to reduce the Mean Squared Error (MSE) between the observed and predicted logarithmic flux values. The loss function is written as:\begin{equation} \mathcal{L} = \frac{1}{N} \sum_{i=1}^N \left( \log_{10}(\hat{y}_i) - \log_{10}(y_i) \right)^2.
 \end{equation}
where \( N \) is the number of data points, \( \hat{y}_i \) represents the predicted flux, and \( y_i \) represents the observed flux.

 The U-Net with Attention mechanism is well-suited for tasks involving complex temporal structures like GRB LCs, where capturing long-range dependencies and local variations is critical \citep{perslev2019u}.

\subsubsection{Long Short Term Memory Neural Network}

Long Short-Term Memory (LSTM; \citealt{hochreiter1997long}) is a form of Recurrent neural network \citep{recurrent2001} that performs well in sequence prediction through its ability to capture long-term dependencies. Unlike traditional neural networks, LSTM incorporates feedback connections, allowing it to process data sequences rather than individual data points. This distinctive architecture equips LSTM with the ability to effectively learn and retain crucial information over time, making it a competent deep learning tool well-suited for applications in speech recognition, language translation, time series forecasting, and more \citep{lstm2016speech, lstm2016lang, lstm2019time}.

The main feature of an LSTM is its memory cell, which is managed by three essential gates: the Input gate, the Forget gate, and the Output gate. These gates determine which information is added, removed, and output from the memory cell to learn long-range relationships within the data effectively.

In an LSTM network, a hidden state function as the short-term memory, capturing information across time steps. This combination allows the LSTM to effectively maintain context and adjust its memory in response to new inputs, facilitating its ability to learn complex patterns across long sequences.

LSTM networks have proven valuable in astrophysics, particularly for time-series data analysis of dynamic celestial phenomena. An example is the LSTM-FCNN model, which combines LSTM and a fully connected neural network (FCNN) to measure time delays in strongly lensed Type Ia supernovae, crucial for measuring the Hubble constant  \citep{huber2024holismokes}. Furthermore, LSTMs have been applied for solar flare prediction from active regions  \citep{liu2019predicting}, effectively predicting them using time-series analysis. These examples show that LSTMs can outperform traditional methods in certain astrophysical time-series tasks.

Our model uses Bi-LSTM \citep{graves2005bidirectional} to reconstruct the GRB LCs. The Bi-LSTM architecture extends the standard LSTM by processing the input data sequence in both forward and reverse directions, allowing it to capture temporal dependencies more effectively. This is particularly useful for GRB LCs, which feature complex temporal patterns such as flares and breaks. Bi-LSTMs also excel at handling missing data by learning from surrounding observations, enabling accurate imputation. The basic structure of the LSTM is described below and shown in Fig.\ref{fig:methods}c.
Table \ref{tab:lstm_variables} defines all the variables and symbols.

\begin{table}[htbp] 
    \centering
    \begin{tabular}{|c|l|}
    \hline
    \textbf{Symbol} & \textbf{Variable} \\ \hline
        \(t\)            & Time  \\ \hline
        \(x_t'\)         & Input sequence \\ \hline
        \(h_t'\)         & Hidden state \\ \hline
        \(h_{t-1}'\)     & Previous hidden state \\ \hline
        \(C_t\)         & Cell state \\ \hline
        \(C_{t-1}'\)     & Previous cell state \\ \hline
        \(f_t'\)         & Forget gate \\ \hline
        \(i_t'\)         & Input gate \\ \hline
        \(o_t'\)         & Output gate \\ \hline
        \(W_f'\)         & Forget gate weights \\ \hline
        \(W_i'\)         & Input gate weights \\ \hline
        \(W_o'\)         & Output gate weights \\ \hline
        \(b_f'\)         & Forget gate bias \\ \hline
        \(b_i'\)         & Input gate bias \\ \hline
        \(b_o'\)         & Output gate bias \\ \hline
    \end{tabular}
    \caption{List of variables and symbols used in the Bi-LSTM model.}
    \label{tab:lstm_variables}
\end{table}

\begin{itemize}
  \item \textbf{Input Gate:} It controls which new information should be incorporated in the cell state. The input is denoted with $i$. It computes a gating signal ($i_{t}'$) using a sigmoid activation function applied to the weighted combination of current input ($x_{i}'$) and the previously hidden stage ($h_{t-1}'$), where $t$ denotes the time, along with a trainable weight ($W_i$') and bias ($b_{i}'$):

\begin{equation}
    i_t' = \sigma(W_i' \cdot [h_{t-1}', x_t'] + b_i').
\end{equation}

 \item \textbf{Forget Gate:} It determines the amount of the previous memory cell ($C_{t-1}$') should be discarded or retained.
 The forget gate is denoted with $f$. This is also computed using a sigmoid activation function. The gate output is between 0 and 1, where a value close to 0 indicates that the information must be forgotten:

\begin{equation}
    f_t' = \sigma(W_f' \cdot [h_{t-1}', x_t'] + b_f').
\end{equation}

 \item \textbf{Cell State Update:} It combines the retained information from the previous cell state and the new candidate memory. This update is governed by both the input and forget gates. The forget gate determines how much of the previous memory ($C_{(t-1)}'$) is preserved, while the input gate controls the amount of new information to be added:

\begin{equation}
    C_t' = f_t' \cdot C_{t-1}' + i_t' \cdot \tilde{C}_t'.
\end{equation}

 \item \textbf{Output Gate:} It specifies the portion of the updated cell state to be passed as the hidden state at the current time step. This is also computed using the sigmoid activation function:

 \begin{equation}
    o_t' = \sigma(W_o' \cdot [h_{t-1}', x_t'] + b_o'),
\end{equation}

where $o'$ denotes the output gate.

Finally, the updated cell state is passed through the hyperbolic tangent function, and the final hidden state is computed as:

\textbf{
\begin{equation}
h_t' = o_t' \cdot \tanh{C_t'}.
\end{equation}
}

\end{itemize}

In Bi-LSTM, the input sequence is processed in forward and reverse directions. The forward LSTM processes the set from the first to the last time step, learning dependencies from the past, while the backward LSTM processes the set from the end, retrieving information from future time steps. Each LSTM layer independently computes the gating signals, hidden and cell states, as described earlier. The outputs from the forward and backward LSTMs are then merged at each time instance to form the final output:

\begin{equation}
  h_t'^{\text{Bi-LSTM}} = \left[h_t'^{\text{forward}},h_t'^{\text{backward}} \right].
\end{equation}



For hyperparameter optimization, \textit{Optuna} was employed to fine-tune essential parameters, including the number of Bi-LSTM layers and the count of hidden units. The architecture consists of Bi-LSTM layers to capture both forward and backward temporal dependencies. In our model, we use 4 layers of Bidirectional LSTM, each with 100 hidden units, followed by a dense layer with Leaky Rectified Linear Unit (\textit{LeakyReLU}) activation to help the LSTM model learn complex features of flares and breaks in the LC. \textit{He-uniform} initialization was applied to initialize the network weights, promoting stable convergence.  \textit{ReduceLROnPlateau}, a callback function in Keras is used to reduce the learning rate when the validation loss plateaus to optimize the learning process further.
The learning rate is set to the default value of 0.001, and the number of training epochs is fixed at 100. Once the optimal hyperparameters were determined, we performed another 5-fold CV using these optimized settings to obtain the final performance, reported as the train and test MSE in Table \ref{tab:reconstruction_results}. The timestep is taken as 1.
The model was developed and trained utilizing the TensorFlow and Keras libraries.

\subsubsection{Bi-Mamba Model}

Mamba \citep{gu2024mambalineartimesequencemodeling} is a novel sequence modeling architecture based on state space models (SSMs), designed to efficiently capture long-range dependencies while reducing computational complexity. Unlike long short-term memory networks (LSTMs), which rely on explicit gating mechanisms to regulate memory updates, Mamba employs a structured state space framework that enables selective long-range memory retention. This structured recurrence mechanism allows Mamba to model temporal dependencies effectively, achieving linear time complexity, making it particularly suitable for large-scale astronomical datasets such as GRB LCs, which exhibit intricate temporal variations.

The fundamental operation of Mamba is governed by its continuous-time state space representation, where the hidden state $s_t$ evolves according to the equation:

\begin{equation} s_t = A s_{t-1} + B x_t, \end{equation}

Here $A \in \mathbb{R}^{N \times N}$ and $B \in \mathbb{R}^{1 \times N}$ are learnable matrices, controlling the state transition dynamics and scaling the contribution of the input $x_t$, respectively. The final output at each time step is obtained using a learned transformation matrix $C \in \mathbb{R}^{1 \times N}$:

\begin{equation} y_t = C s_t. \end{equation}

Since GRB LCs are discrete time-series data, the Mamba is adapted to such inputs using a Zero-Order Hold (ZOH) method -- as described in \citep{10.1093/imamci/dnac005} and \citep{gu2024mambalineartimesequencemodeling} -- which converts discrete inputs into piecewise constant signals. The resulting discrete-time formulation is given by:

\begin{equation} h_t = \bar{A} h_{t-1} + \bar{B} x_t, \end{equation}

\begin{equation} y_t = \bar{C} h_t. \end{equation}

where $\bar{A}$, $\bar{B}$, and $\bar{C}$ are the discrete counterparts of the continuous model parameters.

To enhance temporal context modeling in GRB LC reconstruction, Bi-Mamba model is used. Similar to Bi-LSTM, Bi-Mamba processes input sequences in both forward and reverse directions, ensuring that each time step benefits from contextual information from both past and future observations. This particularly benefits GRB LC reconstruction, where missing data segments require contextual information from surrounding observations. Bi-Mamba consists of two parallel Mamba models: a forward Mamba that processes the sequence in a standard temporal order and a backward Mamba that processes it in reverse. The outputs of both directions are concatenated at each time step to form the final bidirectional representation:

\begin{equation} h_t'^{\text{Bi-Mamba}} = \left[h_t'^{\text{forward}}, h_t'^{\text{backward}} \right]. \end{equation}

Each directional Mamba model employs a selective memory retention mechanism, filtering out irrelevant noise while preserving meaningful historical context. Unlike LSTMs, which rely on discrete gating operations, Mamba's continuous-time state transition model enables a more flexible representation of temporal dependencies, particularly beneficial for modeling the complex structure of GRB LCs, including flares and breaks.

We used a modified version of the \textit{MambaModel} from \cite{thielmann2024mambular} (shown in Fig. \ref{fig:methods}d), which contains an output activation layer with \textit{LeakyReLU} activation function and a linear mapping at the end. The model was built using the \textit{PyTorch} framework with hyperparameter tuning using \textit{Optuna}. The data (D) was scaled to a range of [0, 1] using \textit{MinMaxScaler}, given in the Eq. \ref{eq:minmax}. The hidden state dimension (d\_state) was set to 32, while the hidden layer expansion factor (d\_expand) was configured to 80. A dropout rate of 0.1 was applied to mitigate overfitting. Training was conducted over 120 epochs with a batch size of 64, using the Adam optimizer with a learning rate of 1e-4.

\subsubsection{Gaussian Process - Random Forest Model}\label{sec:GP_RF}
To overcome the challenges of traditional machine learning models in representing localized patterns within complex LC, we build a hybrid model that combines the strengths of GPs and RFs. 


In this method, the GP serves as the base model that captures continuous, smooth trends. At the same time, the RF acts as an embedded model to model the intricate, non-linear patterns (e.g., flares and breaks) in the LCs that the GP might overlook. By analyzing the residuals, defined as the deviation between the predictions by the GP model and the observed values, the RF effectively learns to model the flares left unaddressed by the GP, allowing it to complement the GP’s global predictive capabilities with its localized expertise. This hybrid approach offers a reliable solution for high predictive accuracy and reliability. As the base model, GP provides a smooth curve, and RF, as the meta-model, decreases the MSE.

GP forms the foundation of the hybrid model. 
The GP can capture global, non-linear relationships as a probabilistic, non-parametric model while providing smooth predictions and uncertainty.
The GP was trained on the input data to generate initial predictions representing the broad underlying structure of the LCs. The GP model was optimized using five random restarts, controlled by the parameter \textit{n\_restarts\_optimizer}=5, to improve the chances of finding the global optimum and reduce the likelihood of the model converging to local minima. These initial GP predictions serve as the baseline for the subsequent hybrid model, upon which further refinements and corrections are made.

Although the GP effectively models global trends, it exhibits limitations in capturing finer, localized variations in the data. To address this, the residuals, defined as the difference between the observed values and the GP predictions, were calculated:
\begin{equation}
Residuals=y_{observed} - y_{predicted}.
\end{equation}
 
Here, $y_\text{observed}$
are the actual flux values corresponding to the data and 
\(y_\text{predicted}\)
are the flux values predicted by the GP model.

These residuals were then modeled using RFs, a powerful ensemble method known for its ability to handle non-linear relationships and complex interactions. RFs were proposed by Leo Breiman \citep{breiman2001random} to build a predictor ensemble with a set of decision trees \citep{breiman2017classification} growing in randomly selected data subspaces. The model is shown in Fig. \ref{fig:methods}e.

The RF component was optimized through a systematic hyperparameter tuning process using \textit{RandomizedSearchCV} with 10 iterations and 3-fold CV. The hyperparameters were carefully defined to balance model complexity and computational efficiency, encompassing key parameters such as the number of estimators (ranging from 100 to 200 trees), maximum tree depth (10, 20, or unrestricted), minimum samples required for node splitting (2 or 5), and minimum samples per leaf node (1 or 2). This comprehensive parameter optimization strategy ensures reliable model performance while maintaining adaptability to local features in the LCs. The model was implemented using the Scikit-learn (sklearn) library.

The final prediction of the hybrid model is obtained by combining the outputs of the GP and the RF:
\begin{equation}
y_{final} = y_{GP} + y_{RF}.
\end{equation}



\subsubsection{Conditional Generative Adversarial Networks}
GANs \cite{goodfellow2014generative} provide an innovative approach to reconstruct the LCs. Unlike traditional models that rely on Markov chains—where the current state is dependent on the previous one—GANs bypass this sequential dependency during the learning phase \citep{mirza2014conditionalgenerativeadversarialnets}, incorporating it only during gradient backpropagation. These networks are referred to as generative models and are particularly useful when dealing with complex data distributions, reducing overfitting by incorporating a dual-model approach. 


These adversarial networks have a generator model $G(z)$ and a discriminator model $D(x)$ working harmoniously. Both the generator and the discriminator try to improve their results. If the generator constructs more real samples $x$ from random noise $z$, it becomes challenging for the discriminator to distinguish between the actual data and the synthetic data. Their adversarial relationship is mathematically described as:

\begin{equation}
    \label{eq:GAN}
    \min_G \max_D V(D, G) = \mathbb{E}_{x \sim q(x)} \left[ \log D(x) \right] + \\\mathbb{E}_{z \sim r(z)} \left[ \log(1 - D(G(z))) \right],
\end{equation}

where the $\mathbb{E}_{x \sim q(x)} \left[ \log D(x) \right]$ term denotes the expected value, $\mathbb{E}$, of $\log D(x)$, where $x$ is sampled from the observed data distribution $q(x)$. Likewise, $\mathbb{E}_{z \sim r(z)} \left[ \log(1 - D(G(z))) \right]$ is the expected value of $\log(1 - D(G(z)))$, where $z$ is sampled from the input noise distribution $r(z)$ of the generator $G(z)$. 

The CGANs \cite{mirza2014conditionalgenerativeadversarialnets} extend this architecture by incorporating a condition set $y$ into the generator and discriminator, guiding the data generation process. The architecture of CGAN (as seen in Fig. \ref{fig:methods}f) comprises a generator $G(z|y)$ with two inputs: a condition set $y$ and a random noise vector $z$ and a conditioned discriminator $D(x|y)$ on $y$. The generator aims to create new samples that closely resemble the original samples by using the noise and the condition set as input. After the generation of samples, these, along with the real samples, are fed into the discriminator. The objective for the discriminator is to effectively classify between the generated and the real samples, retrospectively training the generator and discriminator in the process. It is mathematically represented as:

\begin{equation}
    \label{eq:CGAN}
    \min_G \max_D V(D, G) = \mathbb{E}_{x \sim q(x)} \left[ \log D(x | y) \right] + \\ \mathbb{E}_{z \sim r(z)} \left[ \log(1 - D(G(z|y))) \right]. 
\end{equation}

In our implementation, the generator begins with a dense layer containing 256 neurons and the \textit{Leaky-ReLU} activation function \citep{xu2015empiricalevaluationrectifiedactivations}. It includes a positive slope $m$ and comparatively a tiny negative slope $\alpha m$, preventing the occurrences of dead neurons. In our case, $\alpha$ is taken to be $1e-2$. Following the dense layer, we utilized two Convolution-1D layers with 64 and 16 filter sizes, respectively. Paired with \textit{ReLU} activation, these Convolution-1D layers enhance the feature extraction process, ensuring distinct, high-quality features. The flux and time values are preprocessed by converting them to a symmetric logarithmic scale. The time values act as condition sets for the CGAN, the flux values provide the actual data, and the noise set contains random uniform noise samples from 0 to 1 covering the whole domain. All layers in the discriminator, except the final binary classification layer, mirror those of the generator. This design ensures a balanced model complexity, preventing either network from overpowering the other. The discriminator loss is computed as the sum of binary cross-entropy losses for correctly classifying real samples as real and generated samples as fake. The generator loss is defined as the binary cross-entropy between the discriminator’s predictions on generated samples and the target label for real samples, thereby encouraging the generator to produce more realistic outputs. This results in reliable data-augmentation capabilities for the reconstruction of time series data \citep{fu2019time, brophy2023generative}.

The proposed method uses TensorFlow and Keras libraries for its implementation, with 5000 epochs and a batch size of 256. The Adam optimizer was used to train the model, optimizing the loss function at a learning rate of $1e-4$.

\subsubsection{Kolmogorov-Arnold Networks}


The KAN (Kolmogorov-Arnold Network) model was first introduced by \cite{liu2024kan} and has since been studied extensively. One of its key advantages is its ability to recognize complex patterns and relationships in data. 

A distinguishing feature of KANs is their activation mechanism. Instead of applying activation functions at the nodes (as in standard neural networks), KANs place nonlinear functions along the connections between nodes. These functions are composed of adjustable splines and basis functions, allowing for finer control over how information flows through the network.\\

One of the most significant advantages of KANs in scientific contexts is their interpretability. Because each learnable function is univariate and localized, it can be visualized and analyzed directly, offering insights into each input variable's individual contribution, often lost in traditional black-box neural networks. This makes KANs well-suited for analyzing sparse or structured datasets, such as those encountered in GRB LC reconstructions.\\
Pros:
\begin{itemize}
    \item Greater interpretability through visual inspection of learned functions.
    \item Adaptive flexibility due to spline-based transformations.
    \item Potential for better generalization on structured or low-sample-size data. 
\end{itemize}
Cons:
\begin{itemize}
    \item Increased computational complexity in training, especially in high dimensions.
    \item Tooling and ecosystem are new compared to MLPs and CNNs, and lack support and reliability.
    \item High sensitivity to hyperparameter, particularly regarding the number and placement of spline knots.
\end{itemize}
However, many of these limitations have been mitigated, mainly in our case, due to the one-dimensional nature of the data. Furthermore, we employed Optuna-based hyperparameter tuning to ensure the selection of optimal parameters for model efficiency. The implementation was rigorously validated within the PyTorch framework to ensure correctness and stability.


Let the input layer consist of $n_0$ nodes, and the $l$-th layer have $n_l$ nodes. For any node $j$ in the $(l+1)$-th layer, the activation $x_{l+1,j}$ is computed as:
\begin{equation}
    x_{l+1,j} = \sum_{i=1}^{n_l} \phi_{l,j,i}(x_{l,i}),
\label{eq:act}
\end{equation}

where $\phi_{l,j,i}(\cdot)$ denotes the learnable univariate function connecting node $i$ of layer $l$ to node $j$ of layer $l+1$. 

This approach allows KANs to capture intricate data patterns with high precision, making them well-suited for reconstructing missing or incomplete flux data in GRB observations.



The overall computation for a KAN with $l$ layers is expressed as:
\begin{equation}
    KAN(x) = (\Phi_{l-1} \circ \Phi_{l-2} \circ \cdots \circ \Phi_0)(x),
    \label{eq:KAN-activation}
\end{equation}

where $\Phi_l$ represents the activation function matrix for the $l$-th layer. This structure supports deeper networks by stacking layers, analogous to how depth enhances expressiveness in traditional neural networks. In our case, we have used only a single layer of neural network in KAN.

To optimize the spline-based activation functions, each univariate function is represented as a combination of basis splines as:
\begin{equation}
    \phi(x) = w_b b(x) + w_s \text{spline}(x).
    \label{eq:KAN-compute}
\end{equation}

The basis function $b(x)$ is expressed as:
\begin{equation}
b(x) = \text{silu}(x) = \frac{x}{(1 + e^{-x})},
\end{equation}

\begin{equation}
\text{spline}(x) = \sum_i c_i B_i(x), 
\end{equation} 
with $c_i$ as the trainable coefficient that scales the contribution of the i-th basis function and $B_i$ is the i-th B-spline basis function. This parameterization ensures smooth and adaptable transformations while maintaining computational efficiency.

 We construct the KAN model with a width configuration of [1, 5, 1], indicating that the input and output each consist of a single feature vector (or column), while the hidden layer contains five neurons. The model also utilizes a grid size of 5 (shown in Fig. \ref{fig:methods}g).\\
The training process employs the Adam optimizer, with a learning rate of 0.001 and a batch size of 64, trained on 350 epochs. The model's performance is monitored every five epochs, and the best loss value is tracked throughout the training process.

\subsection{The statistical approach}

These models follow the same pre-processing and result calculation steps as mentioned in \ref{section:ML}.

\subsubsection{Fourier Transform}
Transformation in mathematics is a technique to transform or map a function from its original function space to a different function space. The Fourier transform is, therefore, a transformation technique that maps the functions dependent on the time domain into functions dependent on the temporal frequency domain. This breaks down a signal into a sum of sinusoidal functions. The Fourier transform of a function \( f(t) \) is defined by \( \hat{f}(\omega) \), which represents the frequency domain representation of the signal:

\begin{equation}
    \hat{f}(\omega) = \int_{-\infty}^{\infty} f(t) \, e^{-i \omega t} \, dt.
\end{equation}
The Fourier transform helps to understand the data's periodicity and reduce noise.

This concept was implemented to reconstruct GRB LCs by transforming time-domain flux values into frequency-domain coefficients. The initial data was first processed using the Fast Fourier Transform (FFT) \citep{OberstUlrich} on the flux values. The interpolation technique imputes the new flux values, and the uncertainties were predicted using the normal distribution. This process helps us understand how flux values may vary over time, taking uncertainties into account. Fig. \ref{fig:methods}h is the FFT synthesis flow diagram that shows how the FFT works. It shows how the FFT combines two 4-point spectra into an 8-point spectrum. The "$\times S$" symbol represents the signal multiplied by a sinusoid of appropriate frequency. 


Implementing the fourier transform helps deal with noisy and sparse data. This method also aids in understanding the periodic patterns and gives a cleaner representation of the underlying signal by removing unwanted noise. Furthermore, Gaussian smoothing and interpolation techniques enhances the model's performance, increasing its reliability.

\subsection{SARIMAX-based Kalman Smoothing Model}
We researched the Seasonal Autoregressive Integrated Moving Average with Exogenous Regressors (SARIMAX) \citep{inventions7040094} and Kalman smoothing \citep{articleks} and how integrating both models helps reconstruct the GRB LCs. This combination is quite effective since it combines the statistical capability of SARIMAX for time-series modeling with the refinement capabilities of Kalman smoothing, which considers the entire data for the smoothing process. The Kalman smoothing in our implementation is applied through the state-space representation provided by the Statsmodels package.
The time and flux data are preprocessed by transforming them into the logarithmic domain to stabilize variance and enhance the relationship between time and flux. This ensures computational stability. 

The SARIMAX model captures the observed flux data's temporal dependencies, trends, and noise. An ARIMA model \citep{HO1998213} of order (p,d,q) is tuned accordingly to model the LC’s temporal dynamics. Here, p is the number of autoregressive terms, d is the number of non-seasonal differences, and q is the number of lagged forecast errors in the prediction equation. An order of (1,1,1) was used that often captures basic temporal structure with fewer parameters, avoiding overfitting while reducing autocorrelation in residuals.
In our model, we applied Kalman smoothing to refine the SARIMAX output. Kalman smoothing works in such a way that it estimates the system's state using measurements from both past and future time steps.
Let $\mathbf{x}_t$ denote the state vector at time $t$, modeled as:

\begin{equation}
\mathbf{x}_t = \mathbf{A} \mathbf{x}_{t-1} + \mathbf{B} \mathbf{u}_t + \mathbf{w}_t,
\end{equation}

where $\mathbf{A}$ is the state transition matrix, $\mathbf{B}$ is the control input matrix, $\mathbf{u}_t$ is the control input, and $\mathbf{w}_t$ represents process noise. In this work, we utilize the SARIMAX framework without incorporating exogenous regressors; hence, the control input term $\mathbf{Bu}_t$ is omitted from our practical implementation.

The observation $\mathbf{y}_t$ is modeled as:

\begin{equation}
\mathbf{y}_t = \mathbf{C} \mathbf{x}_t + \mathbf{v}_t,
\end{equation}

where $\mathbf{C}$ is the observation matrix and $\mathbf{v}_t$ represents measurement noise. However, in our implementation, we have added noise, as mentioned at the start of \S\ref{section:ML}.


The Kalman smoothing algorithm minimizes the mean squared error of state estimates, yielding a refined sequence of flux values $\hat{\mathbf{F}}_t$.
After Kalman smoothing, uncertainties were added (as seen in the model flow Fig. \ref{fig:methods}i).

The model was implemented using Python libraries such as \textit{Statsmodels}, specifically the \textit{SARIMAX} class and \textit{Matplotlib} for visualizing the final results, including the time series plots and confidence bands. The SARIMAX fitting is optimized for computational efficiency, and specifically, the Kalman smoother estimated the latent states (\textit{smoothed\_state[0]}) to refine noisy measurements by incorporating information from both past and future time steps. These smoothed flux values were then interpolated across a dense logarithmic time grid to reconstruct the light curve. The \textit{Scipy} library’s \textit{gaussian\_filter1d} function was applied to the predicted confidence interval bounds.

\section{Attributes of the models}

The attributes of each model, as shown in Table \ref{tab:attributes}, provide insight into their performance. Using these models, the following inferences can be made:

\begin{itemize}
\item \textbf{Bi-Mamba:} Best for sequential GRB data with long-term dependencies. It adapts to different data distributions but requires careful management of computational cost and noise overfitting.  
\item \textbf{MLP:} Flexible across data distributions but prone to noise overfitting without regularization. Less suitable for sparse, irregular, or sequential GRB data.  
\item \textbf{Fourier Transform:} Well-suited for periodic or smooth components in GRB signals, offering efficiency in high-throughput tasks but less effective for non-periodic or non-linear data.  
\item \textbf{GP-RF:} Effective for sparse or noisy GRB data, combining probabilistic modeling and reliable feature inference. High computational cost for large datasets, but handles complex data distributions well.  
\item \textbf{Bi-LSTM:} Highly effective for sequential data due to their ability to retain long-term dependencies and capture non-linear relationships within the data. However, they are less effective when the data is sparse.
\item \textbf{CGAN:}  Excels at capturing complex, non-linear relationships and learning high-dimensional data distributions. However, it can be computationally expensive and require careful training.
\item \textbf{SARIMAX-based Kalman model:} Efficient in handling sparse data and are computationally efficient, making them suitable for data sets with limited information. They effectively capture non-linear and periodic relationships, making them ideal for modeling time-series data.
\item \textbf{KAN:} Excellent for capturing complex non-linear patterns, though the model requires sufficient training data and careful tuning to avoid noise overfitting.  
\item \textbf{Attention U-Net:} Effective for GRB data with complex patterns due to its ability to focus on critical features through an attention mechanism. However, significant computational resources and careful tuning of attention parameters are required to avoid overfitting.
\end{itemize}

These proposed models are detailed in Table \ref{tab:attributes} below.

\begin{table}[htbp]
\centering
\hspace*{-3.5cm}
\resizebox{1.2\textwidth}{!}{
\begin{tabular}{|l|c|c|c|c|c|c|c|c|c|c|}
    \hline
    \textbf{Attributes}    &
    \textbf{GP} & 
    \textbf{Bi-Mamba} & 
    \textbf{MLP} & 
    \textbf{Fourier} & 
    \textbf{GP-RF} & 
    \textbf{Bi-LSTM} &
    \textbf{CGAN} & 
    \textbf{SARIMAX} &
    \textbf{KAN} & 
    \textbf{U-Net} \\ \hline
    High Throughput                      & \checkmark &            & \checkmark & \checkmark &            &            &           & &            & \\ \hline
    Good Sparse Data Performance       & \checkmark &            &            &            & \checkmark &            &           &\checkmark & \checkmark & \\ \hline
    Long-term Memory                     & \checkmark & \checkmark &            &            & \checkmark & \checkmark &           & &            & \\ \hline
    Low Computational Cost               & \checkmark &            & \checkmark & \checkmark &            &            &           &\checkmark &            & \\ \hline
    Resilience to Noise Overfitting      & \checkmark &            &            & \checkmark & \checkmark & \checkmark &\checkmark & &            & \checkmark \\ \hline
    Captures Complex Patterns &            & \checkmark & \checkmark &            &            & \checkmark &\checkmark &\checkmark & \checkmark & \checkmark \\ \hline
    Captures Periodic Components         &            &            &            & \checkmark &            &            &           &\checkmark &            & \\ \hline
    Models Data Distribution             & \checkmark & \checkmark & \checkmark &            & \checkmark & \checkmark &\checkmark & & \checkmark & \checkmark  \\ \hline
    \end{tabular}
}
    \caption{Comparison of models on various attributes.}
    \label{tab:attributes}
\end{table}

\section{Results}\label{section:results}

Decreasing the uncertainty associated with the W07 parameters is an objective of LCR. To evaluate this, the error fractions, symbolized by $EF$, are calculated for every model parameter in both the primary datasets and after reconstructing them using the model. The error fractions for the three W07 parameters as described in \citealt{dainotti2023stochastic} are represented in Eq. \ref{eqn5}, \ref{eqn6}, and \ref{eqn7}:

 \begin{equation}
  EF_{\log_{10}(T_{a})}=\left|\frac{\Delta{\log_{10}(T_{a})}}{\log_{10}(T_{a})}\right|,
  \label{eqn5}
 \end{equation}

 \begin{equation}
  EF_{\log_{10}(F_{a})}=\left|\frac{\Delta{\log_{10}(F_{a})}}{\log_{10}(F_{a})}\right|, 
  \label{eqn6}
 \end{equation}

 \begin{equation}
 EF_{\alpha_{a}}=\left|\frac{\Delta{\alpha_{a}}}{\alpha_{a}}\right|.    
 \label{eqn7}
 \end{equation}
 
We compute the reduction in the percentage of EF to evaluate the enhancement in fit post-reconstruction.

 \begin{equation}
 \%_{DEC}= \frac{\left|EF^{\rm{after}}_{X}\right|-\left|EF^{\rm{before}}_{X}\right|}{\left|EF^{\rm{before}}_{X}\right|}\times 100   
 \label{eqn8}.
 \end{equation}

\subsection{Results from the Willingale Model}\label{sec:funcresults}

\begin{figure*}[htbp]
\begin{center}
\includegraphics[width=.24\textwidth, height=.21\textwidth]{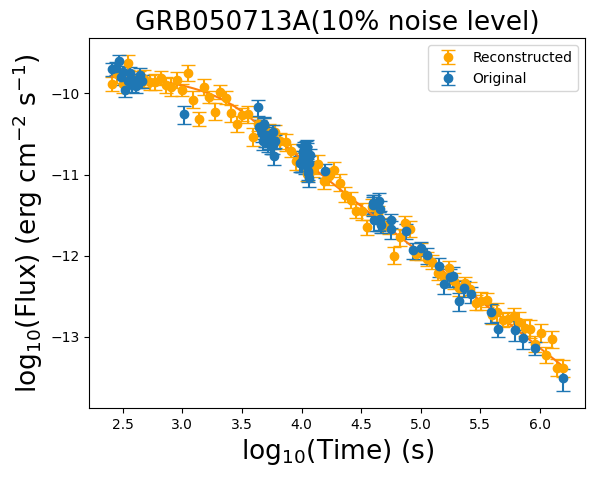}
\includegraphics[width=.24\textwidth, height=.21\textwidth]{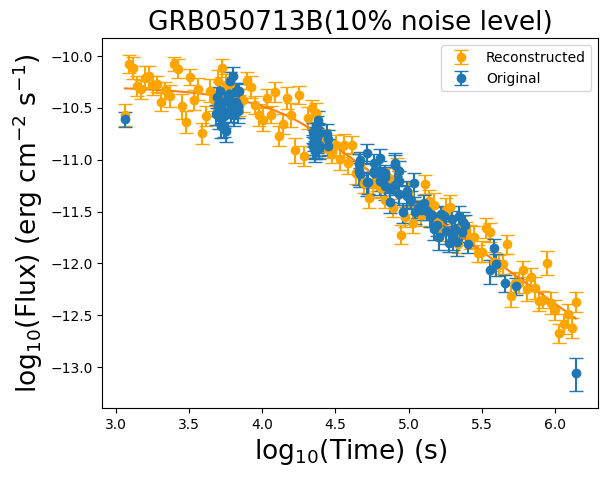}
\includegraphics[width=.24\textwidth, height=.21\textwidth]{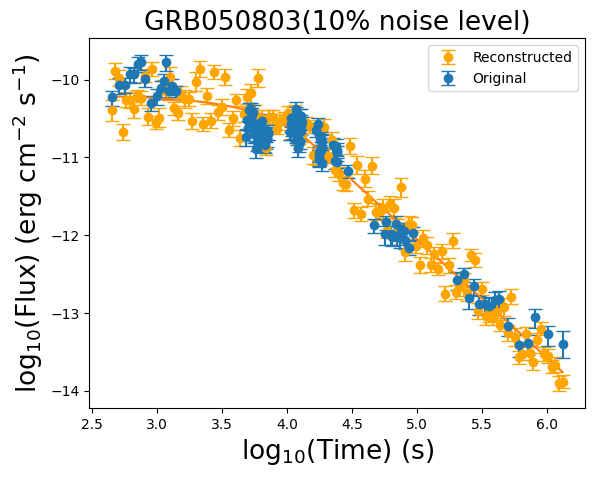}
\includegraphics[width=.24\textwidth, height=.21\textwidth]{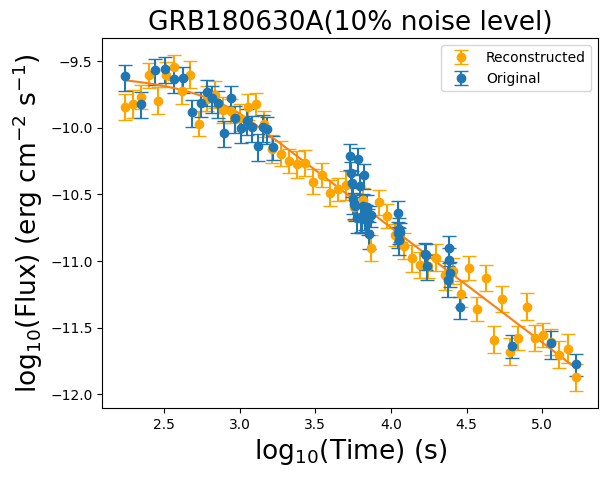}

\includegraphics[width=.24\textwidth, height=.21\textwidth]
{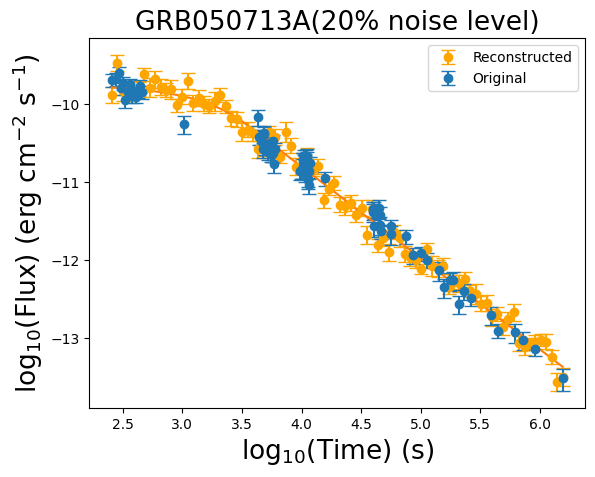}
\includegraphics[width=.24\textwidth, height=.21\textwidth]{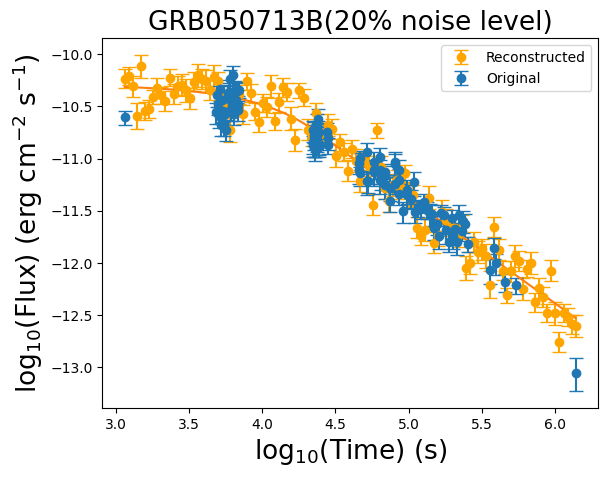}
\includegraphics[width=.24\textwidth, height=.21\textwidth]{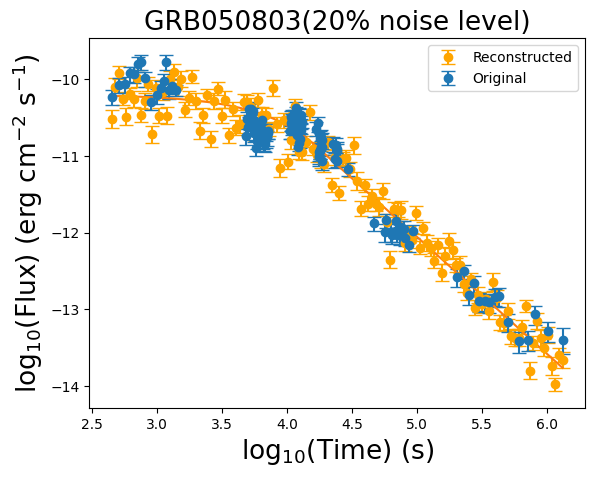}
\includegraphics[width=.24\textwidth, height=.21\textwidth]{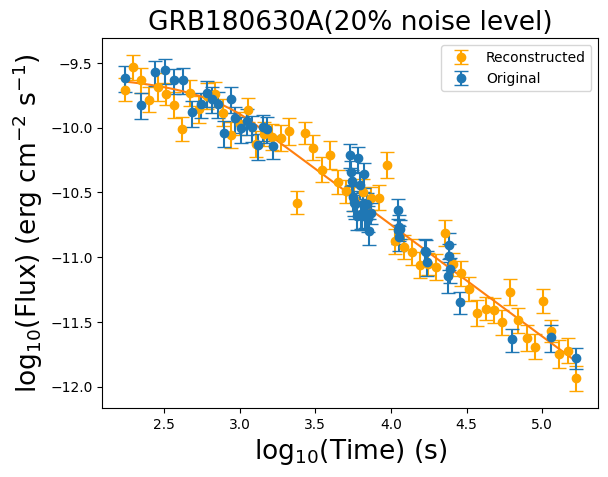}

\end{center}
\caption{Reconstruction of LCs using the W07 function. Row 1 shows the reconstructed LCs at 10\% noise level and Row 2 shows the reconstructed LCs at 20\% noise level for all four classes: i) Good GRBs (Column 1); ii) a break towards the end of the GRB LC (Column 2); iii) flares or bumps in the afterglow (Column 3); iv) flares or bumps with a double break towards the end of the LC (Column 4). }
\label{fig: w_rec} 
\end{figure*}

The reconstruction results for all four classes of GRBs under 10\% and 20\% noise levels are illustrated in Fig. \ref{fig: w_rec}.
For all classes of GRBs, the LCs are reconstructed to demonstrate how increasing noise levels affect the distribution of data points near the W07 fit. As anticipated, increasing the noise level generally leads to higher uncertainty due to the broader uncertainty around
the best-fit line. However, in our case, the decrease in uncertainty does not follow this pattern with increasing noise levels. This is because we are considering the entire sample. The data points deviate significantly from the model function for GRBs exhibiting flares and breaks. As a result, with more significant noise levels, the decrease in uncertainty is less pronounced.
Additionally, the EFs for the W07 model applied to 521 good GRBs at both 10\% and 20\% noise levels are summarized in Table \ref{tab:ALL_Table}. The results include the histogram distribution of the percentage decrease for all three W07 parameters at these noise levels.

A reduction of 24.5\% for $\log T_a$, 25.7\% for $\log F_a$, and 36.2\% for $\alpha$ is observed at a 10\% noise level. However, for the subset containing only 207 good GRBs, it performs the best showcasing 33.3\% for $\log T_a$, 35.0\% for $\log F_a$\% and 43.3\% for $\alpha$ with no additional outliers. As expected, the reductions were marginally smaller, with 21.2\% for $\log T_a$, 22.9\% for $\log F_a$, and 34.7\% for $\alpha$, for a 20\% noise level.

\subsection{Results from the ML and statistical methods}\label{sec:GPresults}

\begin{figure*}[htbp]
\begin{center}

    \includegraphics[width=.24\textwidth, height=.21\textwidth]{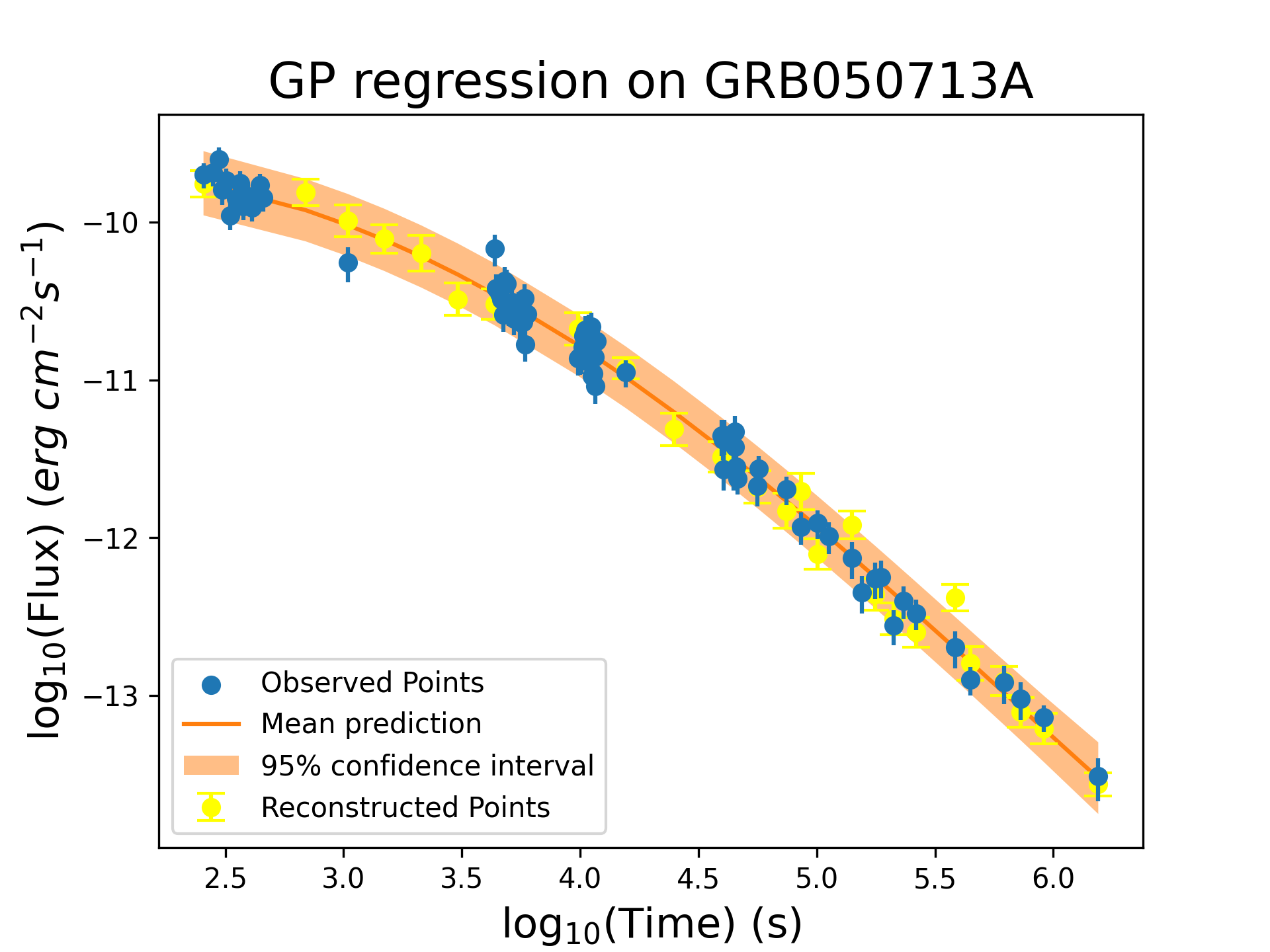}
    \includegraphics[width=.24\textwidth, height=.21\textwidth]{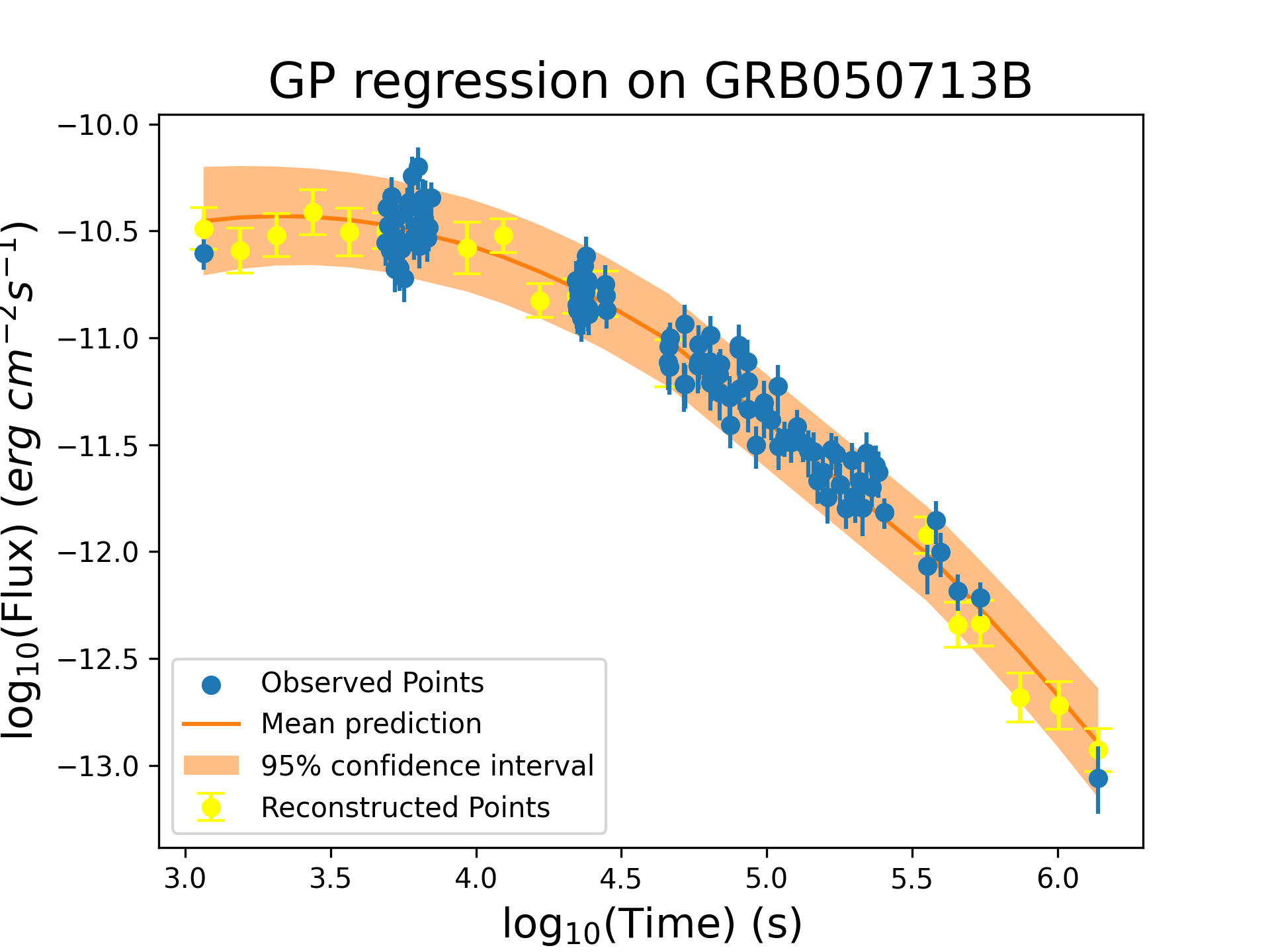}
    \includegraphics[width=.24\textwidth, height=.21\textwidth]{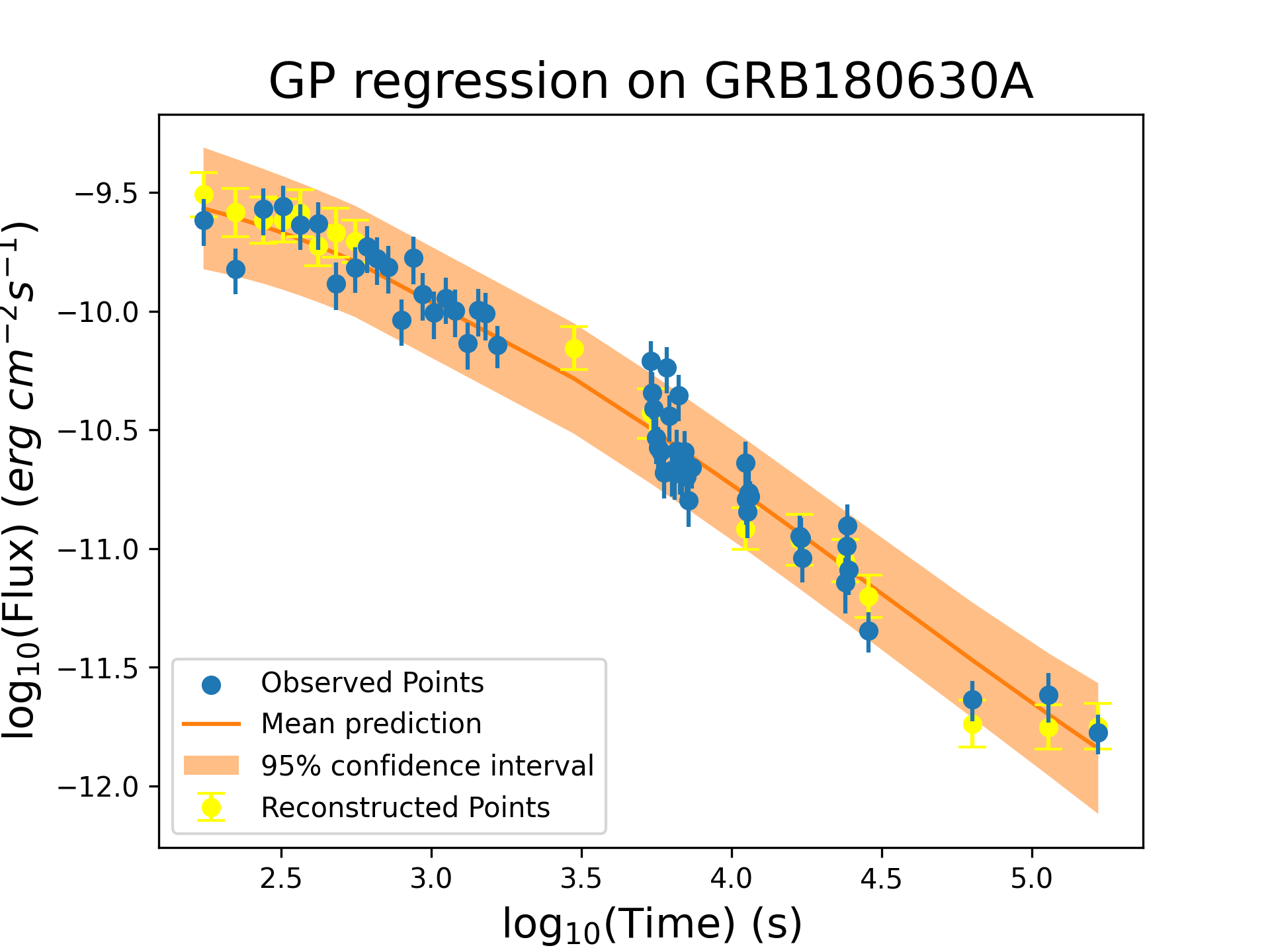}
    \includegraphics[width=.24\textwidth, height=.21\textwidth]{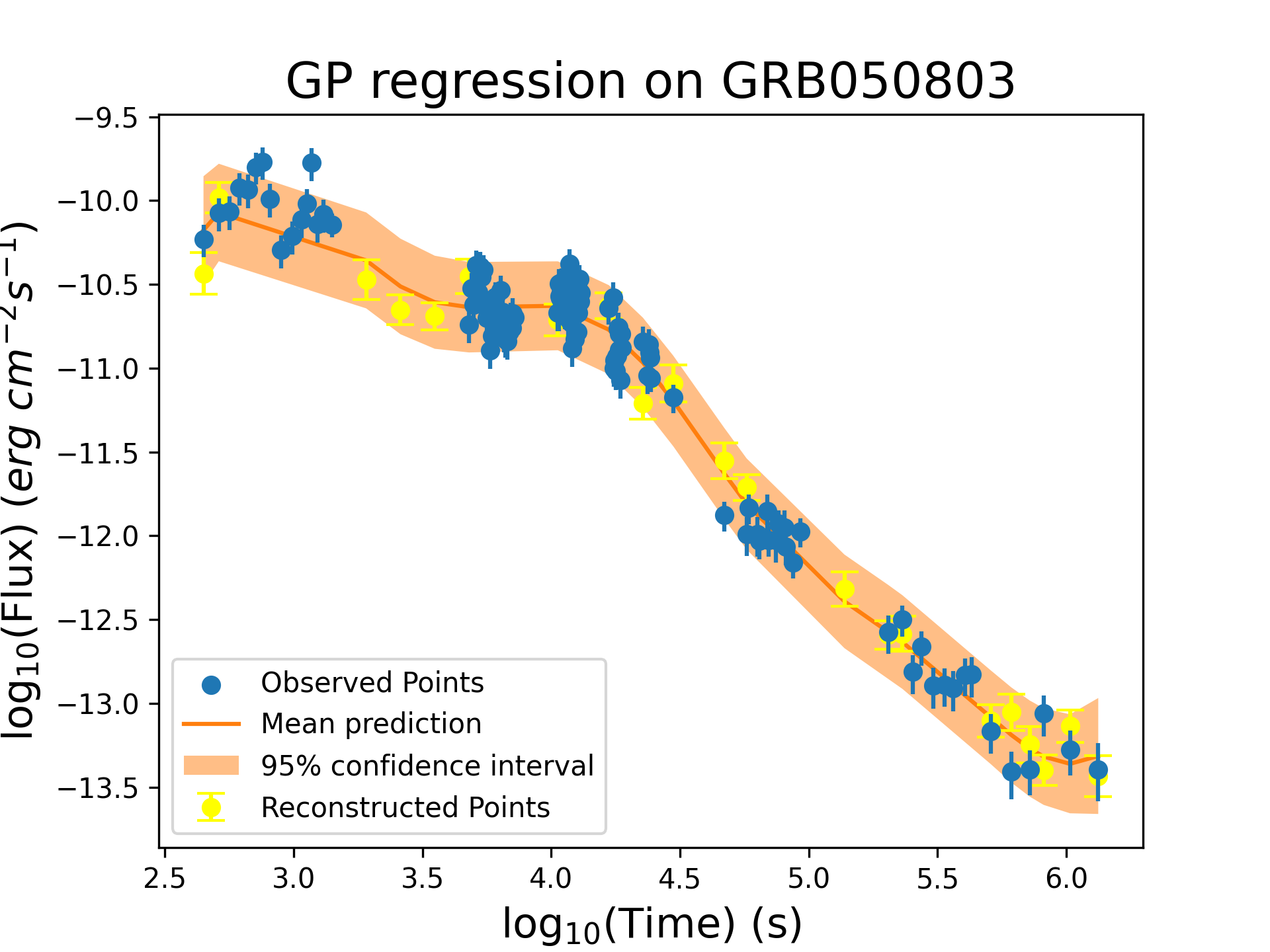}

    \includegraphics[width=.24\textwidth, height=.21\textwidth]{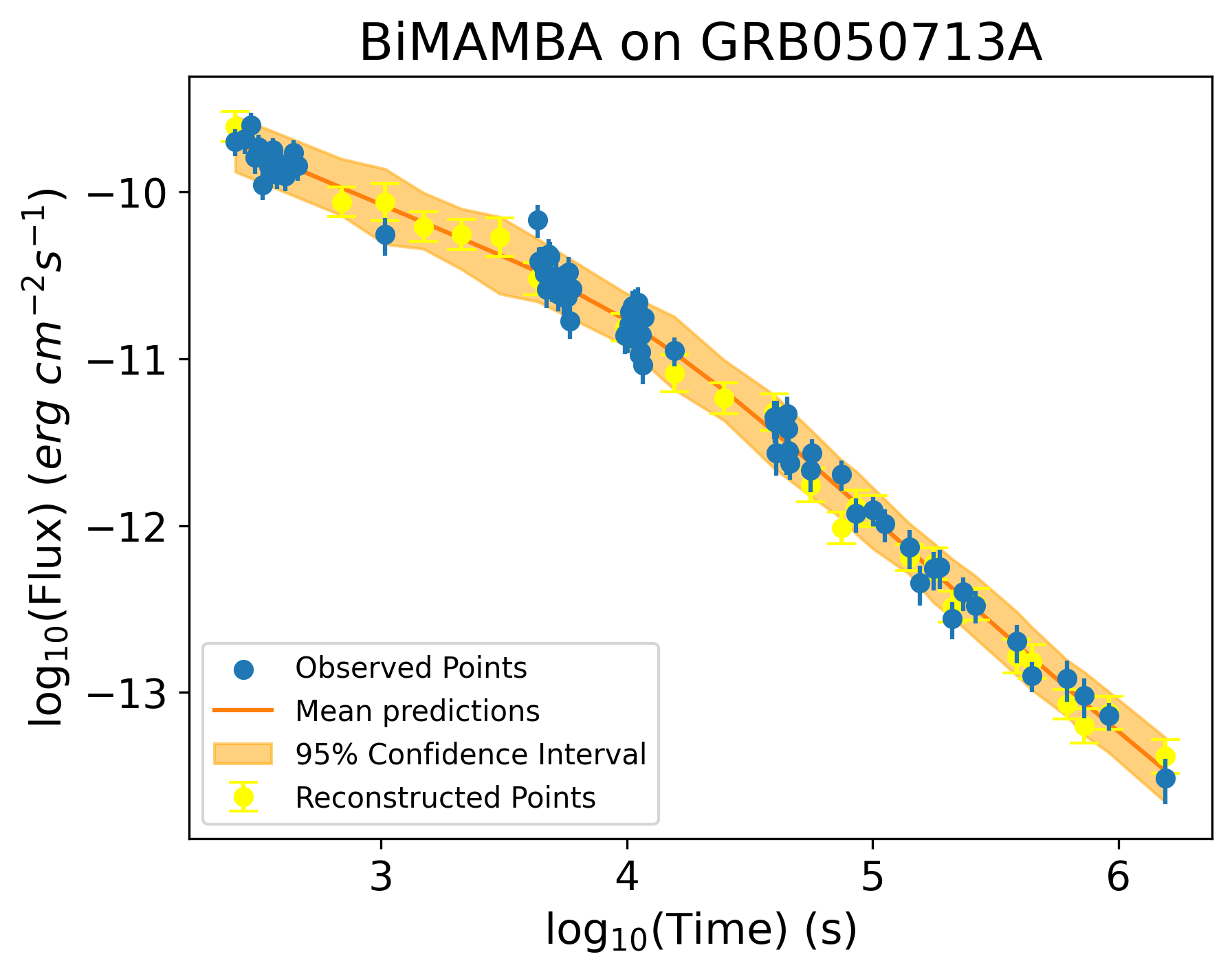}
    \includegraphics[width=.24\textwidth, height=.21\textwidth]{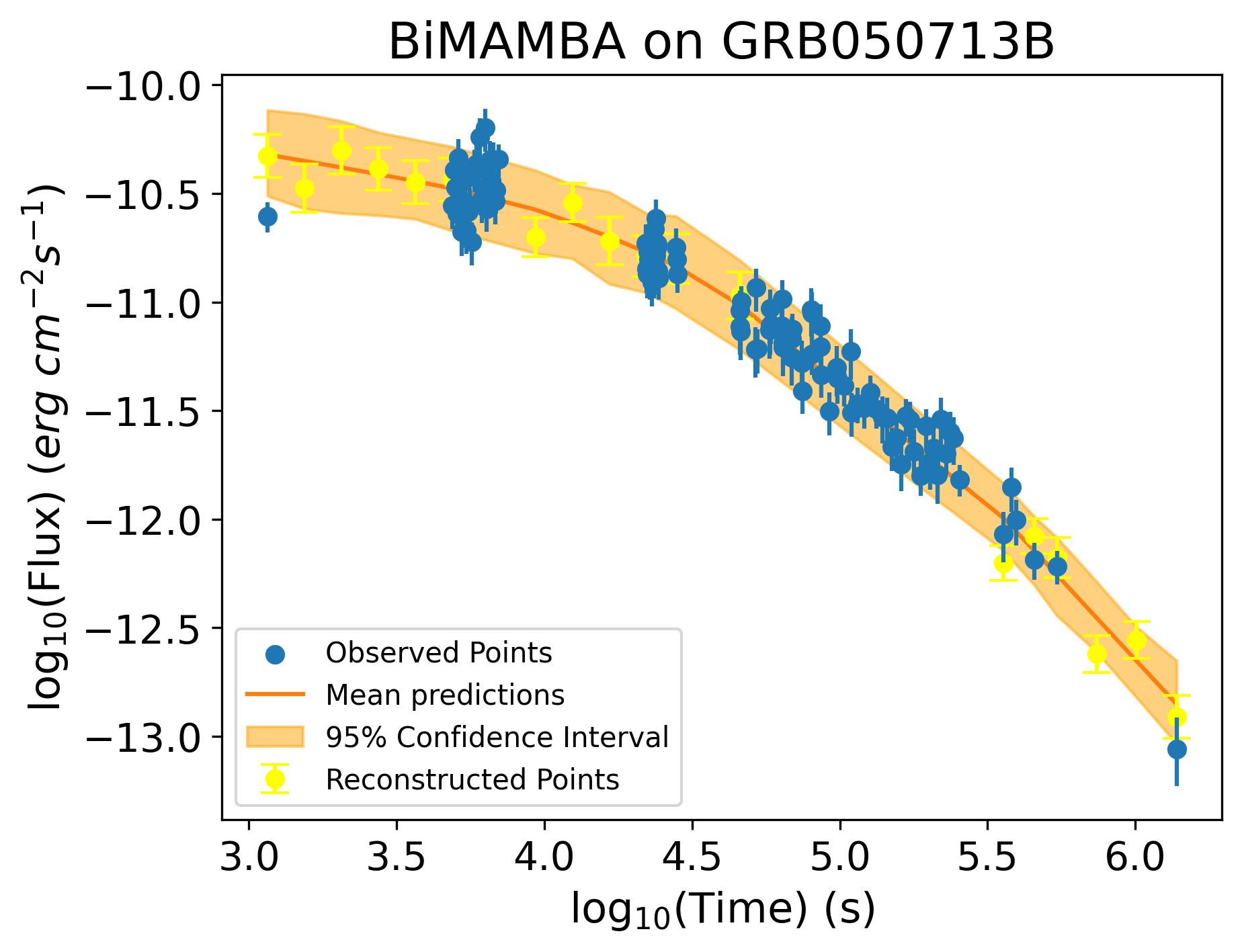}
    \includegraphics[width=.24\textwidth, height=.21\textwidth]{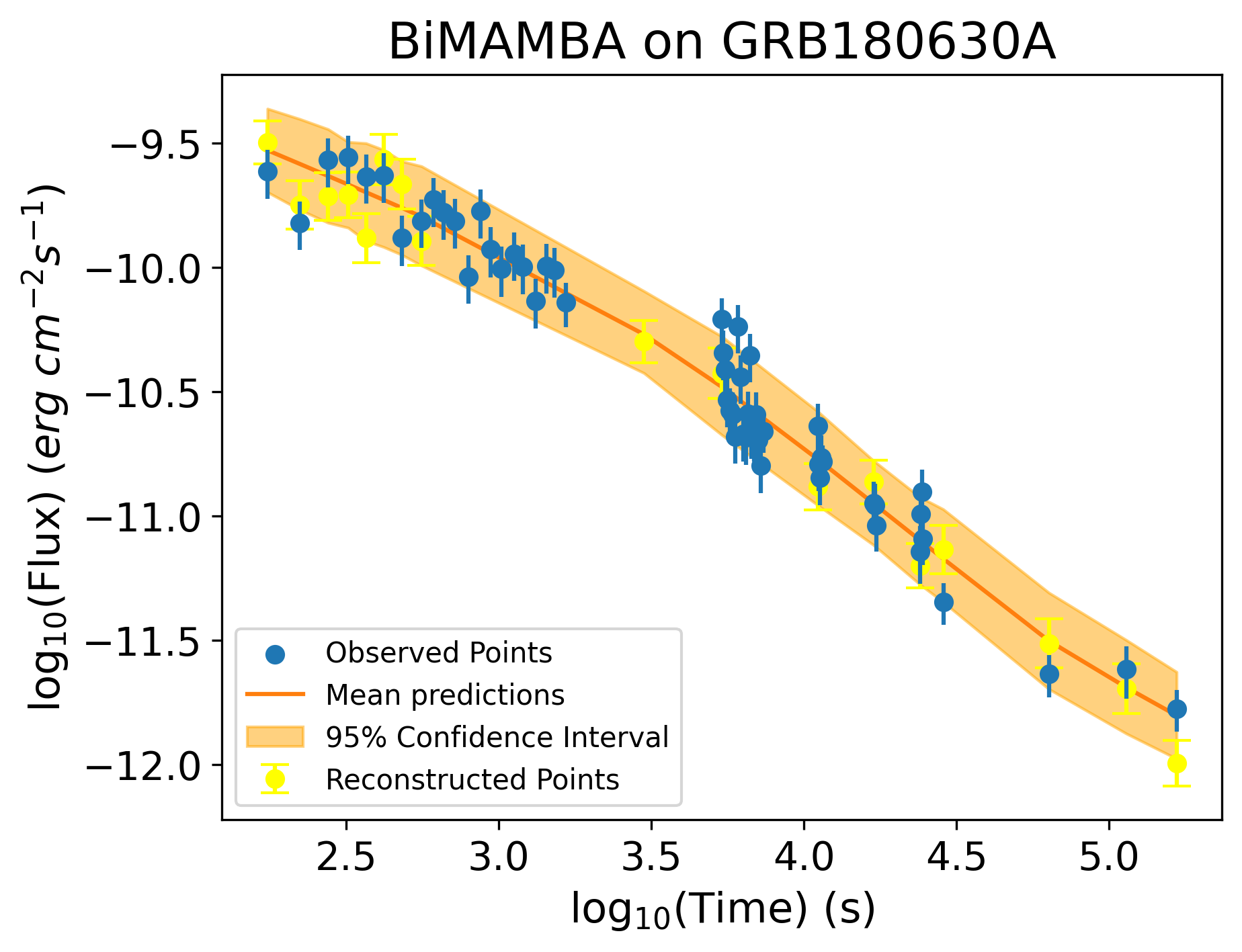}
    \includegraphics[width=.24\textwidth, height=.21\textwidth]{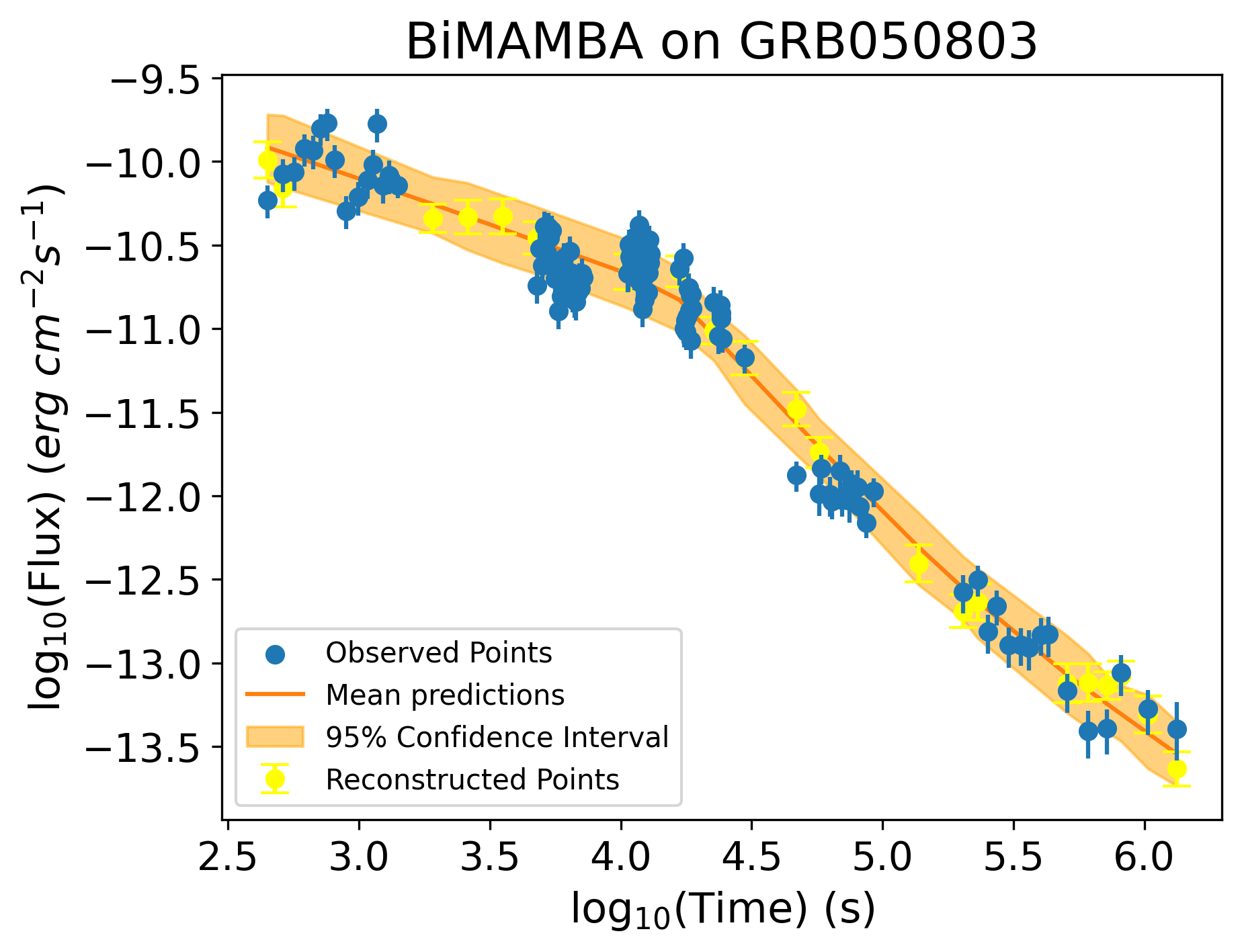}

    \includegraphics[width=.24\textwidth, height=.21\textwidth]{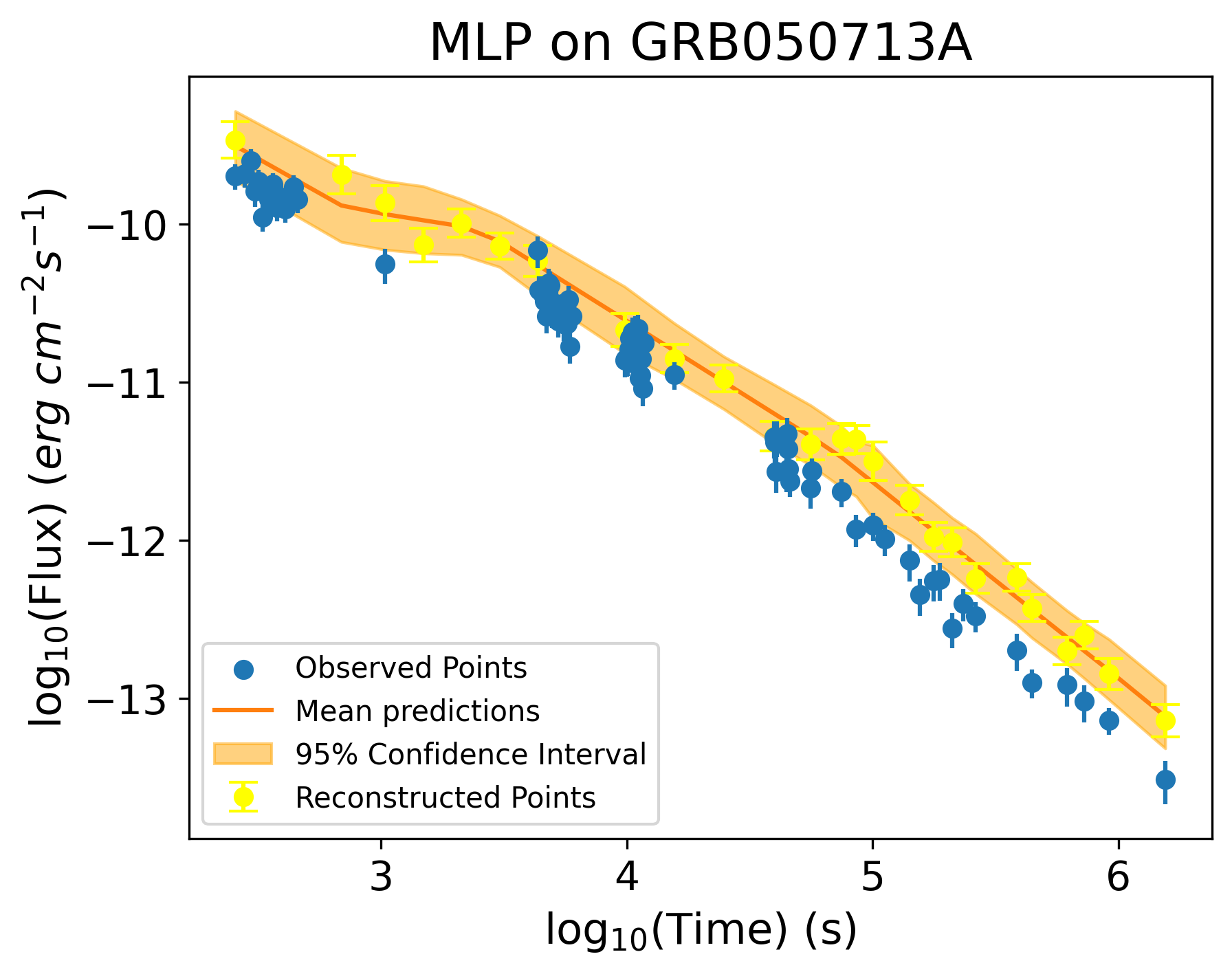}
    \includegraphics[width=.24\textwidth, height=.21\textwidth]{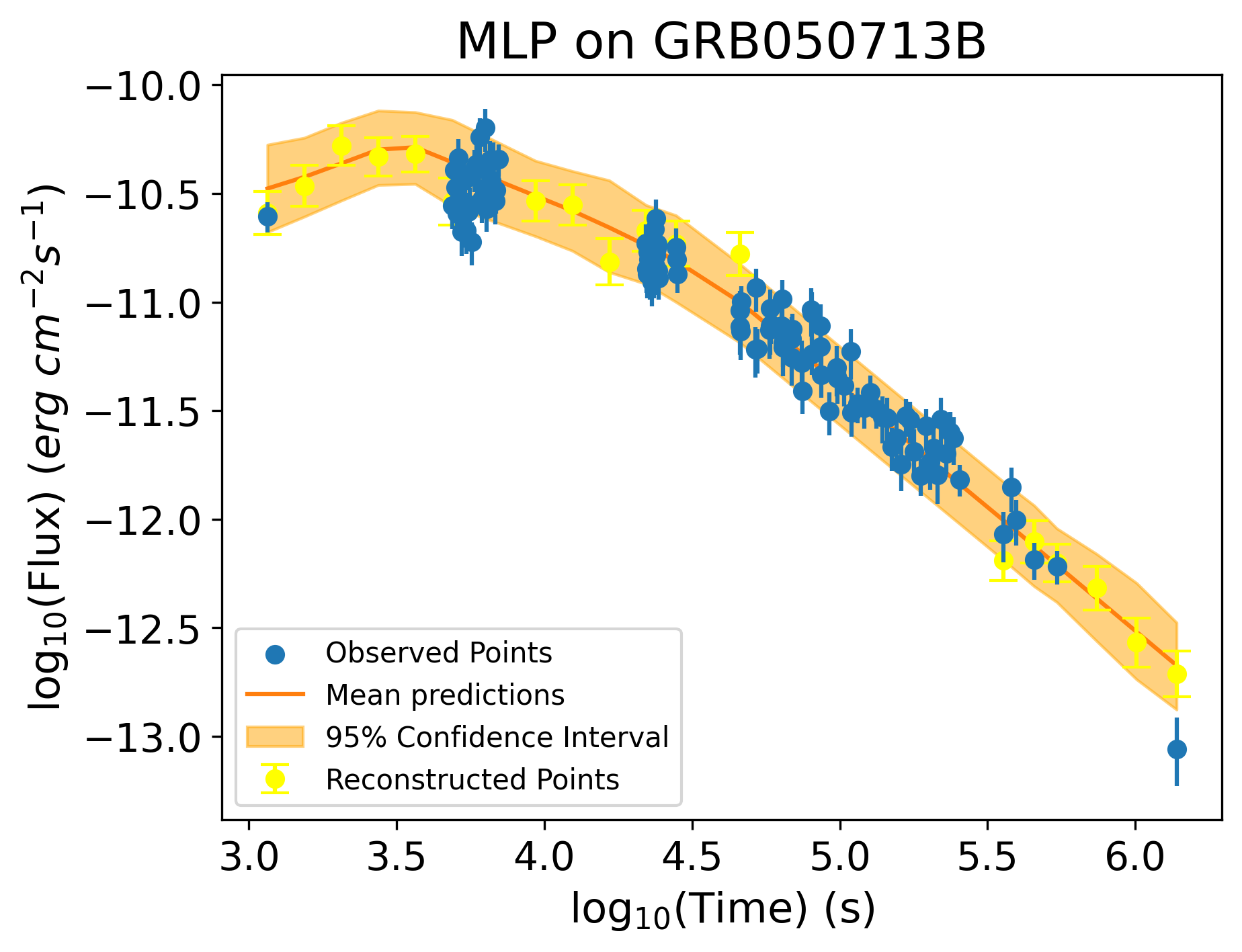}
    \includegraphics[width=.24\textwidth, height=.21\textwidth]{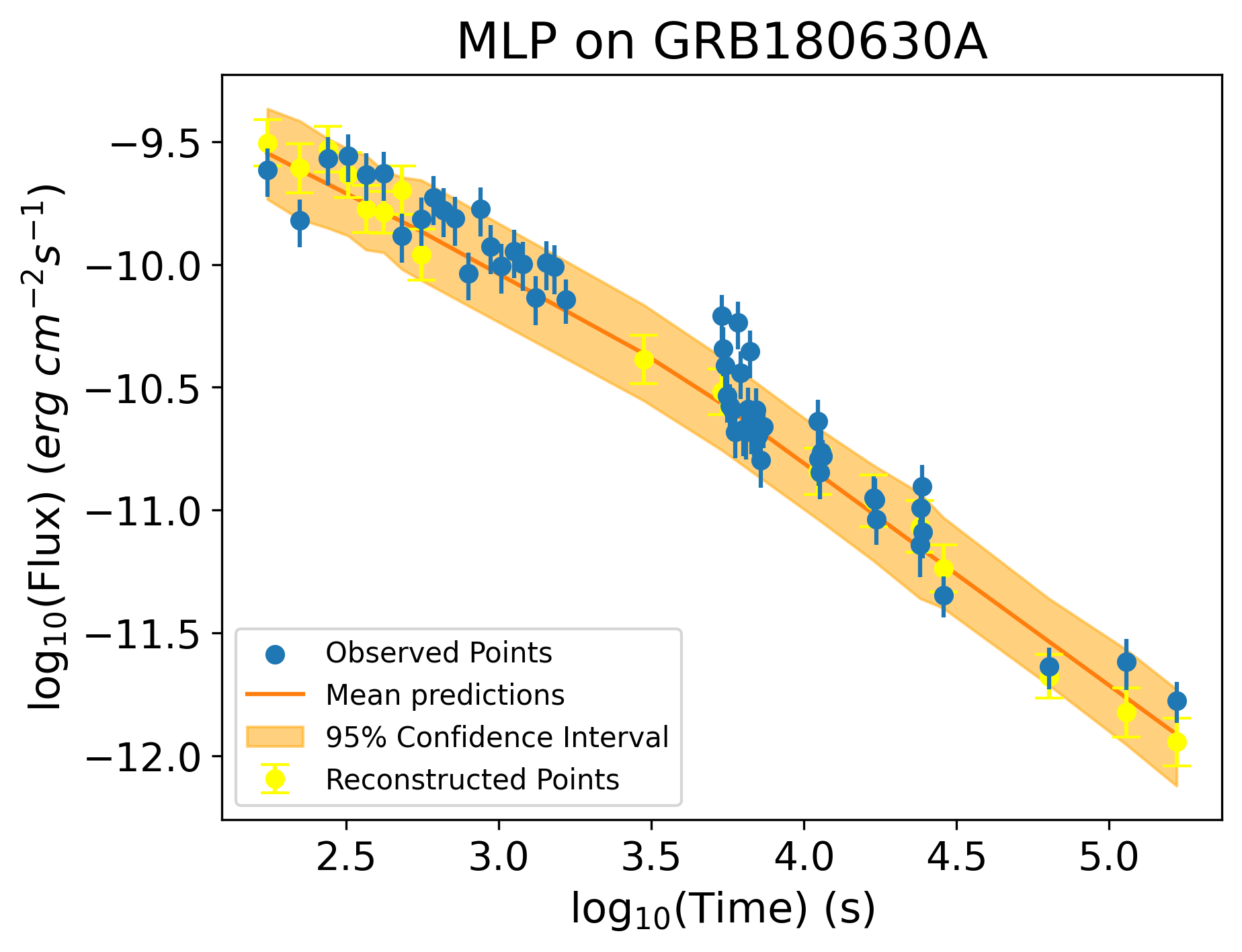}
    \includegraphics[width=.24\textwidth, height=.21\textwidth]{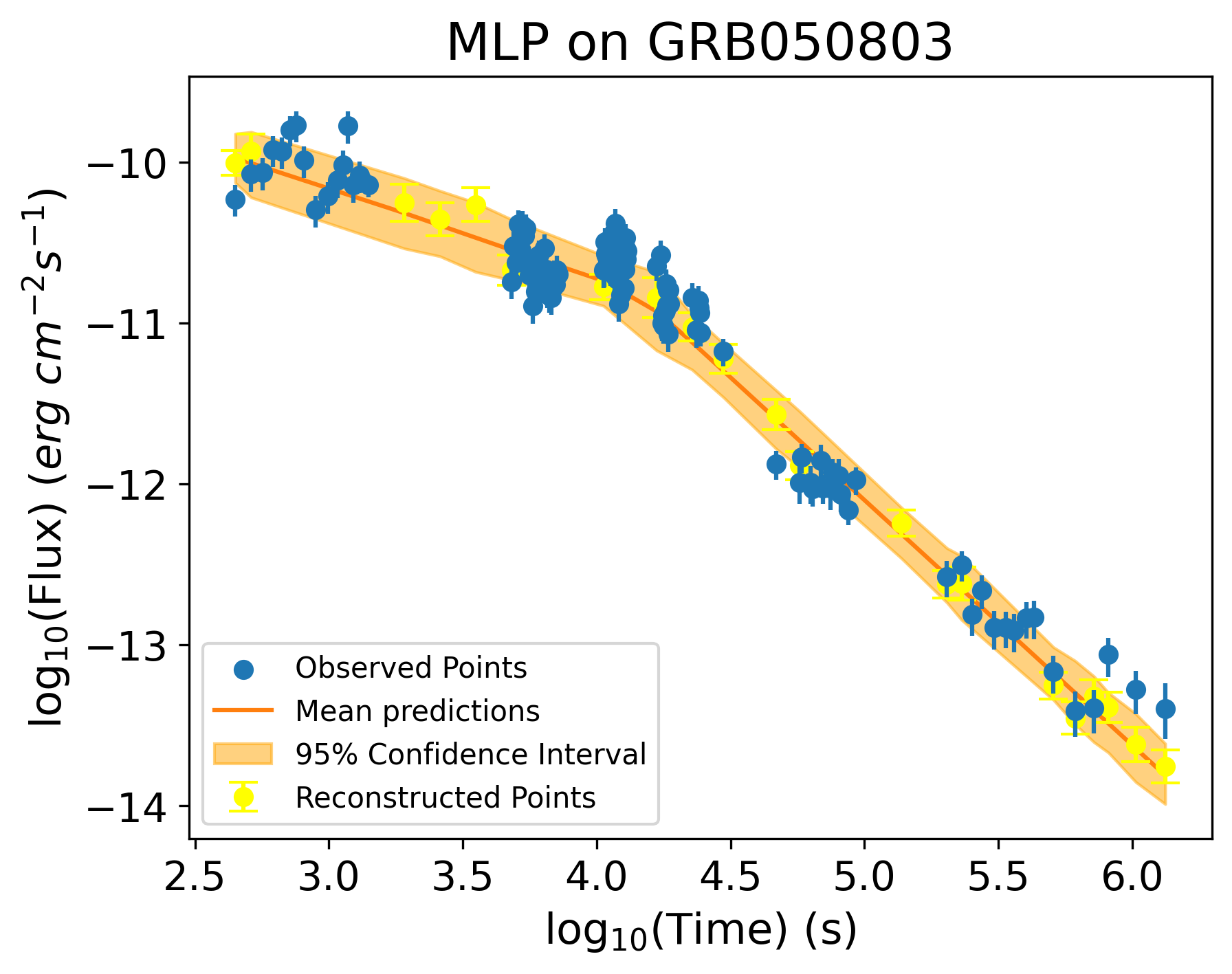}

    \includegraphics[width=.24\textwidth, height=.21\textwidth]{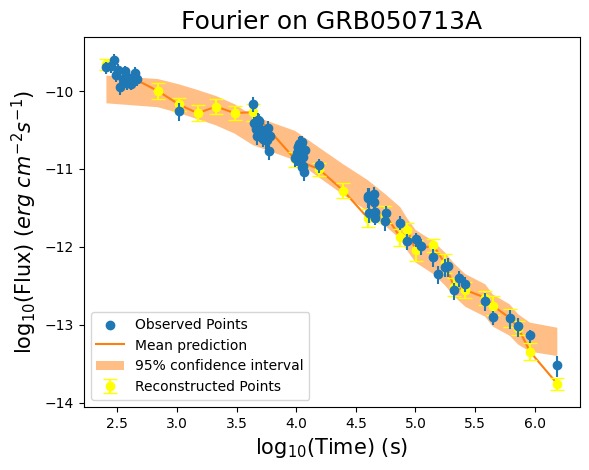}
    \includegraphics[width=.24\textwidth, height=.21\textwidth]{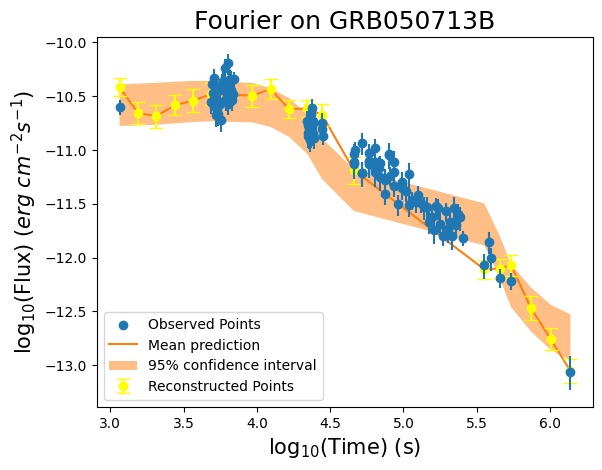}
    \includegraphics[width=.24\textwidth, height=.21\textwidth]{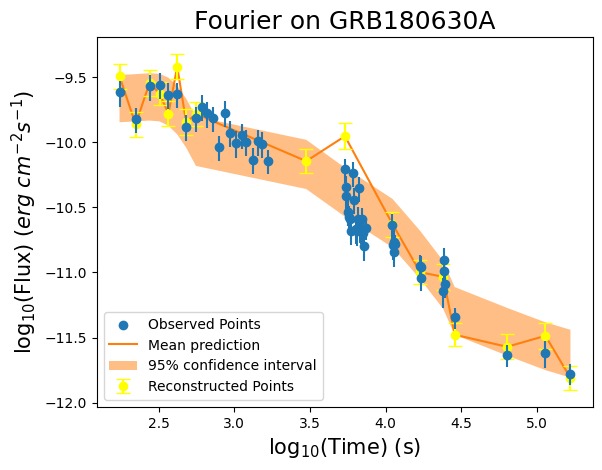}
    \includegraphics[width=.24\textwidth, height=.21\textwidth]{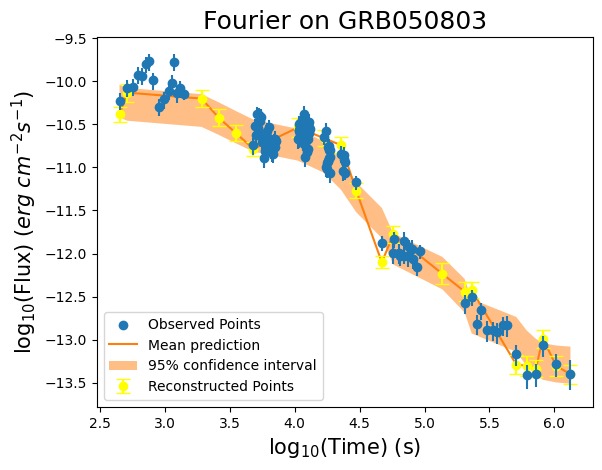}

    \includegraphics[width=.24\textwidth, height=.21\textwidth]{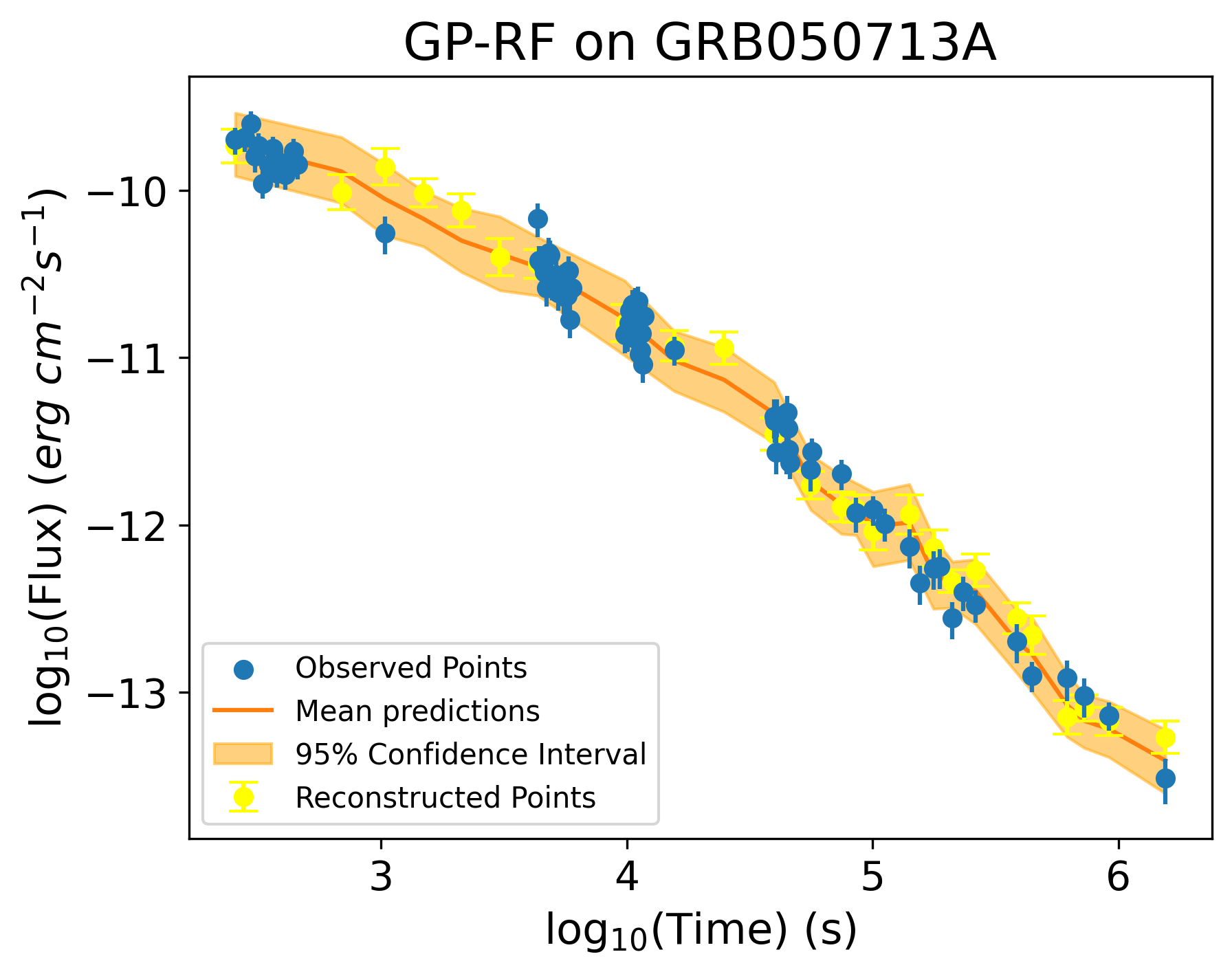}
    \includegraphics[width=.24\textwidth, height=.21\textwidth]{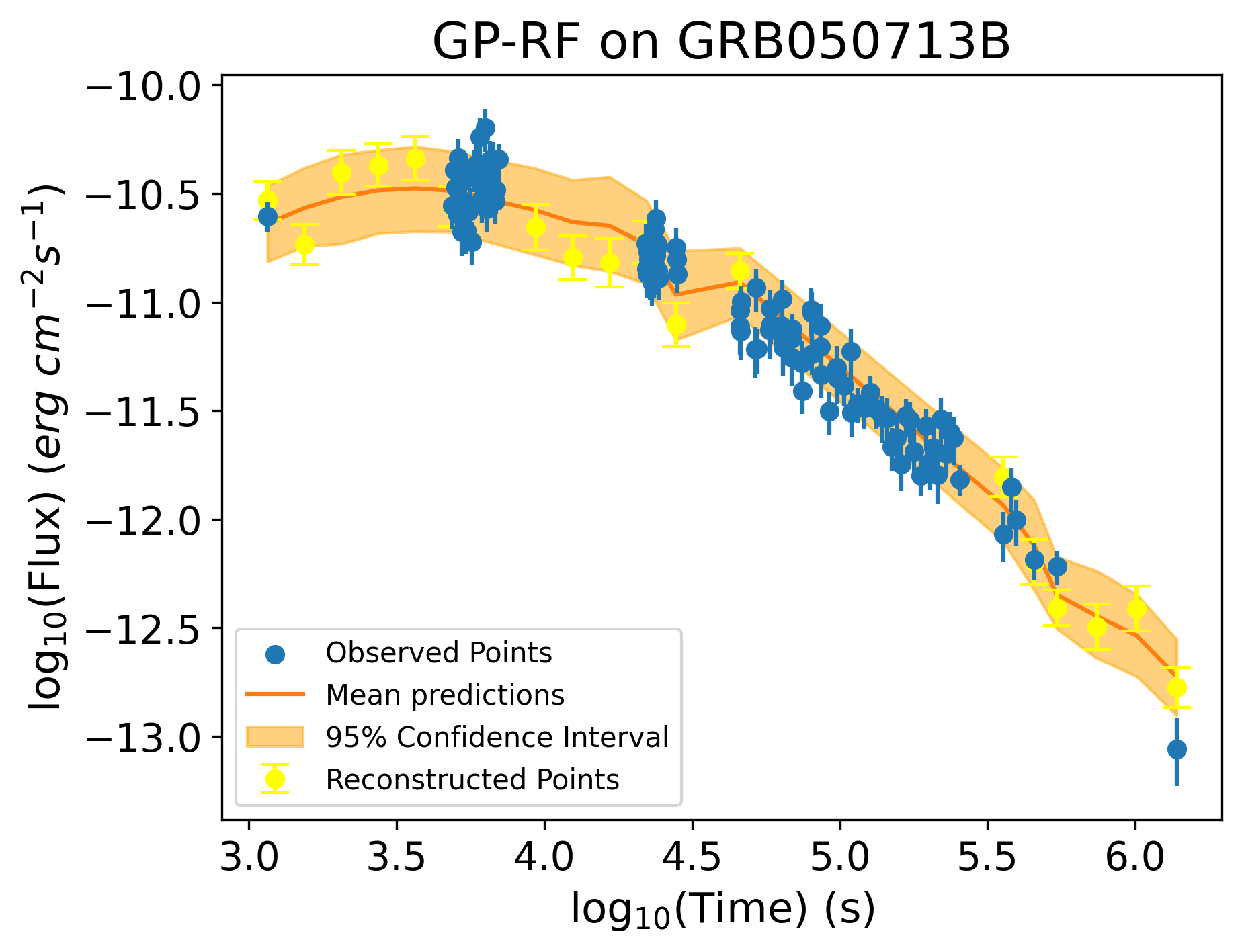}
    \includegraphics[width=.24\textwidth, height=.21\textwidth]{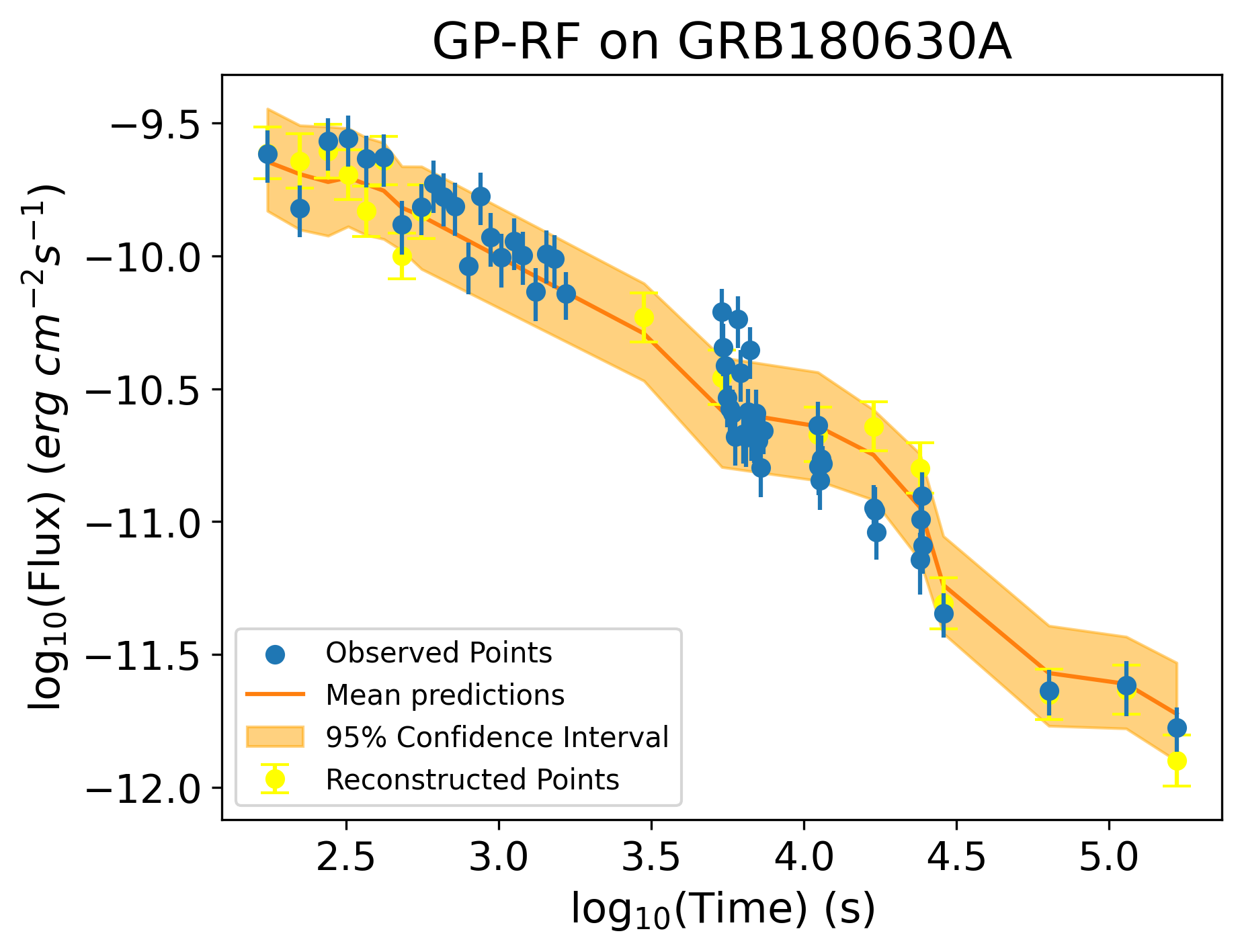}
    \includegraphics[width=.24\textwidth, height=.21\textwidth]{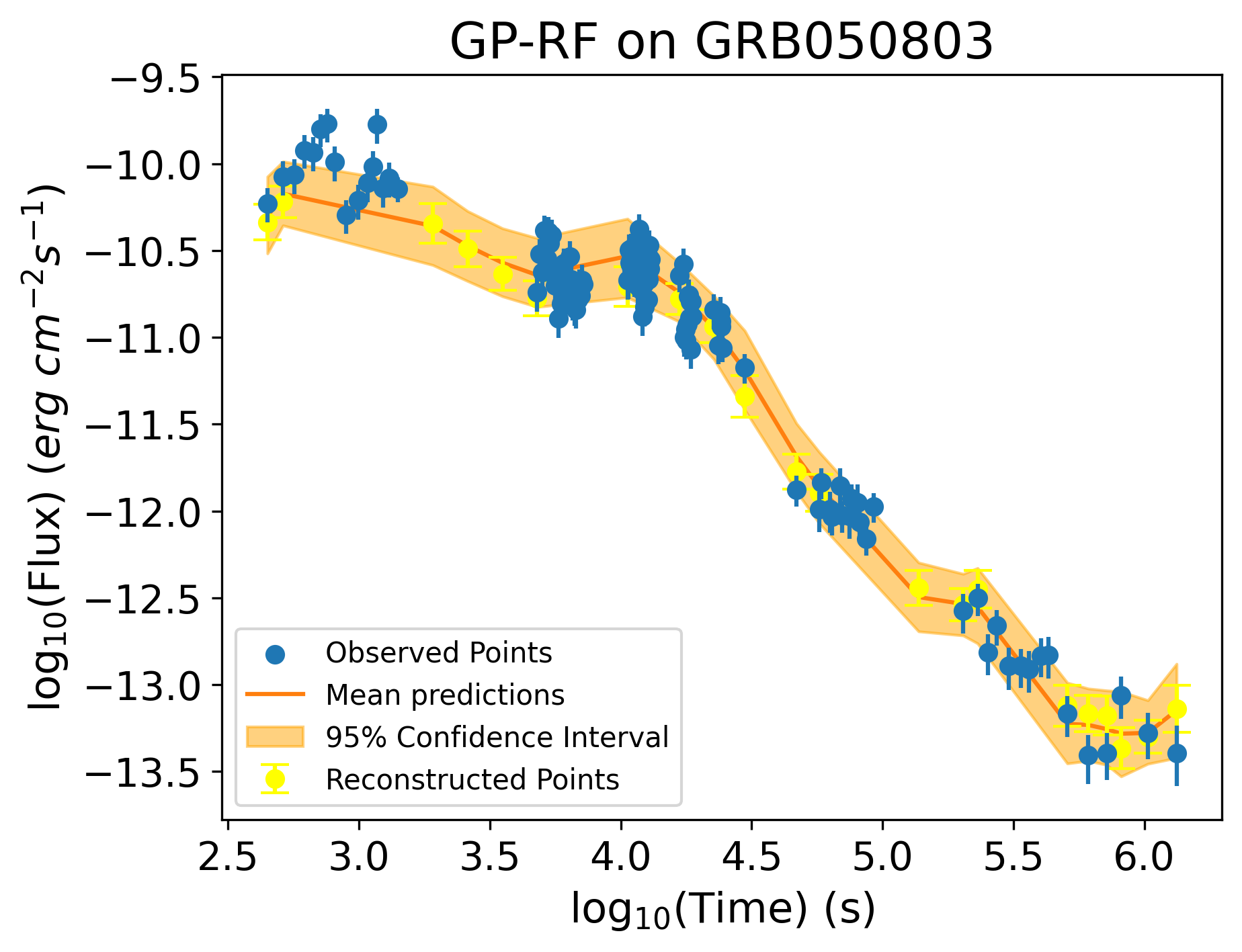} 

    \caption{Reconstruction of LCs for all four varieties of GRBs are shown in a grid with four types of GRBs (left to right): i) Good GRBs (Column 1); ii) a GRB LC with a break towards the end (Column 2); iii) flares or bumps in the afterglow (Column 3); iv) flares or bumps with a double break towards the end of the LC (Column 4) and the models (top to bottom): i) GP (Row 1); ii) Bi-Mamba Model (Row 2); iii) MLP model (Row 3); iv) Fourier Transform (Row 4); v) GP+RF model (Row 5).}
\label{fig: ALL-reconstruction-1}
\end{center}
\end{figure*}

\begin{figure*}[htbp]
\begin{center}

    \includegraphics[width=.24\textwidth, height=.21\textwidth]{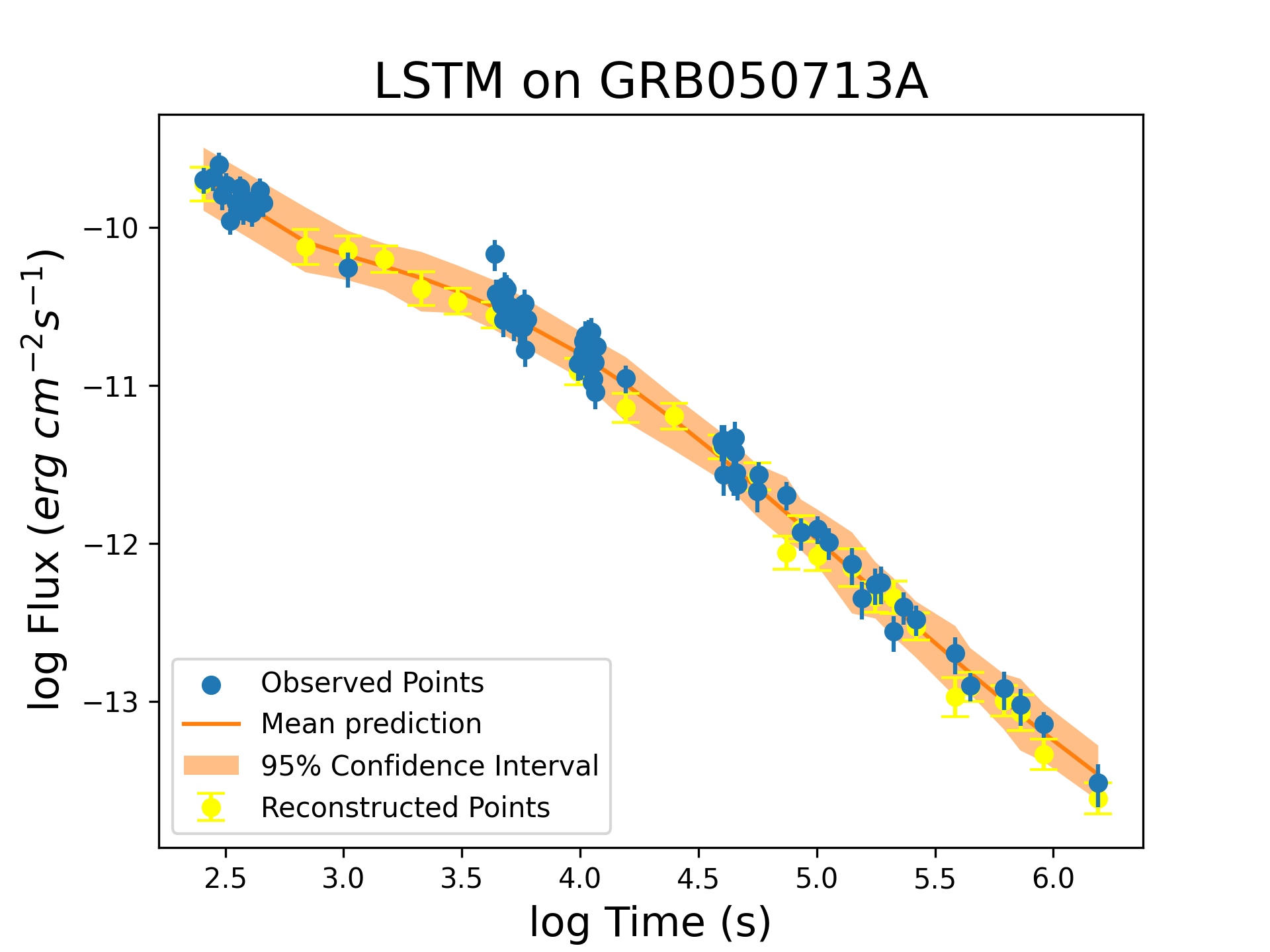}
    \includegraphics[width=.24\textwidth, height=.21\textwidth]{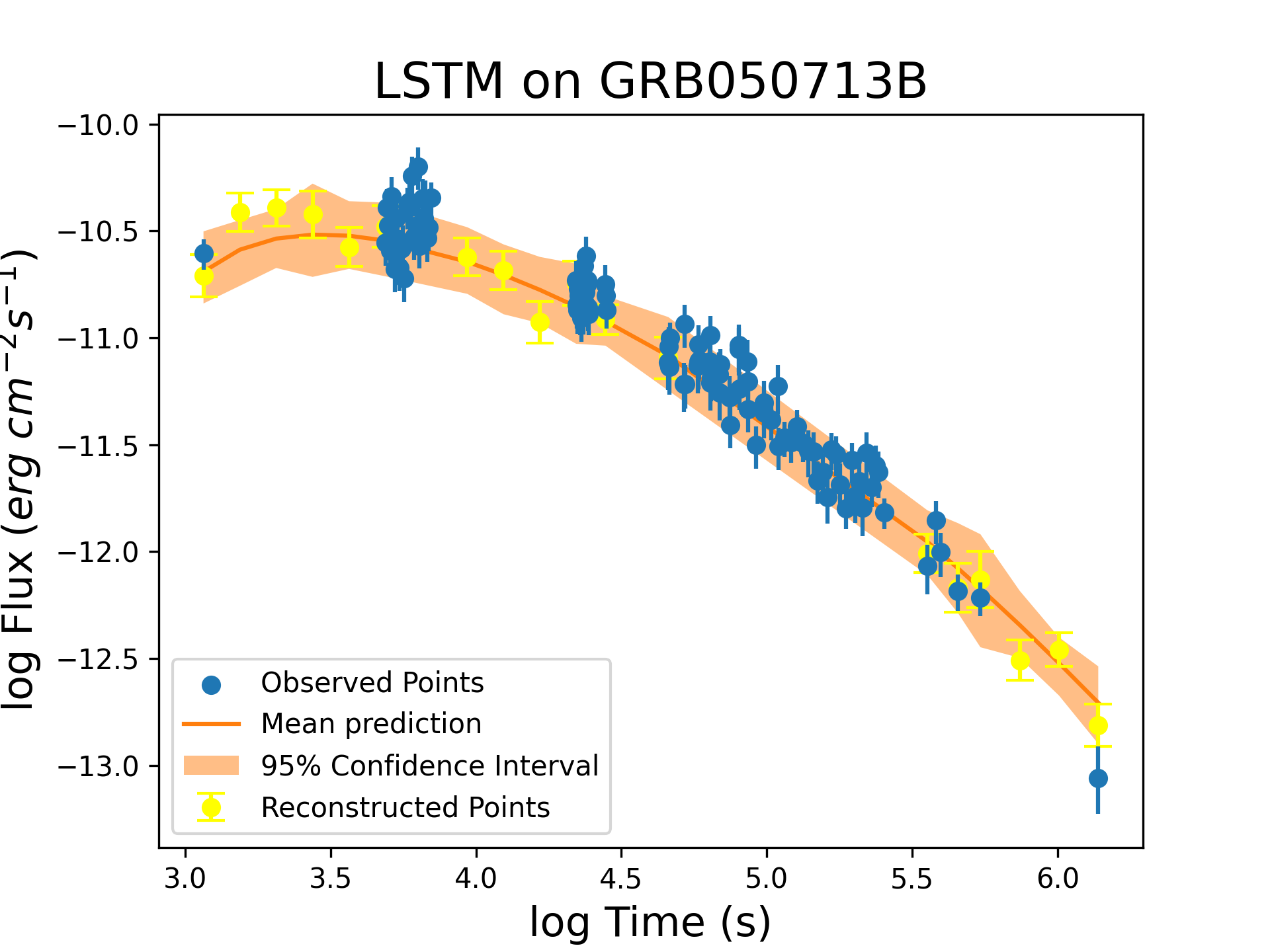}
    \includegraphics[width=.24\textwidth, height=.21\textwidth]{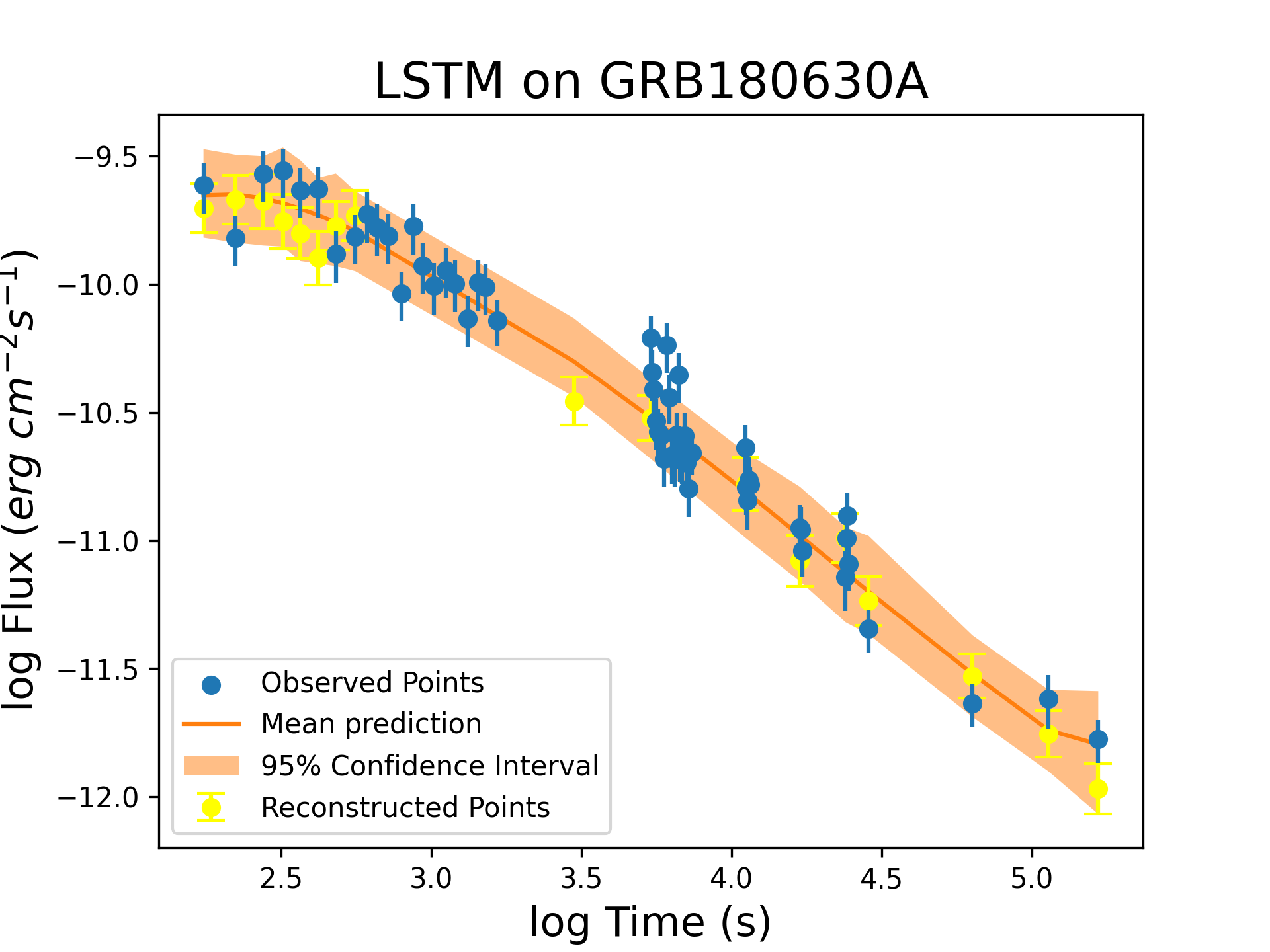}
    \includegraphics[width=.24\textwidth, height=.21\textwidth]{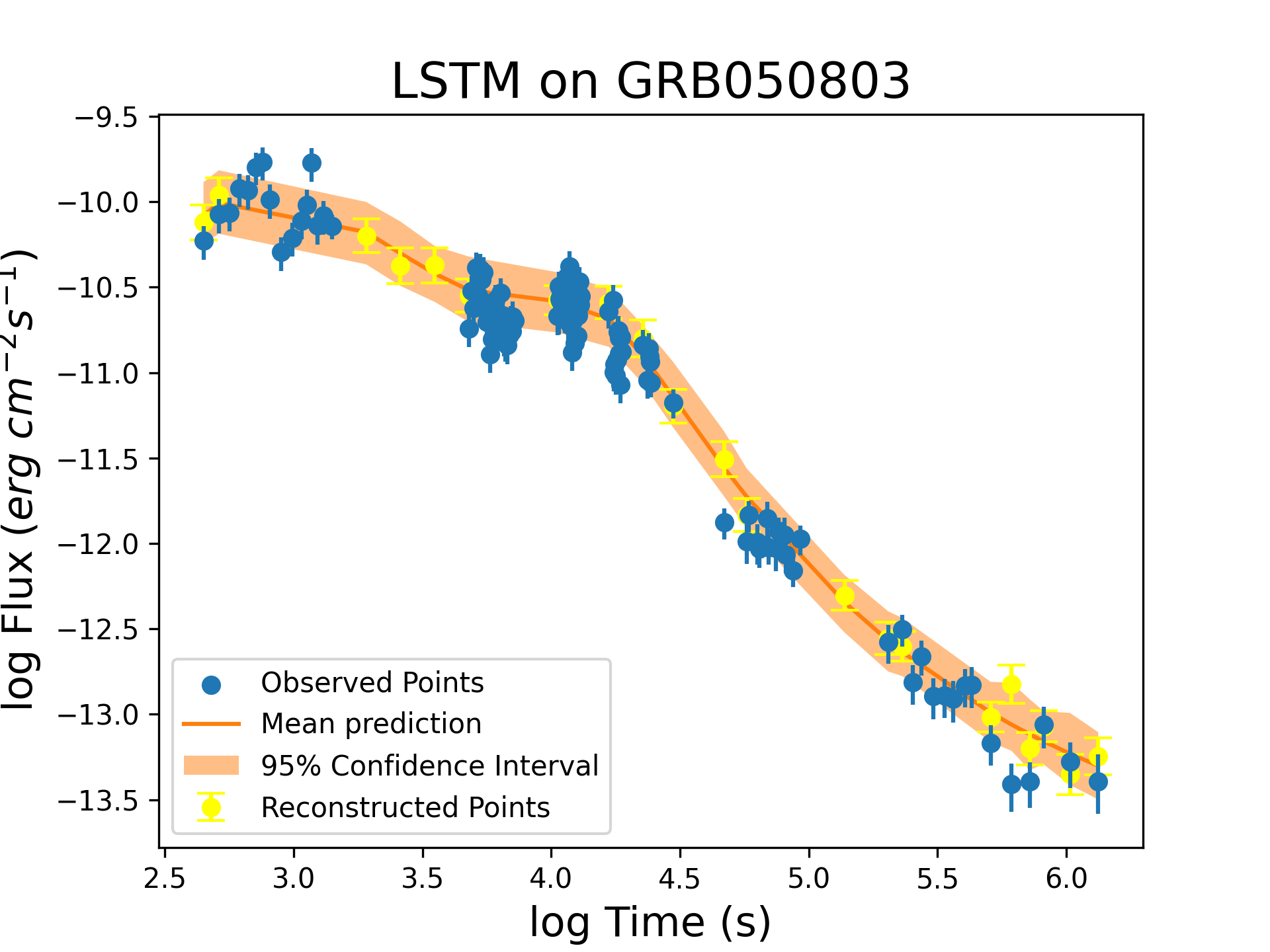} 

    \includegraphics[width=.24\textwidth, height=.21\textwidth]{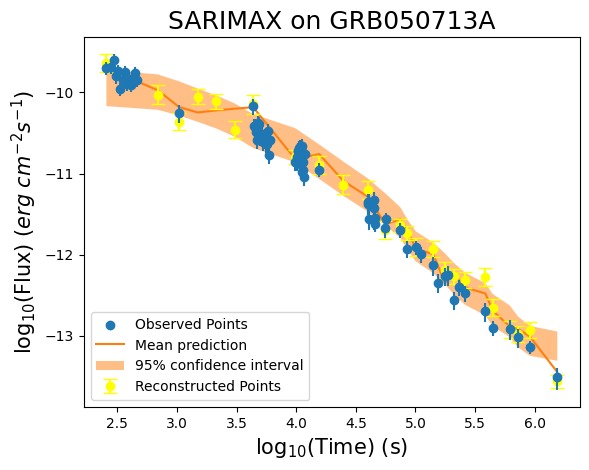}
    \includegraphics[width=.24\textwidth, height=.21\textwidth]{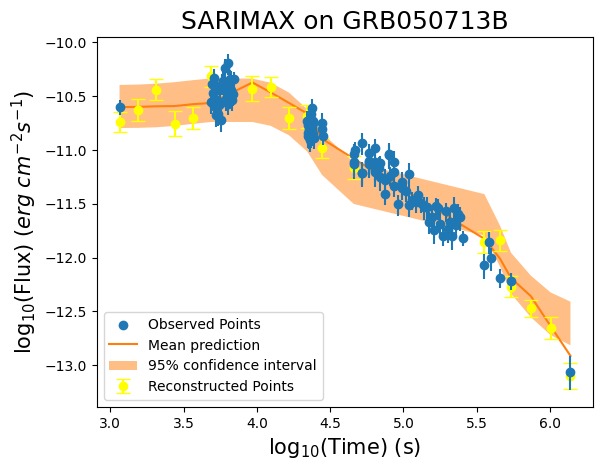}
    \includegraphics[width=.24\textwidth, height=.21\textwidth]{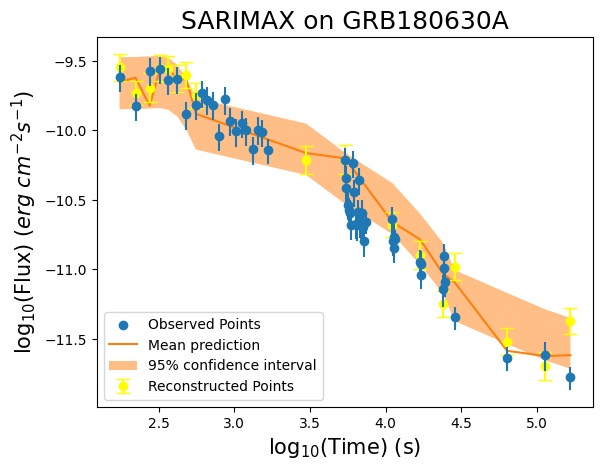}
    \includegraphics[width=.24\textwidth, height=.21\textwidth]{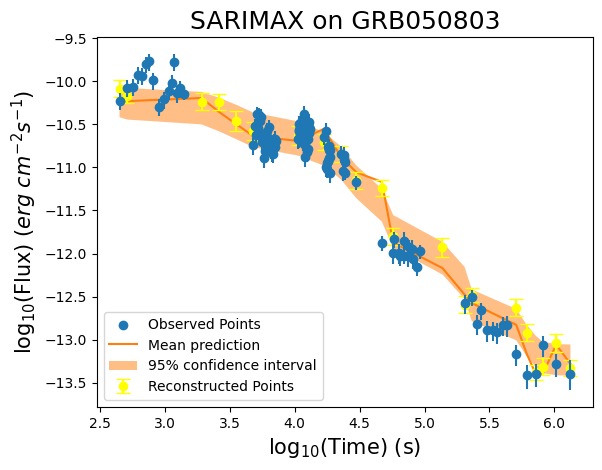}

    \includegraphics[width=.24\textwidth, height=.21\textwidth]{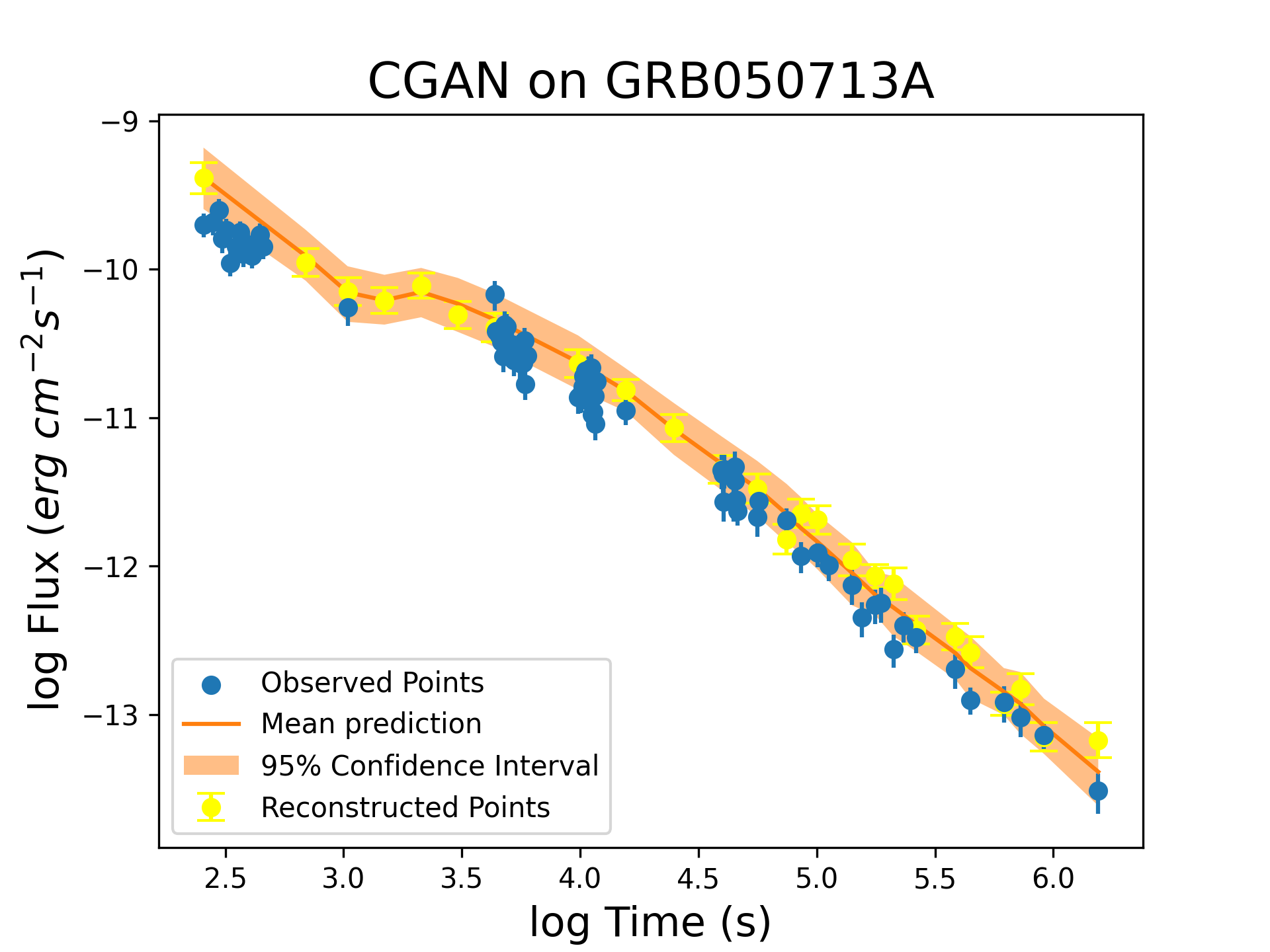}
    \includegraphics[width=.24\textwidth, height=.21\textwidth]{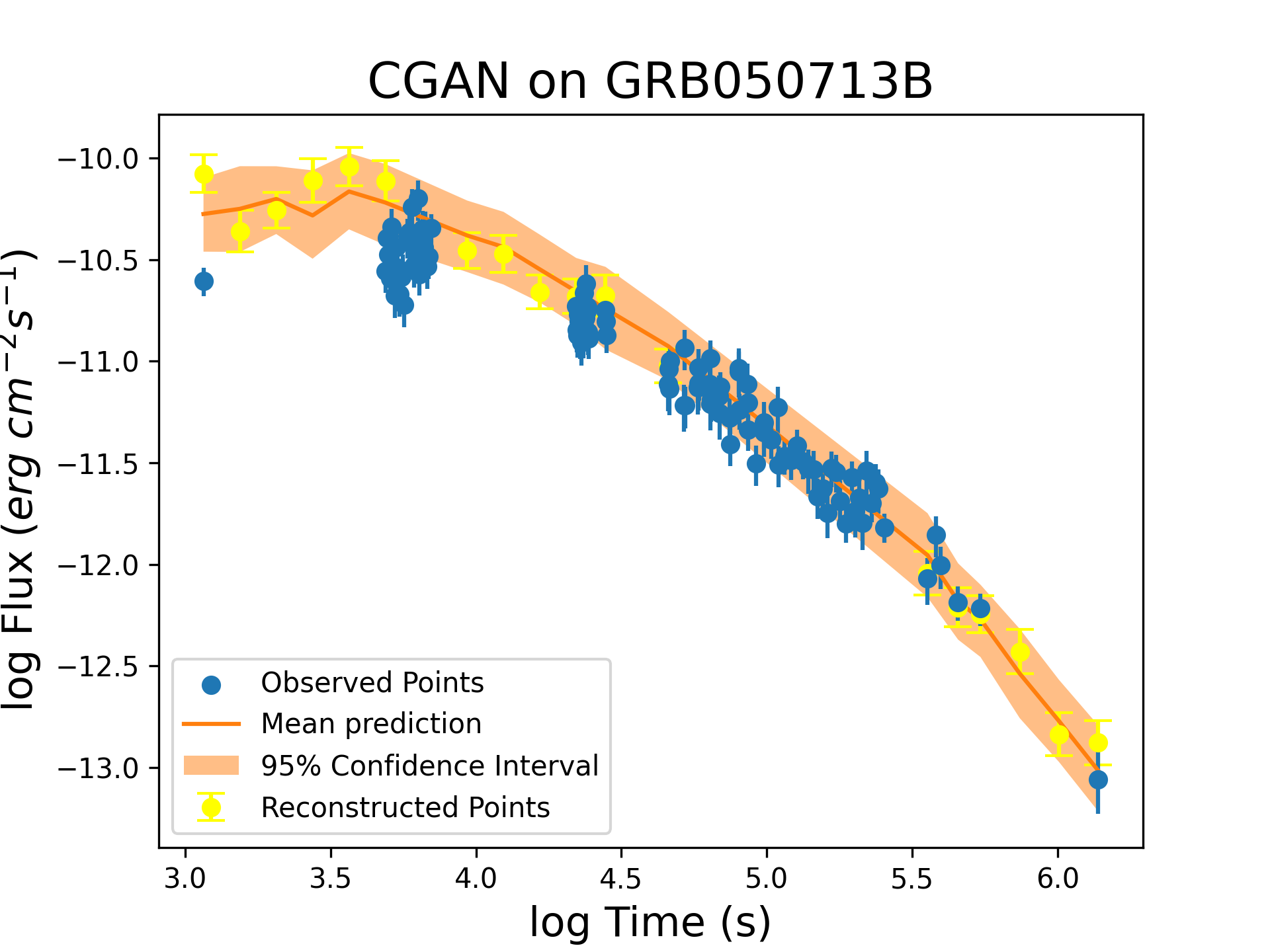}
    \includegraphics[width=.24\textwidth, height=.21\textwidth]{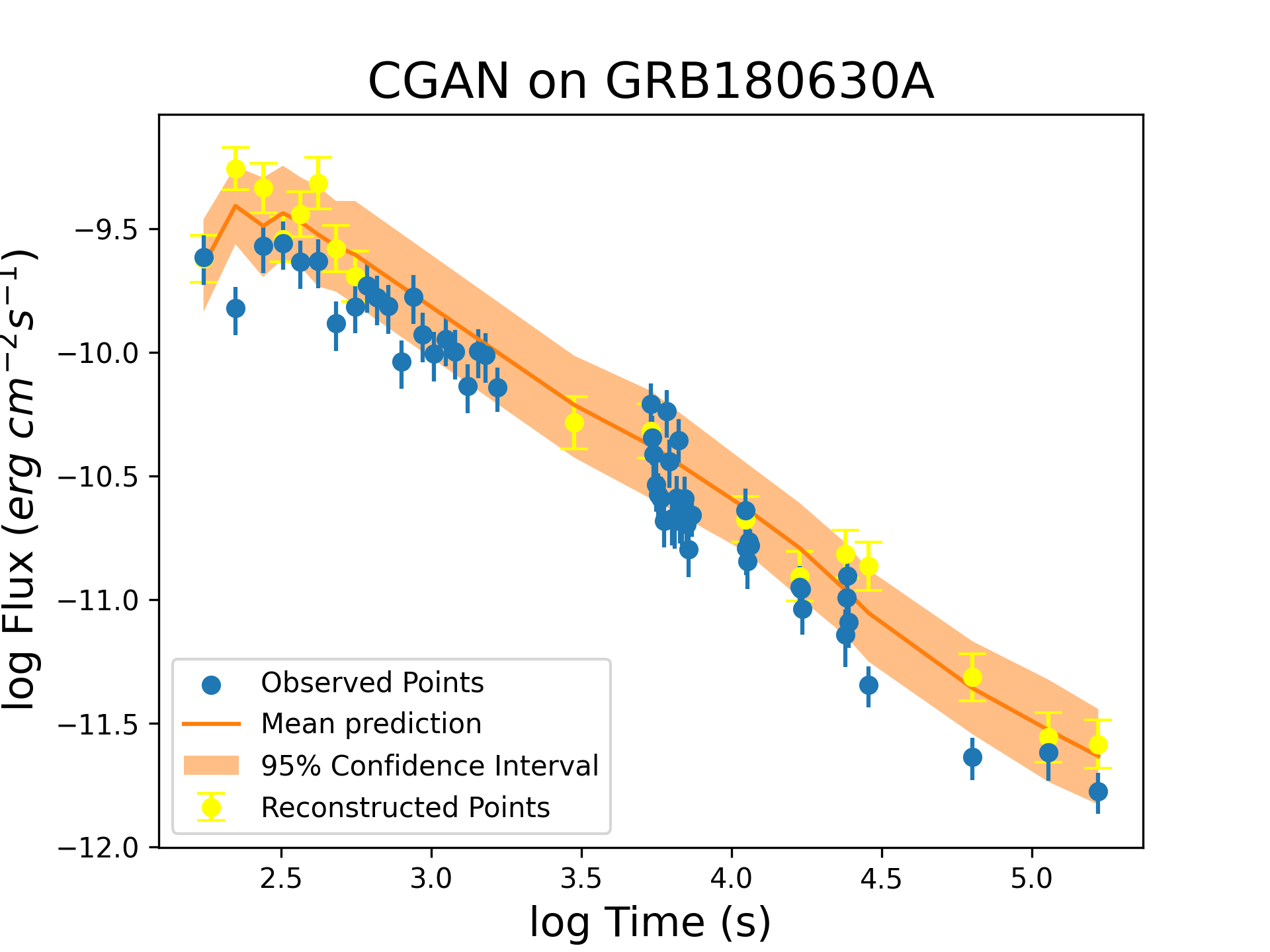}
    \includegraphics[width=.24\textwidth, height=.21\textwidth]{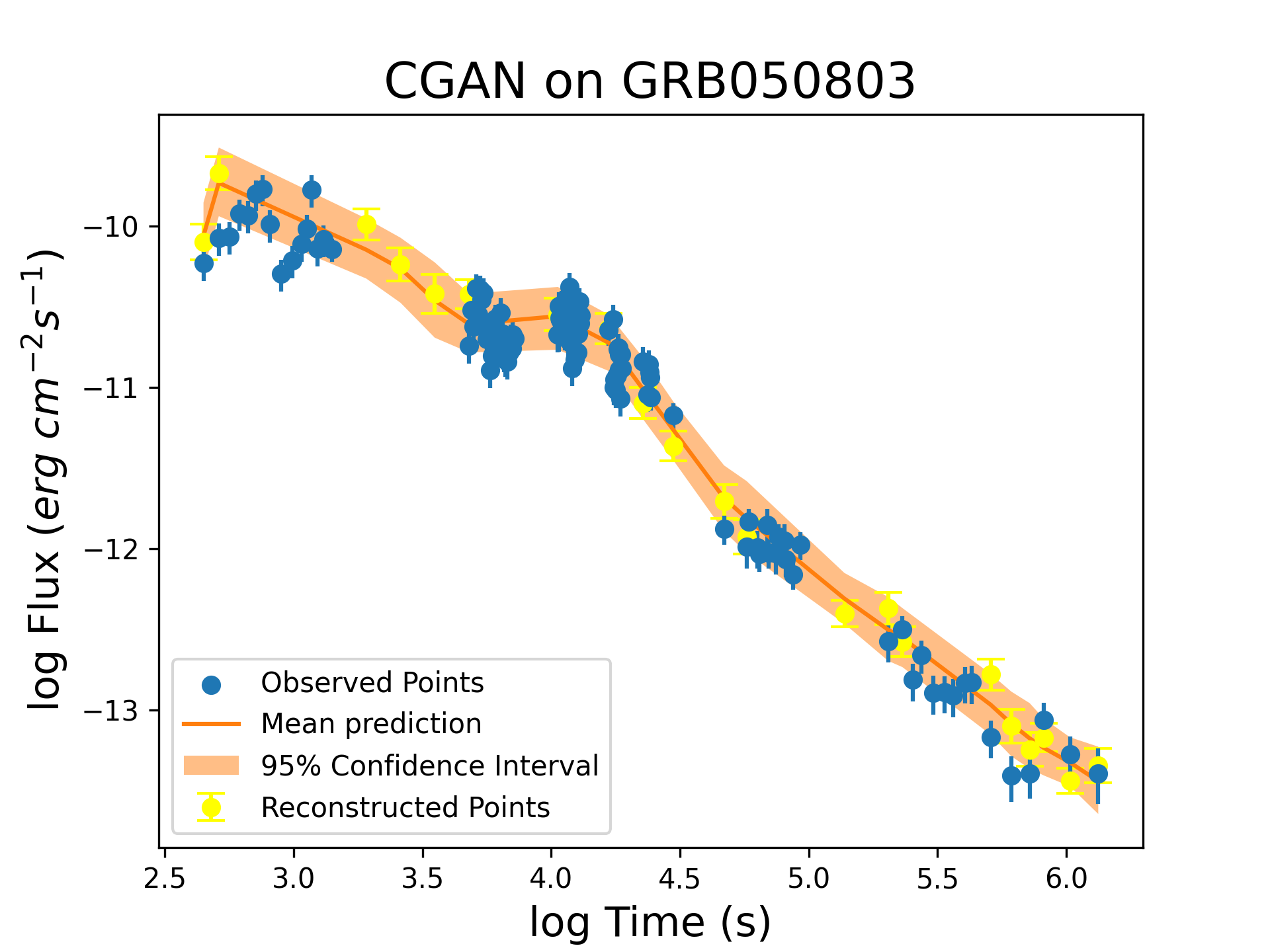}
    
    \includegraphics[width=.24\textwidth, height=.21\textwidth]{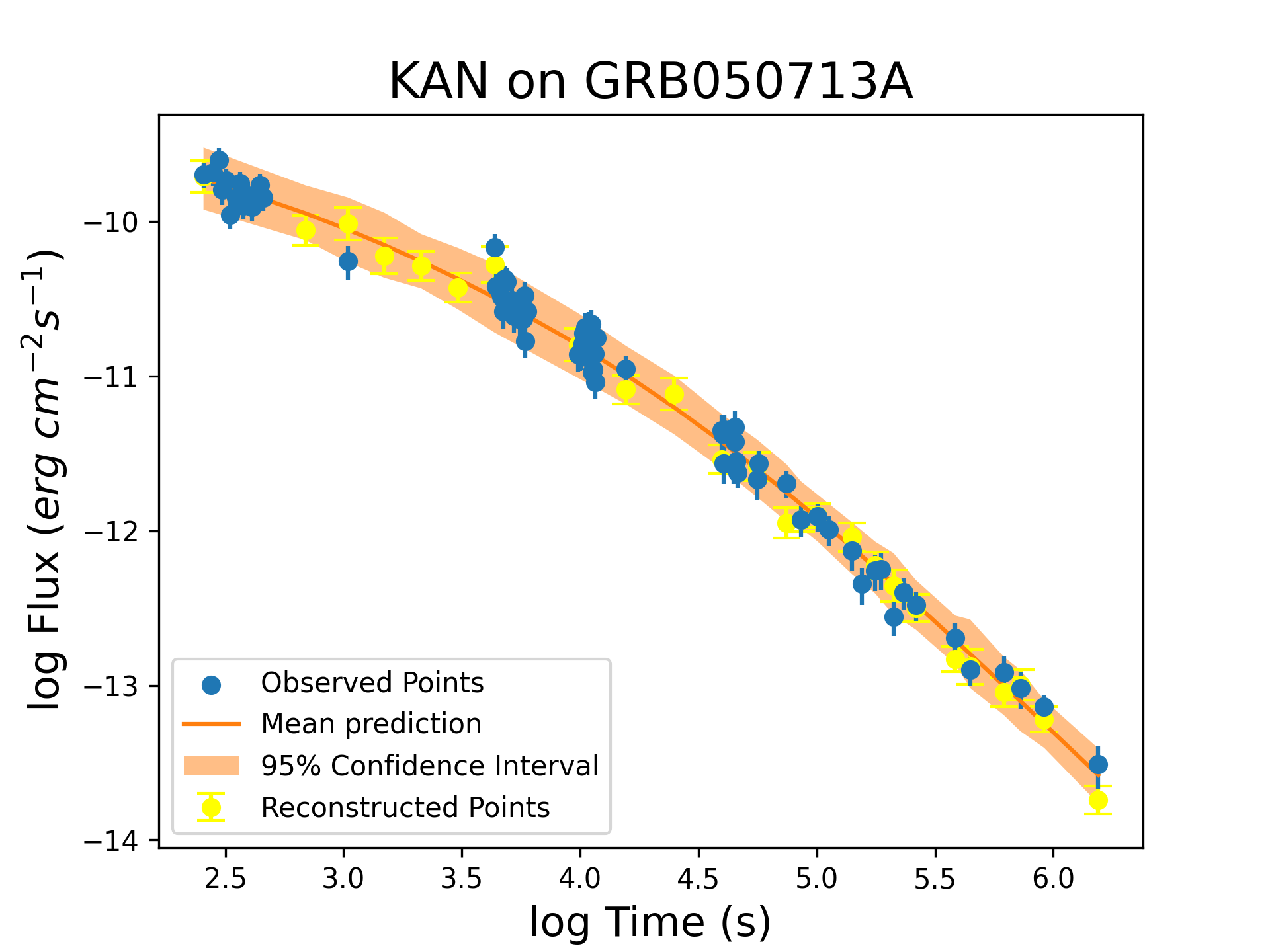}
    \includegraphics[width=.24\textwidth, height=.21\textwidth]{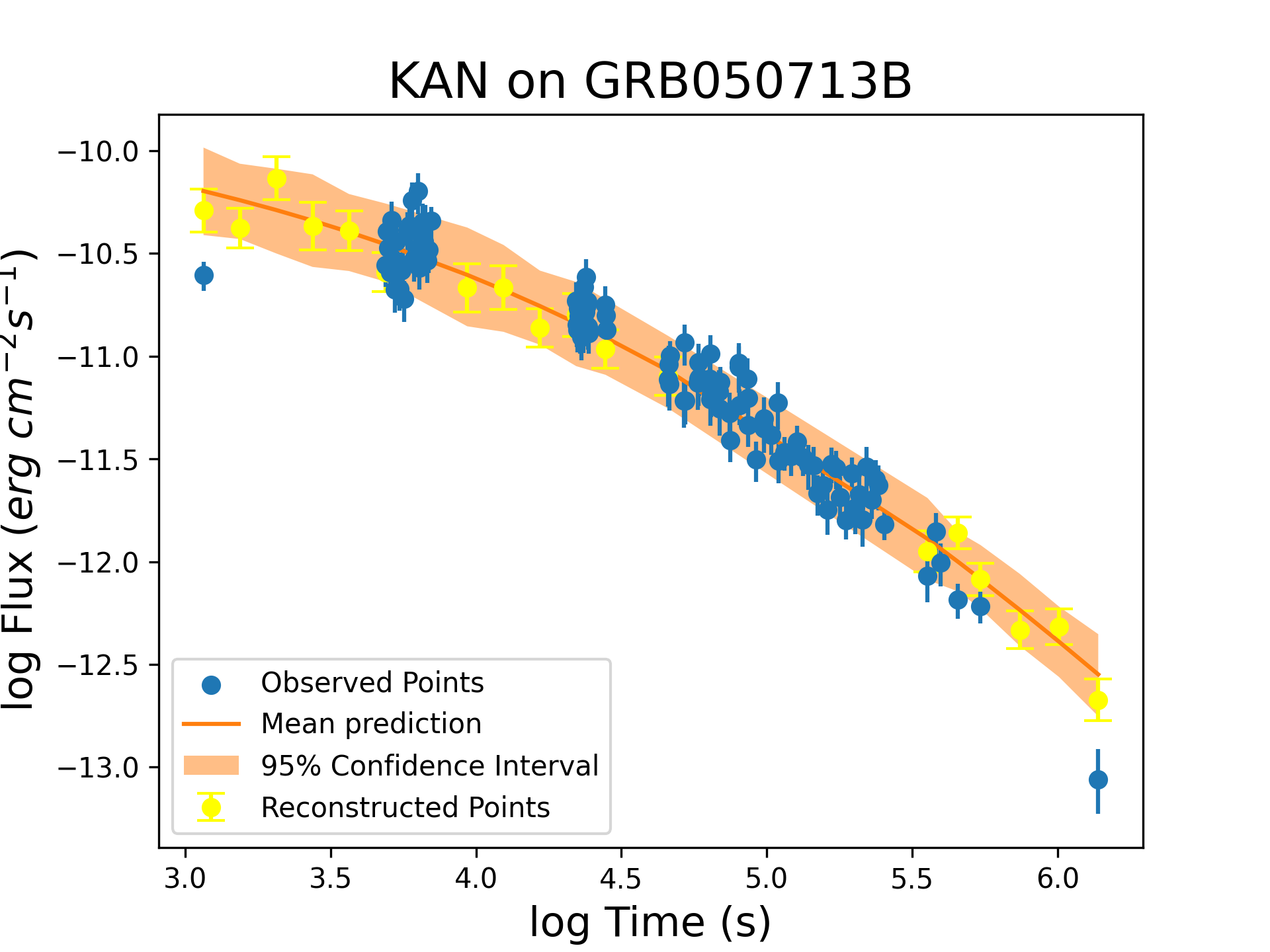}
    \includegraphics[width=.24\textwidth, height=.21\textwidth]{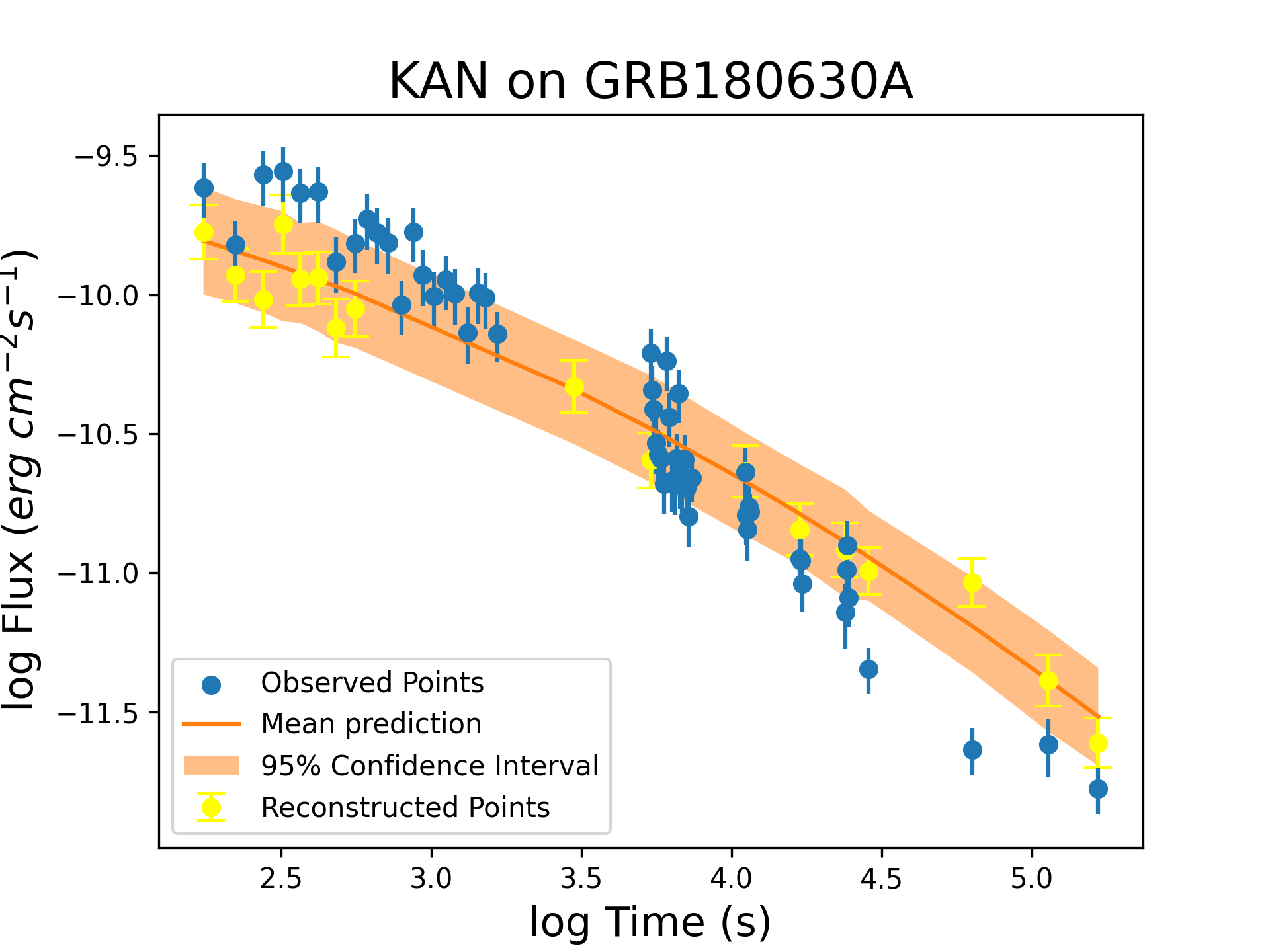}
    \includegraphics[width=.24\textwidth, height=.21\textwidth]{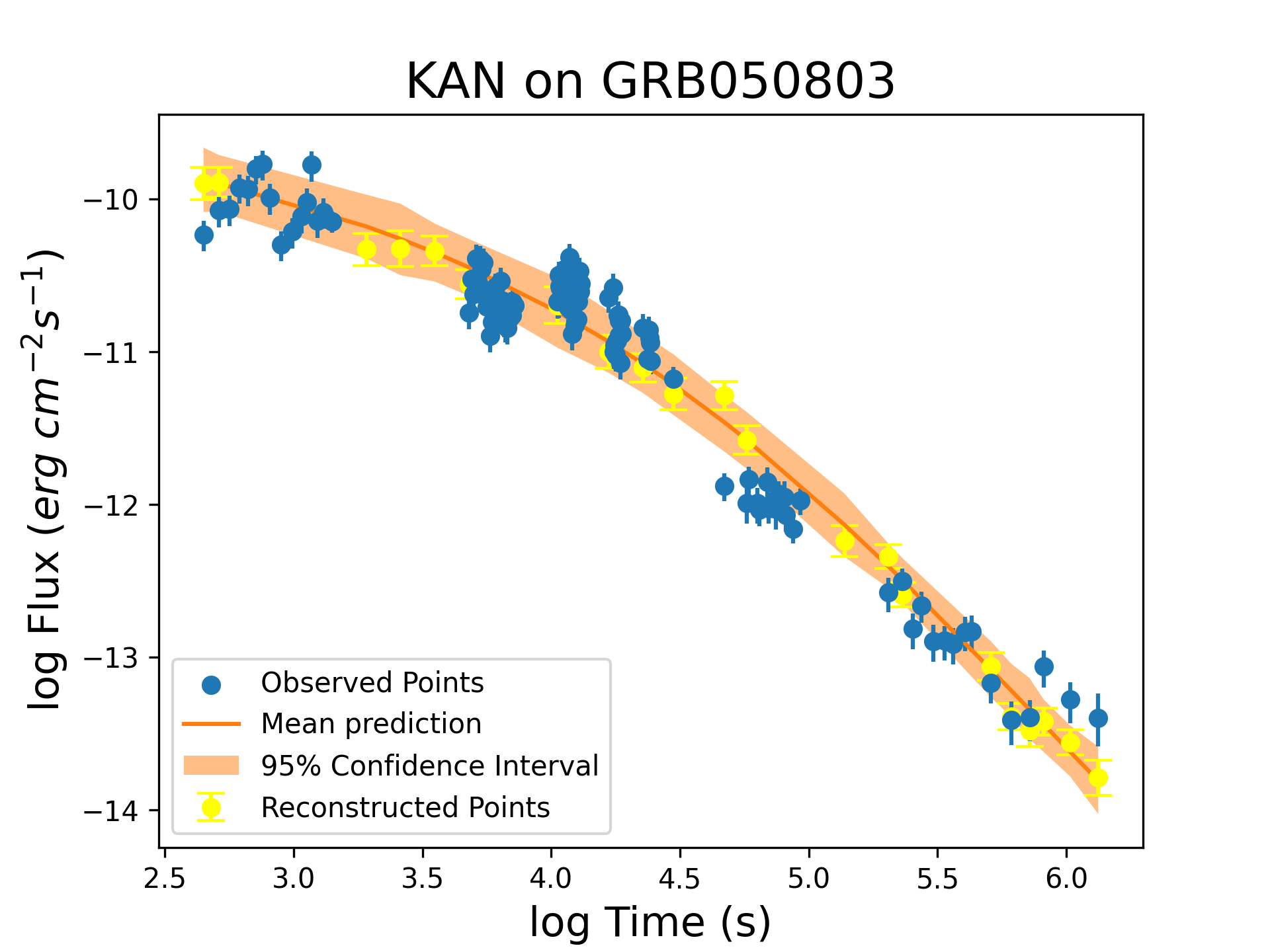}

    \includegraphics[width=.24\textwidth, height=.21\textwidth]{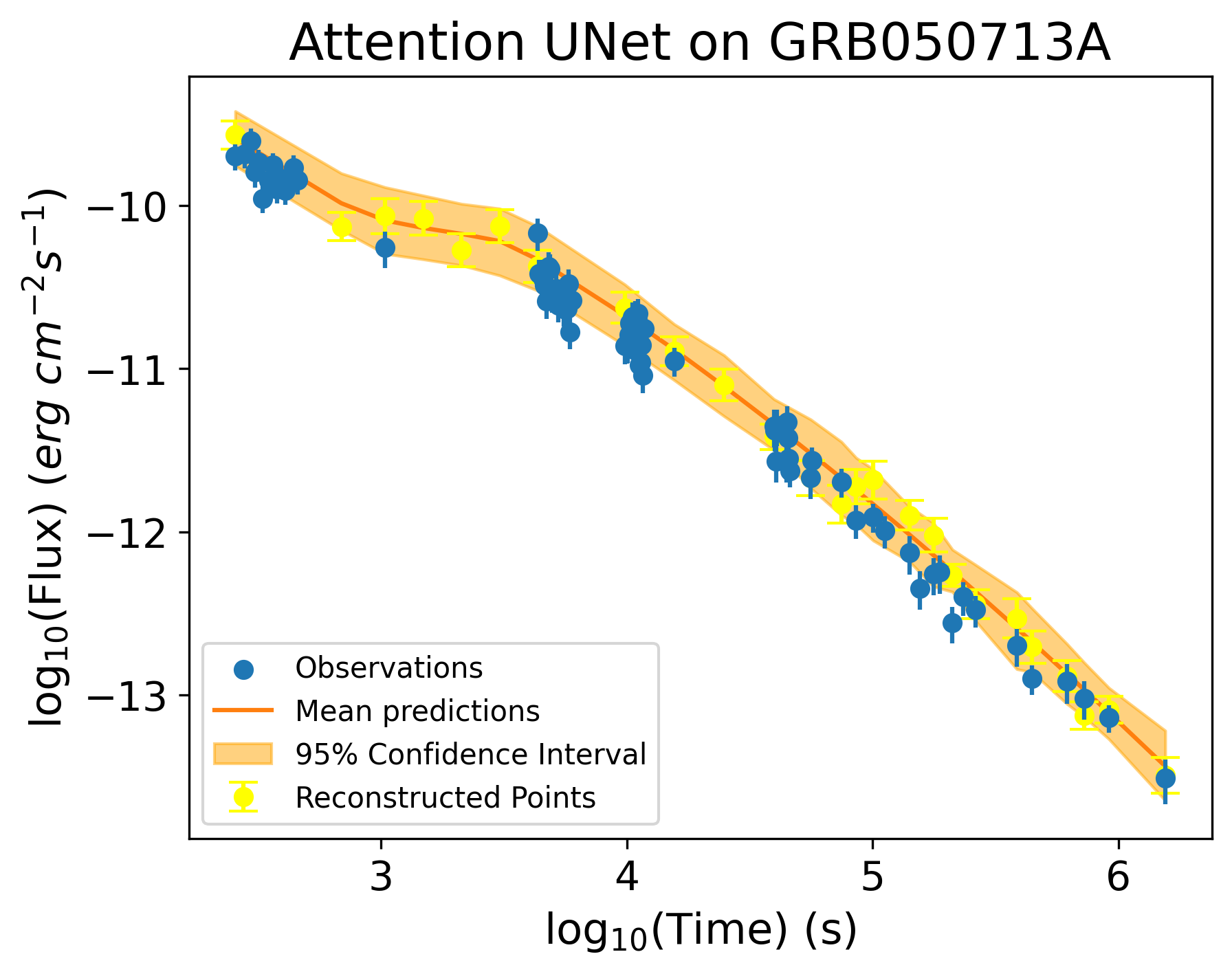}
    \includegraphics[width=.24\textwidth, height=.21\textwidth]{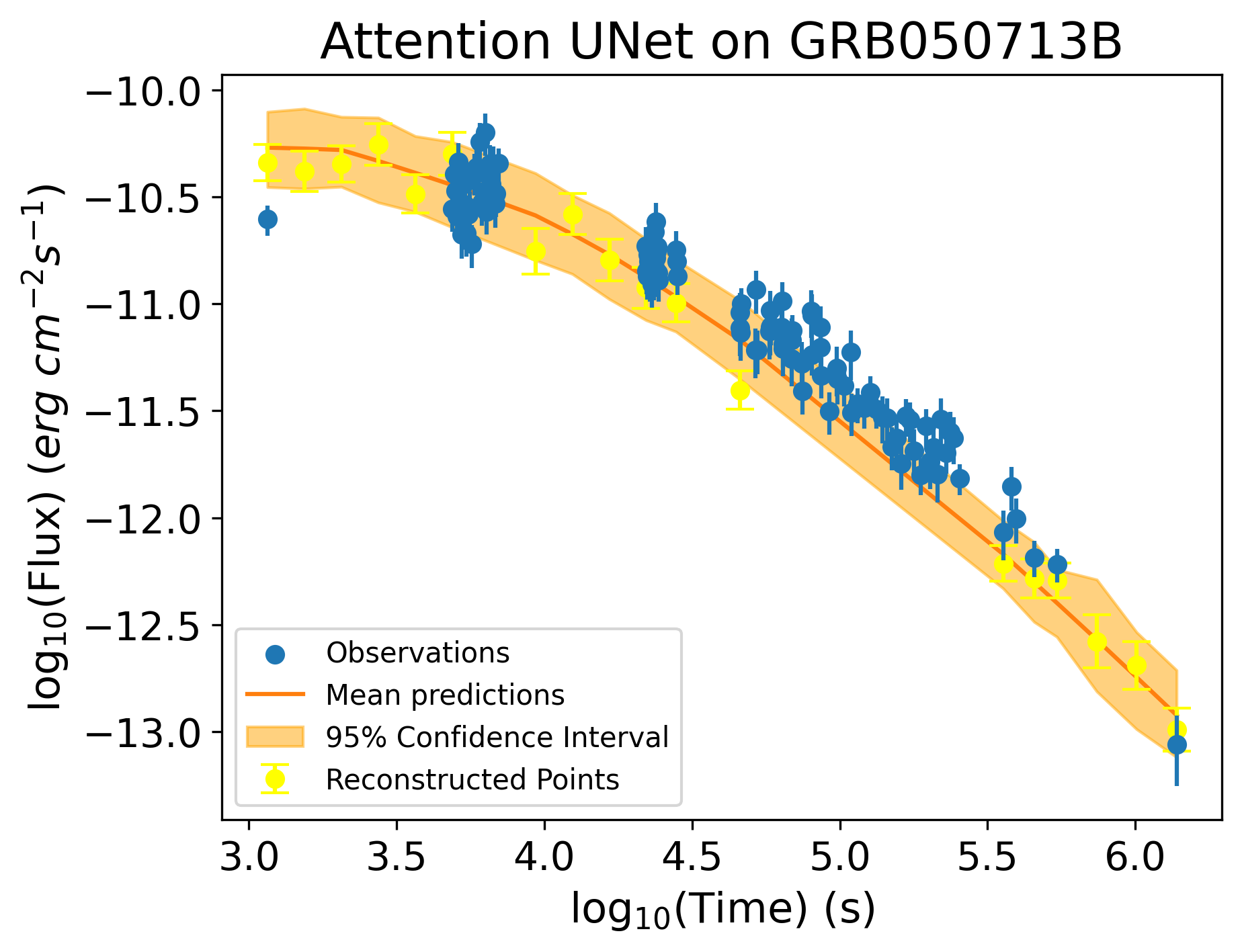}
    \includegraphics[width=.24\textwidth, height=.21\textwidth]{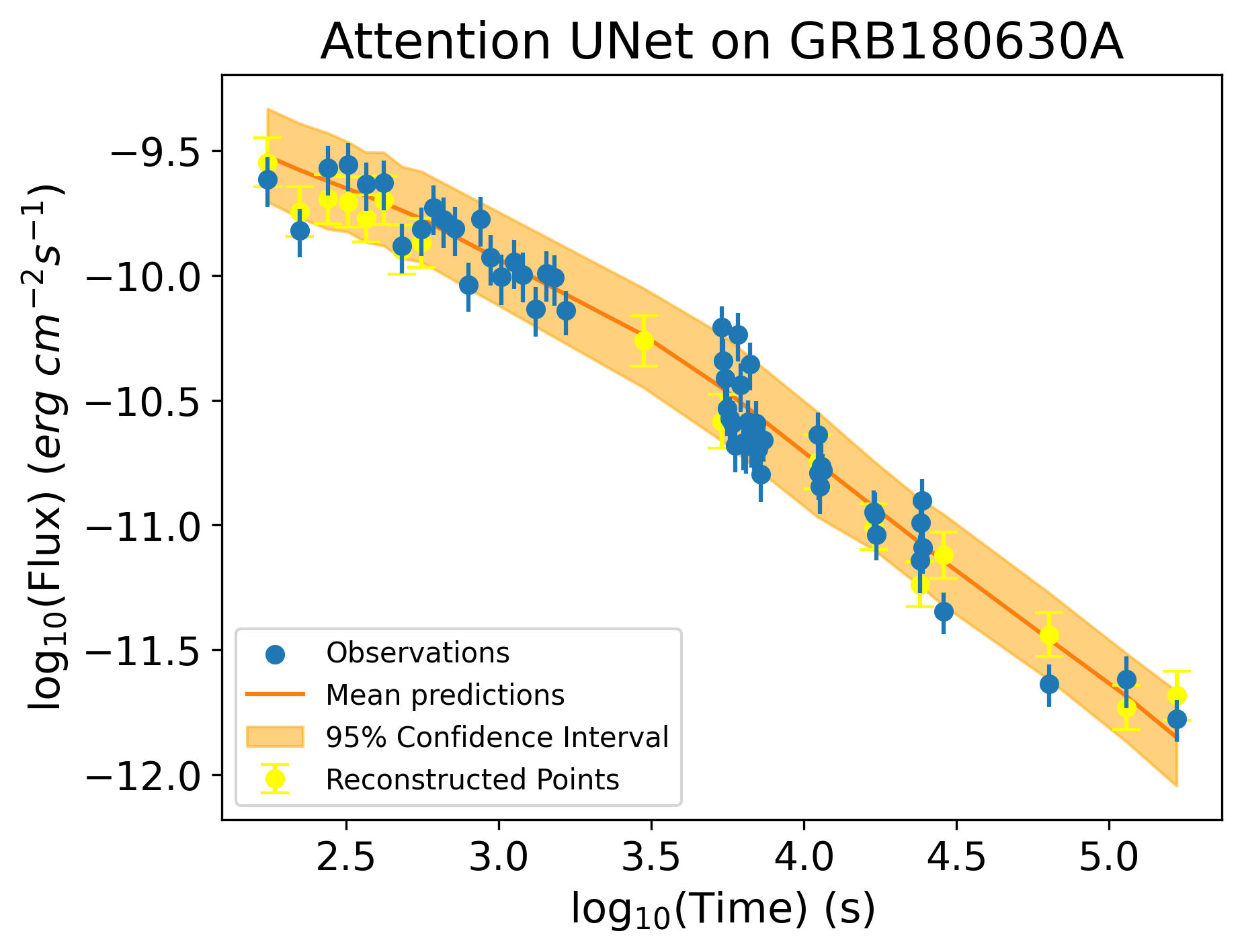}
    \includegraphics[width=.24\textwidth, height=.21\textwidth]{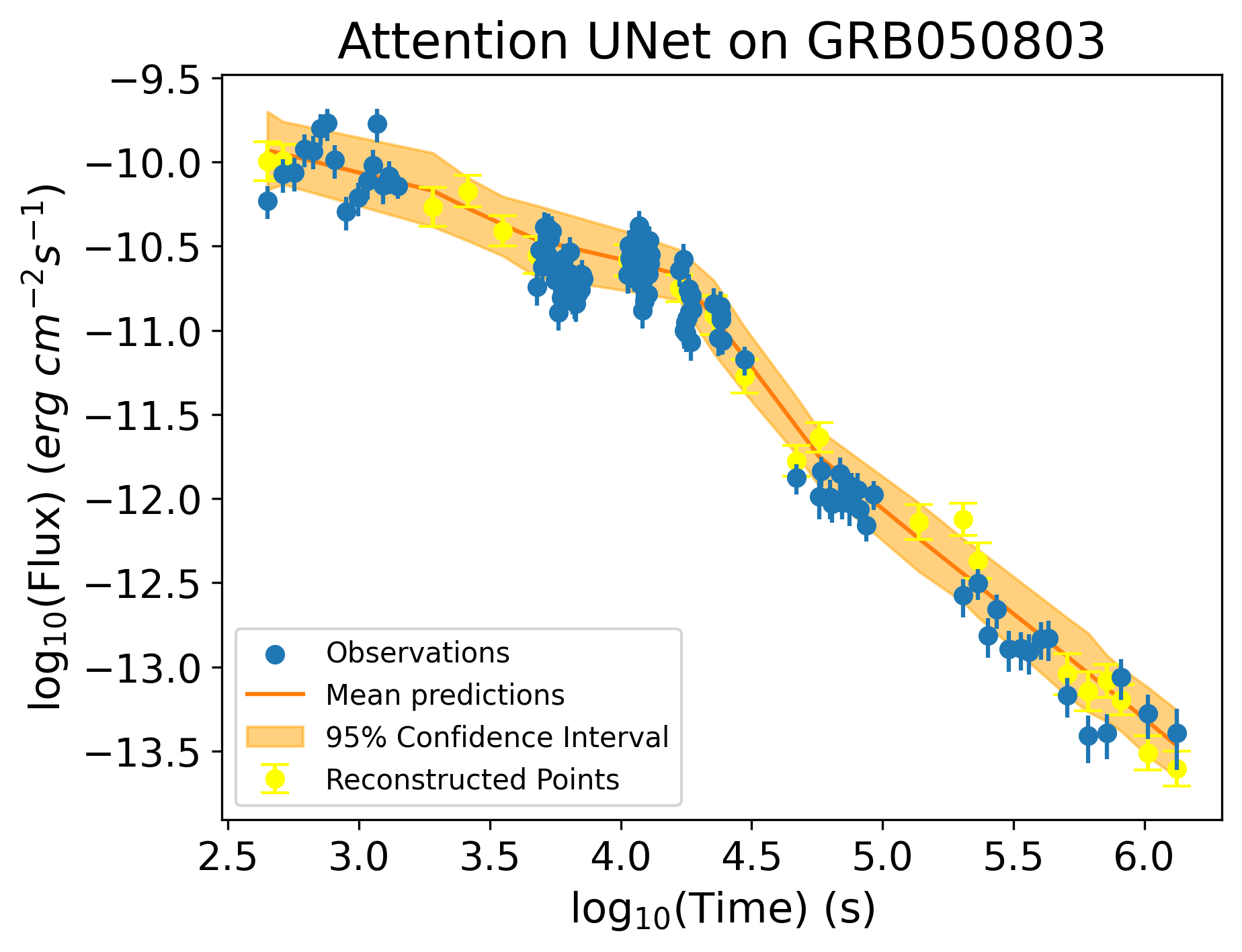}

\end{center}
    \caption{Reconstruction of LCs for all four varieties of GRBs are shown in a grid with four types of GRBs (left to right): i) Good GRBs (Column 1); ii) a GRB LC with a break towards the end (Column 2); iii) flares or bumps in the afterglow (Column 3); iv) flares or bumps with a double break towards the end of the LC (Column 4) and the models (top to bottom): i) Bi-LSTM Model(Row 1); ii) SARIMA-based Kalman Model (Row 2); iii) CGAN (Row 3); iv) KAN model (Row 4); v) Attention U-Net model (Row 5).}
    \label{fig: ALL-reconstruction-2}
\end{figure*}

\begin{figure*}[htbp]
\begin{center}

\includegraphics[width=.27\textwidth, height=.19\textwidth]{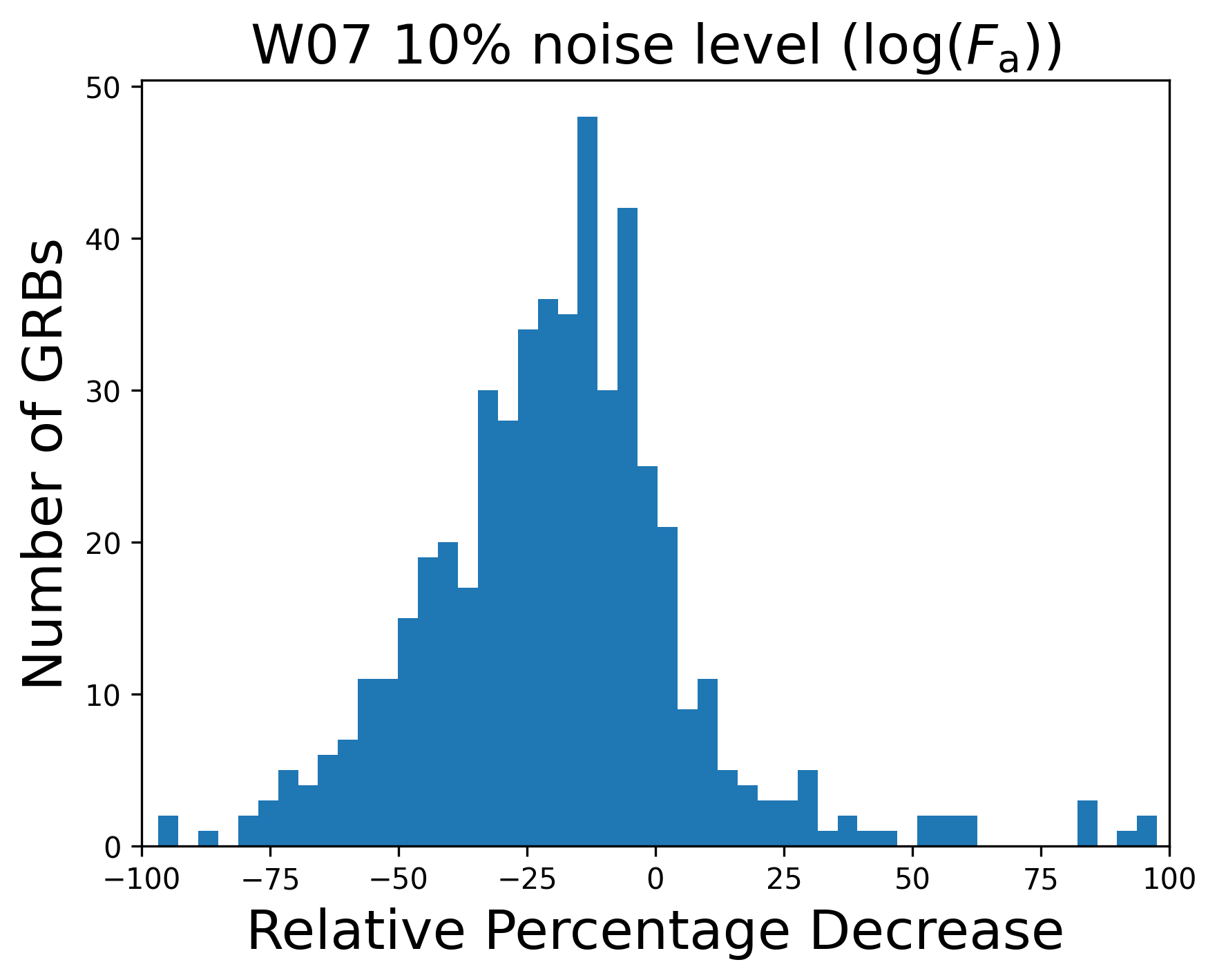}
\includegraphics[width=.27\textwidth, height=.19\textwidth]{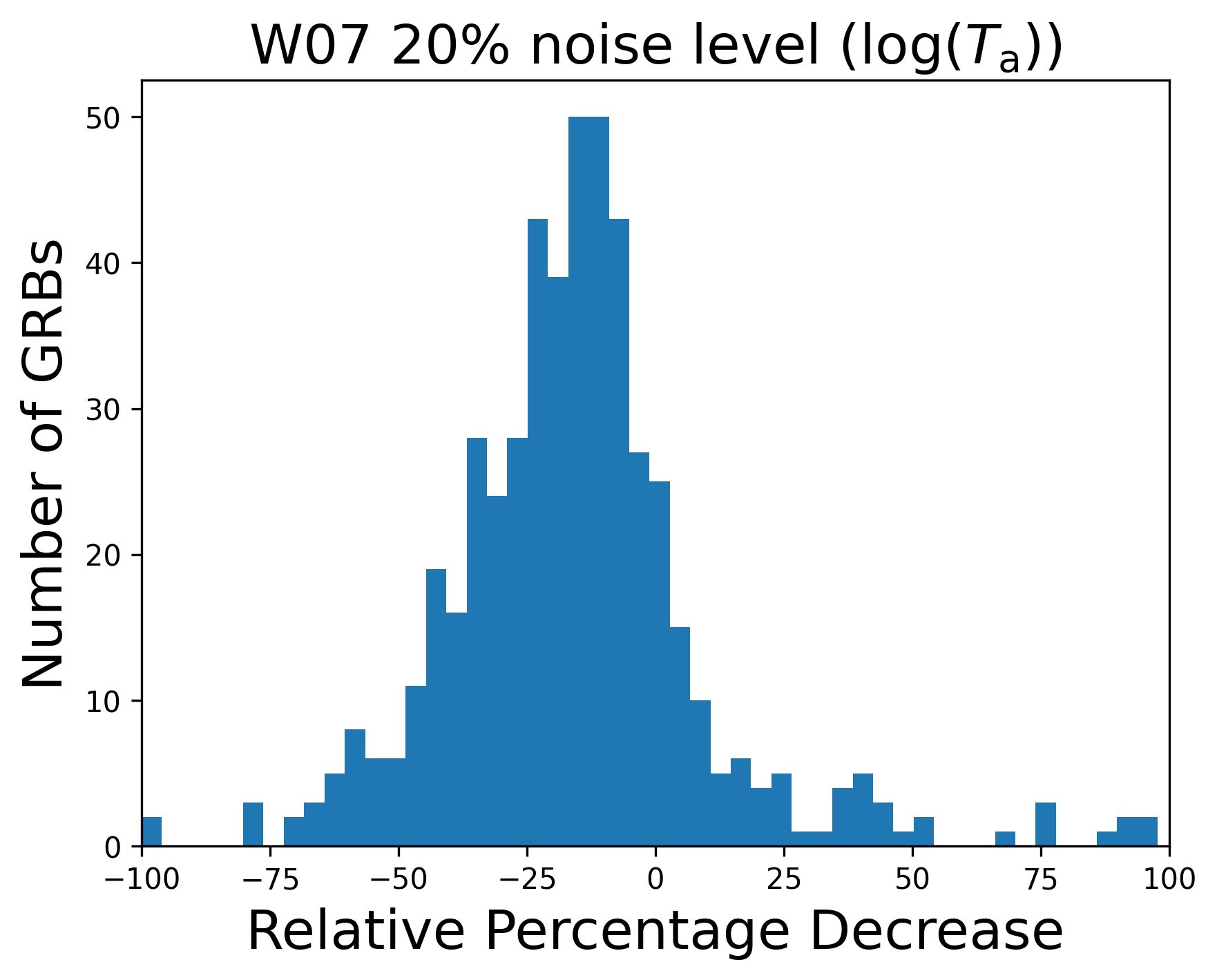}
\includegraphics[width=.27\textwidth, height=.19\textwidth]{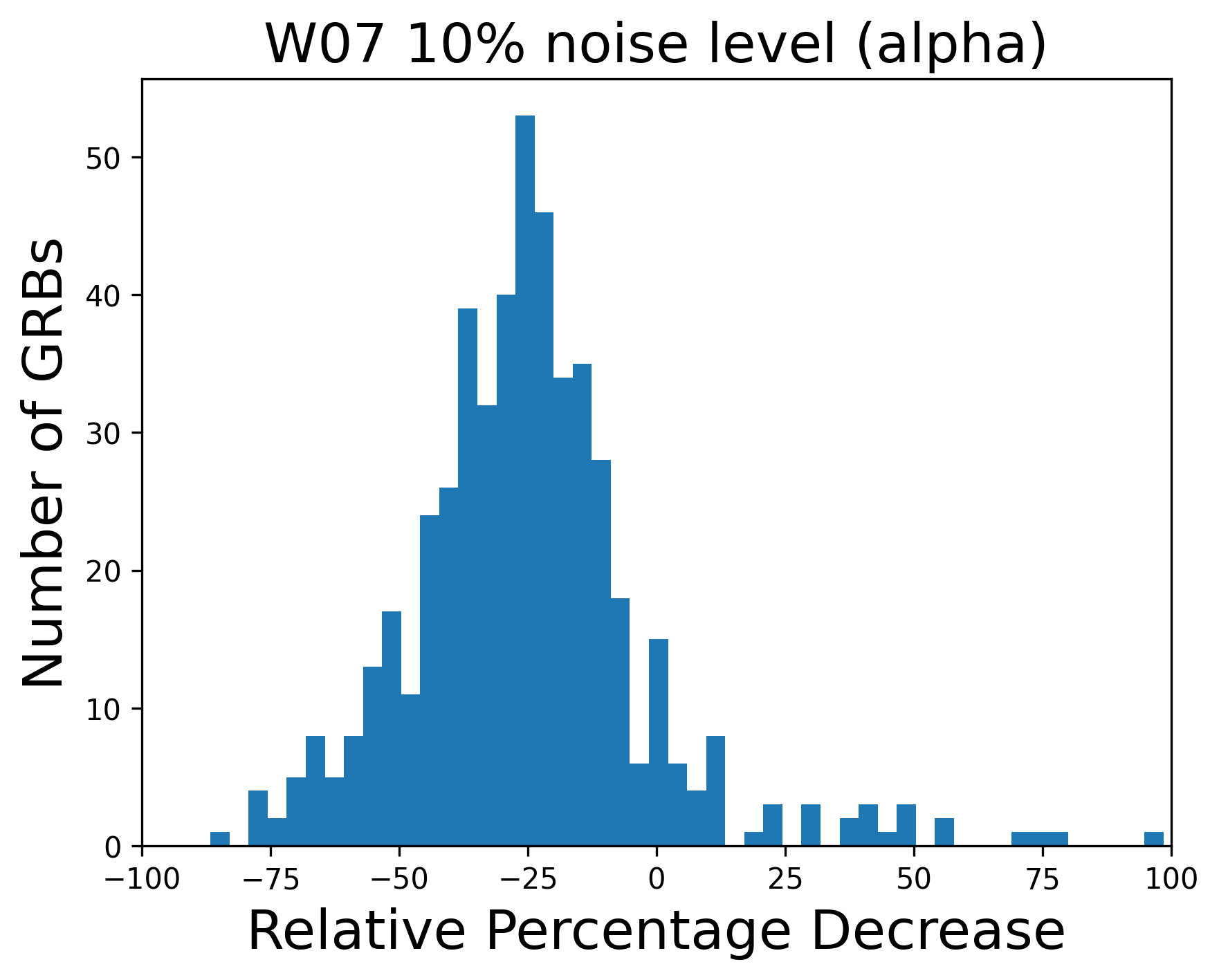}

\includegraphics[width=.27\textwidth, height=.19\textwidth]{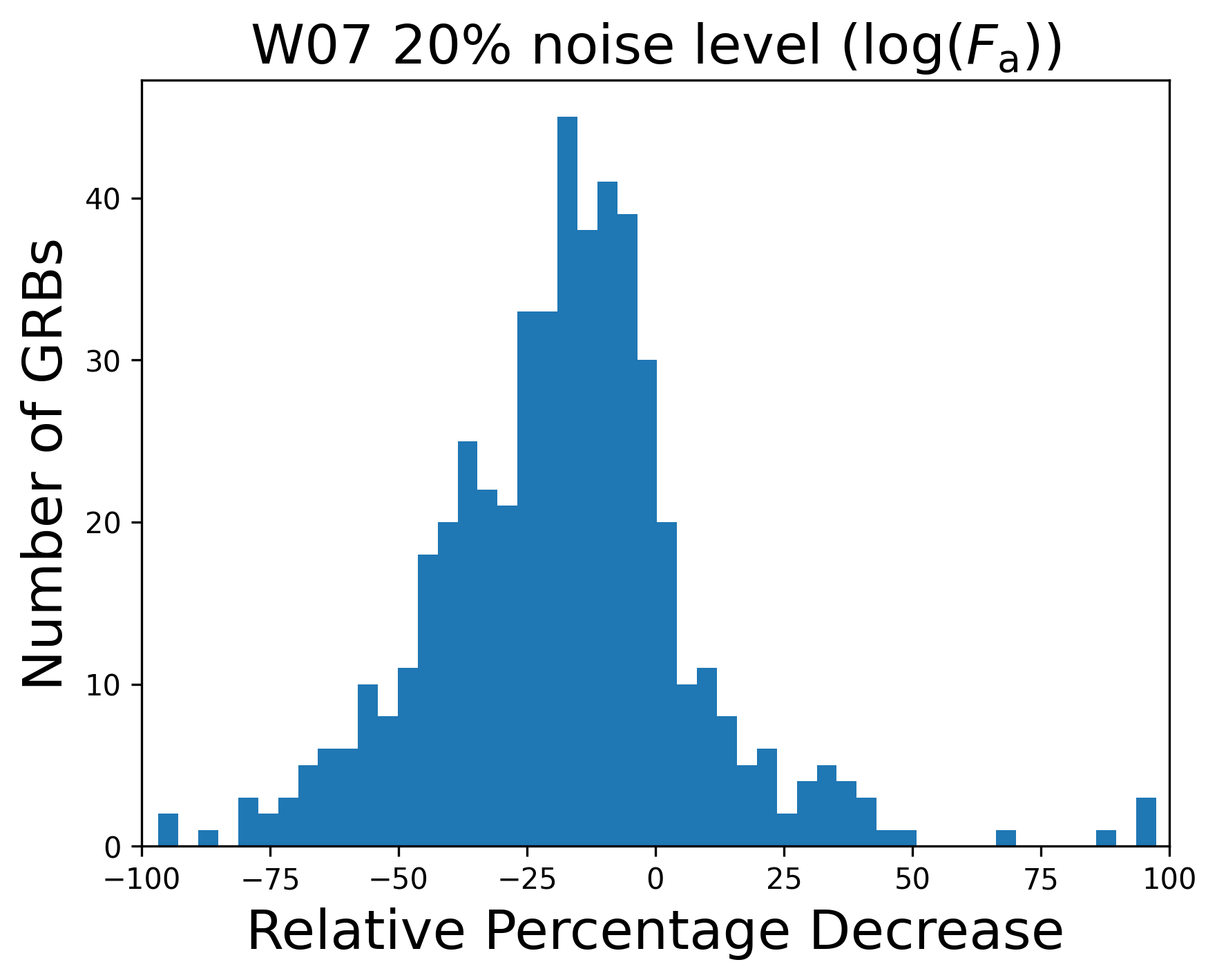}
\includegraphics[width=.27\textwidth, height=.19\textwidth]{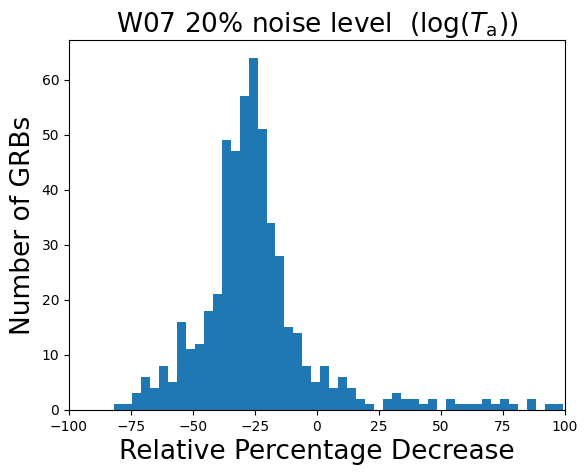}
\includegraphics[width=.27\textwidth, height=.19\textwidth]{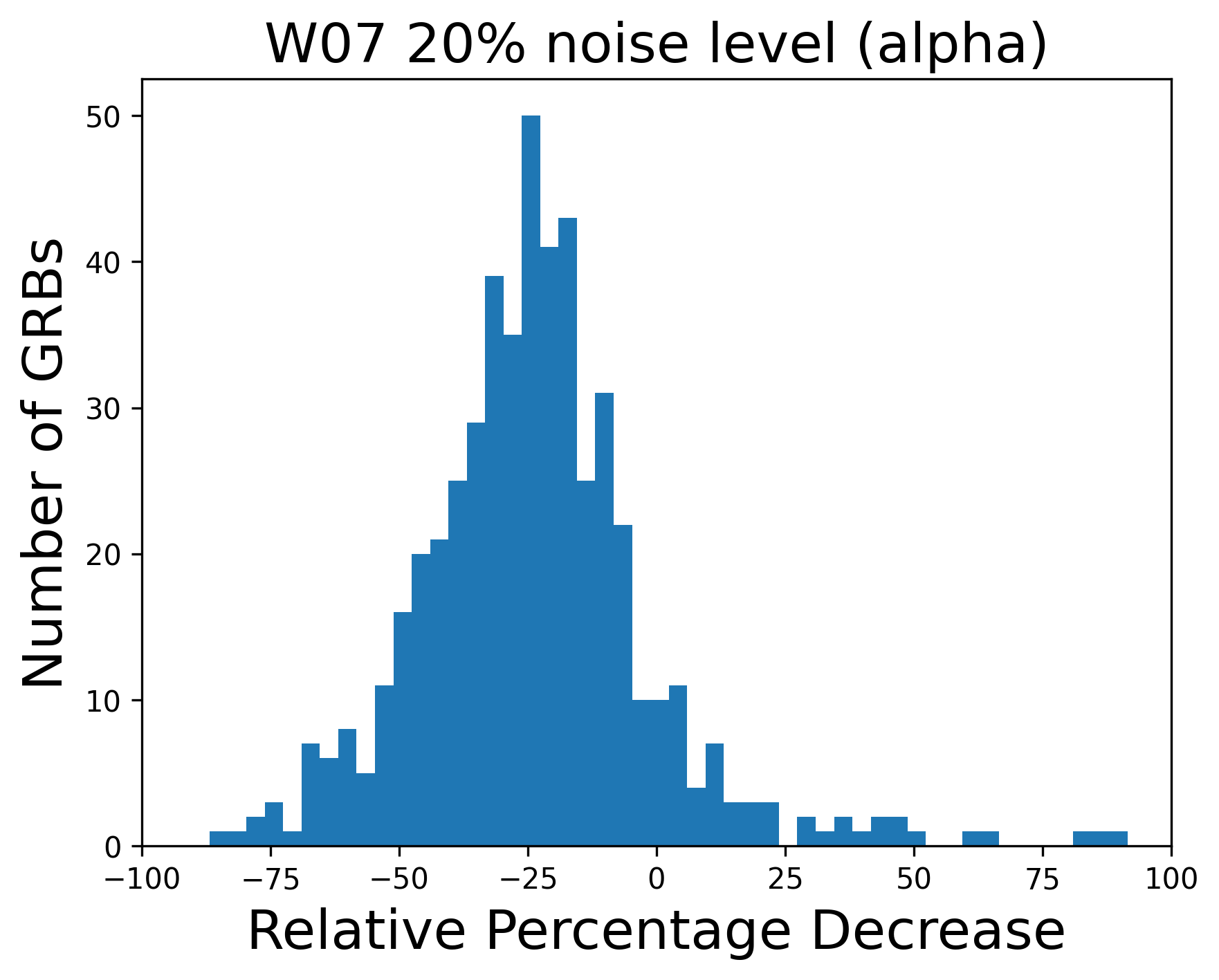}

\includegraphics[width=.27\textwidth, height=.19\textwidth]{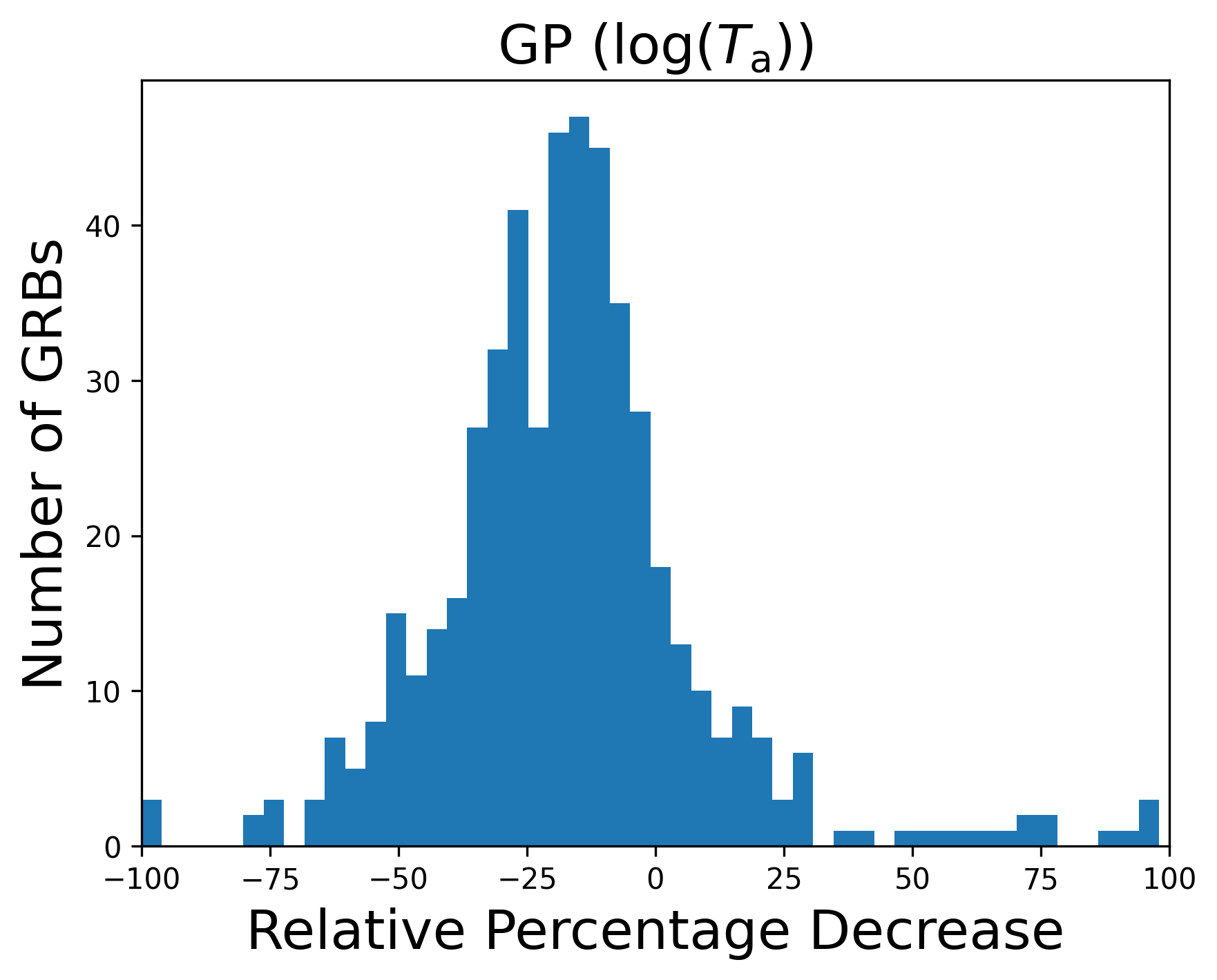}
\includegraphics[width=.27\textwidth, height=.19\textwidth]{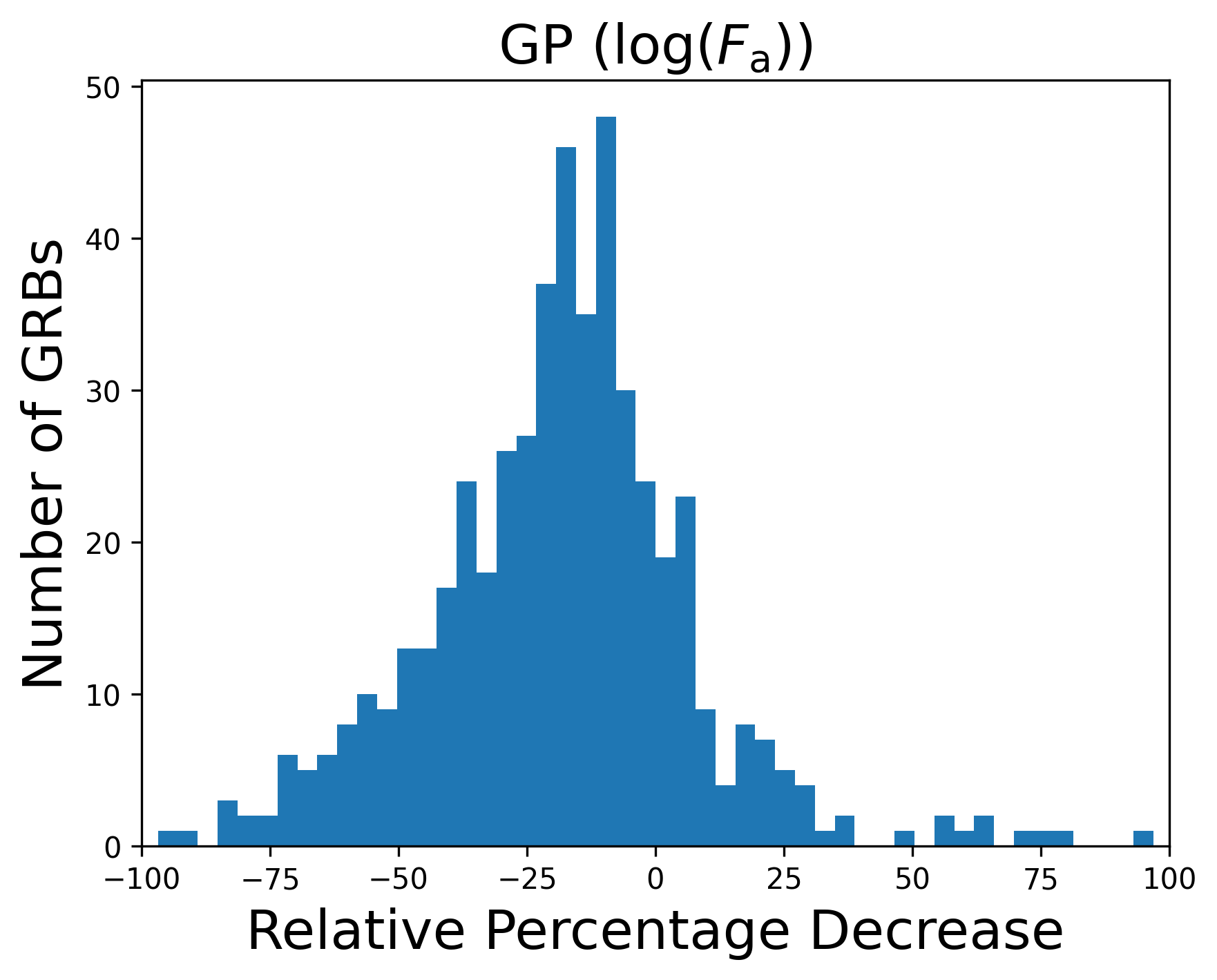}
\includegraphics[width=.27\textwidth, height=.19\textwidth]{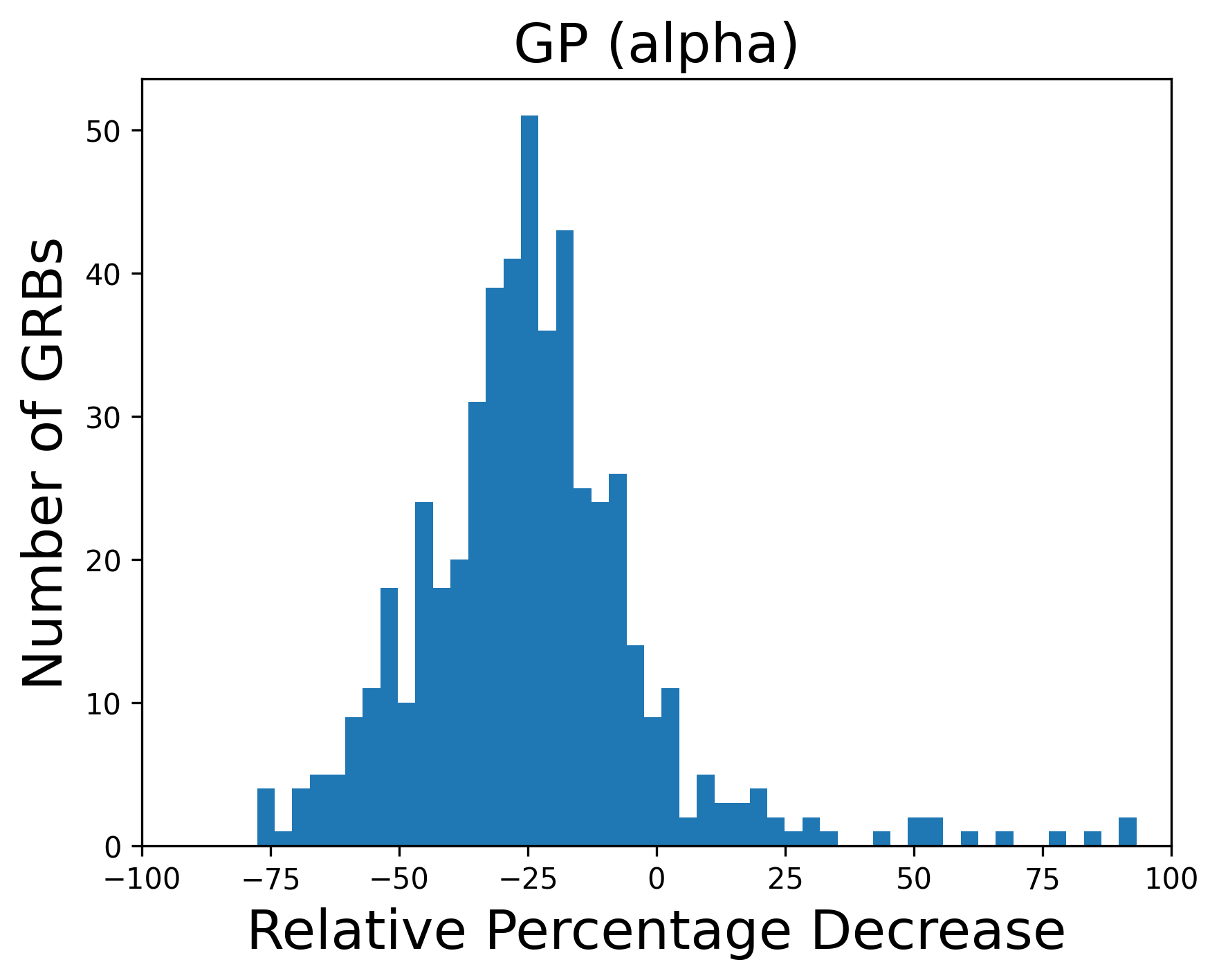}

\includegraphics[width=.27\textwidth, height=.19\textwidth]{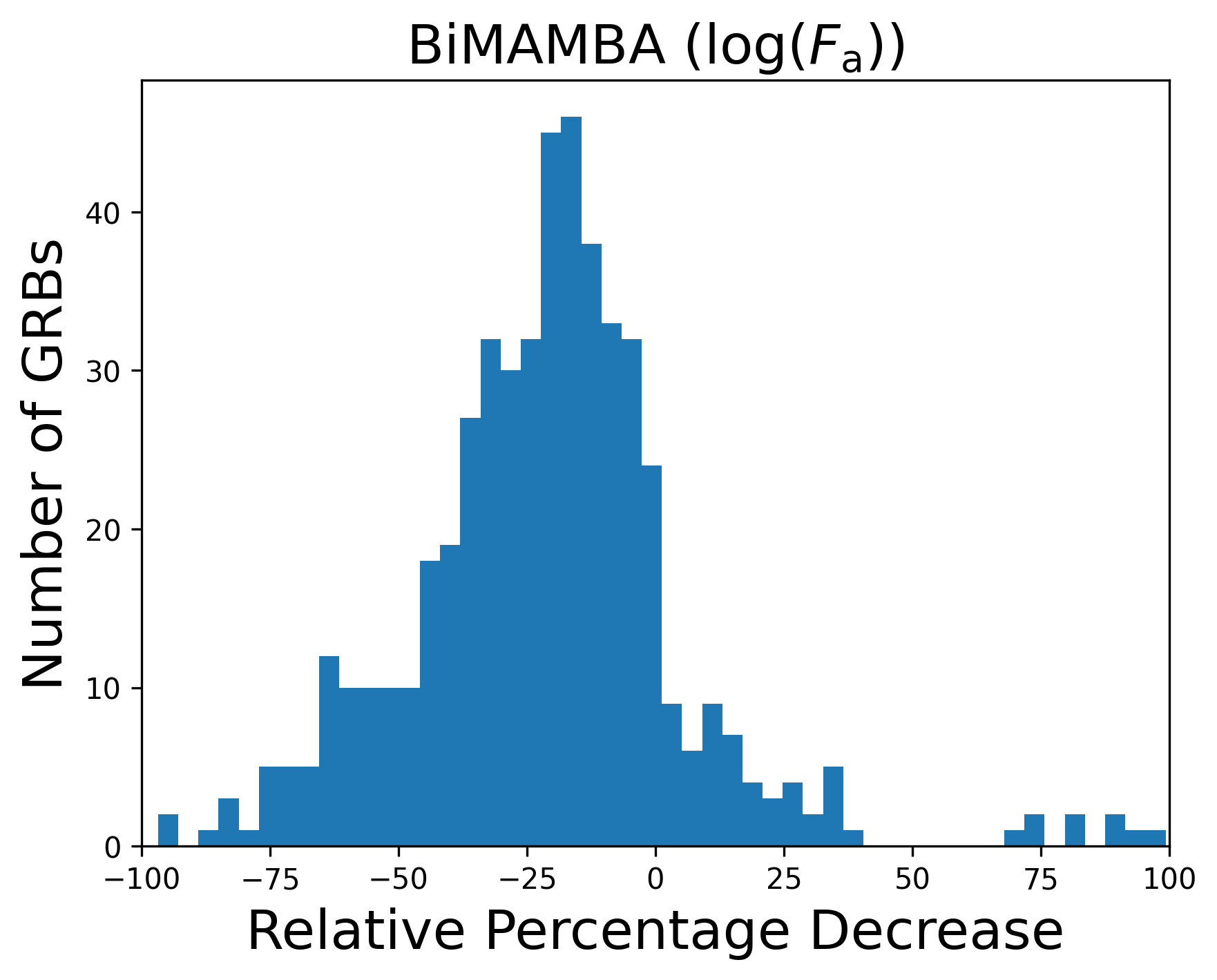}
\includegraphics[width=.27\textwidth, height=.19\textwidth]{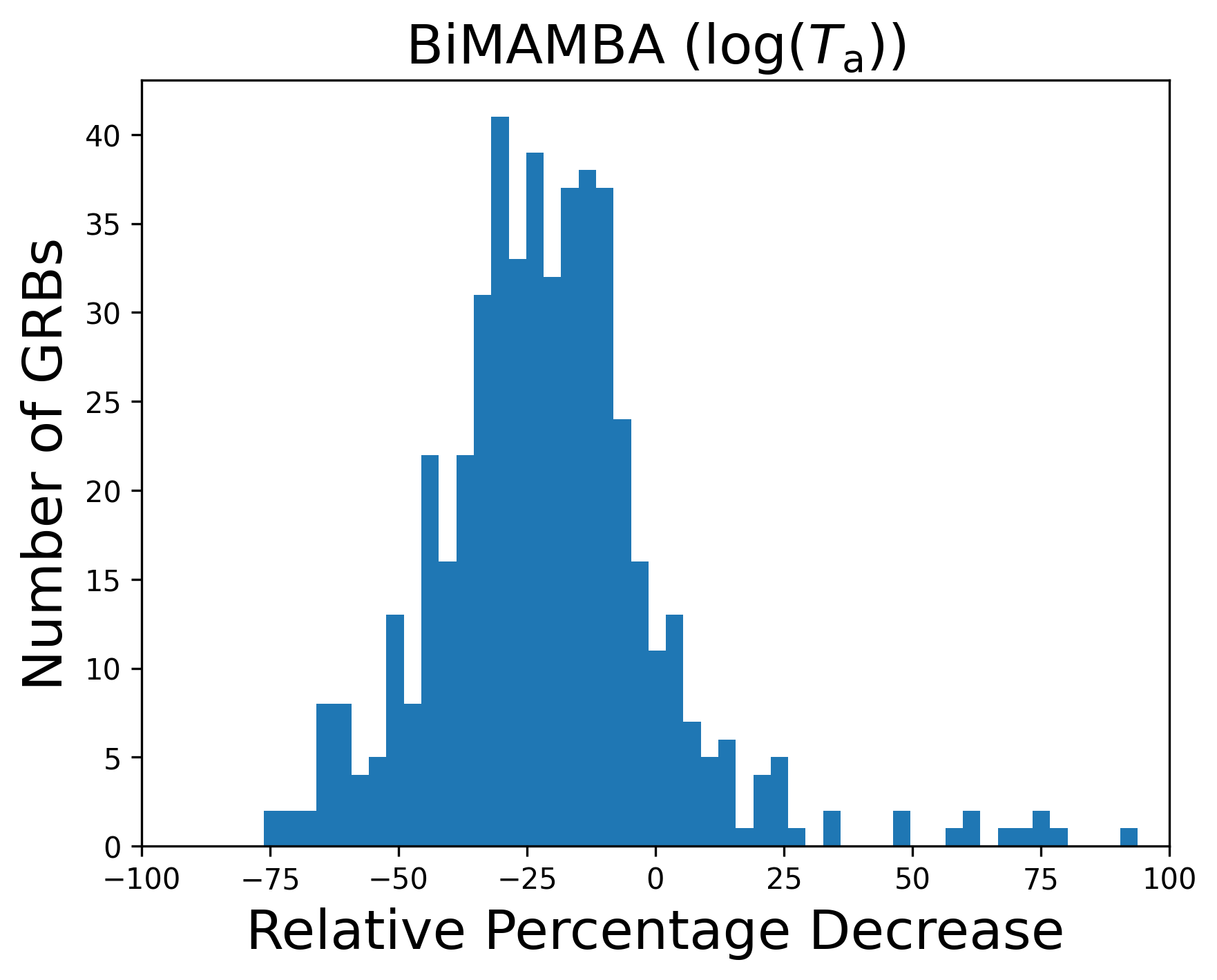}
\includegraphics[width=.27\textwidth, height=.19\textwidth]{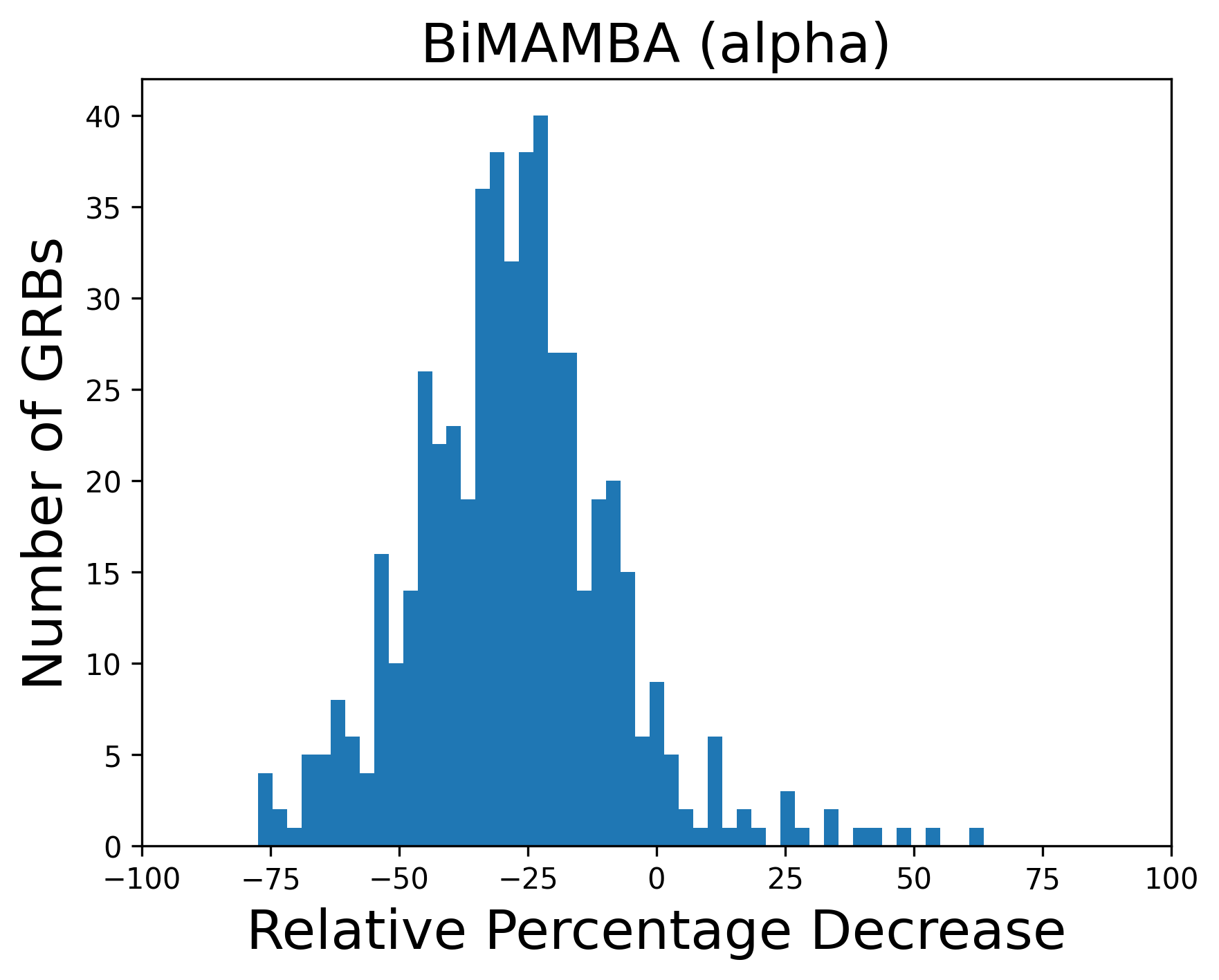}

\includegraphics[width=.27\textwidth, height=.19\textwidth]{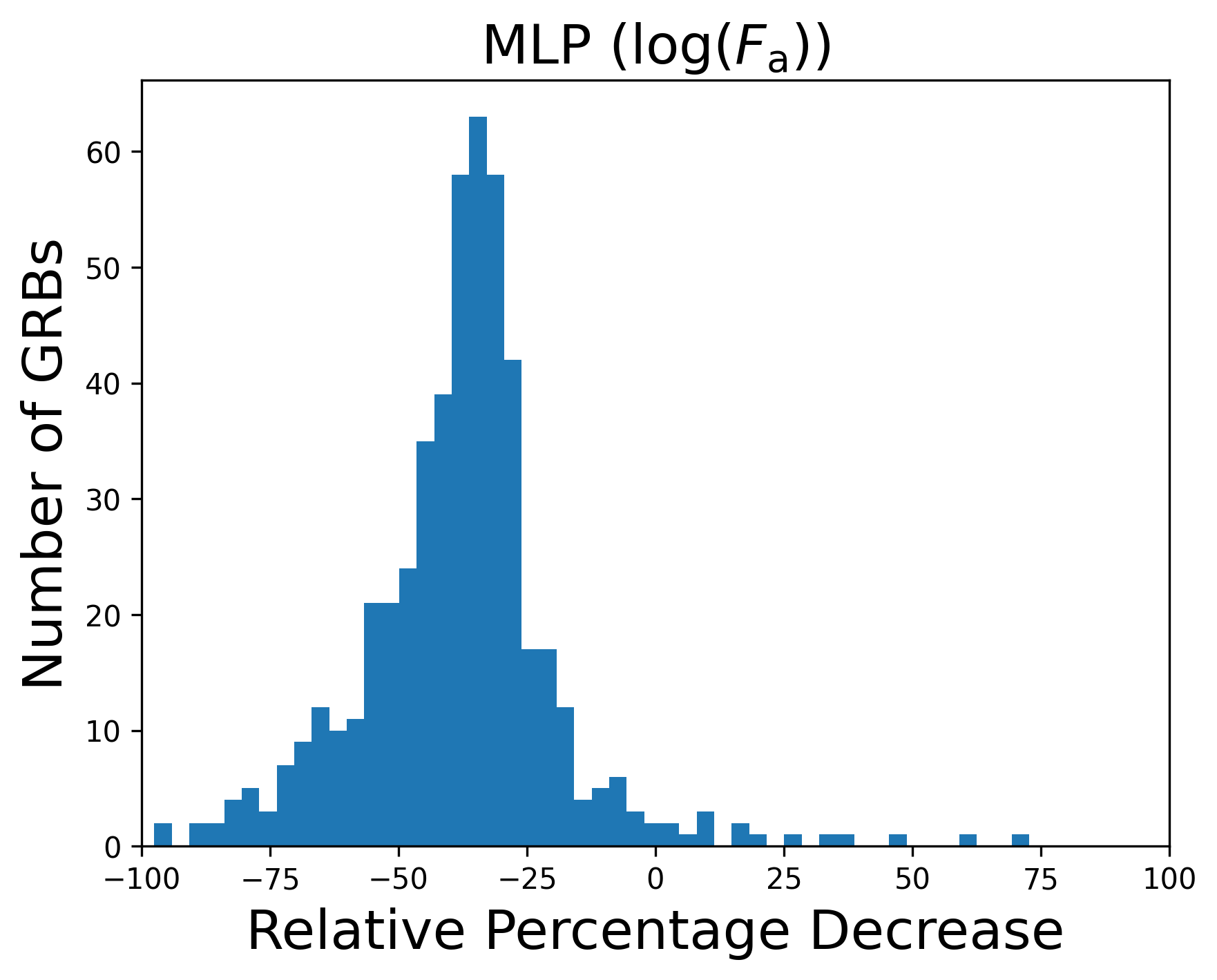}
\includegraphics[width=.27\textwidth, height=.19\textwidth]{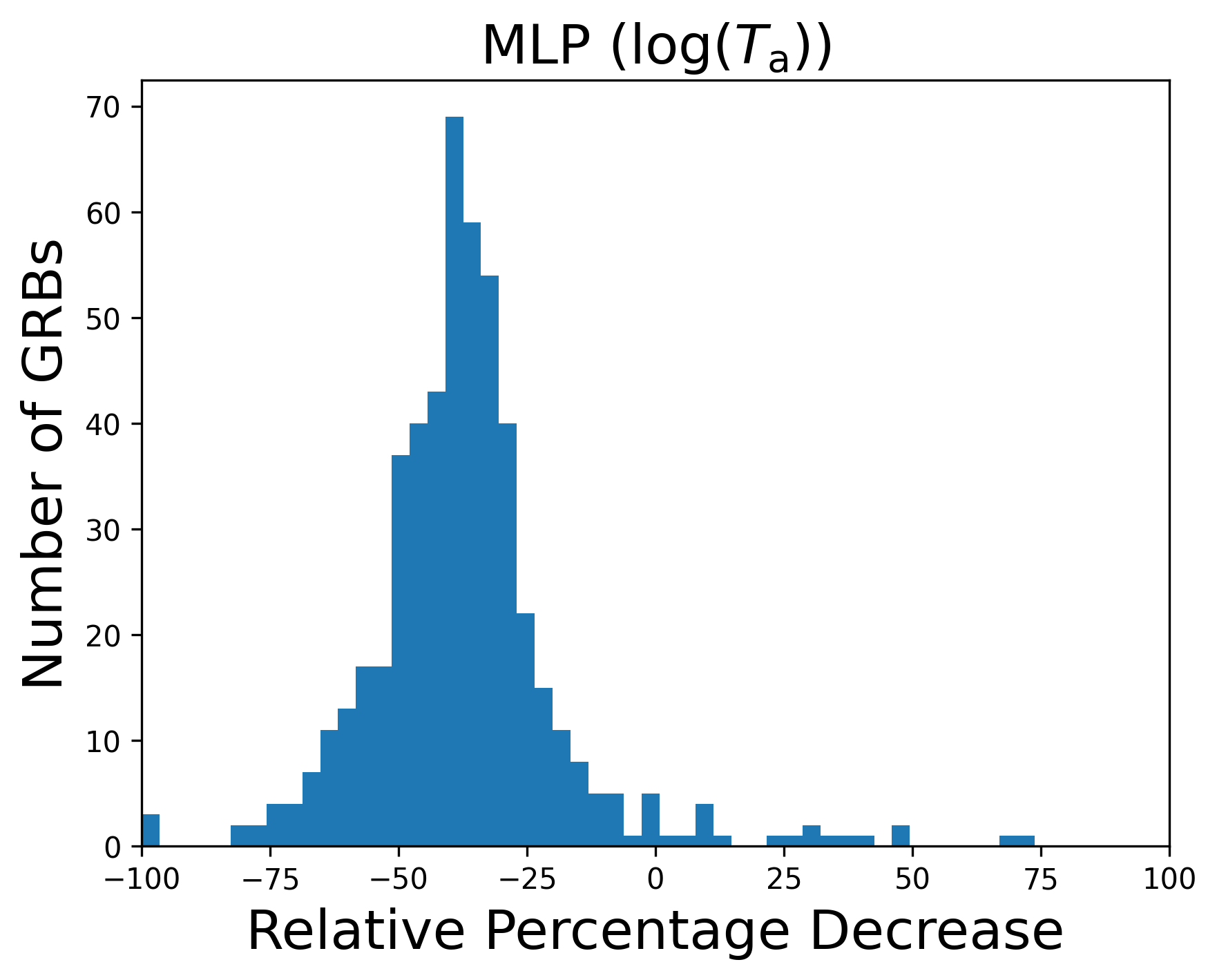}
\includegraphics[width=.27\textwidth, height=.19\textwidth]{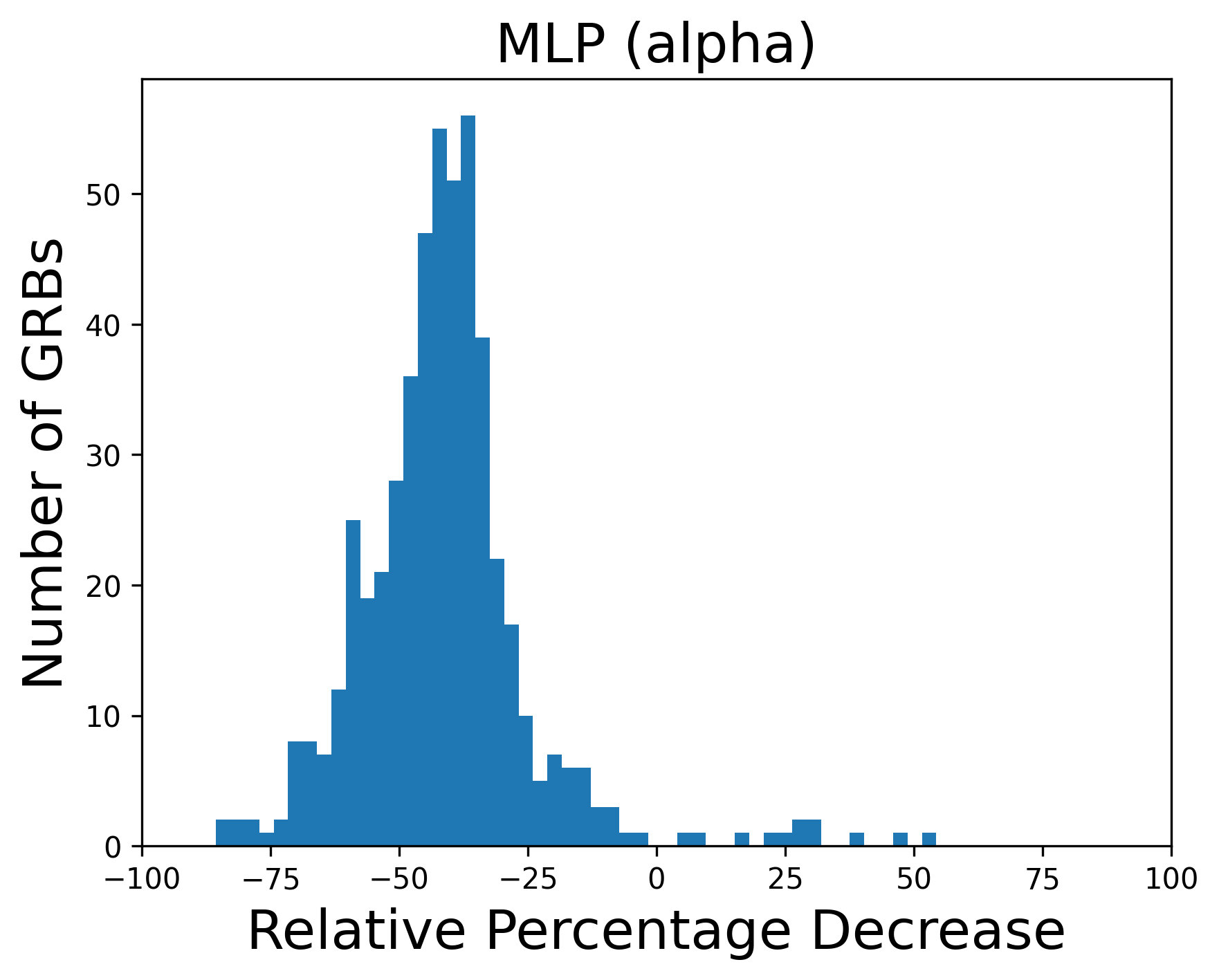}

\includegraphics[width=.27\textwidth, height=.19\textwidth]{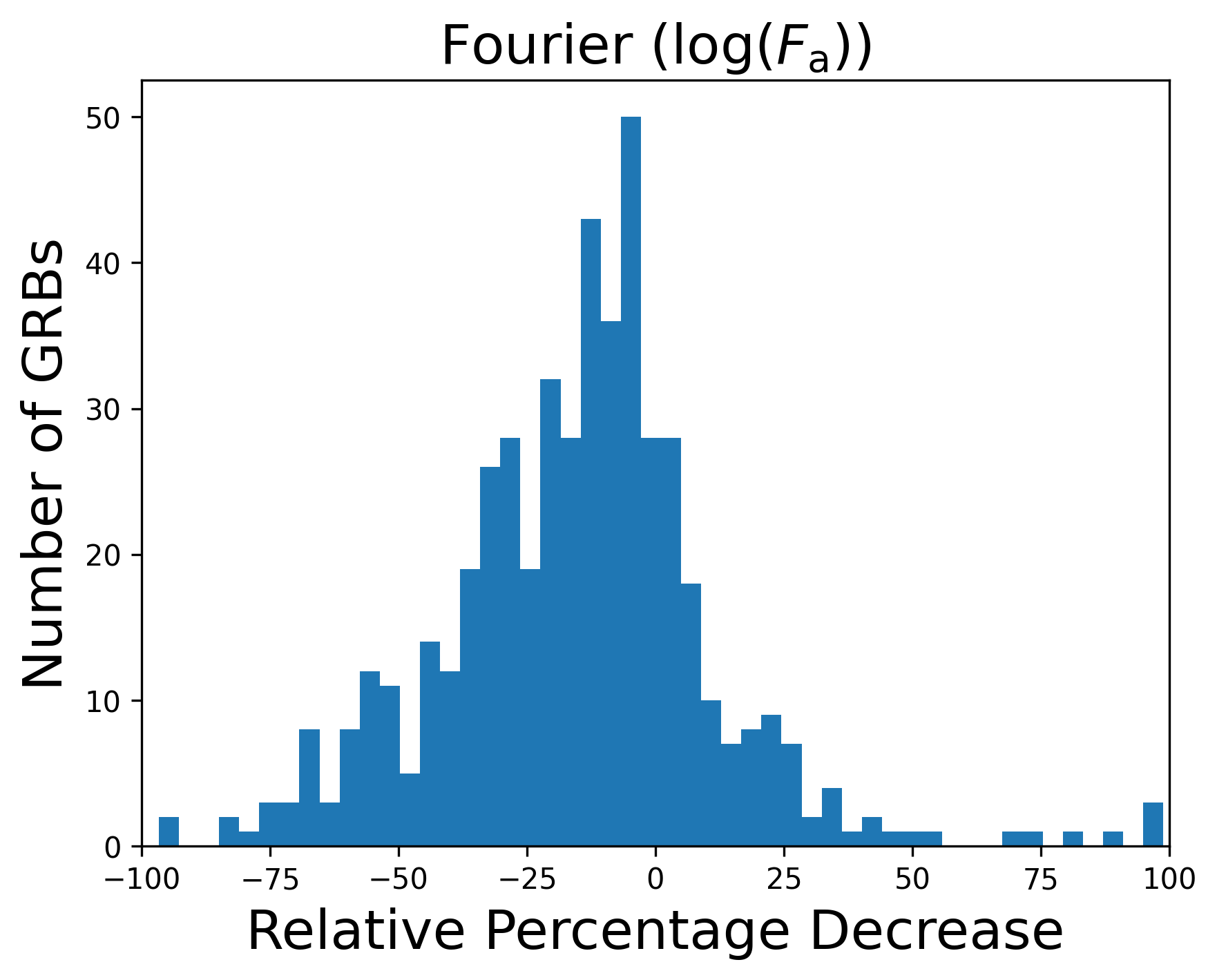}
\includegraphics[width=.27\textwidth, height=.19\textwidth]{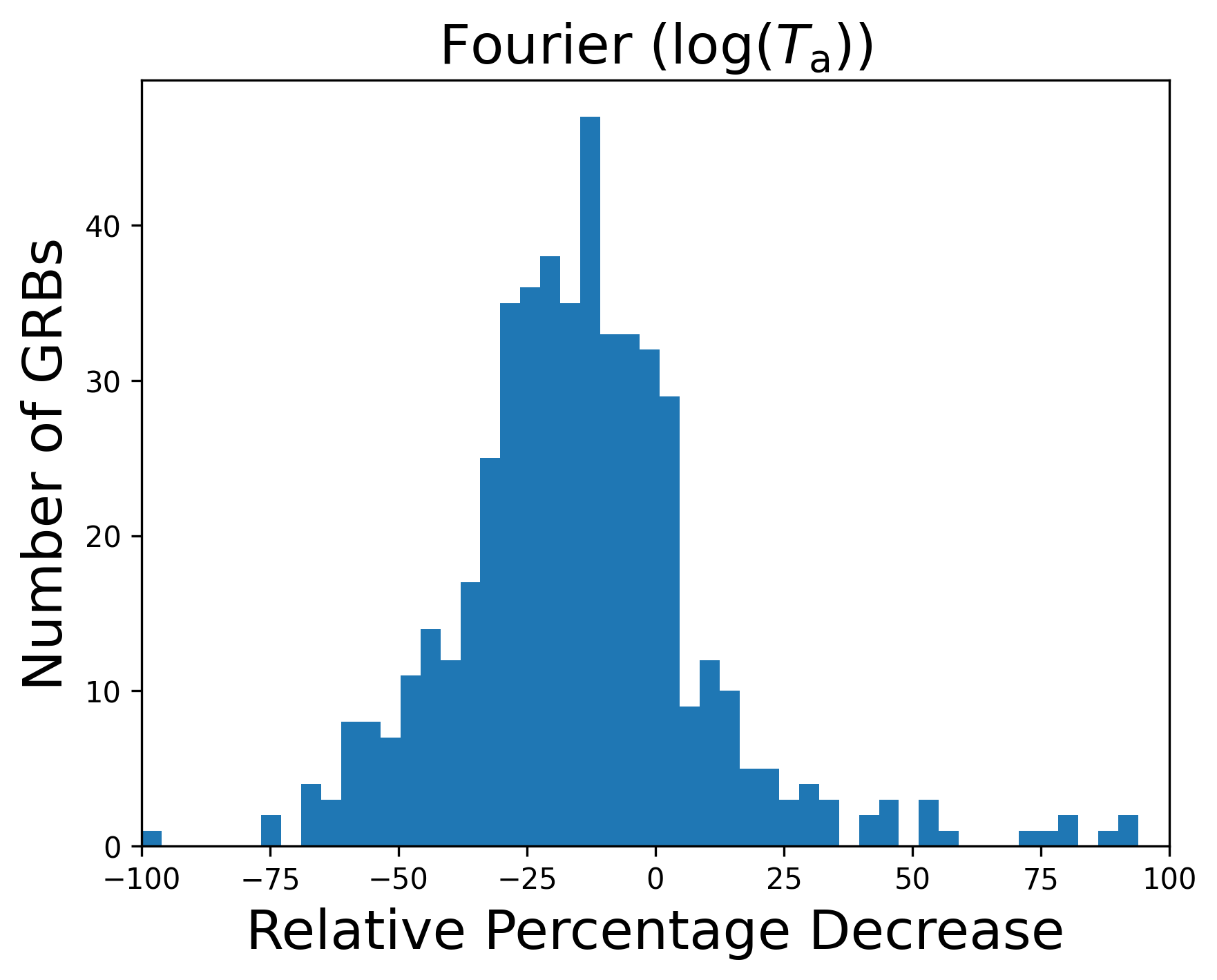}
\includegraphics[width=.27\textwidth, height=.19\textwidth]{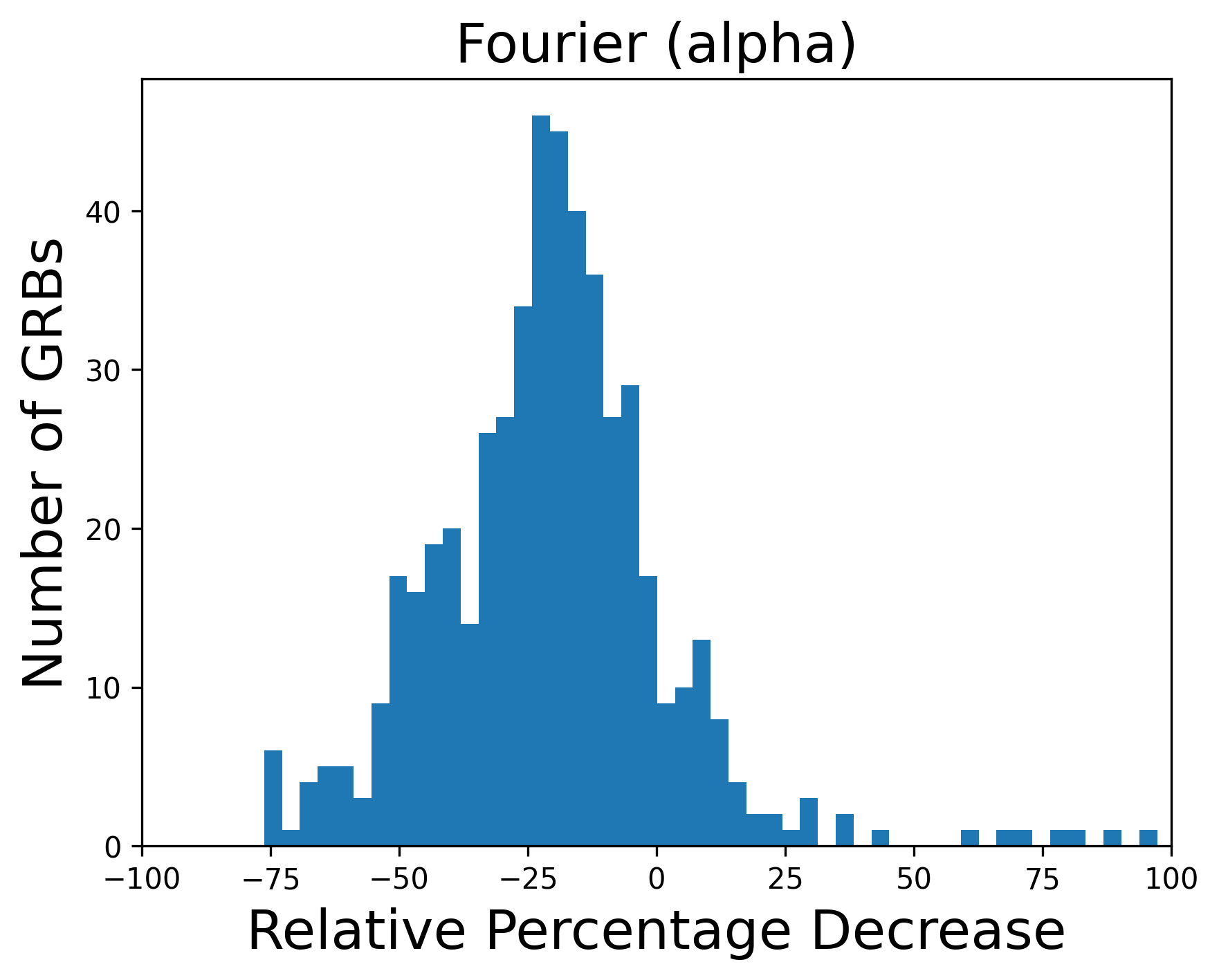}
\end{center}
\caption{Distribution plot of all three W07 parameters in a grid with parameters (left to right): i) $\log F_a$ (Column 1) ii) $\log T_a$ (Column 2) iii) $\alpha$ (Column 3) and the models (top to bottom): i) Willingale (10\% noise) (Row 1); ii) W07 (20\% noise) (Row 2); iii) GP Model(Row 3); iv) Bi-Mamba Model (Row 4); v) MLP Model (Row 5); vi) Fourier (Row 6).}

\label{fig: ALL-results-1} 
\end{figure*}

\begin{figure*}[htbp]
\begin{center}

\includegraphics[width=.27\textwidth, height=.19\textwidth]{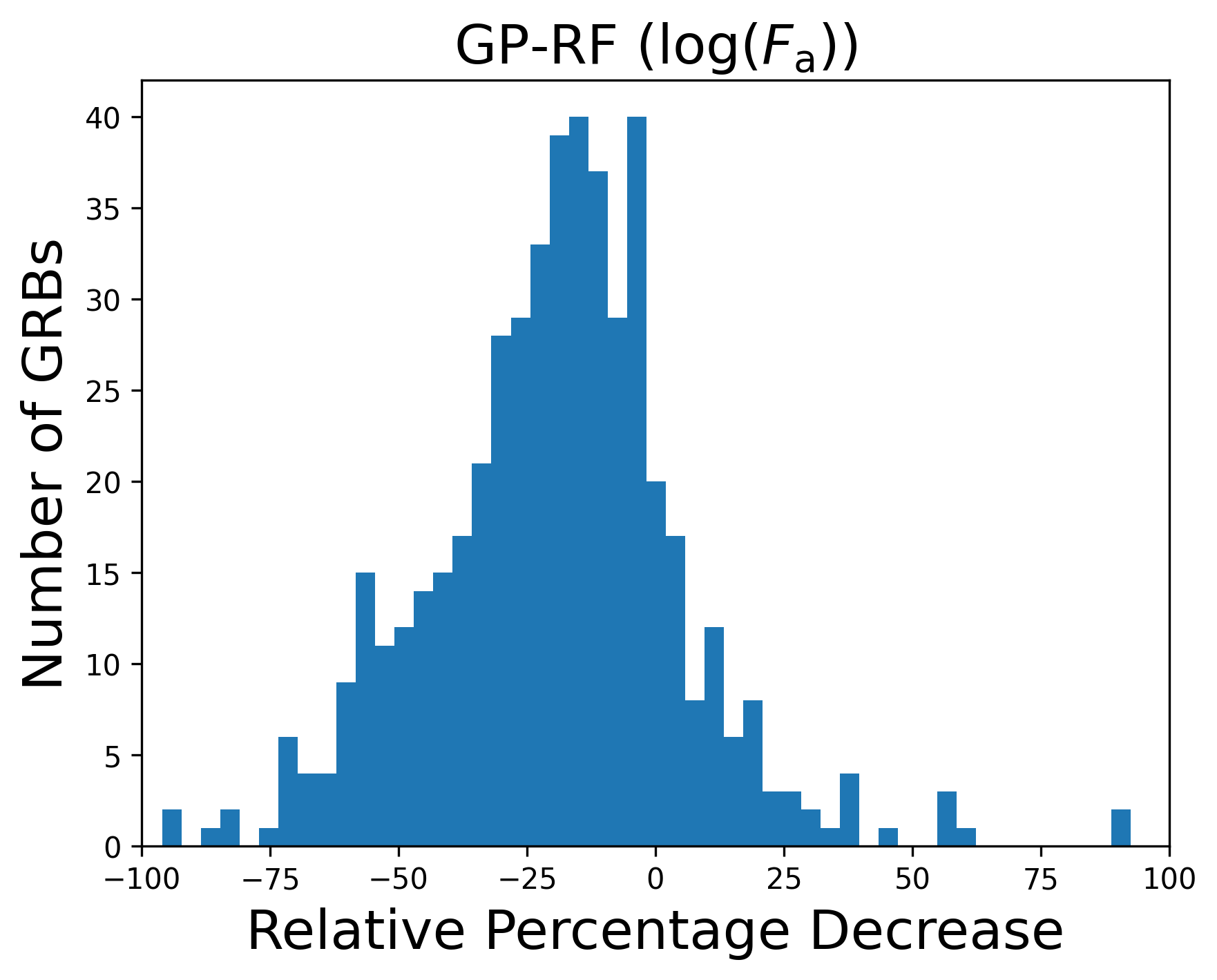}
\includegraphics[width=.27\textwidth, height=.19\textwidth]{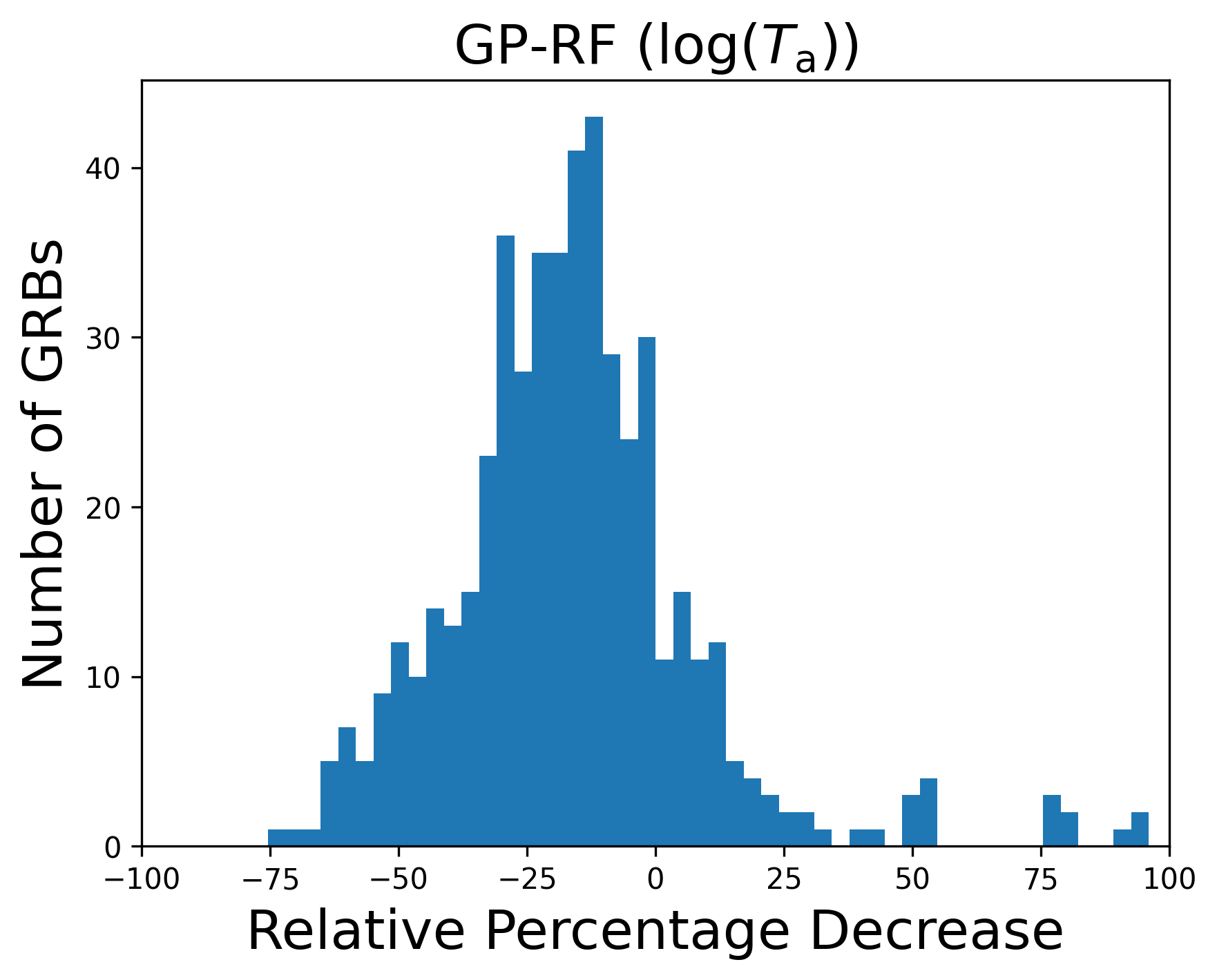}
\includegraphics[width=.27\textwidth, height=.19\textwidth]{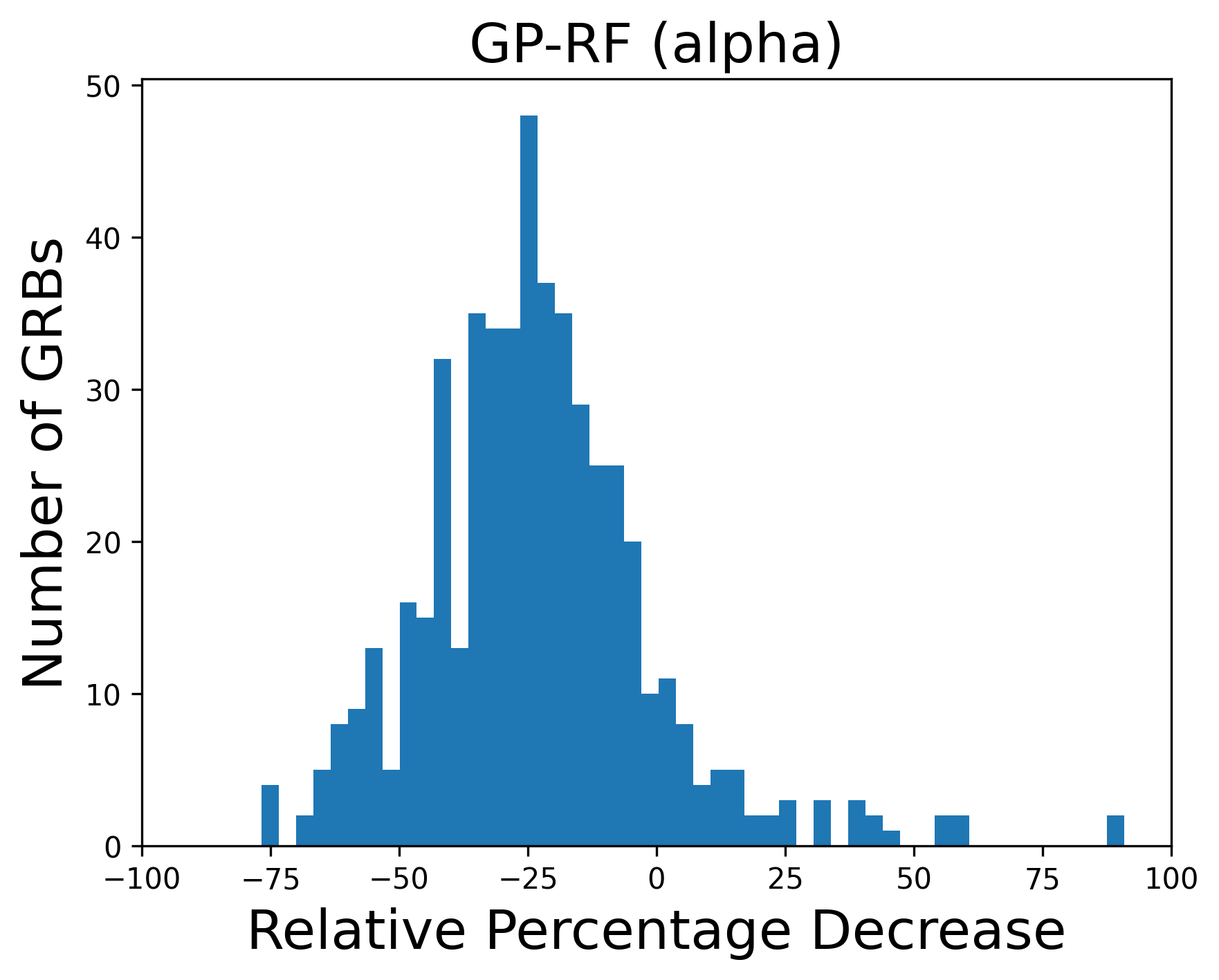}

\includegraphics[width=.27\textwidth, height=.19\textwidth]{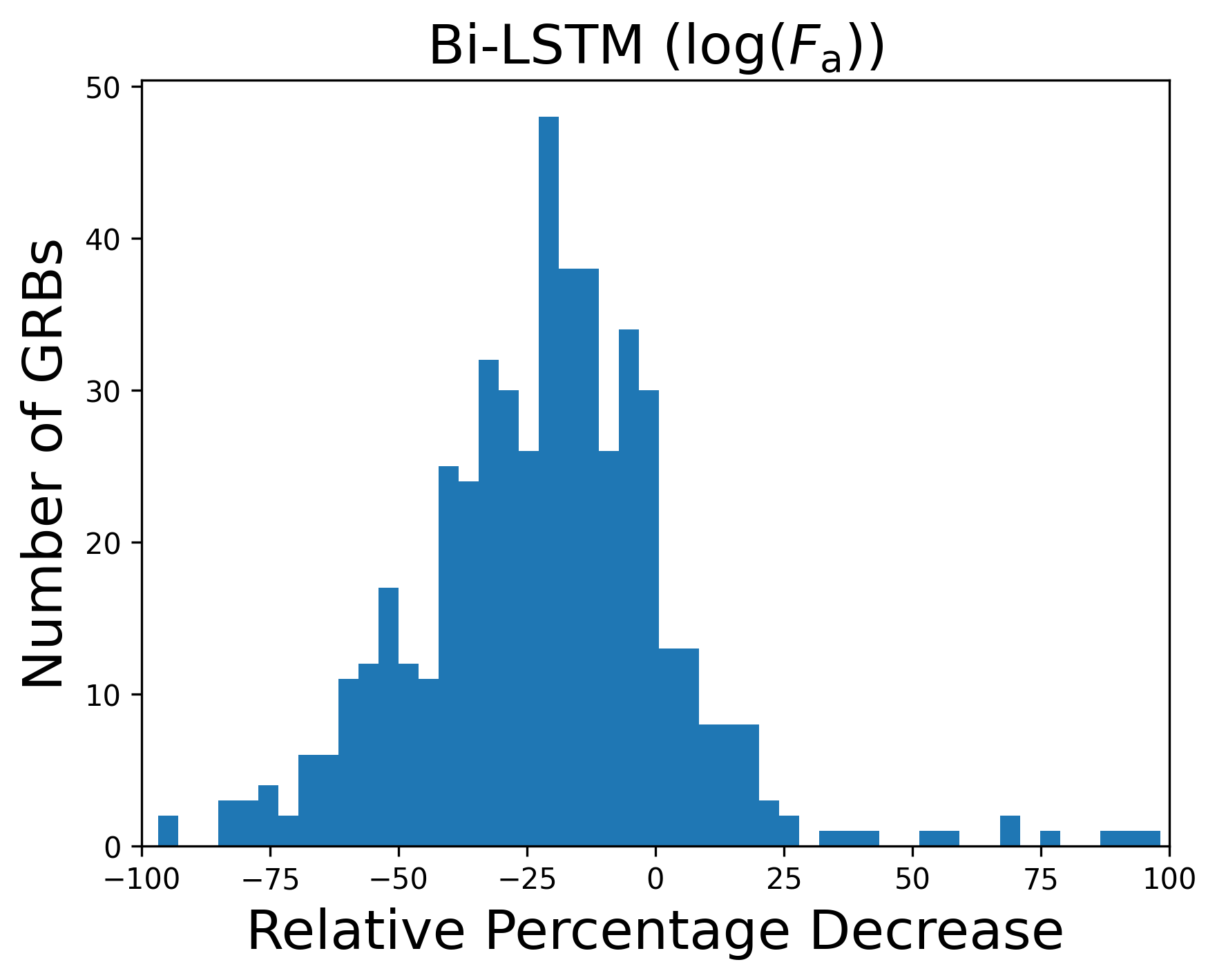}
\includegraphics[width=.27\textwidth, height=.19\textwidth]{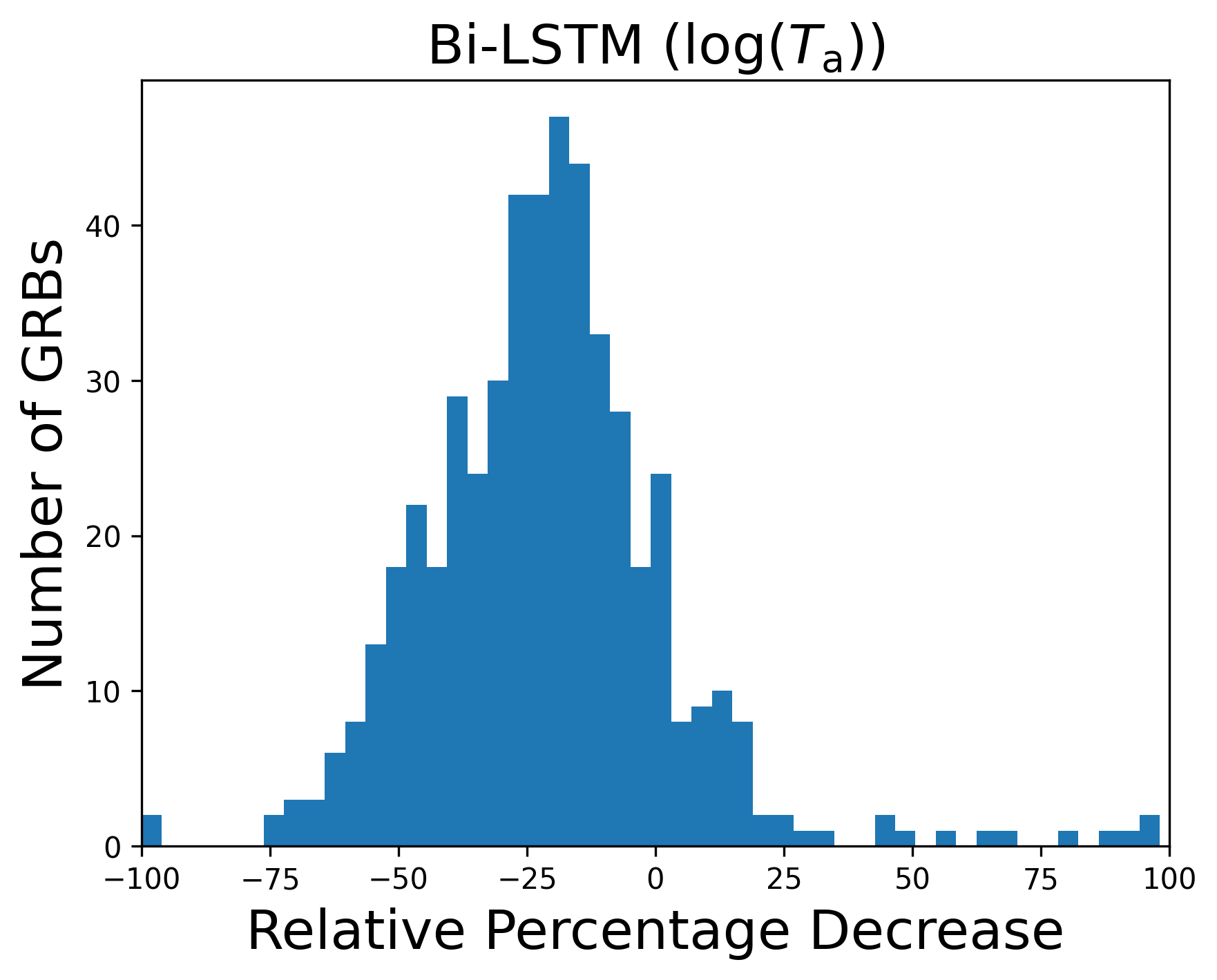}
\includegraphics[width=.27\textwidth, height=.19\textwidth]{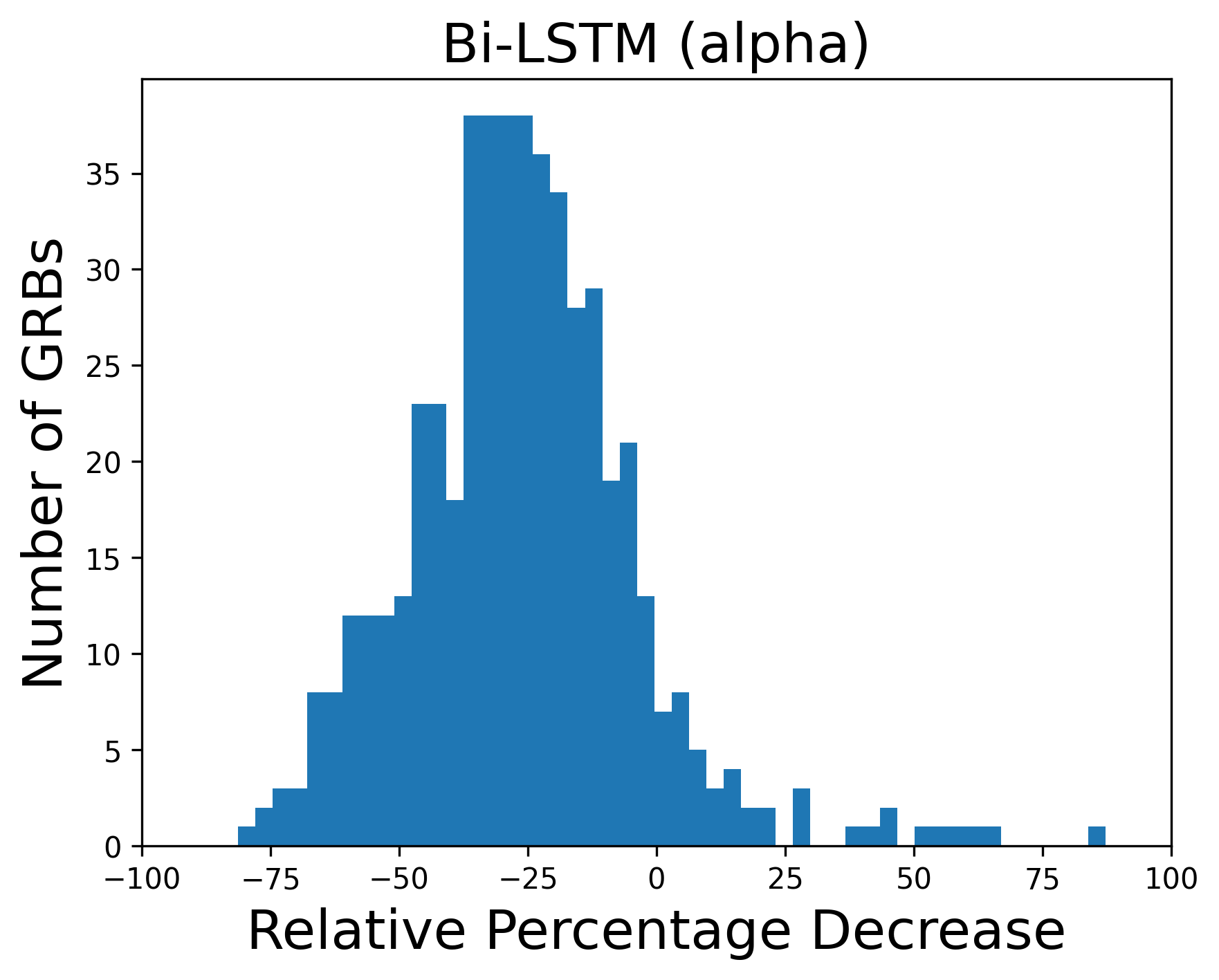}

\includegraphics[width=.27\textwidth, height=.19\textwidth]{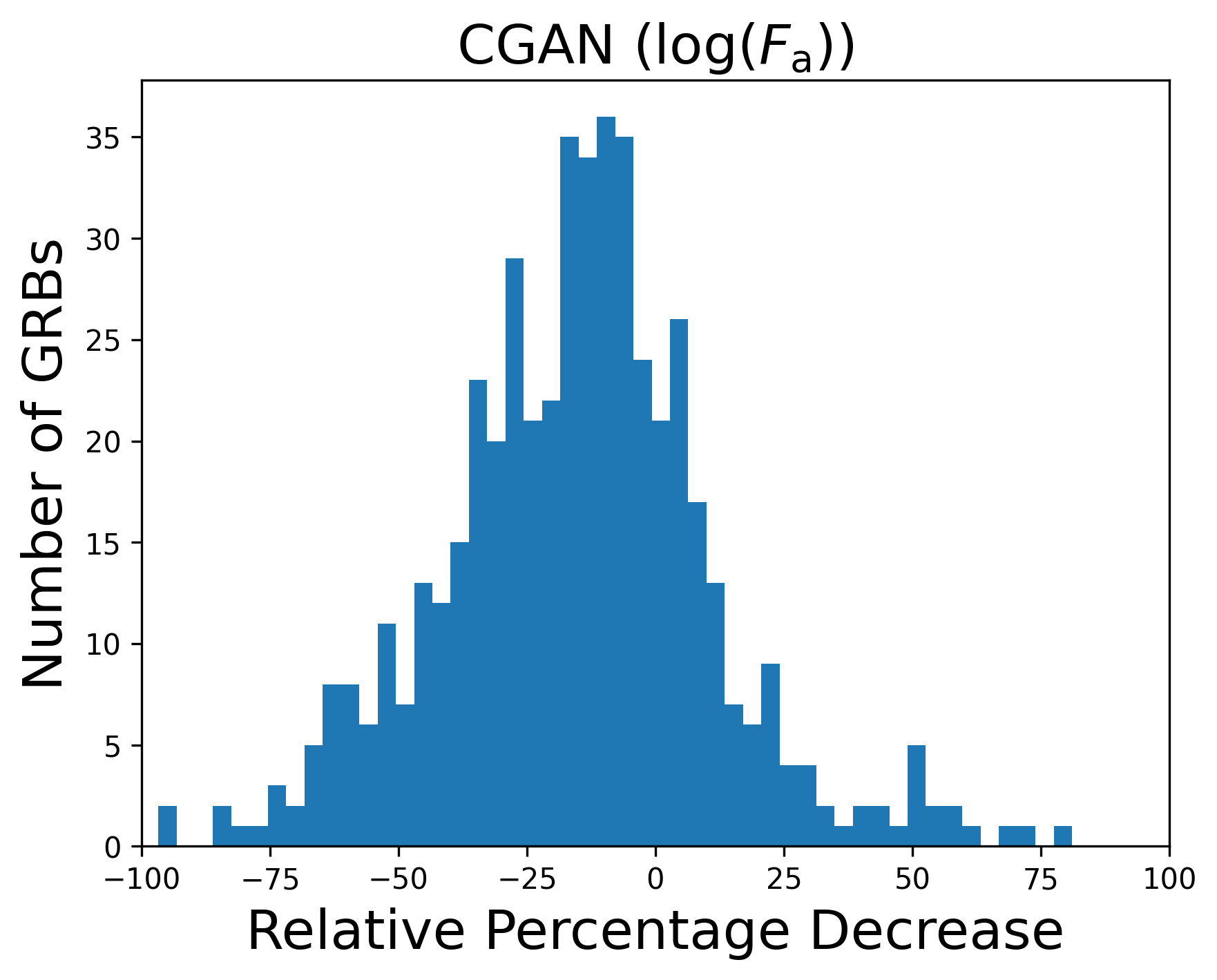}
\includegraphics[width=.27\textwidth, height=.19\textwidth]{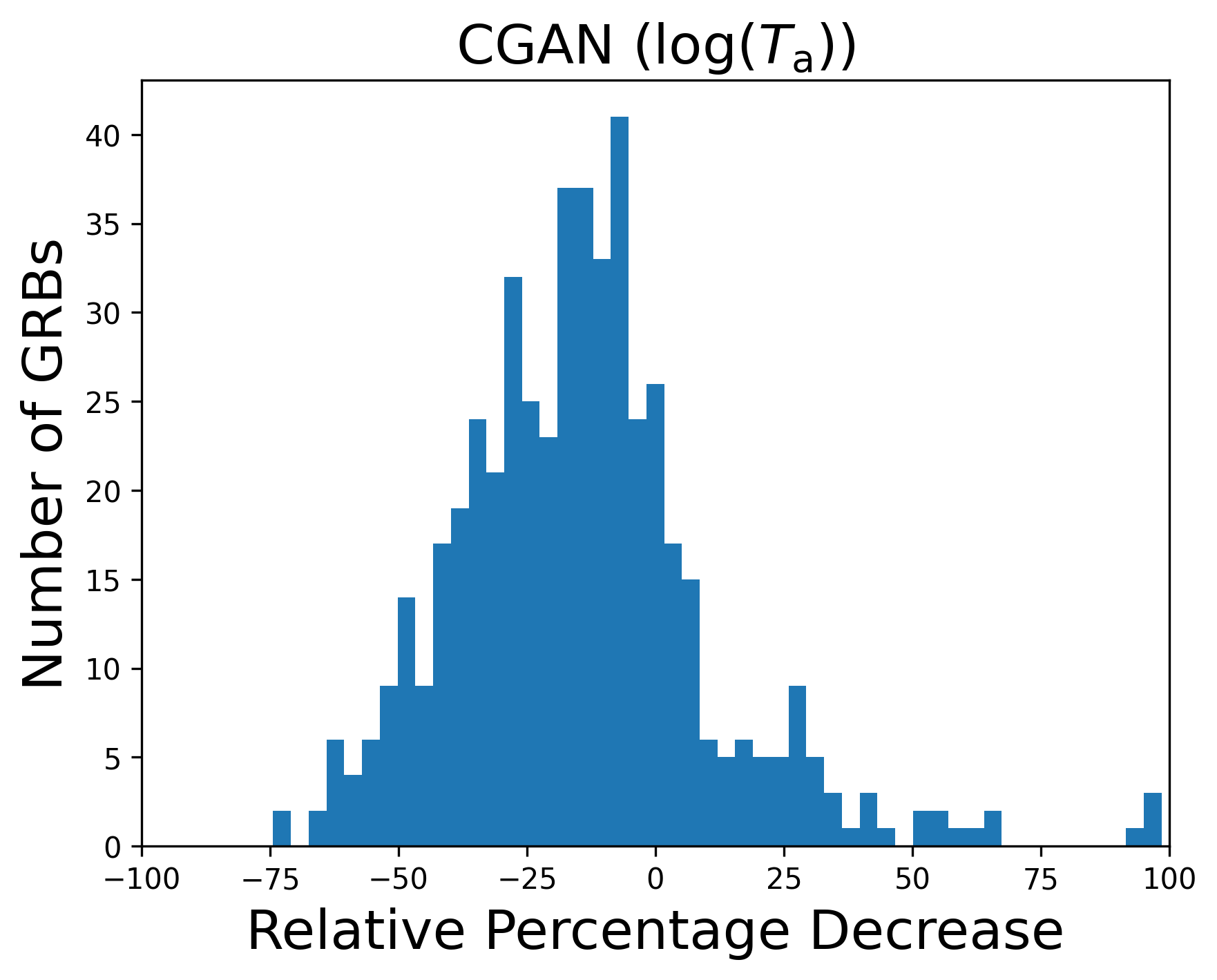}
\includegraphics[width=.27\textwidth, height=.19\textwidth]{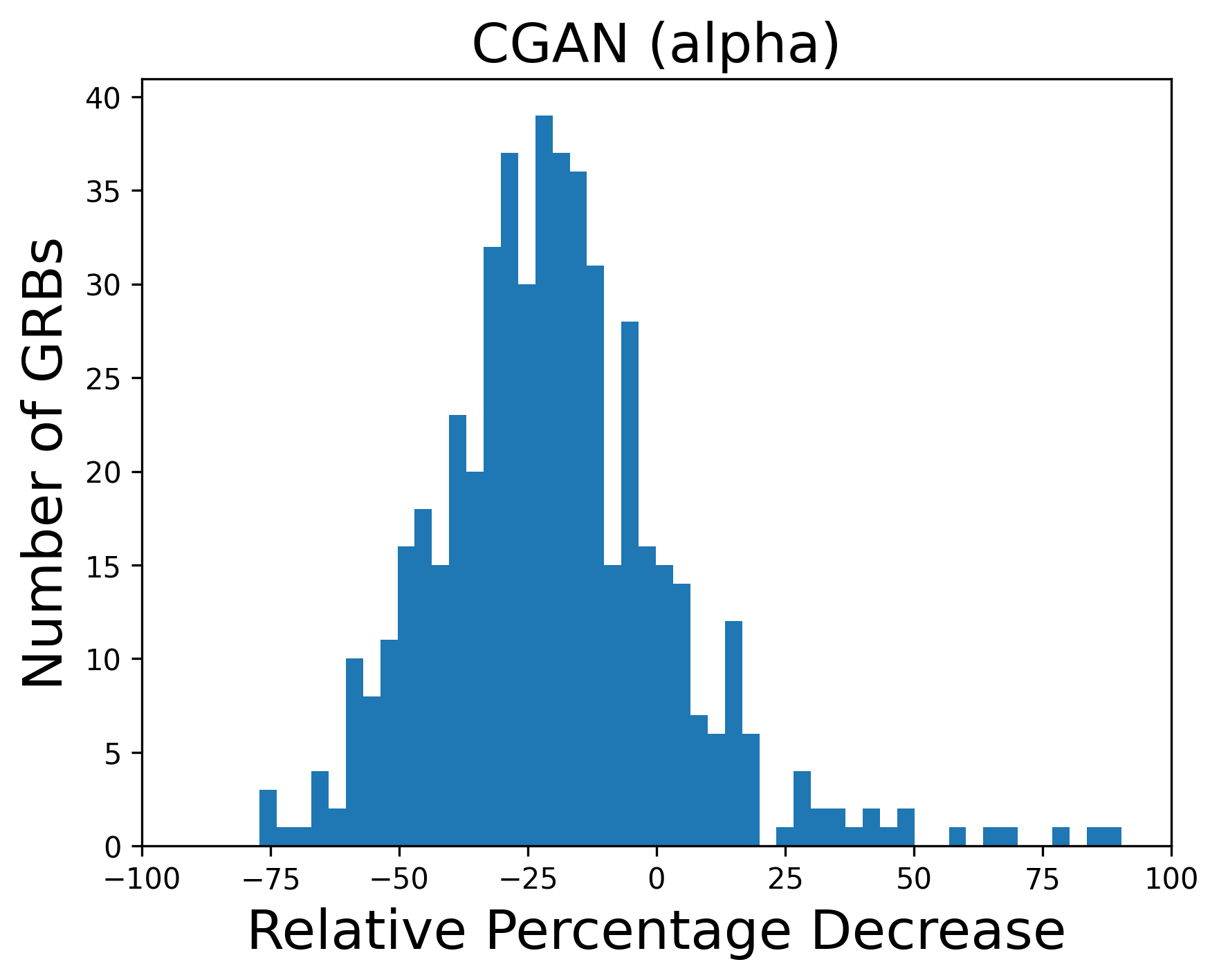}

\includegraphics[width=.27\textwidth, height=.19\textwidth]{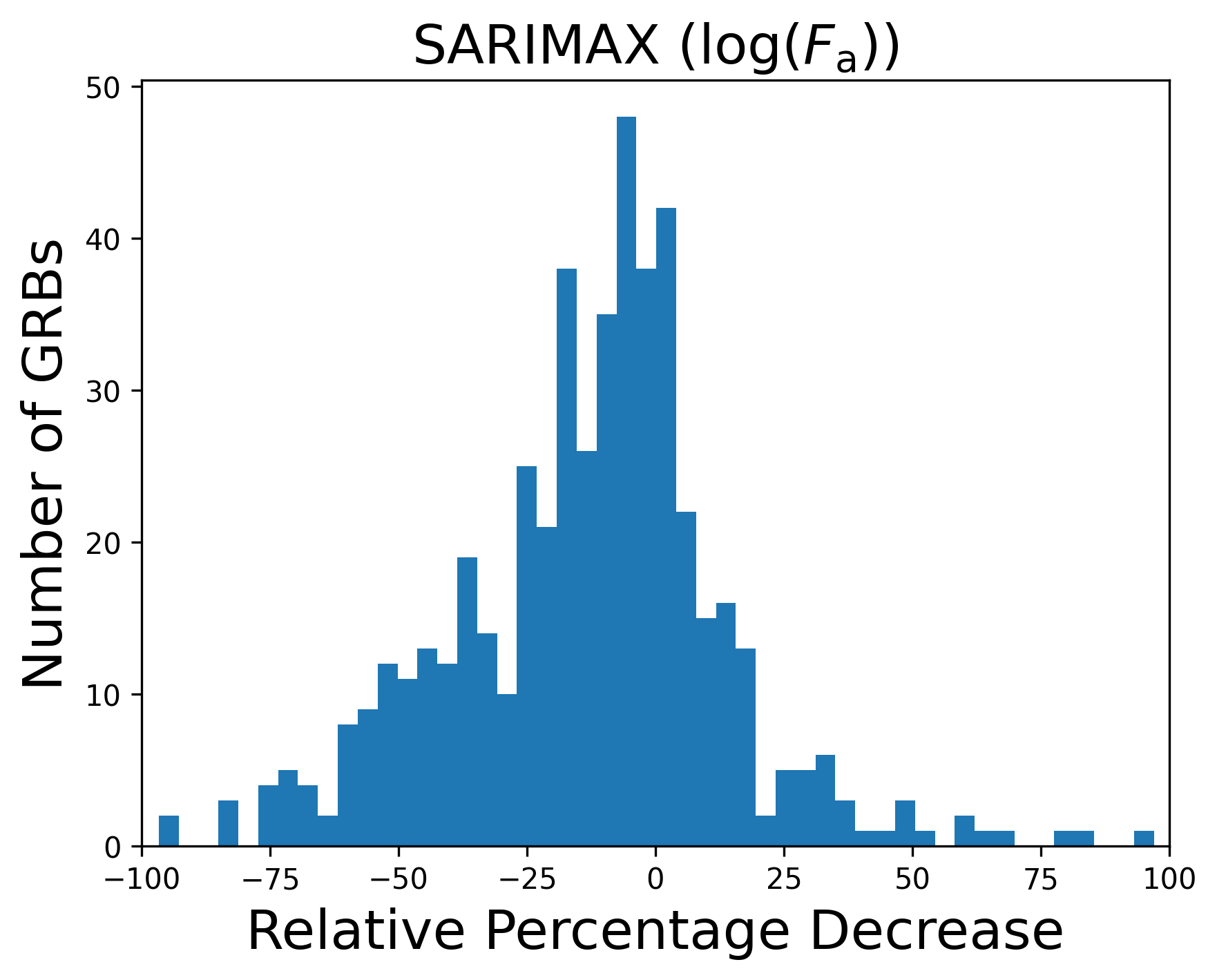}
\includegraphics[width=.27\textwidth, height=.19\textwidth]{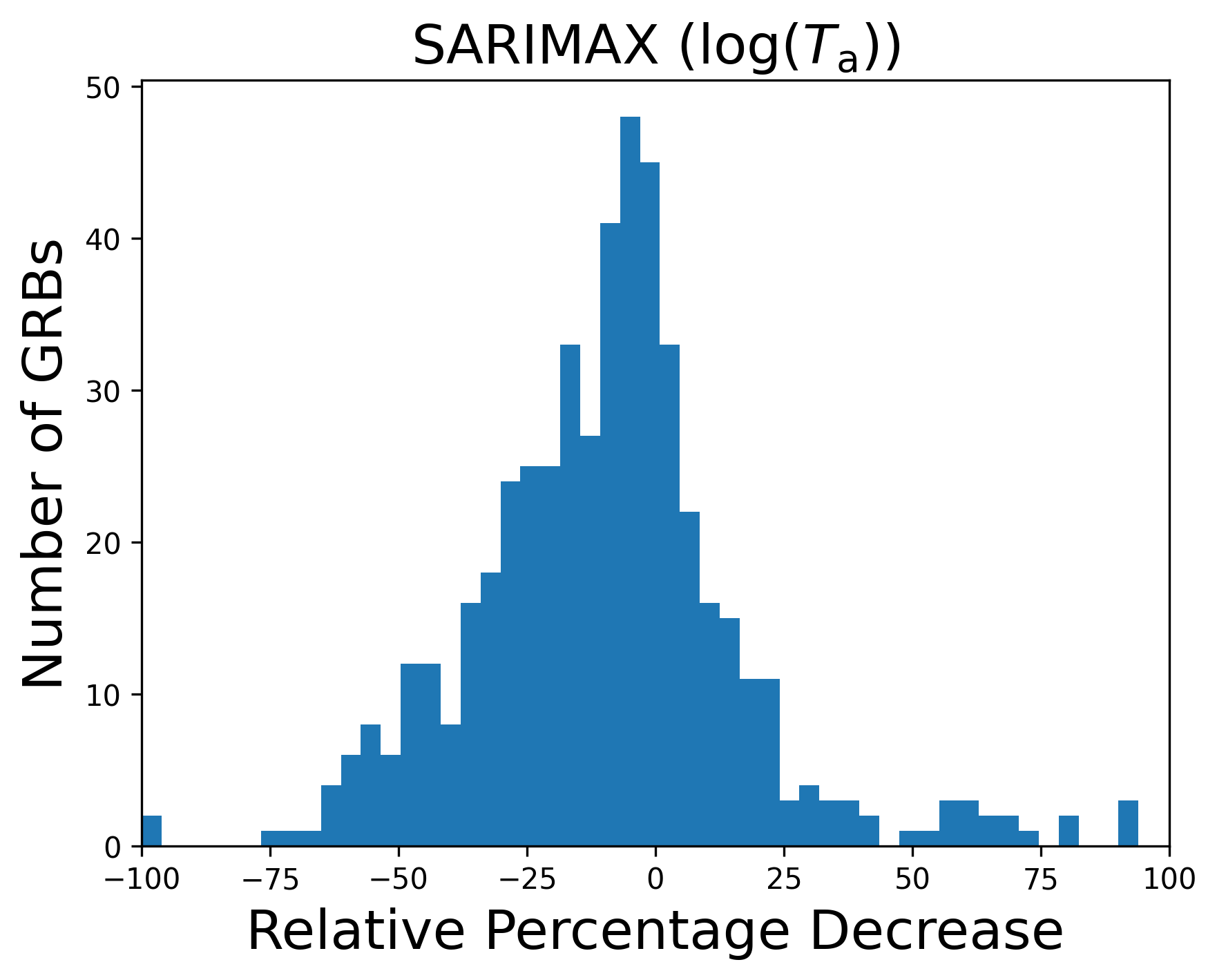}
\includegraphics[width=.27\textwidth, height=.19\textwidth]{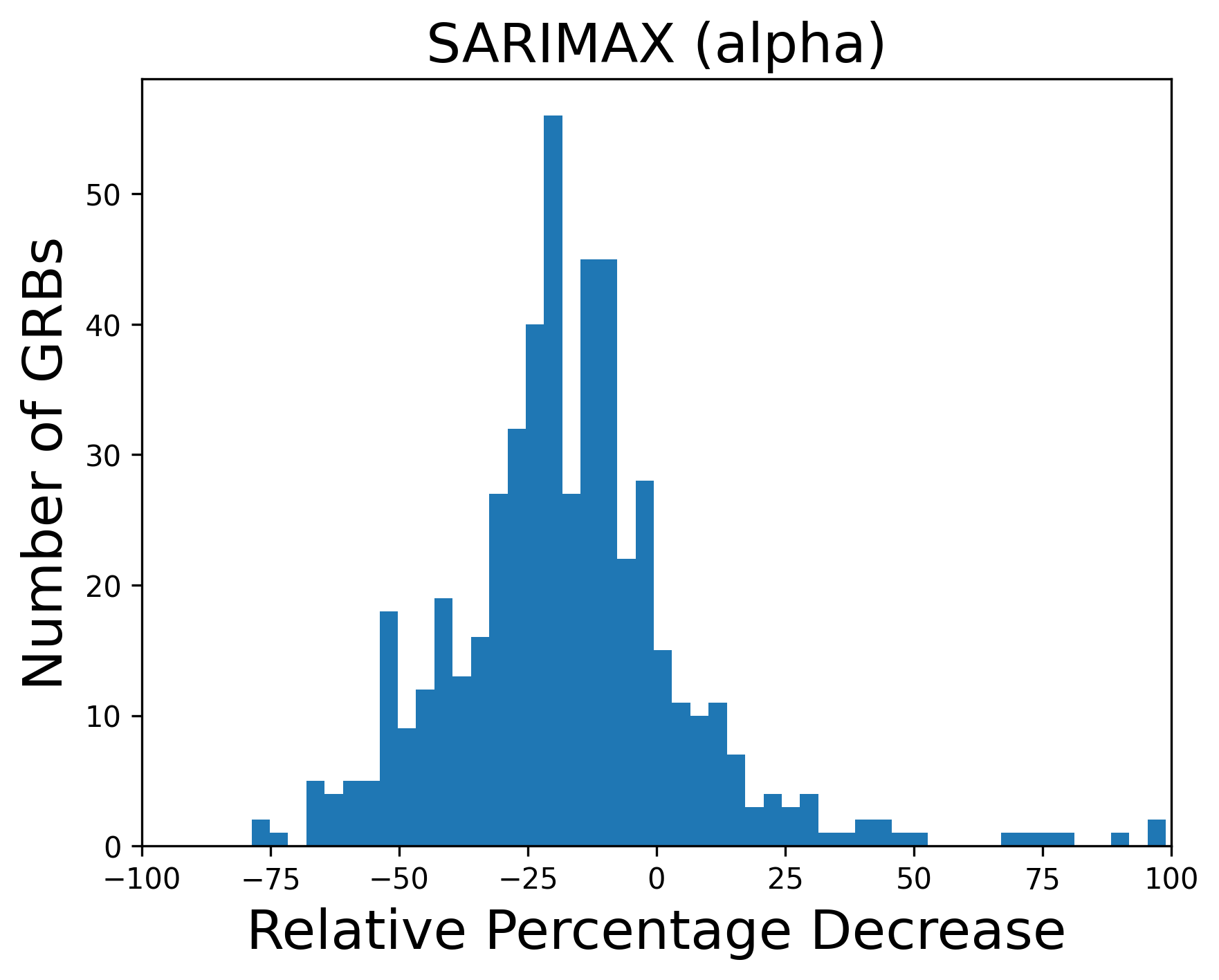}

\includegraphics[width=.27\textwidth, height=.19\textwidth]{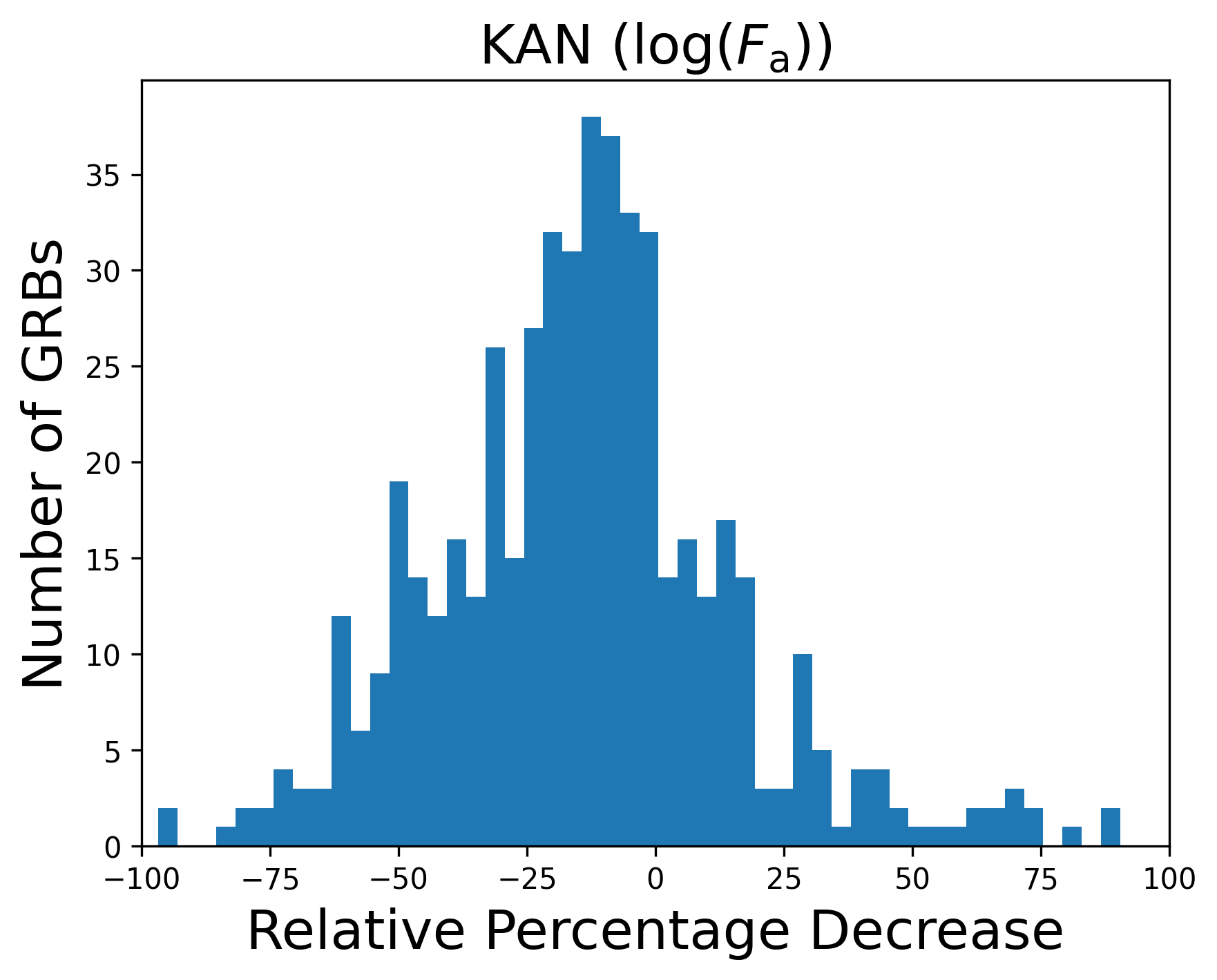}
\includegraphics[width=.27\textwidth, height=.19\textwidth]{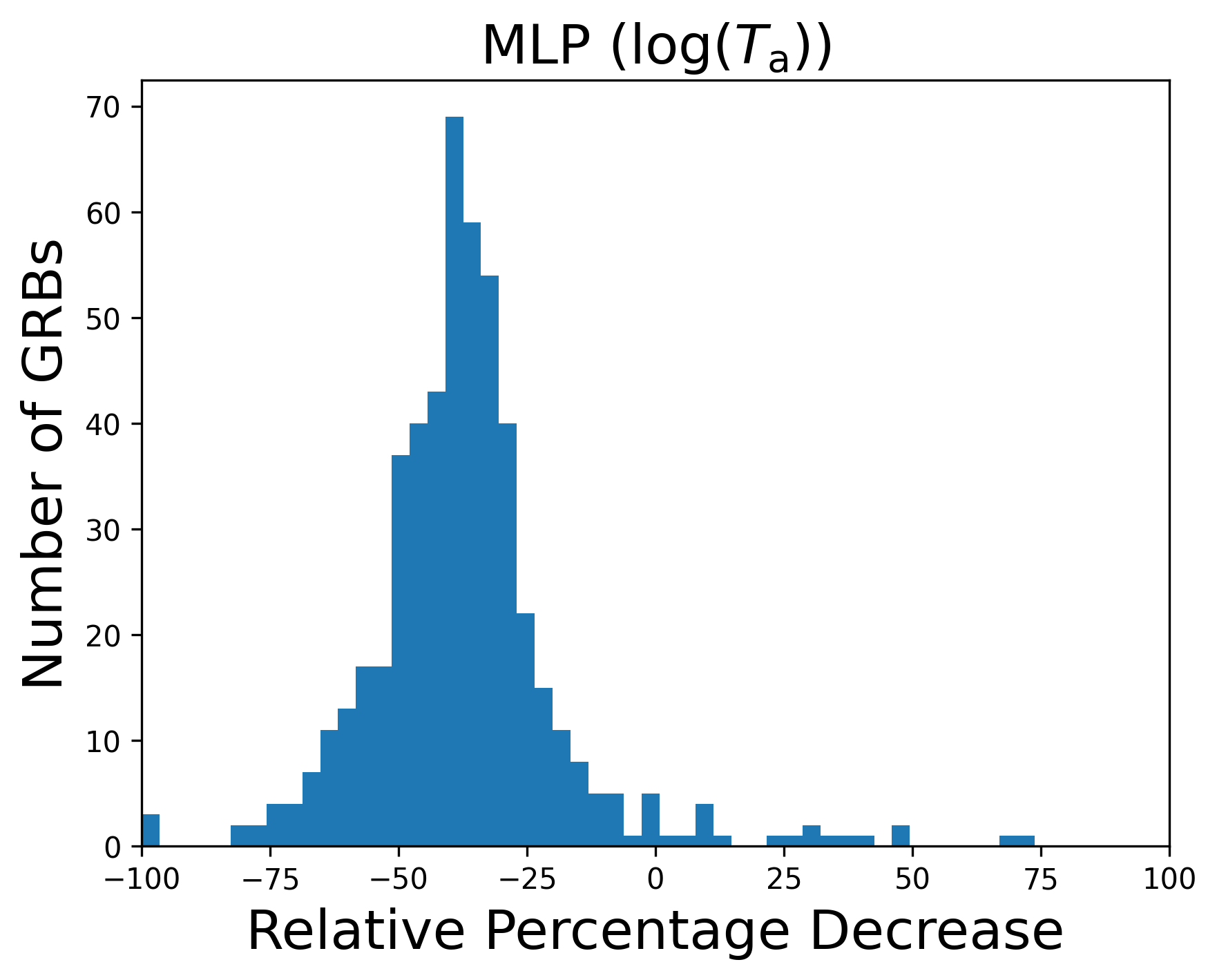}
\includegraphics[width=.27\textwidth, height=.19\textwidth]{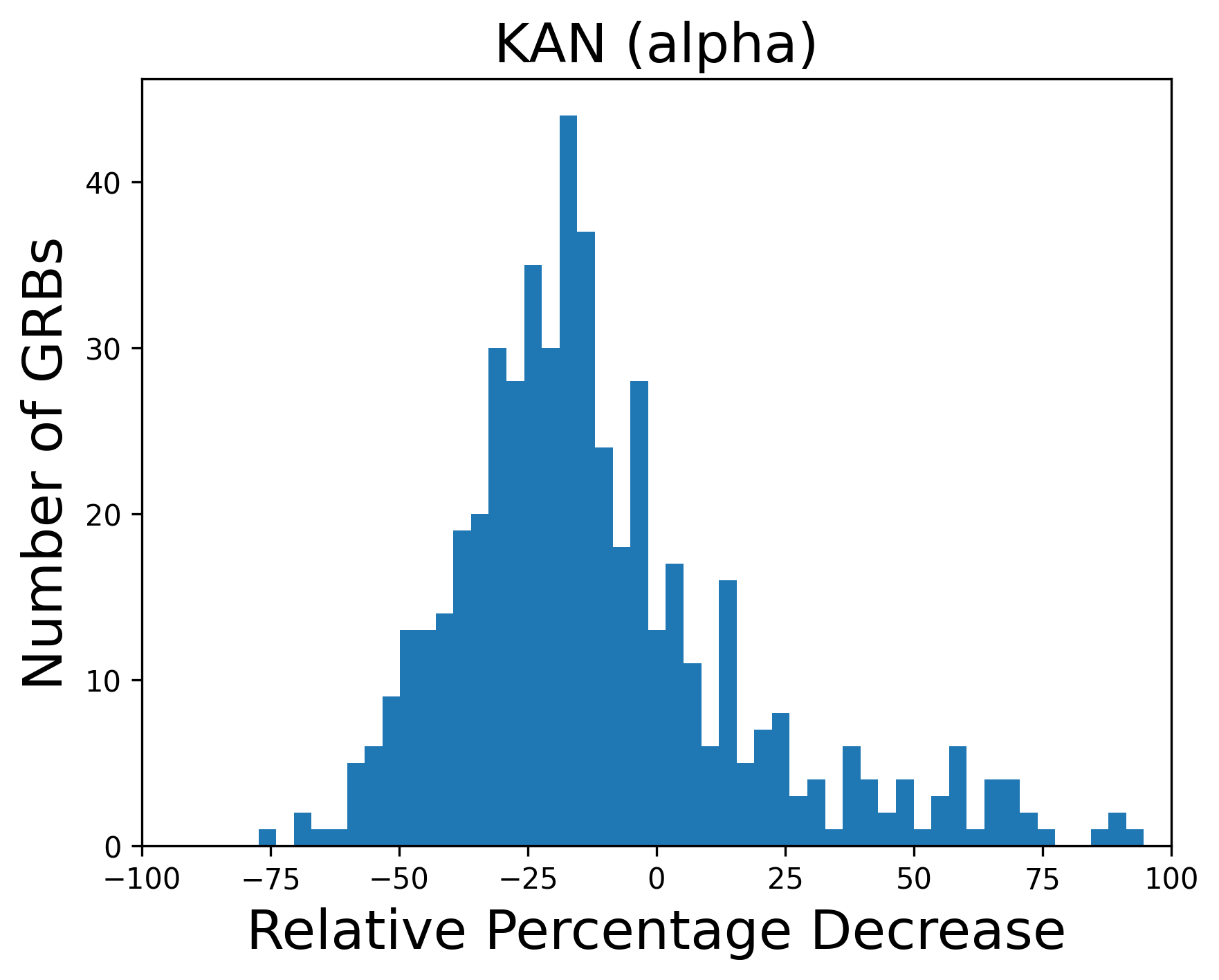}

\includegraphics[width=.27\textwidth, height=.19\textwidth]{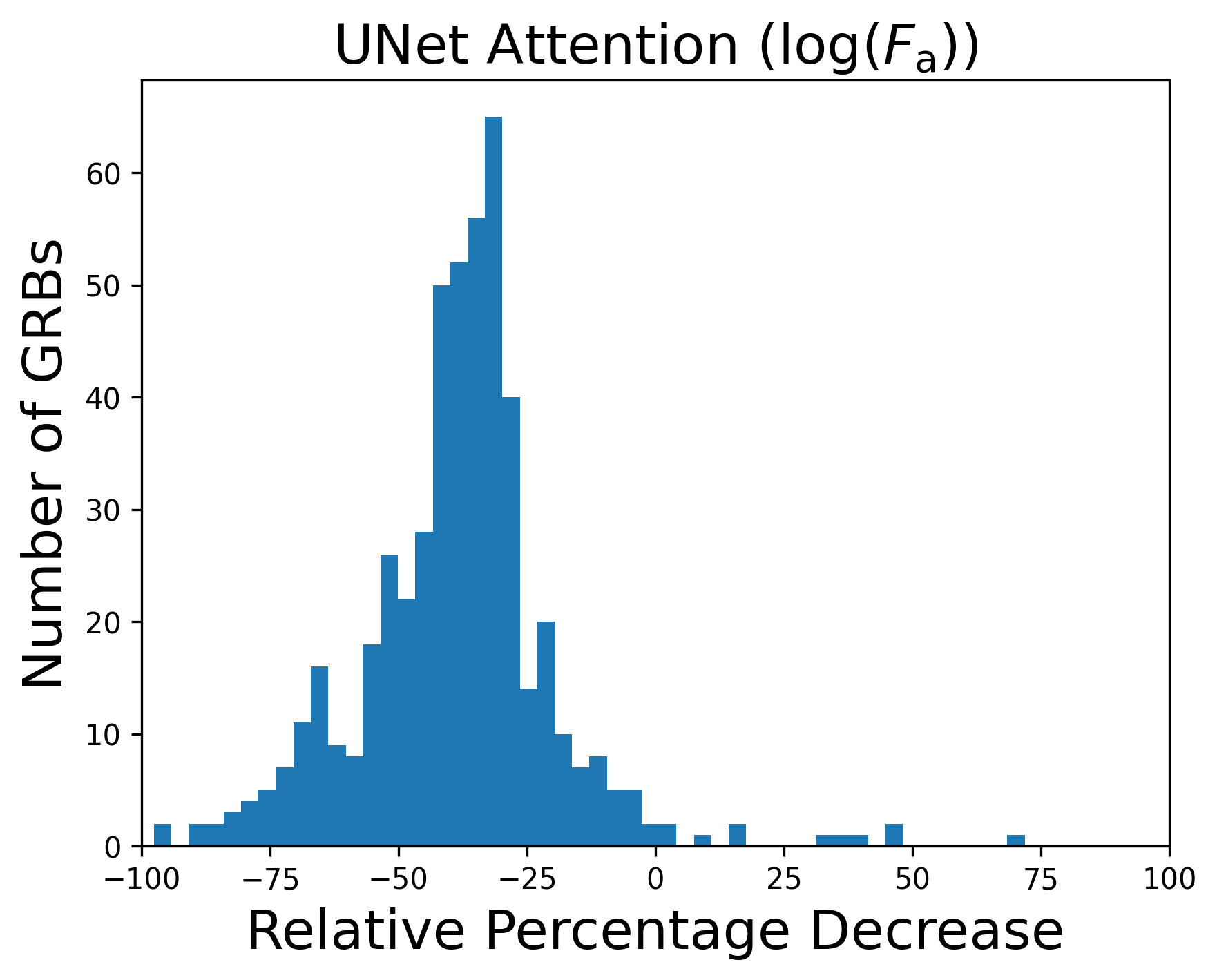}
\includegraphics[width=.27\textwidth, height=.19\textwidth]{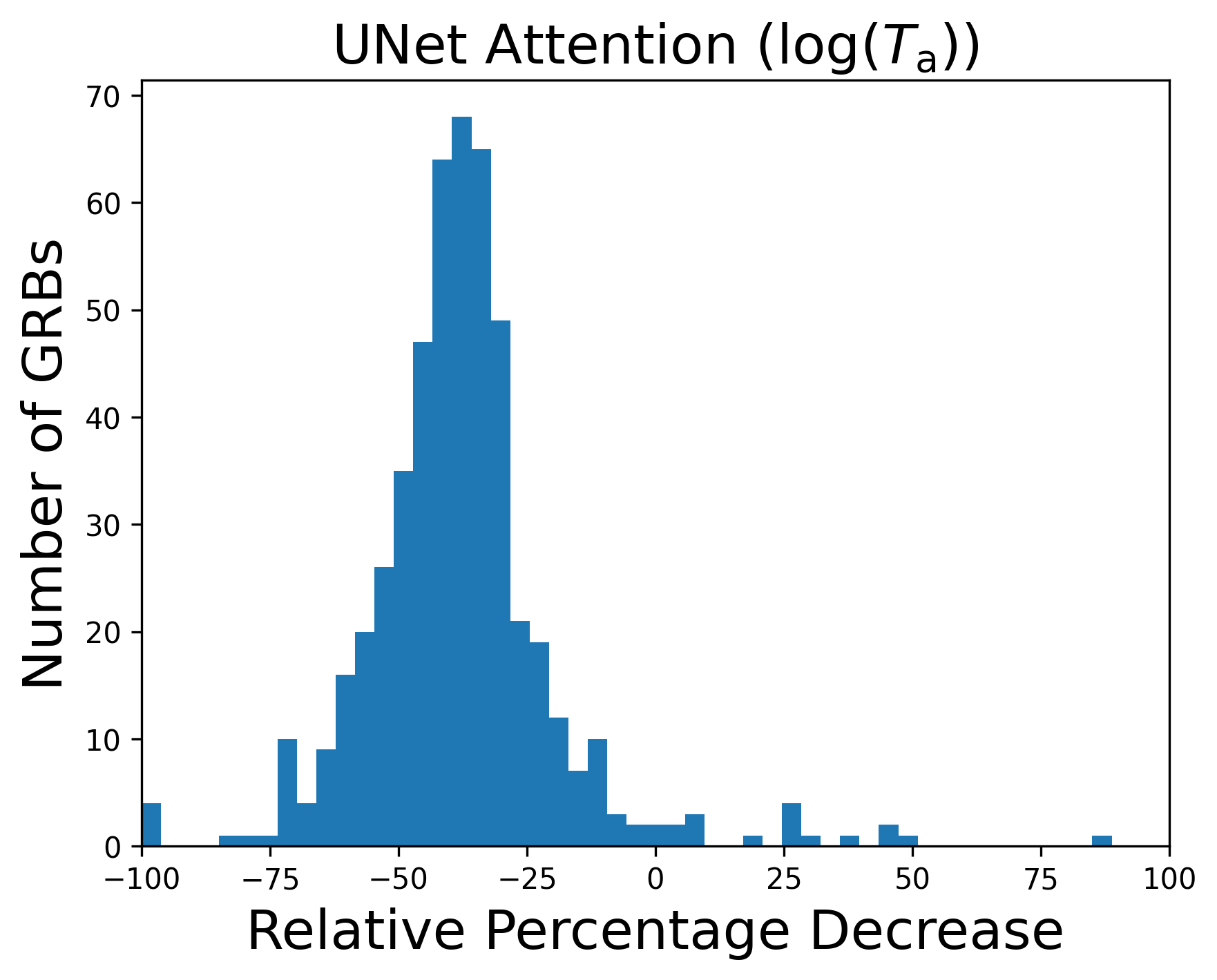}
\includegraphics[width=.27\textwidth, height=.19\textwidth]{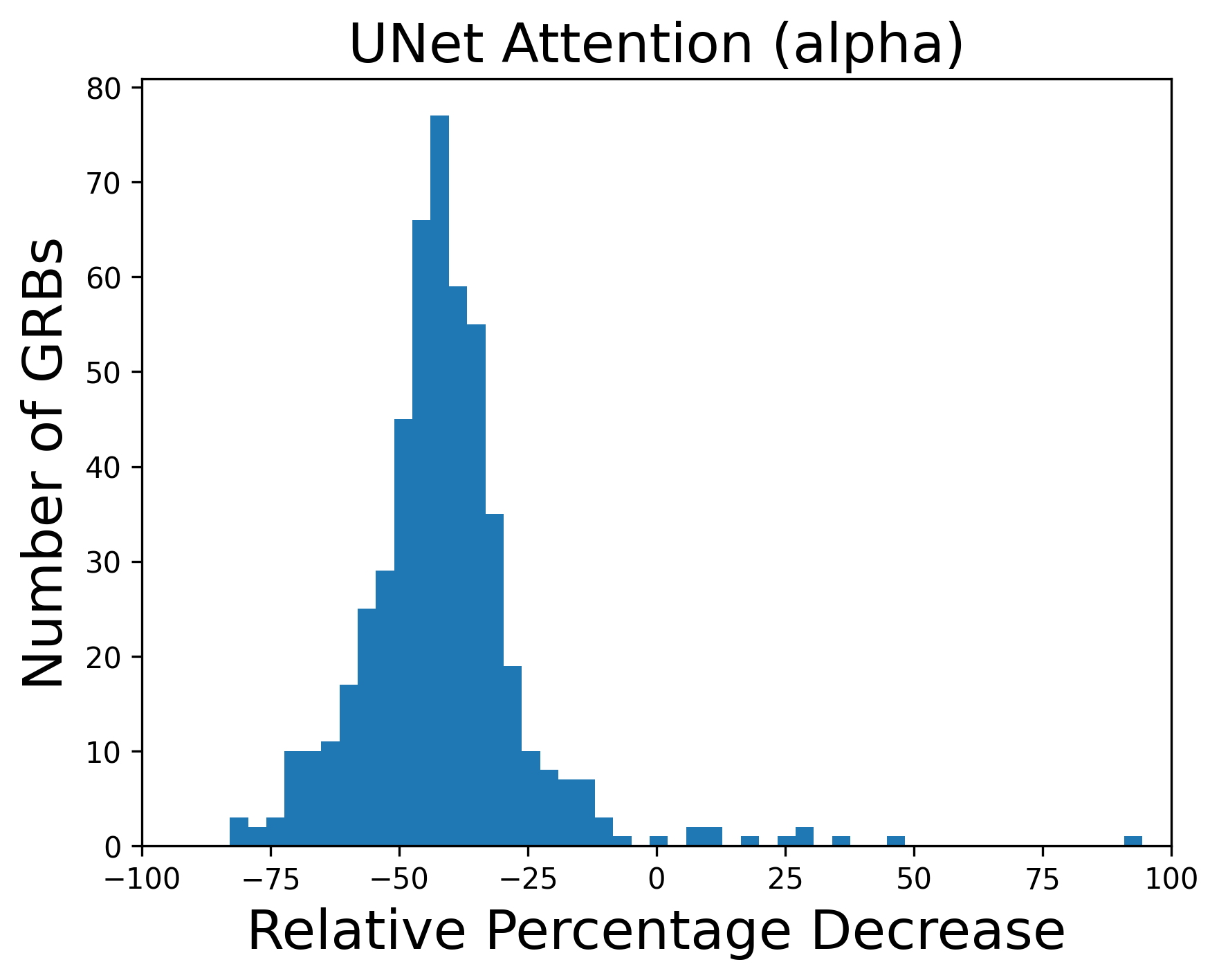}

\caption{Distribution plot of all three W07 parameters in a grid with parameters (left to right): i) $\log F_a$ (Column 1) ii) $\log T_a$ (Column 2) iii) $\alpha$ (Column 3) and the models (top to bottom): i) GP-RF (Row 1); ii) Bi-LSTM (Row 2); iii) CGAN (Row 3); iv) SARIMAX-based Kalman (Row 4); v) KAN (Row 5); vi) Attention U-Net  (Row 6).}
\label{fig: ALL-results-2} 
\end{center}

\end{figure*}

The reconstruction for each category of GRBs for all models is shown in Fig. \ref{fig: ALL-reconstruction-1} and Fig. \ref{fig: ALL-reconstruction-2} and the histogram distribution of the relative percentage decrease for the three W07 parameters, computed using the entire GRB sample, is illustrated in Fig. \ref{fig: ALL-results-1} and Fig. \ref{fig: ALL-results-2}. Table \ref{tab:ALL_Table} shows the EF (before and after LCR) and the percentage reduction of EF for all models across the entire GRB sample.
Along with these, we conducted a 5-Fold CV for all the models to calculate the train and test MSE, ensuring the reliability of our approach.
Table \ref{tab:reconstruction_results} compares the performance of all the ML models.

\vspace{32pt}

\setlength{\LTcapwidth}{\textwidth}

\clearpage

\begin{longtable}[c]{|c|c|c|c|c|c|c|c|c|c|}

\hline
\textbf{GRB ID} & $EF_{\log_{10}(T_i)}$ & $EF_{\log_{10}(F_i)}$ & $EF_{\alpha_i}$ & $EF_{\log_{10}(T_i)}$ RC & $EF_{\log_{10}(F_i)}$ RC & $EF_{\alpha_i}$ RC & $\%\log_{10}(T_i)$ & $\%\log_{10}(F_i)$ & $\%\alpha_i$ \\
\hline
\multicolumn{10}{|c|}{\textbf{ W07 10\% noise}} \\
\hline
050315 & 0.00928  & 0.00285  & 0.0444  & 0.00894  & 0.00268  & 0.0431  & -3.61  & -6.05  & -2.90  \\
050318 & 0.0106   & 0.00554  & 0.0458  & 0.0108   & 0.00540  & 0.0417  &  1.78  & -2.65  & -9.11  \\
050319 & 0.0166   & 0.00389  & 0.0400  & 0.0161   & 0.00370  & 0.0319  & -2.89  & -4.84  & -20.2  \\
050401 & 0.00907  & 0.00202  & 0.0265  & 0.00772  & 0.00174  & 0.0154  & -14.9  & -13.5  & -41.8  \\
050730 & 0.00439  & 0.00263  & 0.0225  & 0.00332  & 0.00205  & 0.0153  & -24.5  & -22.1  & -32.1  \\
050802 & 0.00896  & 0.00267  & 0.0226  & 0.00751  & 0.00239  & 0.0147  & -16.1  & -10.6  & -35.2  \\
050803 & 0.0113   & 0.00450  & 0.0320  & 0.0106   & 0.00419  & 0.0228  & -6.38  & -7.03  & -28.9  \\

\hline
\multicolumn{10}{|c|}{\textbf{W07 20\% noise}} \\
\hline
050315 & 0.00928  & 0.00285  & 0.0444  & 0.00883  & 0.00267  & 0.0424  & -4.78  & -6.42  & -4.40  \\
050318 & 0.0106   & 0.00554  & 0.0458  & 0.0107   & 0.00562  & 0.0462  &  0.85  &  1.37  &  0.720 \\
050319 & 0.0166   & 0.00389  & 0.0400  & 0.0161   & 0.00377  & 0.0325  & -3.06  & -3.24  & -18.9  \\
050401 & 0.00907  & 0.00202  & 0.0265  & 0.00779  & 0.00178  & 0.0156  & -14.0  & -11.7  & -41.3 \\
050730 & 0.00439  & 0.00263  & 0.0225  & 0.00332  & 0.00206  & 0.0154  & -24.4  & -21.9  & -31.9  \\
050802 & 0.00896  & 0.00267  & 0.0226  & 0.00786  & 0.00248  & 0.0151  & -12.2  & -7.32  & -33.1  \\
050803 & 0.0113   & 0.00450  & 0.0320  & 0.0113   & 0.00437  & 0.0245  & -0.370 & -3.04  & -23.4  \\

\hline

\multicolumn{10}{|c|}{\textbf{GP}} \\
\hline
050315 & 0.00928  & 0.00285  & 0.0444  & 0.0105   & 0.00288  & 0.0403  & 13.3   & 0.940  & -9.14  \\
050318 & 0.0106   & 0.00554  & 0.0458  & 0.0105   & 0.00539  & 0.0428  & -1.12  & -2.79  & -6.71  \\
050319 & 0.0166   & 0.00389  & 0.0400  & 0.0124   & 0.00356  & 0.0355  & -24.9  & -8.49  & -11.5  \\
050401 & 0.00907  & 0.00202  & 0.0265  & 0.00671  & 0.00167  & 0.0149  & -26.0  & -17.1  & -43.7  \\
050730 & 0.00439  & 0.00263  & 0.0225  & 0.00326  & 0.00205  & 0.0152  & -25.7  & -22.2  & -32.7  \\
050802 & 0.00896  & 0.00267  & 0.0226  & 0.00856  & 0.00260  & 0.0156  & -4.41  & -2.81  & -31.0  \\
050803 & 0.0113   & 0.00450  & 0.0320  & 0.0106   & 0.00402  & 0.0239  & -5.97  & -10.7  & -25.3  \\

\hline
\multicolumn{10}{|c|}{\textbf{Bi-Mamba}} \\
\hline
050315 & 0.00928  & 0.00285  & 0.0444  & 0.00826  & 0.00260  & 0.0400  & -10.9  & -8.76  & -9.75  \\
050318 & 0.0106   & 0.00554  & 0.0458  & 0.0103   & 0.00536  & 0.0443  & -2.63  & -3.26  & -3.36  \\
050319 & 0.0166   & 0.00389  & 0.0400  & 0.0123   & 0.00371  & 0.0387  & -25.7  & -4.74  & -3.34  \\
050401 & 0.00907  & 0.00202  & 0.0265  & 0.00683  & 0.00164  & 0.0146  & -24.7  & -18.5  & -44.9  \\
050730 & 0.00439  & 0.00263  & 0.0225  & 0.00325  & 0.00207  & 0.0152  & -26.0  & -21.5  & -32.7 \\
050802 & 0.00896  & 0.00267  & 0.0226  & 0.00771  & 0.00241  & 0.0147  & -13.9  & -9.90  & -35.1  \\
050803 & 0.0113   & 0.00450  & 0.0320  & 0.0102   & 0.00391  & 0.0226  & -9.63  & -13.1  & -29.4  \\

\hline
\multicolumn{10}{|c|}{\textbf{MLP Model}} \\
\hline
050315 & 0.00928  & 0.00285  & 0.0444  & 0.00658  & 0.00202  & 0.0308  & -29.1  & -29.3  & -30.7  \\
050318 & 0.0106   & 0.00554  & 0.0458  & 0.00795  & 0.00402  & 0.0319  & -25.0  & -27.6  & -30.3  \\
050319 & 0.0166   & 0.00389  & 0.0400  & 0.00979  & 0.00262  & 0.0272  & -40.9  & -32.8  & -32.0  \\
050401 & 0.00907  & 0.00202  & 0.0265  & 0.00494  & 0.00121  & 0.0121  & -45.6  & -39.8  & -54.4  \\
050730 & 0.00439  & 0.00263  & 0.0225  & 0.00239  & 0.00151  & 0.0113  & -45.6  & -42.7  & -49.7  \\
050802 & 0.00896  & 0.00267  & 0.0226  & 0.00559  & 0.00175  & 0.0116  & -37.6  & -34.5  & -48.5  \\
050803 & 0.0113   & 0.00450  & 0.0320  & 0.00757  & 0.00295  & 0.0175  & -33.1  & -34.5  & -45.4  \\

\hline
\multicolumn{10}{|c|}{\textbf{Fourier Transform}} \\
\hline
050315 & 0.00928  & 0.00285  & 0.0444  & 0.00940  & 0.00279  & 0.0420  &  1.29  & -2.29  & -5.27  \\
050318 & 0.0106   & 0.00554  & 0.0458  & 0.0118   & 0.00600  & 0.0493  & 11.6   &  8.28  &  7.66  \\
050319 & 0.0166   & 0.00389  & 0.0400  & 0.0105   & 0.00350  & 0.0389  & -36.8  & -10.0  & -2.78  \\
050401 & 0.00907  & 0.00202  & 0.0265  & 0.00698  & 0.00173  & 0.0155  & -23.0  & -14.2  & -41.6 \\
050730 & 0.00439  & 0.00263  & 0.0225  & 0.00332  & 0.00210  & 0.0155  & -24.4  & -20.2  & -31.2  \\
050802 & 0.00896  & 0.00267  & 0.0226  & 0.00803  & 0.00253  & 0.0154  & -10.4  & -5.48  & -32.0  \\
050803 & 0.0113   & 0.00450  & 0.0320  & 0.0108   & 0.00413  & 0.0244  & -4.50  & -8.25  & -23.7  \\

\hline
 \multicolumn{10}{|c|}{\textbf{GP-RF Model}} \\
\hline
050315 & 0.00928  & 0.00285  & 0.0444  & 0.00965  & 0.00272  & 0.0410  &  4.03  & -4.60  & -7.51  \\
050318 & 0.0106   & 0.00554  & 0.0458  & 0.0103   & 0.00529  & 0.0432  & -3.09  & -4.56  & -5.81 \\
050319 & 0.0166   & 0.00389  & 0.0400  & 0.0117   & 0.00342  & 0.0380  & -29.3  & -12.0  & -5.08  \\
050401 & 0.00907  & 0.00202  & 0.0265  & 0.00681  & 0.00171  & 0.0152  & -24.9  & -15.1  & -42.5  \\
050730 & 0.00439  & 0.00263  & 0.0225  & 0.00679  & 0.00416  & 0.0314  & 54.5   & 58.1   & 39.4   \\
050802 & 0.00896  & 0.00267  & 0.0226  & 0.00830  & 0.00254  & 0.0154  & -7.30  & -5.02  & -32.0  \\
050803 & 0.0113   & 0.00450  & 0.0320  & 0.0108   & 0.00407  & 0.0238  & -4.58  & -9.69  & -25.6  \\

\hline
 \multicolumn{10}{|c|}{\textbf{Bi-LSTM Model}} \\
\hline
050315 & 0.00928 & 0.00285 & 0.0444 & 0.00814 & 0.00271 & 0.0414 & -12.3 & -5.04 & -6.69 \\
050318 & 0.0106 & 0.00554 & 0.0458 & 0.00996 & 0.00539 & 0.0443 & -5.98 & -2.84 & -3.28 \\
050319 & 0.0166 & 0.00389 & 0.0400 & 0.0102 & 0.00369 & 0.0410 & -38.2 & -5.26 & 2.52 \\
050401 & 0.00907 & 0.00202 & 0.0265 & 0.00680 & 0.00166 & 0.0150 & -25.0 & -17.6 & -43.3 \\
050730 & 0.00439 & 0.00263 & 0.0225 & 0.00328 & 0.00204 & 0.0152 & -25.4 & -22.6 & -32.7 \\
050802 & 0.00896 & 0.00267 & 0.0226 & 0.00853 & 0.00259 & 0.0158 & -4.77 & -3.26 & -30.1 \\
050803 & 0.0113 & 0.00450 & 0.0320 & 0.0103 & 0.00394 & 0.0225 & -9.13 & -12.5 & -29.6 \\

\hline
 \multicolumn{10}{|c|}{\textbf{CGAN Model}} \\
\hline
050315 & 0.00928 & 0.00285 & 0.0444 & 0.00900 & 0.00283 & 0.0430 & -3.05 & -0.83 & -3.18 \\
050318 & 0.0106 & 0.00554 & 0.0458 & 0.0100 & 0.00508 & 0.0404 & -5.24 & -8.27 & -11.9 \\
050319 & 0.0166 & 0.00389 & 0.0400 & 0.00961 & 0.00353 & 0.0399 & -41.9 & -9.24 & -0.280 \\
050401 & 0.00907 & 0.00202 & 0.0265 & 0.00701 & 0.00171 & 0.0154 & -22.7 & -15.2 & -41.8 \\
050730 & 0.00439 & 0.00263 & 0.0225 & 0.00328 & 0.00208 & 0.0154 & -25.4 & -21.1 & -31.7 \\
050802 & 0.00896 & 0.00267 & 0.0226 & 0.00730 & 0.00238 & 0.0145 & -18.5 & -11.0 & -35.7 \\
050803 & 0.0113 & 0.00450 & 0.0320 & 0.0105 & 0.00412 & 0.0235 & -6.99 & -8.42 & -26.6 \\

\hline 
 \multicolumn{10}{|c|}{\textbf{SARIMAX+Kalman Model}} \\
\hline
050315 & 0.00928 & 0.00285 & 0.0444 & 0.00845 & 0.00268 & 0.0417 & -8.87 & -6.01 & -6.11 \\
050318 & 0.0106 & 0.00554 & 0.0458 & 0.0112 & 0.00596 & 0.0472 & 5.36 & 7.46 & 3.02 \\
050319 & 0.0166 & 0.00389 & 0.0400 & 0.0127 & 0.00361 & 0.0395 & -23.6 & -7.27 & -1.30 \\
050401 & 0.00907 & 0.00202 & 0.0265 & 0.00822 & 0.00188 & 0.0169 & -9.39 & -6.75 & -36.2 \\
050730 & 0.00439 & 0.00263 & 0.0225 & 0.00334 & 0.00208 & 0.0155 & -24.1 & -21.0 & -31.3 \\
050802 & 0.00896 & 0.00267 & 0.0226 & 0.00856 & 0.00262 & 0.0160 & -4.48 & -1.96 & -29.2 \\
050803 & 0.0113 & 0.00450 & 0.0320 & 0.0108 & 0.00408 & 0.0248 & -4.50 & -9.29 & -22.5 \\

\hline
 \multicolumn{10}{|c|}{\textbf{KAN Model}} \\
\hline
050315 & 0.00928 & 0.00285 & 0.0444 & 0.00810 & 0.00254 & 0.0391 & -12.7 & -10.9 & -11.8 \\
050318 & 0.0106 & 0.00554 & 0.0458 & 0.00970 & 0.00526 & 0.0456 & -8.46 & -5.18 & -0.500 \\
050319 & 0.0166 & 0.00389 & 0.0400 & 0.0135 & 0.00331 & 0.0286 & -18.3 & -14.9 & -28.5 \\
050401 & 0.00907 & 0.00202 & 0.0265 & 0.00695 & 0.00165 & 0.0149 & -23.4 & -18.0 & -43.8 \\
050730 & 0.00439 & 0.00263 & 0.0225 & 0.00330 & 0.00209 & 0.0153 & -25.0 & -20.9 & -32.1 \\
050802 & 0.00896 & 0.00267 & 0.0226 & 0.00730 & 0.00233 & 0.0145 & -18.5 & -12.8 & -35.8 \\
050803 & 0.0113 & 0.00450 & 0.0320 & 0.0102 & 0.00395 & 0.0222 & -9.83 & -12.2 & -30.5 \\

\hline
 \multicolumn{10}{|c|}{\textbf{Attention U-Net}} \\
\hline
050315 & 0.00928 & 0.00285 & 0.0444 & 0.00785 & 0.00214 & 0.0294 & -15.4 & -25.2 & -33.8 \\
050318 & 0.0106 & 0.00554 & 0.0458 & 0.00818 & 0.00417 & 0.0330 & -22.8 & -24.7 & -28.0 \\
050319 & 0.0166 & 0.00389 & 0.0400 & 0.00984 & 0.00275 & 0.0298 & -40.6 & -29.3 & -25.5 \\
050401 & 0.00907 & 0.00202 & 0.0265 & 0.00504 & 0.00123 & 0.0123 & -44.4 & -38.9 & -53.5 \\
050730 & 0.00439 & 0.00263 & 0.0225 & 0.00240 & 0.00149 & 0.0113 & -45.4 & -43.3 & -49.9 \\
050802 & 0.00896 & 0.00267 & 0.0226 & 0.00579 & 0.00179 & 0.0118 & -35.4 & -33.1 & -47.7 \\
050803 & 0.0113 & 0.00450 & 0.0320 & 0.00756 & 0.00295 & 0.0174 & -33.1 & -34.4 & -45.6 \\

\hline
\caption{The W07 $\log T_a$, $\log F_a$ and $\alpha$ error fractions before and after reconstruction (with a decrease in relative percentage in error for all three parameters). Columns first, second, and third provide error fractions for the original W07 fit; the successive columns fourth, fifth, and sixth show the error fraction for the W07 fit after LCs reconstruction. The seventh, eighth, and ninth columns show a percentage decrease in the error fraction after the reconstruction. The full table will be available online.}

\label{tab:ALL_Table}
\end{longtable}

\begin{table*}[htbp] 
\begin{center}
\begin{tabular}{|l|c|c|c|c|c|c|c|c|}
\hline 
\textbf{Reconstruction Model} & \multicolumn{3}{c|}{\textbf{Uncertainty Decrease}} & \multicolumn{3}{c|}{\textbf{\% Outliers}} & \multicolumn{2}{c|}{\textbf{5 K-Fold CV}} \\
\hline
 & \% $\log T_a$ & \% $\log F_a$ & \% $\alpha$ & \% $\log T_a$ & \% $\log F_a$ & \% $\alpha$ & Train MSE ($10^{-1}$) & Test MSE ($10^{-1}$) $\downarrow$\\
\hline
&\multicolumn{8}{c|}{\textbf{521 GRBs}}\\
\hline

MLP              & -37.2 & -38.0 & -41.2 & \textbf{1.73} & 2.30 & 1.34 & 0.227 & \textbf{0.275} \\
Fourier          & -14.9 & -15.2 & -20.6 & 4.61 & 4.22 & 2.50 & \textbf{0.0270} & 0.339 \\
CGAN             & -14.5 & -15.4 & -20.0 & 3.26 & 3.45 & 1.34 & 0.260 & 0.429 \\
Bi-LSTM          & -21.2 & -21.5 & -26.1 & 2.50 & 2.88 & 1.34 & 0.231 & 0.532 \\
SARIMAX-based Kalman & -9.70 & -13.3 & -17.3 & 3.26 & 3.84 & 1.34 & 0.814 & 0.825 \\
Bi-Mamba         & -20.8 & -21.2 & -27.6 & 2.88 & 2.30 & \textbf{1.15} & 0.151 & 1.30 \\
GP-RF            & -16.3 & -19.5 & -23.2 & 4.03 & 4.03 & 2.30 & 0.0551 & 1.33 \\
Attention U-Net  & \textbf{-37.9} & \textbf{-38.5} & \textbf{-41.4} & \textbf{1.73} & 2.50 & 1.34 & 0.206 & 1.34 \\
KAN model        & -4.74 & -14.2 & -11.6 & 4.99 & \textbf{1.92} & 1.73 & 0.423 & 1.74 \\
\hline
GP (W07)         & -16.9 & -18.6 & -24.3 & 3.07 & 3.45 & 1.54 & 8.56  & 3.63 \\
W07 model (10\%) & -18.0 & -19.1 & -25.2 & 2.30 & 2.30 & 2.11 & - & - \\
W07 model (20\%) & -15.8 & -17.8 & -23.8 & 2.30 & 2.69 & 2.30 & - & - \\

\hline
&\multicolumn{8}{c|}{\textbf{207 Good GRBs \cite{dainotti2023stochastic}}}\\
\hline
SARIMAX-based Kalman   & -10.9  & -15.9 & -19.8  & 0.966  & 1.45  & 0  & 0.191 & \textbf{0.192} \\
MLP              & -38.1 & -39.3 & -43.9 & 0.483 & 0.483 & \textbf{0} & 0.230  & 0.268 \\
CGAN             & -16.3 & -17.9 & -24.6 & 0.966  & 0.966  & \textbf{0} & 0.230  & 0.383  \\
Fourier          & -15.5 & -17.3 & -24.1 & 1.45  & 1.45  & 0.483 & 0.0900 & 0.480  \\
Bi-LSTM          & -22.7 & -24.0 & -30.1 & 0.483  & 0.483  & \textbf{0} & 0.210  & 0.626  \\
Bi-Mamba         & -22.5 & -24.4 & -31.2 & 0.483 & 0.483 & \textbf{0} & 0.110  & 0.850  \\
GP-RF            & -17.6 & -21.8 & -27.8 & 0.483 & 0.966  & 0.483  & \textbf{0.0468} & 0.920 \\
Attention U-Net  & \textbf{-38.8} & \textbf{-40.3} & \textbf{-44.0} & 0.483 & 0.483 & \textbf{0} & 0.173  & 1.07 \\
KAN model        & -7.06 & -17.3 & -12.8 & 3.38  & \textbf{0}  & \textbf{0}  & 0.336  & 1.61  \\
\hline
GP (W07)         & -17.2 & -20.0 & -28.8 & 1.93 & 1.45 & 0 & 7.66 & 1.81  \\
W07 model (10\%) & -23.0 & -24.9 & -30.8 & 0.966 & 0.966 & \textbf{0} & - & - \\ 
W07 model (20\%) & -21.2 & -23.2 & -28.9 & \textbf{0} & \textbf{0} & \textbf{0} & - & -  \\

\hline
\end{tabular}
\caption{Summary of average decrease in the \% uncertainty along with the \% increase in outliers across different reconstruction processes and their parameters. The train MSE and test MSE for all the models across 5 K-Fold CV are provided. The models are sorted in ascending order based on Test MSE values (top to bottom). GP, W07 (10\%) and W07 (20\%) are shown separately at the bottom for both 521 GRBs and 207 Good GRBs.}
\label{tab:reconstruction_results}
\end{center}
\end{table*}



\section{Summary and Conclusion}\label{section:conclusion}

We analyzed various unique approaches, showcasing the efficacy of reconstructing LCs. Below, we summarize the key findings of our results:

\begin{itemize}

    \item The W07 model (10\% noise level) is most effective for reconstructing 207 Good GRBs, achieving the highest uncertainty reduction across all parameters while eliminating outliers. However, its performance diminishes with flares or breaks in the data, limiting its applicability. The model fundamentally relies on the Willingale function, which is designed to represent a generalized LC structure. As such, it does not explicitly account for complex features like flares or breaks often observed in half of the GRBs. Similarly, the GP approach shows reduced reliability in handling GRBs with flares or breaks. Although it provides a more flexible non-parametric framework compared to the fixed functional form of W07, it still tends to follow the overall smooth trend of the LC and has issues in reconstructing transient deviations accurately.

    \item The MLP model provides the lowest 5-Fold Test MSE of 0.0275 among deep learning models, making it the best option when prioritizing reconstruction accuracy. It also offers a high uncertainty reduction (37.2\% for $\log T_a$, 38.0\% for $\log F_a$, and 41.2\% for $\alpha$) while maintaining low outlier rates across all parameters.
    
    \item The Attention U-Net model achieves a higher test MSE of 0.134 but excels in minimizing outliers, particularly for $\log T_a$ and $\log F_a$. It also exhibits the highest uncertainty reduction across all parameters (37.9\% for $\log T_a$, 38.5\% for $\log F_a$, and 41.4\% for $\alpha$). 

    \item In comparison, the Bi-LSTM and Bi-Mamba models show comparable levels of uncertainty reduction across the three parameters: 21.2\%, 21.5\%, and 26.1\% for Bi-LSTM, and 20.8\%, 21.2\%, and 27.6\% for Bi-Mamba, respectively. They are an effective alternative, especially when reducing computational complexity is a priority. Bi-Mamba also exhibits the lowest outliers for $\alpha$ of 1.15.

    \item For $\log T_a$, the Attention U-Net reduces the outliers by 1.88\% compared to the MLP baseline. Similarly, for $\log F_a$, it achieves a 1.32\% reduction, while for $\alpha$, the reduction amounts to 0.49\%.

    \item The Fourier model, CGAN, and SARIMAX-based Kalman model, compared to the other models, demonstrate lower Test MSE values (ranging from 0.0339 to 0.0825) but are not the best overall choices due to trade-offs in handling uncertainty, outliers, and model overfitting (as indicated by the discrepancy in the Train and Test losses).

    \item The GP-RF and KAN models show higher test MSE and poor outlier control than the other tested models, making them less suitable for this specific reconstruction task.

In summary, while Attention U-Net achieves the highest reduction in outliers and predictive uncertainty, its test MSE (0.134) is higher than that of the MLP. The MLP, in contrast, delivers the lowest test MSE and remains highly competitive in uncertainty reduction, with performance very close to the Attention U-Net (within 0.5–1.9\% across all parameters). This balance between accuracy and stability suggests that the MLP provides a more reliable baseline for GRB LCR, while the Attention U-Net offers advantages in minimizing extreme deviations.

\end{itemize}


In comparison to the prior study by \cite{dainotti2023stochastic}, we have the following results:

\begin{itemize}


\item As highlighted by \cite{dainotti2023stochastic}, a study of 207 Good GRBs revealed an average reduction in uncertainties of 26\% across all parameters with a 10\% noise level and 24\% at a 20\% noise level on modeling with the W07 model. 


\item In our work, we include this exact dataset of 207 GRBs for a fair comparison (see Table \ref{tab:reconstruction_results}). When compared to the W07 model, which achieves uncertainty reductions of 23.0\% for $\log T_a$, 24.9\% for $\log F_a$, and 30.8\% for $\alpha$, the Attention U-Net improves these values substantially to 38.8\%, 40.3\%, and 44.0\%, respectively. This corresponds to relative improvements of 68.7\%, 61.9\%, and 42.9\%, indicating that the Attention U-Net provides a more effective reduction of predictive uncertainty compared to the W07 model. Compared to the GP model, Attention U-Net outperforms for all three parameters. Additionally, the 5-fold CV results show that both training and testing MSE values for Attention U-Net are lower than those of GP, further demonstrating its reliability.
Another advantage of Attention U-Net is that it is model-independent while achieving performance equivalent to W07.

\item This study effectively reduces parameter uncertainty for the subset of 207 well-behaved GRBs. However, when the approach is extended to the full dataset encompassing all four GRB categories, the GP method exhibits reduced performance, as outlined in \cite{dainotti2023stochastic}. In contrast,  the other ML and DL models, particularly Attention U-Net, maintain a similar level of precision in reducing parameter uncertainty. This highlights the strength of our models in reconstructing more complex GRB LCs that include features such as flares or breaks. Furthermore, both training and testing MSE values remain low, demonstrating the continued reliability of the models. Although the number of outliers increases, from an average of 0.13\% for the 207 GRBs to 2.5\% across 521 GRBs, this is expected. This rise is expected as the Willingale function is not designed to model the behavior of more complex LCs.

\item In summary, this work undertakes a more comprehensive and challenging analysis than \cite{dainotti2023stochastic} by applying advanced ML and DL methods to a significantly larger and more diverse GRB dataset. We provide more insight on the strengths and weaknesses of each approach by methodically evaluating a larger variety of GRBs, including those with intricate characteristics like flares and breaks. These advancements not only extend the scope of previous studies but also highlight the reliability and generalizability of our models.

\end{itemize}

The substantial decrease in uncertainties attained through these reconstructions enhances the applicability of the GRB plateau parameters for cosmological tools \citep{dainotti2022g, dainotti2022c, dainotti2023b}. 

Among the different evaluation aspects, the most important parameters in our analysis are $\log T_a$, $\log F_a$, and $\alpha$, as they carry physical significance:
$\alpha$ is crucial for theoretical modeling, as it describes the post-plateau decay behavior. This is relevant for the testing of the standard fireball model \citep{piran1999gamma}, where the reduction of the $\alpha$ parameter is crucial to obtain reduced uncertainties on the closure relationship \citep{closure,dainotti2021closure, dainotti2024analysis}. It refers to the relationship between the spectral and temporal indices of a GRB's afterglow. $\log T_a$ and $\log F_a$ are key for identifying empirical correlations, such as the Dainotti relations in two and three dimensions \citep{Dainotti2008, dainotti2010a, dainotti2011a, dainotti2013determination, dainotti2016fundamental, dainotti2020a, dainotti2020b, dainotti2022b, levine2022}. The 2D Dainotti relation relates the end time of the plateau and its luminosity, while the three-dimensional relation adds to the 2D relation the peak of the prompt emission parameter. These relations are of particular interest in cosmological applications and also for distinguishing among theoretical models.

Therefore, we consider these three parameters as the most important outputs of our reconstruction. Among all models tested, Attention U-Net demonstrates the most significant reduction in error for the three parameters. Thus, we recommend it as the preferred model when accurate estimation of these physically meaningful quantities is the main objective.
The MSE is the second most important metric in our study. While it does not directly capture physical significance, it quantifies the overall performance and accuracy of the ML models, making it essential for model comparison. By this measure, the MLP performs best, delivering the lowest test MSE while still achieving uncertainty reductions that are very close to those of the Attention U-Net. Outlier analysis is complementary and helps interpret cases where reconstruction might be less representative or behave anomalously.

In future research, we intend to explore advanced deep-learning techniques, such as genetic algorithms, Bayesian neural networks, and neural ODEs. Moreover, we aim to investigate transformer models and large language models for their ability to identify complex temporal patterns. These approaches are expected to significantly improve LC reconstruction's accuracy, reliability, and applicability when combined with new datasets, particularly in theoretical model interpretation and cosmology.

This reconstruction framework has been developed exclusively using Swift LCs so far; however, we intend to extend its application to additional current missions such as SVOM \citep{atteia2022svom} and Einstein Probe \citep{yuan2022einstein}, as well as future missions, including THESEUS \citep{amati2018theseus} and HiZ-GUNDAM \citep{yonetoku2024high}. Our extension will also incorporate data across multiple wavelengths. Expanding this work to optical wavelengths is especially timely, given the availability of the most extensive optical catalog to date \citep{dainotti2020optical, dainotti2022b, dainotti2024analysis, dainotti2024optical}.

\section{Appendix} \label{section:appendix}

This section explains the rationale for not using a separate test set to evaluate our model. It also highlights alternative models we experimented with, though they did not produce satisfactory results for our specific use case. However, we acknowledge that the model may hold more potential in different hybrid configurations for future work.

\subsection{Test Set Evaluation}

\begin{figure*}[htbp]
\begin{center}
    \includegraphics[width=.47\textwidth, height=.35\textwidth]{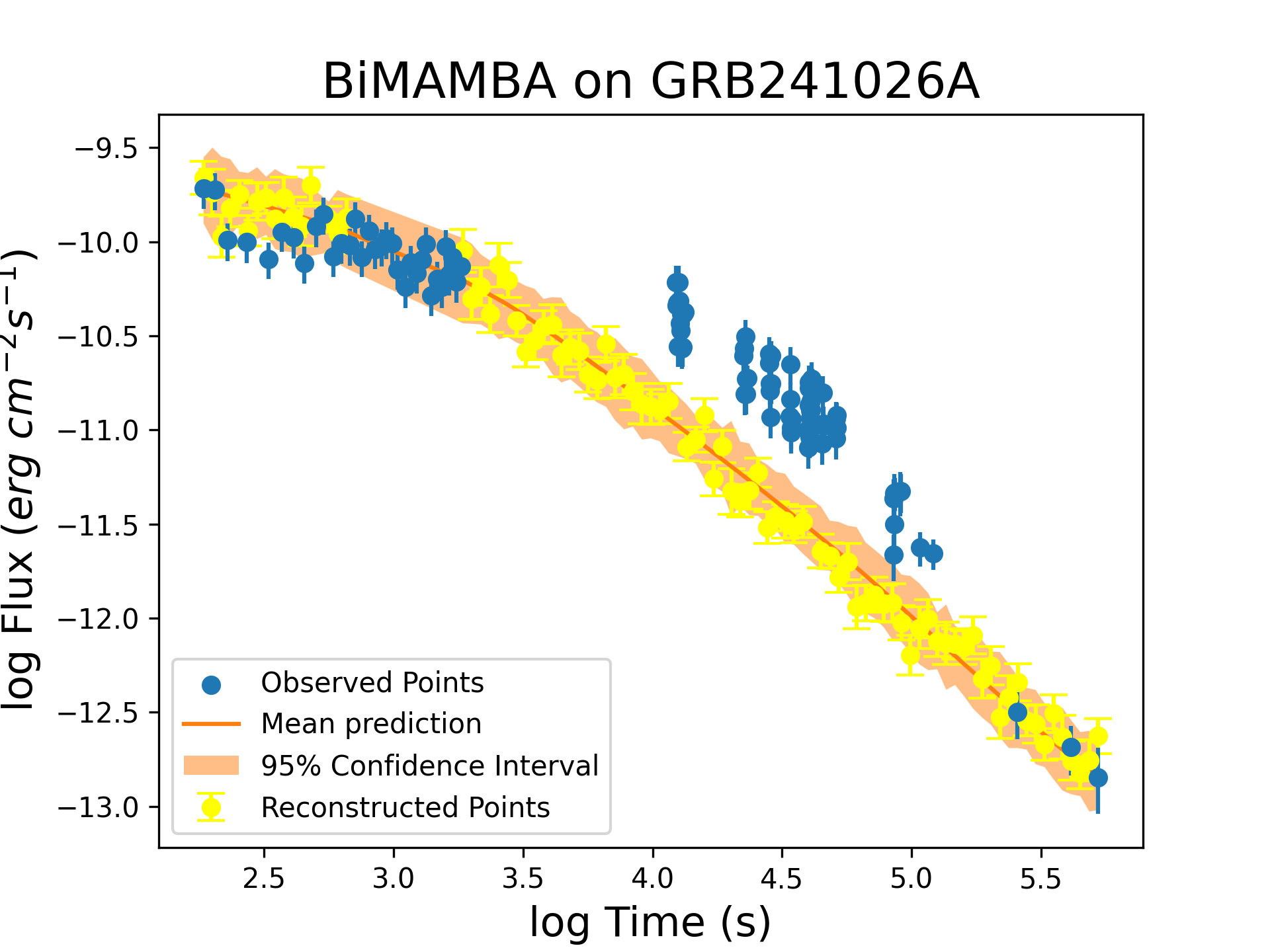}
    \includegraphics[width=.47\textwidth, height=.35\textwidth]{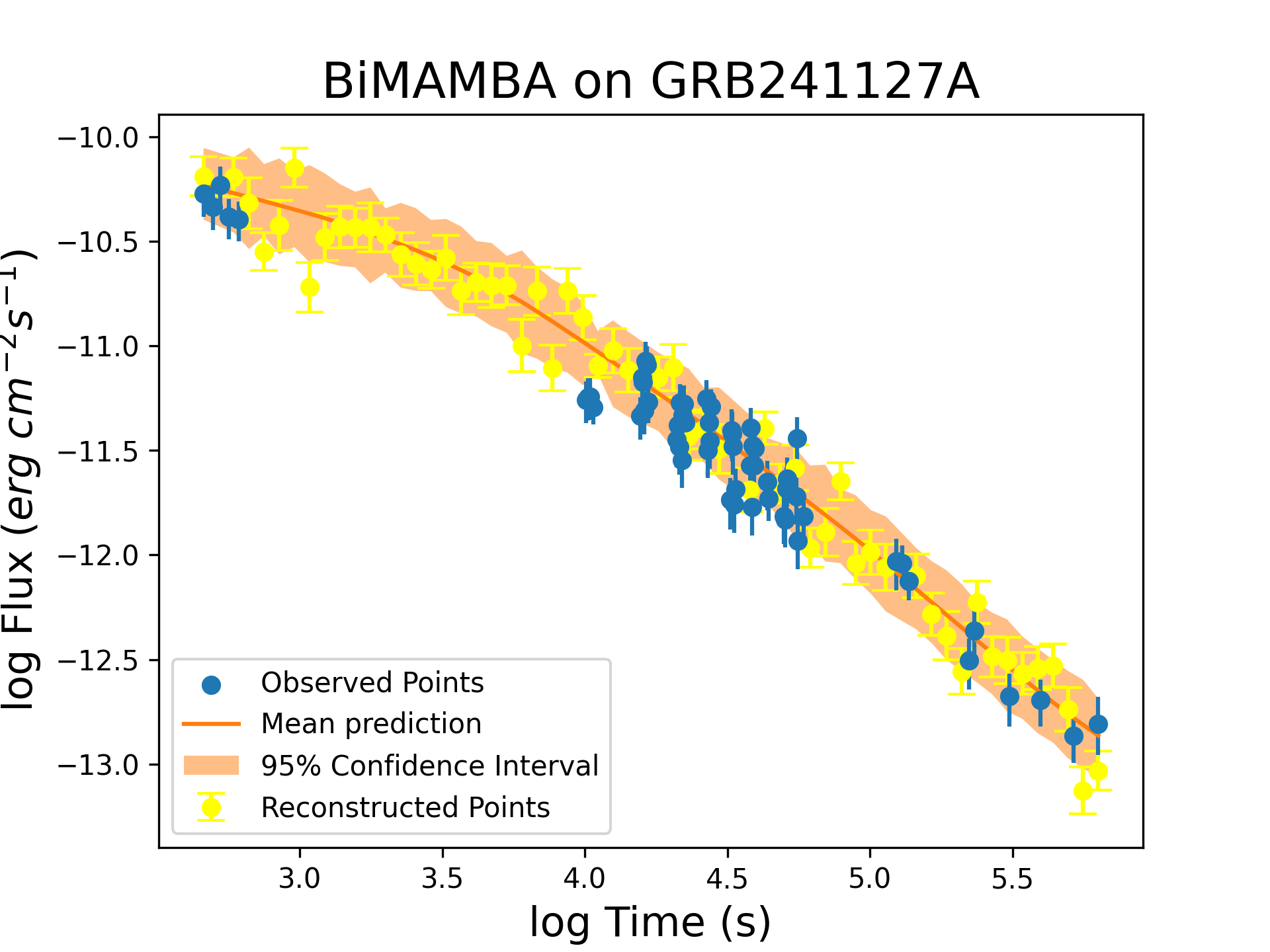}
\end{center}
\caption{Reconstruction of LCs using Bi-Mamba trained on GRB050713A. Left: Reconstructed LC of GRB241026A; Right: Reconstructed LC of GRB241127A.}
\label{fig:appendix-test-set}
\end{figure*}

Due to the nature of our problem being data reconstruction rather than pure regression analysis, a separate test set cannot be effectively used. Each GRB must be fitted individually on the sparse time and flux values available to reconstruct the variability within its curve accurately. Attempts to train on multiple GRBs and reconstruct a separate test set led the model to reproduce a similar GRB structure across all test instances. This occurs because the model learns a single flux value for each corresponding timestamp and tends to generalize when reconstruction specific to each GRB is required. This behavior is evident in Figure \ref{fig:appendix-test-set}, where Bi-Mamba, trained on GRB050713A and tested on a separate set containing GRB241026A and GRB241127A, produces similar reconstructions for both test GRBs. This is because the model has access to sparse training data but only time samples from the test set.

\subsection{TimeAutoDiff}

TimeAutoDiff \citep{timeautodiff2024} is a model designed to synthesize time-series tabular data, addressing the challenges of temporal dependencies and heterogeneous features. It combines a Variational Autoencoder (VAE) \citep{vae2019} with a Denoising Diffusion Probabilistic Model (DDPM) \citep{denoisingdiff2020} to generate synthetic data. 

The architecture of TimeAutoDiff consists of the following components:

\begin{itemize}

    \item \textbf{VAE Component:} This leverages transformers to capture feature correlations and Recurrent Neural Networks (RNNs) to model temporal dependencies in the data.

    \item \textbf{DDPM Component:} This module learns the data's latent distribution, enabling new samples with learned temporal structures.

\end{itemize}

While TimeAutoDiff demonstrated strong performance in reconstructing flux for timestamps it had already observed, it struggled to generalize to unseen timestamps, making it unsuitable for our reconstruction task.

\subsection{Variational Autoencoders (VAE)}

VAEs are generative models that are used for the reconstruction of data. Like other autoencoders, VAEs are made up of a decoder that reconstructs the input data from the latent variables and an encoder that learns to extract important latent variables from training data. The ability of VAEs to encode a continuous, probabilistic representation of the latent space rather than a fixed, discrete one distinguishes them from conventional autoencoders. VAEs can produce new data samples closely resembling the original dataset through variational inference.

This probabilistic approach allows VAEs to explicitly capture and model the uncertainty inherent to the data. For GRB LCR, this capability is particularly advantageous, as it captures the variability and noise present in astrophysical observations while producing reliable reconstructions and confidence intervals for the predictions.

Despite their strengths, the VAEs struggled to accurately capture the temporal dependencies and the variability in the GRB LCs. However, we will continue exploring this method and believe that increasing the model's complexity, such as incorporating recurrent components or deeper architectures, could improve its ability to capture the intricate details of the LCs and enhance its performance.

\subsection{Decision Trees}
Decision Trees (DTs) are a non-parametric supervised learning method mainly used for classification and regression.
We use the CART algorithm, which stands for Classification and Regression Tree algorithm, to build
a tree, and the Python package scikit-learn for implementing Decision Trees for Regression.

Decision trees work by recursively splitting the data based on feature values to minimize the variance within the resulting subgroups. Decision trees are beneficial for LCR because they handle nonlinear relationships well.
However, decision trees tend to overfit the training data, especially when the LC has noisy measurements or complex temporal behavior. Although they perform well in capturing local variations, their predictions often lack the smoothness required for continuous processes like LCR. This limitation leads to step-like predictions rather than a smooth, gradual transition typically expected in astronomical LCs.

\subsection{Random Forest}
Random Forests are an ensemble method that builds upon the foundation of decision trees by creating multiple trees using random subsets of the data and features and then averaging their predictions. This ensemble approach helps mitigate the overfitting issue commonly seen in single decision trees, leading to improved performance.

Random Forests have one significant advantage over individual decision trees: their ability to estimate uncertainty. Although Random Forests are not inherently probabilistic models, uncertainty can be calculated by measuring the variance of predictions across the ensemble of trees. Random Forests provide uncertainty quantification by examining the distribution of predictions from the individual trees.

\begin{figure}[htbp] 
    \centering
    \includegraphics[width=0.45\textwidth]{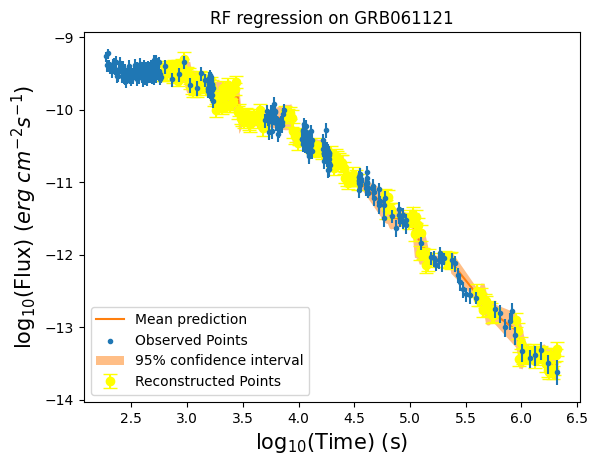}
    \hfill
    \includegraphics[width=0.45\textwidth]{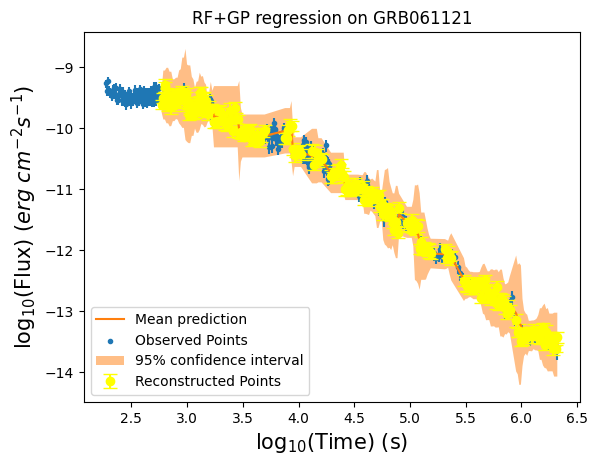}
    \caption{Reconstructed GRB LCs using i) Random Forest model (left); ii) Incorporation of both Random Forest and GP Regression Models (right).}
    \label{fig:rf-comparison}
\end{figure}

However, while Random Forests offer some uncertainty estimation, they are still less precise than the uncertainty provided by fully probabilistic models like GPs. Random Forest uncertainty is derived from the diversity among trees rather than the underlying data noise or model uncertainty, and the predicted intervals are less smooth and more step-like compared to probabilistic methods, as shown in Fig. \ref{fig:rf-comparison}.

\subsection{The Hybrid Model: Random Forest with Gaussian Process}
To counter the step-like nature of Random Forest predictions in LCR, we develop a hybrid model by combining Random Forest with a GP. Here, the Random Forest acts as the base model, trained on the data to capture initial predictions that efficiently handle nonlinear relationships. However, due to the Random Forest’s step-like nature, these predictions lack the smoothness typically required for reconstructing continuous time-series data.

Once the Random Forest produces its predictions, they are passed to a GP for further modeling. The GP is trained on the Random Forest predictions, learning a functional mapping that transforms the step-like predictions into smoother, continuous outputs. GPs are particularly well-suited for this task because they can model uncertainty in a probabilistic framework while naturally producing smooth, continuous predictions. By integrating the GP, we aim to refine the overall prediction, achieving a smoother representation of the LC with a continuity that better reflects its underlying structure.

However, a limitation of this approach is that the error propagation from the Random Forest propagates via the GP, and the step-like uncertainty patterns remain prominent in the hybrid model’s confidence intervals. While the mean predictions become smoother, the uncertainty estimates retain a step-like nature as shown in Fig. \ref{fig:rf-comparison}. Thus, this hybrid model improves the smoothness of the LC reconstruction but still lacks the fully continuous uncertainty representation characteristic of models with an entirely probabilistic approach. In section \ref{section:ML}, we introduce another hybrid approach that reverses the base structure: instead of using RF as the foundation, we employ GP as the base model and combine it with RF. In this, GP is first trained on the observed data to generate smooth, uncertainty-aware predictions. The RF acts as a secondary learner that attempts to correct residual patterns or finer-scale structures that the GP might have smoothed out or underfit. Since the RF is now operates on a smoother input (from GP), the resulting uncertainty propagation becomes more consistent and less step-like.

\section*{Acknowledgements}
We want to thank Biagio De Simone for providing tips on the initial code for the reconstruction of the LC.
The authors also thank Aditya Narendra, Nikita Khatiya, and Dhruv Bal for their insightful comments on the analysis of our models.
We thank Spencer James Gibson and Federico Da Rold for their valuable insights on exploring different new algorithms for the reconstruction process. NF acknowledges financial support from UNAM-DGAPA-PAPIIT through the grant IN112525. M.G.D. acknowledges the support of the JSPS Grant-in-Aid for Scientific Research (KAKENHI) (A), Grant Number JP25H00675.

We thank Dr. Jurgen Mifsud, Dr. Purba Mukherjee, Dr. Konstantinos F. Dialektopoulos, and Prof. Jackson Said for their helpful comments and discussion of our analysis.

\bibliography{LC_reconstruction}

\begin{thebibliography}{}
\expandafter\ifx\csname natexlab\endcsname\relax\def\natexlab#1{#1}\fi
\providecommand{\url}[1]{\href{#1}{#1}}
\providecommand{\dodoi}[1]{doi:~\href{http://doi.org/#1}{\nolinkurl{#1}}}
\providecommand{\doeprint}[1]{\href{http://ascl.net/#1}{\nolinkurl{http://ascl.net/#1}}}
\providecommand{\doarXiv}[1]{\href{https://arxiv.org/abs/#1}{\nolinkurl{https://arxiv.org/abs/#1}}}

\bibitem[{Agarap(2018)}]{agarap2018deep}
Agarap, A. 2018, arXiv preprint arXiv:1803.08375

\bibitem[{Akiba {et~al.}(2019)Akiba, Sano, Yanase, Ohta, \& Koyama}]{2019optuna}
Akiba, T., Sano, S., Yanase, T., Ohta, T., \& Koyama, M. 2019, in Proceedings of the 25th ACM SIGKDD international conference on knowledge discovery \& data mining, 2623--2631

\bibitem[{Alharbi \& Csala(2022)}]{inventions7040094}
Alharbi, F.~R., \& Csala, D. 2022, Inventions, 7, \dodoi{10.3390/inventions7040094}

\bibitem[{Amati {et~al.}(2018)Amati, O’Brien, G{\"o}tz, Bozzo, Tenzer, Frontera, Ghirlanda, Labanti, Osborne, Stratta, {et~al.}}]{amati2018theseus}
Amati, L., O’Brien, P., G{\"o}tz, D., {et~al.} 2018, Advances in Space Research, 62, 191

\bibitem[{Aravkin {et~al.}(2013)Aravkin, Burke, \& Pillonetto}]{articleks}
Aravkin, A., Burke, J., \& Pillonetto, G. 2013

\bibitem[{Atteia {et~al.}(2022)Atteia, Cordier, \& Wei}]{atteia2022svom}
Atteia, J.-L., Cordier, B., \& Wei, J. 2022, International Journal of Modern Physics D, 31, 2230008

\bibitem[{Bargiacchi {et~al.}(2025)Bargiacchi, Dainotti, \& Capozziello}]{bargiacchi2025high}
Bargiacchi, G., Dainotti, M.~G., \& Capozziello, S. 2025, New Astronomy Reviews, 100, 101712

\bibitem[{{Barthelmy} {et~al.}(2005){Barthelmy}, {Barbier}, {Cummings}, {Fenimore}, {Gehrels}, {Hullinger}, {Krimm}, {Markwardt}, {Palmer}, {Parsons}, {Sato}, {Suzuki}, {Takahashi}, {Tashiro}, \& {Tueller}}]{barthelmy2005burst}
{Barthelmy}, S.~D., {Barbier}, L.~M., {Cummings}, J.~R., {et~al.} 2005, Space Science Reviews, 120, 143, \dodoi{10.1007/s11214-005-5096-3}

\bibitem[{{Beskin} {et~al.}(2010){Beskin}, {Karpov}, {Bondar}, {Greco}, {Guarnieri}, {Bartolini}, \& {Piccioni}}]{Beskin2010ApJ}
{Beskin}, G., {Karpov}, S., {Bondar}, S., {et~al.} 2010, The Astrophysical Journal Letters, 719, L10, \dodoi{10.1088/2041-8205/719/1/L10}

\bibitem[{Betoule {et~al.}(2014)Betoule, Kessler, Guy, Mosher, Hardin, Biswas, Astier, El-Hage, Konig, Kuhlmann, Marriner, Pain, Regnault, Balland, Bassett, Brown, Campbell, Carlberg, Cellier-Holzem, Cinabro, Conley, D’Andrea, DePoy, Doi, Ellis, Fabbro, Filippenko, Foley, Frieman, Fouchez, Galbany, Goobar, Gupta, Hill, Hlozek, Hogan, Hook, Howell, Jha, Le~Guillou, Leloudas, Lidman, Marshall, Möller, Mourão, Neveu, Nichol, Olmstead, Palanque-Delabrouille, Perlmutter, Prieto, Pritchet, Richmond, Riess, Ruhlmann-Kleider, Sako, Schahmaneche, Schneider, Smith, Sollerman, Sullivan, Walton, \& Wheeler}]{Betoule}
Betoule, M., Kessler, R., Guy, J., {et~al.} 2014, Astronomy \&; Astrophysics

\bibitem[{{Blake} {et~al.}(2005){Blake}, {Bloom}, {Starr}, {Falco}, {Skrutskie}, {Fenimore}, {Duch{\^e}ne}, {Szentgyorgyi}, {Hornstein}, {Prochaska}, {McCabe}, {Ghez}, {Konopacky}, {Stapelfeldt}, {Hurley}, {Campbell}, {Kassis}, {Chaffee}, {Gehrels}, {Barthelmy}, {Cummings}, {Hullinger}, {Krimm}, {Markwardt}, {Palmer}, {Parsons}, {McLean}, \& {Tueller}}]{Blake2005Natur}
{Blake}, C.~H., {Bloom}, J.~S., {Starr}, D.~L., {et~al.} 2005, Nature, 435, 181, \dodoi{10.1038/nature03520}

\bibitem[{Breiman(2001)}]{breiman2001random}
Breiman, L. 2001, Machine learning, 45, 5

\bibitem[{Breiman {et~al.}(2017)Breiman, Friedman, Olshen, \& Stone}]{breiman2017classification}
Breiman, L., Friedman, J., Olshen, R.~A., \& Stone, C.~J. 2017, Classification and regression trees (Routledge)

\bibitem[{Brophy {et~al.}(2023)Brophy, Wang, She, \& Ward}]{brophy2023generative}
Brophy, E., Wang, Z., She, Q., \& Ward, T. 2023, ACM Computing Surveys, 55, 1

\bibitem[{{Burrows} {et~al.}(2005){Burrows}, {Hill}, {Nousek}, {Kennea}, {Wells}, {Osborne}, {Abbey}, {Beardmore}, {Mukerjee}, {Short}, {Chincarini}, {Campana}, {Citterio}, {Moretti}, {Pagani}, {Tagliaferri}, {Giommi}, {Capalbi}, {Tamburelli}, {Angelini}, {Cusumano}, {Br{\"a}uninger}, {Burkert}, \& {Hartner}}]{burrows2005swift}
{Burrows}, D.~N., {Hill}, J.~E., {Nousek}, J.~A., {et~al.} 2005, Space Science Reviews, 120, 165, \dodoi{10.1007/s11214-005-5097-2}

\bibitem[{{Cao} {et~al.}(2022{\natexlab{a}}){Cao}, {Dainotti}, \& {Ratra}}]{2022MNRAS.512..439C}
{Cao}, S., {Dainotti}, M., \& {Ratra}, B. 2022{\natexlab{a}}, \mnras, 512, 439, \dodoi{10.1093/mnras/stac517}

\bibitem[{{Cao} {et~al.}(2022{\natexlab{b}}){Cao}, {Khadka}, \& {Ratra}}]{2022MNRAS.510.2928C}
{Cao}, S., {Khadka}, N., \& {Ratra}, B. 2022{\natexlab{b}}, \mnras, 510, 2928, \dodoi{10.1093/mnras/stab3559}

\bibitem[{Cardone {et~al.}(2009)Cardone, Capozziello, \& Dainotti}]{cardone2009updated}
Cardone, V.~F., Capozziello, S., \& Dainotti, M.~G. 2009, Monthly Notices of the Royal Astronomical Society, 400, 775, \dodoi{10.1111/j.1365-2966.2009.15456.x}

\bibitem[{Cardone {et~al.}(2010)Cardone, Dainotti, Capozziello, \& Willingale}]{cardone2010constraining}
Cardone, V.~F., Dainotti, M.~G., Capozziello, S., \& Willingale, R. 2010, Monthly Notices of the Royal Astronomical Society, 408, 1181, \dodoi{10.1111/j.1365-2966.2010.17197.x}

\bibitem[{Cheng {et~al.}(2016)Cheng, Dong, \& Lapata}]{lstm2016lang}
Cheng, J., Dong, L., \& Lapata, M. 2016, arXiv preprint arXiv:1601.06733

\bibitem[{{Costa} {et~al.}(1997){Costa}, {Frontera}, {Heise}, {Feroci}, {in't Zand}, {Fiore}, {Cinti}, {Dal Fiume}, {Nicastro}, {Orlandini}, {Palazzi}, {Rapisarda\#}, {Zavattini}, {Jager}, {Parmar}, {Owens}, {Molendi}, {Cusumano}, {Maccarone}, {Giarrusso}, {Coletta}, {Antonelli}, {Giommi}, {Muller}, {Piro}, \& {Butler}}]{costa1997}
{Costa}, E., {Frontera}, F., {Heise}, J., {et~al.} 1997, Nature, 387, 783, \dodoi{10.1038/42885}

\bibitem[{{Cucchiara} {et~al.}(2011){Cucchiara}, {Levan}, {Fox}, {Tanvir}, {Ukwatta}, {Berger}, {Kr{\"u}hler}, {K{\"u}pc{\"u} Yolda{\textcommabelow s}}, {Wu}, {Toma}, {Greiner}, {Olivares}, {Rowlinson}, {Amati}, {Sakamoto}, {Roth}, {Stephens}, {Fritz}, {Fynbo}, {Hjorth}, {Malesani}, {Jakobsson}, {Wiersema}, {O'Brien}, {Soderberg}, {Foley}, {Fruchter}, {Rhoads}, {Rutledge}, {Schmidt}, {Dopita}, {Podsiadlowski}, {Willingale}, {Wolf}, {Kulkarni}, \& {D'Avanzo}}]{Cucchiara2011}
{Cucchiara}, A., {Levan}, A.~J., {Fox}, D.~B., {et~al.} 2011, The Astrophysical Journal, 736, 7, \dodoi{10.1088/0004-637X/736/1/7}

\bibitem[{Dainotti {et~al.}(2024{\natexlab{a}})Dainotti, Bhardwaj, Bissaldi, Fraija, Sourav, \& Galvan-Gamez}]{dainotti2024analysis}
Dainotti, M., Bhardwaj, S., Bissaldi, E., {et~al.} 2024{\natexlab{a}}, arXiv preprint arXiv:2411.10736

\bibitem[{Dainotti {et~al.}(2023{\natexlab{a}})Dainotti, Lenart, Chraya, Sarracino, Nagataki, Fraija, Capozziello, \& Bogdan}]{dainotti2023b}
Dainotti, M., Lenart, A.~{\L}., Chraya, A., {et~al.} 2023{\natexlab{a}}, Monthly Notices of the Royal Astronomical Society, 518, 2201

\bibitem[{Dainotti {et~al.}(2020{\natexlab{a}})Dainotti, Livermore, Kann, Li, Oates, Yi, Zhang, Gendre, Cenko, \& Fraija}]{dainotti2020optical}
Dainotti, M., Livermore, S., Kann, D., {et~al.} 2020{\natexlab{a}}, The Astrophysical Journal Letters, 905, L26

\bibitem[{Dainotti {et~al.}(2024{\natexlab{b}})Dainotti, Taira, Wang, Lehman, Narendra, Pollo, Madejski, Petrosian, Bogdan, Dey, {et~al.}}]{giovanna2024inferring}
Dainotti, M., Taira, E., Wang, E., {et~al.} 2024{\natexlab{b}}, arXiv preprint arXiv:2401.03589

\bibitem[{Dainotti {et~al.}(2024{\natexlab{c}})Dainotti, De~Simone, Mohideen~Malik, Pasumarti, Levine, Saha, Gendre, Kido, Watson, Becerra, {et~al.}}]{dainotti2024optical}
Dainotti, M., De~Simone, B., Mohideen~Malik, R., {et~al.} 2024{\natexlab{c}}, Monthly Notices of the Royal Astronomical Society, 533, 4023

\bibitem[{{Dainotti} {et~al.}(2008){Dainotti}, {Cardone}, \& {Capozziello}}]{Dainotti2008}
{Dainotti}, M.~G., {Cardone}, V.~F., \& {Capozziello}, S. 2008, Monthly Notices of the Royal Astronomical Society, 391, L79, \dodoi{10.1111/j.1745-3933.2008.00560.x}

\bibitem[{Dainotti {et~al.}(2022c)Dainotti, De~Simone, Schiavone, Montani, Rinaldi, Lambiase, Bogdan, \& Ugale}]{dainotti2022c}
Dainotti, M.~G., De~Simone, B., Schiavone, T., {et~al.} 2022c, Galaxies, 10, 24

\bibitem[{{Dainotti} {et~al.}(2015){Dainotti}, {Del Vecchio}, {Shigehiro}, \& {Capozziello}}]{dainotti2015}
{Dainotti}, M.~G., {Del Vecchio}, R., {Shigehiro}, N., \& {Capozziello}, S. 2015, The Astrophysical Journal, 800, 31, \dodoi{10.1088/0004-637X/800/1/31}

\bibitem[{{Dainotti} {et~al.}(2011){Dainotti}, {Fabrizio Cardone}, {Capozziello}, {Ostrowski}, \& {Willingale}}]{dainotti2011a}
{Dainotti}, M.~G., {Fabrizio Cardone}, V., {Capozziello}, S., {Ostrowski}, M., \& {Willingale}, R. 2011, The Astrophysical Journal, 730, 135, \dodoi{10.1088/0004-637X/730/2/135}

\bibitem[{Dainotti {et~al.}(2020{\natexlab{b}})Dainotti, Lenart, Sarracino, Nagataki, Capozziello, \& Fraija}]{dainotti2020x}
Dainotti, M.~G., Lenart, A., Sarracino, G., {et~al.} 2020{\natexlab{b}}, The Astrophysical Journal, 904, 97

\bibitem[{Dainotti {et~al.}(2021)Dainotti, Lenart, Fraija, Nagataki, Warren, De~Simone, Srinivasaragavan, \& Mata}]{dainotti2021closure}
Dainotti, M.~G., Lenart, A.~{\L}., Fraija, N., {et~al.} 2021, Publications of the Astronomical Society of Japan, 73, 970

\bibitem[{{Dainotti} {et~al.}(2020{\natexlab{a}}){Dainotti}, {Lenart}, {Sarracino}, {Nagataki}, {Capozziello}, \& {Fraija}}]{dainotti2020a}
{Dainotti}, M.~G., {Lenart}, A.~{\L}., {Sarracino}, G., {et~al.} 2020{\natexlab{a}}, The Astrophysical Journal, 904, 97, \dodoi{10.3847/1538-4357/abbe8a}

\bibitem[{{Dainotti} {et~al.}(2017){Dainotti}, {Nagataki}, {Maeda}, {Postnikov}, \& {Pian}}]{dainotti2017a}
{Dainotti}, M.~G., {Nagataki}, S., {Maeda}, K., {Postnikov}, S., \& {Pian}, E. 2017, Astronomy and Astrophysics, 600, A98, \dodoi{10.1051/0004-6361/201628384}

\bibitem[{{Dainotti} {et~al.}(2022g){Dainotti}, {Nielson}, {Sarracino}, {Rinaldi}, {Nagataki}, {Capozziello}, {Gnedin}, \& {Bargiacchi}}]{dainotti2022g}
{Dainotti}, M.~G., {Nielson}, V., {Sarracino}, G., {et~al.} 2022g, \mnras, 514, 1828, \dodoi{10.1093/mnras/stac1141}

\bibitem[{{Dainotti} {et~al.}(2013){Dainotti}, {Petrosian}, {Singal}, \& {Ostrowski}}]{dainotti2013}
{Dainotti}, M.~G., {Petrosian}, V., {Singal}, J., \& {Ostrowski}, M. 2013, The Astrophysical Journal, 774, 157, \dodoi{10.1088/0004-637X/774/2/157}

\bibitem[{{Dainotti} {et~al.}(2016){Dainotti}, {Postnikov}, {Hernandez}, \& {Ostrowski}}]{Dainotti2016}
{Dainotti}, M.~G., {Postnikov}, S., {Hernandez}, X., \& {Ostrowski}, M. 2016, The Astrophysical Journal Letters, 825, L20, \dodoi{10.3847/2041-8205/825/2/L20}

\bibitem[{Dainotti {et~al.}(2016)Dainotti, Postnikov, Hernandez, \& Ostrowski}]{dainotti2016fundamental}
Dainotti, M.~G., Postnikov, S., Hernandez, X., \& Ostrowski, M. 2016, The Astrophysical Journal Letters, 825, L20

\bibitem[{Dainotti {et~al.}(2022b)Dainotti, Sarracino, \& Capozziello}]{dainotti2022b}
Dainotti, M.~G., Sarracino, G., \& Capozziello, S. 2022b, Publications of the Astronomical Society of Japan, 74, 1095

\bibitem[{Dainotti {et~al.}(2023{\natexlab{b}})Dainotti, Sharma, Narendra, Levine, Rinaldi, Pollo, \& Bhatta}]{dainotti2023stochastic}
Dainotti, M.~G., Sharma, R., Narendra, A., {et~al.} 2023{\natexlab{b}}, The Astrophysical Journal Supplement Series, 267, 42

\bibitem[{Dainotti {et~al.}(2013)Dainotti, Singal, Ostrowski, {et~al.}}]{dainotti2013determination}
Dainotti, M.~G., Singal, J., Ostrowski, M., {et~al.} 2013, The Astrophysical Journal, 774, 157

\bibitem[{{Dainotti} {et~al.}(2010){Dainotti}, {Willingale}, {Capozziello}, {Fabrizio Cardone}, \& {Ostrowski}}]{dainotti2010a}
{Dainotti}, M.~G., {Willingale}, R., {Capozziello}, S., {Fabrizio Cardone}, V., \& {Ostrowski}, M. 2010, The Astrophysical Journal Letters, 722, L215, \dodoi{10.1088/2041-8205/722/2/L215}

\bibitem[{{Dainotti} {et~al.}(2020{\natexlab{b}}){Dainotti}, {Livermore}, {Kann}, {Li}, {Oates}, {Yi}, {Zhang}, {Gendre}, {Cenko}, \& {Fraija}}]{dainotti2020b}
{Dainotti}, M.~G., {Livermore}, S., {Kann}, D.~A., {et~al.} 2020{\natexlab{b}}, The Astrophysical Journal Letters, 905, L26, \dodoi{10.3847/2041-8213/abcda9}

\bibitem[{{Dainotti} {et~al.}(2022){Dainotti}, {Young}, {Li}, {Levine}, {Kalinowski}, {Kann}, {Tran}, {Zambrano-Tapia}, {Zambrano-Tapia}, {Cenko}, {Fuentes}, {S{\'a}nchez-V{\'a}zquez}, {Oates}, {Fraija}, {Becerra}, {Watson}, {Butler}, {Gonz{\'a}lez}, {Kutyrev}, {Lee}, {Prochaska}, {Ramirez-Ruiz}, {Richer}, \& {Zola}}]{2022ApJS..261...25D}
{Dainotti}, M.~G., {Young}, S., {Li}, L., {et~al.} 2022, \apjs, 261, 25, \dodoi{10.3847/1538-4365/ac7c64}

\bibitem[{Demianenko {et~al.}(2023)Demianenko, Malanchev, Samorodova, Sysak, Shiriaev, Derkach, \& Hushchyn}]{demianenko2023understanding}
Demianenko, M., Malanchev, K., Samorodova, E., {et~al.} 2023, Astronomy \& Astrophysics, 677, A16

\bibitem[{Dereli-B{\'e}gu{\'e} {et~al.}(2024)Dereli-B{\'e}gu{\'e}, Pe'er, B{\'e}gu{\'e}, \& Ryde}]{dereli2024unraveling}
Dereli-B{\'e}gu{\'e}, H., Pe'er, A., B{\'e}gu{\'e}, D., \& Ryde, F. 2024, arXiv preprint arXiv:2412.11533

\bibitem[{Evans {et~al.}(2009)Evans, Beardmore, Page, Osborne, O'Brien, Willingale, Starling, Burrows, Godet, Vetere, {et~al.}}]{evans2009methods}
Evans, P., Beardmore, A., Page, K., {et~al.} 2009, Monthly Notices of the Royal Astronomical Society, 397, 1177

\bibitem[{Fu {et~al.}(2019)Fu, Chen, Zeng, Zhuang, \& Sudjianto}]{fu2019time}
Fu, R., Chen, J., Zeng, S., Zhuang, Y., \& Sudjianto, A. 2019, arXiv preprint arXiv:1904.11419

\bibitem[{{Gehrels} {et~al.}(2009){Gehrels}, {Ramirez-Ruiz}, \& {Fox}}]{Gehrels2009ARA&A}
{Gehrels}, N., {Ramirez-Ruiz}, E., \& {Fox}, D.~B. 2009, Annual Review of Astronomy and Astrophysics, 47, 567, \dodoi{10.1146/annurev.astro.46.060407.145147}

\bibitem[{{Gehrels} {et~al.}(2004){Gehrels}, {Chincarini}, {Giommi}, {Mason}, {Nousek}, {Wells}, {White}, {Barthelmy}, {Burrows}, {Cominsky}, {Hurley}, {Marshall}, {M{\'e}sz{\'a}ros}, {Roming}, {Angelini}, {Barbier}, {Belloni}, {Campana}, {Caraveo}, {Chester}, {Citterio}, {Cline}, {Cropper}, {Cummings}, {Dean}, {Feigelson}, {Fenimore}, {Frail}, {Fruchter}, {Garmire}, {Gendreau}, {Ghisellini}, {Greiner}, {Hill}, {Hunsberger}, {Krimm}, {Kulkarni}, {Kumar}, {Lebrun}, {Lloyd-Ronning}, {Markwardt}, {Mattson}, {Mushotzky}, {Norris}, {Osborne}, {Paczynski}, {Palmer}, {Park}, {Parsons}, {Paul}, {Rees}, {Reynolds}, {Rhoads}, {Sasseen}, {Schaefer}, {Short}, {Smale}, {Smith}, {Stella}, {Tagliaferri}, {Takahashi}, {Tashiro}, {Townsley}, {Tueller}, {Turner}, {Vietri}, {Voges}, {Ward}, {Willingale}, {Zerbi}, \& {Zhang}}]{Gehrels2004ApJ...611.1005G}
{Gehrels}, N., {Chincarini}, G., {Giommi}, P., {et~al.} 2004, The Astrophysical Journal, 611, 1005, \dodoi{10.1086/422091}

\bibitem[{Goodfellow {et~al.}(2014)Goodfellow, Pouget-Abadie, Mirza, Xu, Warde-Farley, Ozair, Courville, \& Bengio}]{goodfellow2014generative}
Goodfellow, I.~J., Pouget-Abadie, J., Mirza, M., {et~al.} 2014, Advances in neural information processing systems, 27

\bibitem[{{Gorbovskoy} {et~al.}(2012){Gorbovskoy}, {Lipunova}, {Lipunov}, {Kornilov}, {Belinski}, {Shatskiy}, {Tyurina}, {Kuvshinov}, {Balanutsa}, {Chazov}, {Kuznetsov}, {Zimnukhov}, {Kornilov}, {Sankovich}, {Krylov}, {Ivanov}, {Chvalaev}, {Poleschuk}, {Konstantinov}, {Gress}, {Yazev}, {Budnev}, {Krushinski}, {Zalozhnich}, {Popov}, {Tlatov}, {Parhomenko}, {Dormidontov}, {Senik}, {Yurkov}, {Sergienko}, {Varda}, {Kudelina}, {Castro-Tirado}, {Gorosabel}, {S{\'a}nchez-Ram{\'\i}rez}, {Jelinek}, \& {Tello}}]{2012MNRAS.421.1874G}
{Gorbovskoy}, E.~S., {Lipunova}, G.~V., {Lipunov}, V.~M., {et~al.} 2012, Monthly Notices of the Royal Astronomical Society, 421, 1874, \dodoi{10.1111/j.1365-2966.2012.20195.x}

\bibitem[{Graves {et~al.}(2005)Graves, Fern{\'a}ndez, \& Schmidhuber}]{graves2005bidirectional}
Graves, A., Fern{\'a}ndez, S., \& Schmidhuber, J. 2005, in International conference on artificial neural networks, Springer, 799--804

\bibitem[{Gu \& Dao(2024)}]{gu2024mambalineartimesequencemodeling}
Gu, A., \& Dao, T. 2024, Mamba: Linear-Time Sequence Modeling with Selective State Spaces.
\newblock \doarXiv{2312.00752}

\bibitem[{He {et~al.}(2015)He, Zhang, Ren, \& Sun}]{7410480}
He, K., Zhang, X., Ren, S., \& Sun, J. 2015, in 2015 IEEE International Conference on Computer Vision (ICCV), 1026--1034, \dodoi{10.1109/ICCV.2015.123}

\bibitem[{Heaton(2018)}]{DeepLearning}
Heaton, J. 2018, Genetic programming and evolvable machines, 19, 305

\bibitem[{Ho {et~al.}(2020)Ho, Jain, \& Abbeel}]{denoisingdiff2020}
Ho, J., Jain, A., \& Abbeel, P. 2020, Advances in neural information processing systems, 33, 6840

\bibitem[{Ho \& Xie(1998)}]{HO1998213}
Ho, S., \& Xie, M. 1998, Computers \& Industrial Engineering, 35, 213, \dodoi{https://doi.org/10.1016/S0360-8352(98)00066-7}

\bibitem[{Hochreiter \& Schmidhuber(1997)}]{hochreiter1997long}
Hochreiter, S., \& Schmidhuber, J. 1997, Neural computation, 9, 1735, \dodoi{10.1162/neco.1997.9.8.1735}

\bibitem[{Hua {et~al.}(2019)Hua, Zhao, Li, Chen, Liu, \& Zhang}]{lstm2019time}
Hua, Y., Zhao, Z., Li, R., {et~al.} 2019, IEEE Communications Magazine, 57, 114

\bibitem[{Huber \& Suyu(2024)}]{huber2024holismokes}
Huber, S., \& Suyu, S. 2024, arXiv preprint arXiv:2403.08029

\bibitem[{Kingma \& Ba(2014)}]{kingma2014adam}
Kingma, D.~P., \& Ba, J. 2014, Adam: A Method for Stochastic Optimization.
\newblock \doarXiv{1412.6980}

\bibitem[{Kingma {et~al.}(2019)Kingma, Welling, {et~al.}}]{vae2019}
Kingma, D.~P., Welling, M., {et~al.} 2019, Foundations and Trends{\textregistered} in Machine Learning, 12, 307

\bibitem[{Kohavi(1995)}]{kohaviCV}
Kohavi, R. 1995, in Proceedings of the 14th International Joint Conference on Artificial Intelligence (IJCAI), Vol.~14

\bibitem[{Kumar \& Duran(2010)}]{kumar2010external}
Kumar, P., \& Duran, R.~B. 2010, Monthly Notices of the Royal Astronomical Society, 409, 226

\bibitem[{{Kumar} \& {Zhang}(2015)}]{2015PhR...561....1K}
{Kumar}, P., \& {Zhang}, B. 2015, \physrep, 561, 1, \dodoi{10.1016/j.physrep.2014.09.008}

\bibitem[{{Levine} {et~al.}(2022){Levine}, {Dainotti}, {Zvonarek}, {Fraija}, {Warren}, {Chandra}, \& {Lloyd-Ronning}}]{levine2022}
{Levine}, D., {Dainotti}, M., {Zvonarek}, K.~J., {et~al.} 2022, The Astrophysical Journal, 925, 15, \dodoi{10.3847/1538-4357/ac4221}

\bibitem[{{Li} {et~al.}(2018){Li}, {Wu}, {Lei}, {Dai}, {Liang}, \& {Ryde}}]{Li2018b}
{Li}, L., {Wu}, X.-F., {Lei}, W.-H., {et~al.} 2018, The Astrophysical Journal Supplement, 236, 26, \dodoi{10.3847/1538-4365/aabaf3}

\bibitem[{{Liang} {et~al.}(2007){Liang}, {Zhang}, \& {Zhang}}]{Liang2007}
{Liang}, E.-W., {Zhang}, B.-B., \& {Zhang}, B. 2007, The Astrophysical Journal, 670, 565, \dodoi{10.1086/521870}

\bibitem[{Liu {et~al.}(2019)Liu, Liu, Wang, \& Wang}]{liu2019predicting}
Liu, H., Liu, C., Wang, J.~T., \& Wang, H. 2019, The Astrophysical Journal, 877, 121

\bibitem[{Liu {et~al.}(2024)Liu, Wang, Vaidya, Ruehle, Halverson, Solja{\v{c}}i{\'c}, Hou, \& Tegmark}]{liu2024kan}
Liu, Z., Wang, Y., Vaidya, S., {et~al.} 2024, arXiv preprint arXiv:2404.19756

\bibitem[{Medsker {et~al.}(2001)Medsker, Jain, {et~al.}}]{recurrent2001}
Medsker, L.~R., Jain, L., {et~al.} 2001, Design and Applications, 5, 2

\bibitem[{Mirza \& Osindero(2014)}]{mirza2014conditionalgenerativeadversarialnets}
Mirza, M., \& Osindero, S. 2014, Conditional Generative Adversarial Nets.
\newblock \doarXiv{1411.1784}

\bibitem[{Narendra {et~al.}(2024)Narendra, Dainotti, Sarkar, Lenart, Bogdan, Pollo, Zhang, Rabeda, Petrosian, \& Kazunari}]{narendra2024grb}
Narendra, A., Dainotti, M., Sarkar, M., {et~al.} 2024, arXiv preprint arXiv:2410.13985

\bibitem[{{Nousek} {et~al.}(2006){Nousek}, {Kouveliotou}, {Grupe}, {Page}, {Granot}, {Ramirez-Ruiz}, {Patel}, {Burrows}, {Mangano}, {Barthelmy}, {Beardmore}, {Campana}, {Capalbi}, {Chincarini}, {Cusumano}, {Falcone}, {Gehrels}, {Giommi}, {Goad}, {Godet}, {Hurkett}, {Kennea}, {Moretti}, {O'Brien}, {Osborne}, {Romano}, {Tagliaferri}, \& {Wells}}]{Nousek2006}
{Nousek}, J.~A., {Kouveliotou}, C., {Grupe}, D., {et~al.} 2006, \apj, 642, 389, \dodoi{10.1086/500724}

\bibitem[{Oberst(2007)}]{OberstUlrich}
Oberst, U. 2007, SIAM J. Control and Optimization, 46, 496, \dodoi{10.1137/060658242}

\bibitem[{{O'Brien} {et~al.}(2006){O'Brien}, {Willingale}, {Osborne}, {Goad}, {Page}, {Vaughan}, {Rol}, {Beardmore}, {Godet}, {Hurkett}, {Wells}, {Zhang}, {Kobayashi}, {Burrows}, {Nousek}, {Kennea}, {Falcone}, {Grupe}, {Gehrels}, {Barthelmy}, {Cannizzo}, {Cummings}, {Hill}, {Krimm}, {Chincarini}, {Tagliaferri}, {Campana}, {Moretti}, {Giommi}, {Perri}, {Mangano}, \& {LaParola}}]{OBrien2006}
{O'Brien}, P.~T., {Willingale}, R., {Osborne}, J., {et~al.} 2006, The Astrophysical Journal, 647, 1213, \dodoi{10.1086/505457}

\bibitem[{Oktay {et~al.}(2018)Oktay, Schlemper, Folgoc, Lee, Heinrich, Misawa, Mori, McDonagh, Hammerla, Kainz, {et~al.}}]{oktay2018attention}
Oktay, O., Schlemper, J., Folgoc, L.~L., {et~al.} 2018, arXiv preprint arXiv:1804.03999

\bibitem[{Panaitescu \& Kumar(2000)}]{panaitescu2000analytic}
Panaitescu, A., \& Kumar, P. 2000, The Astrophysical Journal, 543, 66

\bibitem[{Pechlivanidou \& Karampetakis(2022)}]{10.1093/imamci/dnac005}
Pechlivanidou, G., \& Karampetakis, N. 2022, IMA Journal of Mathematical Control and Information, 39, 708, \dodoi{10.1093/imamci/dnac005}

\bibitem[{Perslev {et~al.}(2019)Perslev, Jensen, Darkner, Jennum, \& Igel}]{perslev2019u}
Perslev, M., Jensen, M., Darkner, S., Jennum, P.~J., \& Igel, C. 2019, Advances in neural information processing systems, 32

\bibitem[{Piran(1999)}]{piran1999gamma}
Piran, T. 1999, Physics Reports, 314, 575

\bibitem[{{Piro} {et~al.}(1998){Piro}, {Amati}, {Antonelli}, {Butler}, {Costa}, {Cusumano}, {Feroci}, {Frontera}, {Heise}, {in 't Zand}, {Molendi}, {Muller}, {Nicastro}, {Orlandini}, {Owens}, {Parmar}, {Soffitta}, \& {Tavani}}]{Piro1998}
{Piro}, L., {Amati}, L., {Antonelli}, L.~A., {et~al.} 1998, Astronomy and Astrophysics, 331, L41.
\newblock \doarXiv{astro-ph/9710355}

\bibitem[{Postnikov {et~al.}(2014)Postnikov, Dainotti, Hernandez, \& Capozziello}]{postnikov2014nonparametric}
Postnikov, S., Dainotti, M.~G., Hernandez, X., \& Capozziello, S. 2014, The Astrophysical Journal, 783, 126

\bibitem[{Racusin {et~al.}(2009)Racusin, Liang, Burrows, Falcone, Sakamoto, Zhang, Zhang, Evans, \& Osborne}]{racusin2009jet}
Racusin, J., Liang, E., Burrows, D.~N., {et~al.} 2009, The Astrophysical Journal, 698, 43

\bibitem[{{Racusin} {et~al.}(2009){Racusin}, {Liang}, {Burrows}, {Falcone}, {Sakamoto}, {Zhang}, {Zhang}, {Evans}, \& {Osborne}}]{Racusin2009}
{Racusin}, J.~L., {Liang}, E.~W., {Burrows}, D.~N., {et~al.} 2009, \apj, 698, 43, \dodoi{10.1088/0004-637X/698/1/43}

\bibitem[{Rasmussen(2003)}]{GP2003gaussian}
Rasmussen, C.~E. 2003, in Summer school on machine learning (Springer), 63--71

\bibitem[{{Rea} {et~al.}(2015){Rea}, {Gull{\'o}n}, {Pons}, {Perna}, {Dainotti}, {Miralles}, \& {Torres}}]{Rea2015}
{Rea}, N., {Gull{\'o}n}, M., {Pons}, J.~A., {et~al.} 2015, The Astrophysical Journal, 813, 92, \dodoi{10.1088/0004-637X/813/2/92}

\bibitem[{{Roming} {et~al.}(2005){Roming}, {Kennedy}, {Mason}, {Nousek}, {Ahr}, {Bingham}, {Broos}, {Carter}, {Hancock}, {Huckle}, {Hunsberger}, {Kawakami}, {Killough}, {Koch}, {McLelland}, {Smith}, {Smith}, {Soto}, {Boyd}, {Breeveld}, {Holland}, {Ivanushkina}, {Pryzby}, {Still}, \& {Stock}}]{Roming2005}
{Roming}, P. W.~A., {Kennedy}, T.~E., {Mason}, K.~O., {et~al.} 2005, Space Science Review, 120, 95, \dodoi{10.1007/s11214-005-5095-4}

\bibitem[{Ronneberger {et~al.}(2015)Ronneberger, Fischer, \& Brox}]{ronneberger2015u}
Ronneberger, O., Fischer, P., \& Brox, T. 2015, in Medical image computing and computer-assisted intervention--MICCAI 2015: 18th international conference, Munich, Germany, October 5-9, 2015, proceedings, part III 18, Springer, 234--241

\bibitem[{{Rowlinson} {et~al.}(2014){Rowlinson}, {Gompertz}, {Dainotti}, {O'Brien}, {Wijers}, \& {van der Horst}}]{Rowlinson2014}
{Rowlinson}, A., {Gompertz}, B.~P., {Dainotti}, M., {et~al.} 2014, Monthly Notices of the Royal Astronomical Society, 443, 1779, \dodoi{10.1093/mnras/stu1277}

\bibitem[{Ryan {et~al.}(2020)Ryan, Van~Eerten, Piro, \& Troja}]{ryan2020gamma}
Ryan, G., Van~Eerten, H., Piro, L., \& Troja, E. 2020, The Astrophysical Journal, 896, 166

\bibitem[{{Sakamoto} {et~al.}(2007){Sakamoto}, {Hill}, {Yamazaki}, {Angelini}, {Krimm}, {Sato}, {Swindell}, {Takami}, \& {Osborne}}]{Sakamoto2007}
{Sakamoto}, T., {Hill}, J.~E., {Yamazaki}, R., {et~al.} 2007, The Astrophysical Journal, 669, 1115, \dodoi{10.1086/521640}

\bibitem[{Srinivasaragavan {et~al.}(2020)Srinivasaragavan, Dainotti, Fraija, Hernandez, Nagataki, Lenart, Bowden, \& Wagner}]{srinivasaragavan2020investigation}
Srinivasaragavan, G.~P., Dainotti, M.~G., Fraija, N., {et~al.} 2020, The Astrophysical Journal, 903, 18

\bibitem[{{Srinivasaragavan} {et~al.}(2020{\natexlab{a}}){Srinivasaragavan}, {Dainotti}, {Fraija}, {Hernandez}, {Nagataki}, {Lenart}, {Bowden}, \& {Wagner}}]{2020ApJ...903...18S}
{Srinivasaragavan}, G.~P., {Dainotti}, M.~G., {Fraija}, N., {et~al.} 2020{\natexlab{a}}, \apj, 903, 18, \dodoi{10.3847/1538-4357/abb702}

\bibitem[{{Srinivasaragavan} {et~al.}(2020{\natexlab{b}}){Srinivasaragavan}, {Dainotti}, {Fraija}, {Hernandez}, {Nagataki}, {Lenart}, {Bowden}, \& {Wagner}}]{closure}
---. 2020{\natexlab{b}}, \apj, 903, 18, \dodoi{10.3847/1538-4357/abb702}

\bibitem[{{Stratta} {et~al.}(2018){Stratta}, {Dainotti}, {Dall'Osso}, {Hernandez}, \& {De Cesare}}]{Stratta2018}
{Stratta}, G., {Dainotti}, M.~G., {Dall'Osso}, S., {Hernandez}, X., \& {De Cesare}, G. 2018, The Astrophysical Journal, 869, 155, \dodoi{10.3847/1538-4357/aadd8f}

\bibitem[{Suh {et~al.}(2024)Suh, Yang, Hsieh, Luan, Xu, Zhu, \& Cheng}]{timeautodiff2024}
Suh, N., Yang, Y., Hsieh, D.-Y., {et~al.} 2024, arXiv preprint arXiv:2406.16028

\bibitem[{{Tagliaferri} {et~al.}(2005){Tagliaferri}, {Goad}, {Chincarini}, {Moretti}, {Campana}, {Burrows}, {Perri}, {Barthelmy}, {Gehrels}, {Krimm}, {Sakamoto}, {Kumar}, {M{\'e}sz{\'a}ros}, {Kobayashi}, {Zhang}, {Angelini}, {Banat}, {Beardmore}, {Capalbi}, {Covino}, {Cusumano}, {Giommi}, {Godet}, {Hill}, {Kennea}, {Mangano}, {Morris}, {Nousek}, {O'Brien}, {Osborne}, {Pagani}, {Page}, {Romano}, {Stella}, \& {Wells}}]{Tagliaferri2005}
{Tagliaferri}, G., {Goad}, M., {Chincarini}, G., {et~al.} 2005, Nature, 436, 985, \dodoi{10.1038/nature03934}

\bibitem[{Tak {et~al.}(2019)Tak, Omodei, Uhm, Racusin, Asano, \& McEnery}]{tak2019closure}
Tak, D., Omodei, N., Uhm, Z.~L., {et~al.} 2019, The Astrophysical Journal, 883, 134

\bibitem[{{Tang} {et~al.}(2019){Tang}, {Huang}, {Geng}, \& {Zhang}}]{Tang2019ApJS..245....1T}
{Tang}, C.-H., {Huang}, Y.-F., {Geng}, J.-J., \& {Zhang}, Z.-B. 2019, \apjs, 245, 1, \dodoi{10.3847/1538-4365/ab4711}

\bibitem[{Thielmann {et~al.}(2024)Thielmann, Kumar, Weisser, Reuter, S{\"a}fken, \& Samiee}]{thielmann2024mambular}
Thielmann, A.~F., Kumar, M., Weisser, C., {et~al.} 2024, arXiv preprint arXiv:2408.06291

\bibitem[{{Troja} {et~al.}(2007){Troja}, {Cusumano}, {O'Brien}, {Zhang}, {Sbarufatti}, {Mangano}, {Willingale}, {Chincarini}, {Osborne}, {Marshall}, {Burrows}, {Campana}, {Gehrels}, {Guidorzi}, {Krimm}, {La Parola}, {Liang}, {Mineo}, {Moretti}, {Page}, {Romano}, {Tagliaferri}, {Zhang}, {Page}, \& {Schady}}]{Troja2007}
{Troja}, E., {Cusumano}, G., {O'Brien}, P.~T., {et~al.} 2007, The Astrophysical Journal, 665, 599, \dodoi{10.1086/519450}

\bibitem[{{van Paradijs} {et~al.}(1997){van Paradijs}, {Groot}, {Galama}, {Kouveliotou}, {Strom}, {Telting}, {Rutten}, {Fishman}, {Meegan}, {Pettini}, {Tanvir}, {Bloom}, {Pedersen}, {N{\o}rdgaard-Nielsen}, {Linden-V{\o}rnle}, {Melnick}, {Van der Steene}, {Bremer}, {Naber}, {Heise}, {in't Zand}, {Costa}, {Feroci}, {Piro}, {Frontera}, {Zavattini}, {Nicastro}, {Palazzi}, {Bennett}, {Hanlon}, \& {Parmar}}]{vanParadijs1997}
{van Paradijs}, J., {Groot}, P.~J., {Galama}, T., {et~al.} 1997, Nature, 386, 686, \dodoi{10.1038/386686a0}

\bibitem[{{Vaswani} {et~al.}(2017){Vaswani}, {Shazeer}, {Parmar}, {Uszkoreit}, {Jones}, {Gomez}, {Kaiser}, \& {Polosukhin}}]{2017arXiv170603762V}
{Vaswani}, A., {Shazeer}, N., {Parmar}, N., {et~al.} 2017, arXiv e-prints, arXiv:1706.03762, \dodoi{10.48550/arXiv.1706.03762}

\bibitem[{{Vestrand} {et~al.}(2005){Vestrand}, {Wozniak}, {Wren}, {Fenimore}, {Sakamoto}, {White}, {Casperson}, {Davis}, {Evans}, {Galassi}, {McGowan}, {Schier}, {Asa}, {Barthelmy}, {Cummings}, {Gehrels}, {Hullinger}, {Krimm}, {Markwardt}, {McLean}, {Palmer}, {Parsons}, \& {Tueller}}]{Vestrand2005Natur}
{Vestrand}, W.~T., {Wozniak}, P.~R., {Wren}, J.~A., {et~al.} 2005, Nature, 435, 178, \dodoi{10.1038/nature03515}

\bibitem[{{Vestrand} {et~al.}(2014){Vestrand}, {Wren}, {Panaitescu}, {Wozniak}, {Davis}, {Palmer}, {Vianello}, {Omodei}, {Xiong}, {Briggs}, {Elphick}, {Paciesas}, \& {Rosing}}]{2014Sci...343...38V}
{Vestrand}, W.~T., {Wren}, J.~A., {Panaitescu}, A., {et~al.} 2014, Science, 343, 38, \dodoi{10.1126/science.1242316}

\bibitem[{Wang {et~al.}(2022)Wang, Hu, Zhang, \& Dai}]{Wang:2021hcx}
Wang, F.~Y., Hu, J.~P., Zhang, G.~Q., \& Dai, Z.~G. 2022, Astrophys. J., 924, 97, \dodoi{10.3847/1538-4357/ac3755}

\bibitem[{Wang(2003)}]{ANN}
Wang, S.-C. 2003, in Interdisciplinary computing in java programming (Springer), 81--100

\bibitem[{Willingale {et~al.}(2007)Willingale, O’brien, Osborne, Godet, Page, Goad, Burrows, Zhang, Rol, Gehrels, {et~al.}}]{willingale2007testing}
Willingale, R., O’brien, P., Osborne, J., {et~al.} 2007, The Astrophysical Journal, 662, 1093

\bibitem[{Xu {et~al.}(2015)Xu, Wang, Chen, \& Li}]{xu2015empiricalevaluationrectifiedactivations}
Xu, B., Wang, N., Chen, T., \& Li, M. 2015, Empirical Evaluation of Rectified Activations in Convolutional Network.
\newblock \doarXiv{1505.00853}

\bibitem[{Yonetoku {et~al.}(2024)Yonetoku, Doi, Mihara, Matsuhara, Sakamoto, Tsumura, Ioka, Arimoto, Enoto, Fujimoto, {et~al.}}]{yonetoku2024high}
Yonetoku, D., Doi, A., Mihara, T., {et~al.} 2024, in Space Telescopes and Instrumentation 2024: Ultraviolet to Gamma Ray, Vol. 13093, SPIE, 618--626

\bibitem[{Yuan {et~al.}(2022)Yuan, Zhang, Chen, \& Ling}]{yuan2022einstein}
Yuan, W., Zhang, C., Chen, Y., \& Ling, Z. 2022, in Handbook of X-ray and Gamma-ray Astrophysics (Springer), 1--30

\bibitem[{{Zhang} {et~al.}(2006){Zhang}, {Fan}, {Dyks}, {Kobayashi}, {M{\'e}sz{\'a}ros}, {Burrows}, {Nousek}, \& {Gehrels}}]{Zhang2006}
{Zhang}, B., {Fan}, Y.~Z., {Dyks}, J., {et~al.} 2006, The Astrophysical Journal, 642, 354, \dodoi{10.1086/500723}

\bibitem[{{Zhang} \& {M{\'e}sz{\'a}ros}(2001)}]{Zhang2001}
{Zhang}, B., \& {M{\'e}sz{\'a}ros}, P. 2001, The Astrophysical Journal Letters, 552, L35, \dodoi{10.1086/320255}

\bibitem[{{Zhang} {et~al.}(2007){Zhang}, {Zhang}, {Liang}, {Gehrels}, {Burrows}, \& {M{\'e}sz{\'a}ros}}]{Zhang2007ApJ...655L..25Z}
{Zhang}, B., {Zhang}, B.-B., {Liang}, E.-W., {et~al.} 2007, The Astrophysical Journal Letters, 655, L25, \dodoi{10.1086/511781}

\bibitem[{Zhang {et~al.}(2016)Zhang, Chen, Yu, Yao, Khudanpur, \& Glass}]{lstm2016speech}
Zhang, Y., Chen, G., Yu, D., {et~al.} 2016, in 2016 IEEE international conference on acoustics, speech and signal processing (ICASSP), IEEE, 5755--5759

\bibitem[{{Zhao} {et~al.}(2019){Zhao}, {Zhang}, {Gao}, {Lan}, {L{\"u}}, \& {Zhang}}]{Zhao2019ApJ...883...97Z}
{Zhao}, L., {Zhang}, B., {Gao}, H., {et~al.} 2019, \apj, 883, 97, \dodoi{10.3847/1538-4357/ab38c4}

\end{thebibliography}
\bibliographystyle{aasjournal}



\end{document}